\documentclass[letterpaper,12pt]{article}

\usepackage{csm-thesis}

\usepackage{bm}

\usepackage[numbers]{natbib}

\usepackage{hyperref}

\usepackage{listings}

\usepackage{pdflscape} 

\usepackage{amsmath}
\usepackage{amssymb}
\usepackage{mathrsfs}
\usepackage{perpage}
\MakePerPage{footnote}

\makeatletter
\renewcommand*{\@dotsep}{1}
\makeatother

	\title{Gradient Bundle Analysis: A Full Topological Approach to Chemical Bonding}

\degreetitle{Doctor of Philosophy}
\discipline{Applied Chemistry}
\department{Chemistry and Geochemistry}

\author{Amanda Morgenstern}
\advisor{Dr. Mark E. Eberhart}
\dpthead{Dr. David Wu}{Professor and Head}

\begin{document}



\frontmatter


\maketitle
\newpage




\makesubmittal
\newpage


\begin{abstract}
The ``chemical bond" is a central concept in molecular sciences, but there is no consensus as to what a bond actually is.
Therefore, a variety of bonding models have been developed, each defining the structure of molecules in a different manner with the goal of explaining and predicting chemical properties.
While many twentieth century bonding models provide useful information for a variety of chemical systems, these models are sometimes less insightful for more lofty goals such as designing metalloenzymes. 
The design process of novel catalysts could be improved if more predictive and accurate models of chemical bonding are created.
One recently developed bonding model based on the topology of the electron charge density is the quantum theory of atoms in molecules (QTAIM).
QTAIM defines bonding interactions as one-dimensional ridges of electron density, $\rho(\bm{r})$, which are known as bond paths.
As with any bonding model, there are instances where bond paths do not adequately describe properties of interest, such as in the analysis of histone deacetylase.

This thesis describes the initial development of gradient bundle analysis (GBA), a chemical bonding model that creates a higher resolution picture of chemical interactions within the charge density framework. 
GBA is based on concepts from QTAIM, but uses a more complete picture of the topology and geometry of $\rho(\bm{r})$ to understand and predict bonding interactions. 
Gradient bundles are defined as volumes bounded by zero-flux surfaces (ZFSs) in the gradient of the charge density with well-defined energies.
The structure of gradient bundles provides an avenue for detecting the locations of valence electrons, which correspond to reactive regions in a molecule.
The number of electrons in bonding regions, which are defined by the lowering of kinetic energy in gradient bundles, is found to correlate to bond dissociation energy in diatomic molecules.
Furthermore, site reactivity can be understood and predicted by observing the motion of ZFSs bounding gradient bundles and calculating condensed gradient bundle Fukui functions. 
Using only the ground state charge density, I present preliminary results for a method that predicts which regions in a molecule are most likely to undergo nucleophilic and electrophilic attack, effectively locating the HOMO and LUMO.

\end{abstract}

\newpage


\tableofcontents
\newpage


\listoffiguresandtables
\newpage





\listofsymbols*
\newpage



\listofabbreviations*
\newpage







\bodymatter



\chapter{Introduction---Topological Chemical Bonding Models}
\label{cha:introduction}

The evolution of chemical bonding models and creation of entirely new models is imperative for the progression of the field of chemistry, cultivating a better understanding and control of interactions between atoms.
Most models of the chemical bond used today are rooted in century old ideas developed by Lewis, London, and Heitler.
Few foundational changes have been made to they way we think about chemical bonds since the work of Pauling, Mulliken, and Slater in the 1930s. 
While some unique models have been proposed, such as the quantum theory of atoms in molecules (QTAIM) \cite{baderbook} and conceptual density functional theory (CDFT) reactivity indicators \cite{CDFT}, these viewpoints are not widely utilized and have yet to find their way into most introductory chemistry texts.
By radically changing our view of atomic interactions there is potential to revolutionize how we design new molecules and materials.
\addabbreviation{Conceptual density functional theory}{CDFT}

One area where an innovative view of chemical bonding could be especially useful is in the design of novel proteins.
The design of new enzymes to catalyze non-native reactions can lead to more environmentally friendly syntheses and increased production of medicines and materials.
Enzyme catalysis is not a fully understood process, and using computational modeling to simulate the affect of mutations to enzymes is a difficult task \cite{Mayo_review, Houk_enzyme_review_2013, Warshel_design_review_2014, Zanghellini_design_review_2014}.
Even newer chemical bonding models such as QTAIM can not always properly explain the way chemical bonds will transform in designed enzymes, as is shown in chapter \ref{cha:enzymes}.

This thesis presents the foundational work in the development of gradient bundle analysis (GBA). 
GBA is a potentially transformative bonding model that is based on quantum mechanics, recovers observed properties of chemicals, and is predictive in nature.
Gradient bundles are defined using the gradient field of the electron charge density, $\rho(\bm{r})$, and provide a higher resolution picture of bonding interactions than was previously possible with standard QTAIM methods.
The overarching goal of this work is to provide a useful bonding model that is accessible and applicable to chemists specializing in any field.
Two areas where we are currently working on applying GBA are in enzyme design and to complex metallurgical problems such as understanding brittle failure in iridium (see section \ref{sec:closed}).

The creation of any new model necessarily begins by demonstrating its ability to recover known properties in simple molecules. 
To this end, the first test of GBA is presented in chapter \ref{cha:intrinsic}, where the locations of valence electrons are recovered using values calculated directly from the ground state charge density.
Valence shell electron pair repulsion (VSEPR) diagrams for a set of homonuclear diatomic molecules are reproduced by partitioning the charge density and kinetic energy into gradient bundles, which are volumes bounded by zero-flux surfaces (ZFSs) in the gradient of the charge density, $\nabla \rho(\bm{r})$.  
The usefulness of this structure is elucidated in Chapter \ref{cha:BDE}, with the discovery of a quantitative relationship between experimentally determined bond dissociation energies of diatomic molecules and the lowering of kinetic energy in bonding regions defined using gradient bundles. 
\addabbreviation{Highest occupied molecular orbital}{HOMO}
\addabbreviation{Lowest unoccupied molecular orbital}{LUMO}

Once the framework of GBA is in place, we can begin the search for answers to more complex chemical problems.
Chapter \ref{cha:ZFS} presents a method for visualizing chemical reactivity based on the motion of ZFSs in $\nabla \rho(\bm{r})$. 
This visualization method provides an alternative to drawing electron pushing arrows or using electron transfer between frontier orbitals to picture chemical reactions occurring.
The results from this study provide motivation for a method of answering the open question originally posed by Slater, ``where are the HOMO and LUMO in the ground state charge density?".
The first theorem of density functional theory (DFT) states that all properties are determined by the charge density of a system.
Therefore, it should be possible to predict where electrophilic or nucleophilic attack will occur in a molecule based purely on the ground state charge density, which should also generally match where the HOMO and LUMO are located in a molecule, respectively. 
We have begun to answer this question using the size and shape of gradient bundles, and the preliminary results of this study are presented in chapter \ref{cha:conclusion}. 
To put the usefulness and complexity of GBA into context, the remainder of this chapter briefly reviews topological bonding models, with a focus on identifying specific properties that can recovered using each definition of chemical structure. 
\addabbreviation{Valence shell electron pair repulsion}{VSEPR}
\addabbreviation{Zero-flux surface}{ZFS}
\addsymbol{Electron charge density}{$\rho(\bm{r})$}
\addabbreviation{Quantum theory of atoms in molecules}{QTAIM}
\addabbreviation{Bond bundle}{BB}
\addabbreviation{Gradient bundle}{GB}
\addabbreviation{Gradient bundle analysis}{GBA}

\subsection{Quantum Theory of Atoms in Molecules}

The quantum theory of atoms in molecules uses the topology of the $\rho(\bm{r})$ to define and analyze atomic properties and bonding interactions. The main applications of QTAIM to chemical bonding have been on characterizing types of bonding as well as better understanding the strength of interactions and therefore the stability of molecules. This theory has been applied with much success to  a wide variety of chemical systems including small molecules \cite{Bader_hydrocarbons, Popelier_similarity}, biological systems \cite{Matta_AA, Matta_protein, Popelier_pka}, and solids \cite{wigner-seitz, pendas1, eberhart_shear}. QTAIM was originally developed by Professor Richard Bader in the 1970s. The motivation for this theory is given by the following two questions posed by Bader \cite{baderbook, personal}

\begin{enumerate}
\item Does an atom in a molecule exist, and if so, how do you recognize one?  
\item Are bonds physical observables and how do you define them?
\end{enumerate}

To discuss the physical existence of something you must have some way of observing the item in question. While wave functions, $\psi(\bm{r})$, give great insight into chemical interactions, they are not quantum-mechanical observables. They are mathematical constructs that approximate reality and are often complex valued functions. Charge density, on the other hand, is a measurable feature and always a real valued function. With the advancement of x-ray diffraction techniques, chemists are now able to image the electron density. It has been argued that an accurate bonding model must be rooted in something real \cite{fuzzybond, Bickelhaupt1}. This is why the electron density was an obvious starting point for QTAIM. The original theory of QTAIM gives insight into bonding interactions in terms of two topological features: bond critical points in $\rho(\bm{r})$ (and their corresponding bond paths) as well as the electron exchange between atomic basins. 
\addsymbol{Wave function}{$\psi(\bm{r})$}

\subsubsection{Critical Points}

Chemical bonding information can often be recovered by examining critical points (CPs) in the charge density. 
There are four types of CPs in a 3D scalar field such as $ \rho(\bm{r})$:~local minima, local maxima, and two types of saddle points. A CP in 3D space is defined as
\addabbreviation{Critical point}{CP}

\begin{equation}
\label{eqn: CP}
\nabla \rho(\bm{r})  = \mathbf{i} \frac{\mathrm{d \rho}}{\mathrm{d} x} + \mathbf{j} \frac{\mathrm{d \rho}}{\mathrm{d} y} + \mathbf{k} \frac{\mathrm{d \rho}}{\mathrm{d} z} \quad \rightarrow
 \left\{
  \begin{array}{l l}
   = \bm{0} & \quad \textup{at critical points and } \infty\\
   \neq \bm{0} & \quad \textup{at all other points.}\\
  \end{array} 
  \right. 
\end{equation}

\noindent CPs are denoted by the number of dimensions of the space and the net number of positive curvatures. For example, at a minimum, the curvatures in all principal directions are positive; therefore, this is a (3, +3) CP. 

The ground state charge density at an atomic nucleus is always a maximum, a (3, $-3$) CP, hence it is called a nuclear CP (within the coulomb approximation the density is actually cusp at a nuclear CP since the finite size of the nuclei are neglected in quantum mechanical calculations, but the point still acts acts as a maximum, topologically \cite{Dyall199327, andrae}). 
The simplest topological connection results from a shared (3, $-1$) saddle point between two nuclear CPs and is indicative of a charge density ridge originating at the (3, $-1$) CP and terminating at the nuclear CPs (see \ref{fig:cubane}).  
This ridge possesses the topological properties imagined for a chemical bond and is referred to as a bond path. 
The (3, $-1$) CP is thus called a bond CP.  
Other CPs provide additional information about chemical structure.  
A (3, +1) CP is required at the center of ring structures (rings of bond paths), earning its designation as a ring CP. 
Cage structures must enclose a single (3, +3) CP so these points are called cage CPs.  

\csmfigure{cubane}{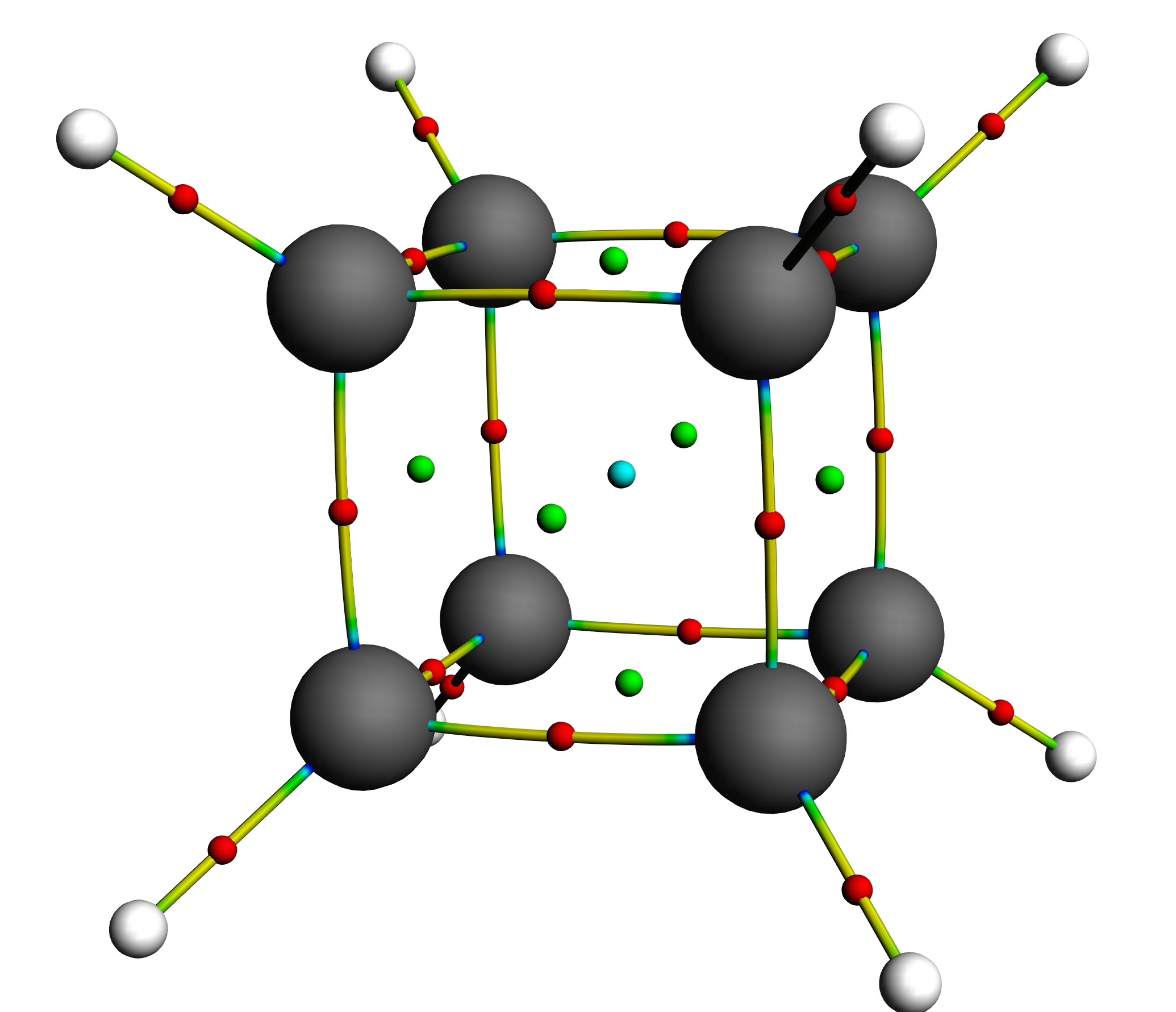}{.5\textwidth}{The CPs and bond paths in a cubane molecule. The image is colored as follows: C-grey, H-white, bond CPs-red, ring CPs-green, cage CPs-blue. Critical points and bond paths were calculated using the Amsterdam Density Functional package \cite{ADF}.}

The amount of charge density at a bond CP has  been used to determine bond strength \cite{baderbook}. 
Oftentimes, bond order (BO) can be determined by
\addabbreviation{Bond order}{BO}

\begin{equation}
\label{eqn:BO}
\textup{BO}=e^{A(\rho_b - B)}
\end{equation} 

\noindent where A and B are constants that are dependent on the bond in question and $\rho_b$ is the value of the charge density at the bond CP. 
The correlation between bond strength and electron density at bond CPs has been especially useful in characterizing hydrogen bonds \cite{popelier_dihydrogen}.  
\addsymbol{Charge density at a bond critical point}{$\rho_b$}
\addsymbol{Laplacian}{$\nabla^2 \rho(\bm{r})$}

The Hessian of $ \rho(\bm{r})$ determines the curvature at critical points. 
Its diagonalized form is

\begin{equation}
\label{eqn: Hessian}
\Lambda =
\begin{bmatrix} \frac{\partial^2 \rho}{\partial x^2} &  0 & 0 \\ 
0 & \frac{\partial^2 \rho}{\partial y^2} & 0\\
0 & 0 & \frac{\partial^2 \rho}{\partial z^2}
\end{bmatrix}
=\begin{bmatrix} \lambda_1 &0 & 0 \\
0 & \lambda_2 & 0\\
0 & 0 & \lambda_3
 \end{bmatrix}
\end{equation}
\addsymbol{Diagonalized Hessian}{$\Lambda$}

\noindent where $\lambda_1$, $\lambda_2$, and $\lambda_3$ are the curvatures of the density in 3 principle directions. 
Its trace, the Laplacian, $\nabla^2  \rho(\bm{r})$, is the sum of the three diagonal elements of the Hessian and has been used to characterize bonding interactions. 
At a bond CP, one of the eigenvectors of the Hessian is parallel to the bond path and the other two are perpendicular. 
Since $ \rho(\bm{r})$ is a maximum perpendicular to the bond path at a bond CP, these two eigenvalues ($\lambda_1$ and $\lambda_2$) are negative while the component parallel to the bond path ($\lambda_3$) is always positive. 
\addsymbol{Curvature of charge density in principle direction $i$}{$\lambda_i$}

Bader argued that a positive value of the Laplacian indicates closed-shell bonding, such as ionic or van der Waals interactions \cite{baderbook}. 
A positive value of $\nabla^2 \rho_b$ occurs when the curvature along the bond path is greater than the sum of the curvatures perpendicular, indicating a depletion of charge density along the bond path. 
On the other hand, a negative value of $\nabla^2 \rho_b$ is associated with open shell (covalent) bonding since there is an accumulation of charge density at the bond CP \cite{Mattabook}.  

Grabowski calculated the amount of charge density and the value of the Laplacian at bond critical points in systems with a large variety of hydrogen bonding interactions ranging from extremely strong (F--H--F)$^-$ to extremely weak HCCH--$\pi$(HCCH) interactions in different chemical environments \cite{grabowski_hbond}. 
He found that parameters purely from the proton-donating part of the H-bond interaction (such as $\rho_b$ and $\nabla \rho_b$) correlate well to hydrogen bond strength, so there is no need to analyze the proton-accepting part of the H-bond or the environment around the bond. 

Ford recently performed a study analyzing values at bond critical points for lithium-bonded complexes such as LiCl$\cdot$NH$_3$ and LiF$\cdot$H$_2$S \cite{ford_li}. 
He was able to predict which structures formed rings, for example, LiF$\cdot$H$_2$O, from structures that were lacking ring CPs, such as LiF$\cdot$H$_2$S. 
Correlation was also found between the amount of charge density and the values of the Laplacian at bond CPs with binding energy.

Further analysis of bonding interactions can be performed using the ratios of individual components of the Hessian. 
The average directionality of a bond is defined by the curvature of $\rho(\bm{r})$ at bond CPs \cite{Eberhart:1996}

\begin{equation}
\label{eqn:directionality}
<\tan(\theta)>=\sqrt{\frac{\lambda_1 + \lambda_2}{2 \lambda_3}}.
\end{equation}

\noindent Directionality can be used to quantify the shape of the electron density at any CP. 
Large values of directionality indicate that the charge density is close to a radial distribution around the CP. 
The charge density is more elliptical if directionality decreases, which is indicative of bonding interactions. 
Values of $\theta$ in eqn (\ref{eqn:directionality}) have been used to explain material properties such as Cauchy pressure \cite{eberhart_cauchy}, elasticity \cite{Eberhart:1996, Eberhart:Giamei, eberhart_shear}, and brittleness \cite{Jones_brittle}.
\addsymbol{Average Directionality}{$<\tan(\theta)>$}

In a similar fashion, ellipticity is defined using the two negative values of $\lambda$ at a bond CP

\begin{equation}
\label{eqn:ellipticity}
\varepsilon = \frac{\lambda_1}{\lambda_2} -1
\end{equation}

\noindent where $|\lambda_1| \geq |\lambda_2|$. 
A value of $\varepsilon$ at a bond CP close to 0 indicates a spherical distribution of density, such as in a single bond. 
The ellipticity will reach a maximum in nonlinear molecules as the bonding interaction becomes more like a standard double bond and will then drop back to zero as a triple bond is reached (since a triple bond is cylindrically symmetric around the bond path). 
For the series ethane, benzene, ethene, and ethyne, the ellipticity values at the C--C bond CPs are around 0, 0.19, 0.30, and 0, respectively.
\addsymbol{Ellipticity}{$\varepsilon$}

Popkov and Breza were able to explain the selectivity of mono- vs. bi-alkylation of a chiral Ni(II) complex using the ellipticity values at bond CPs. 
They found that higher mechanical strain led to increased bond CP ellipticities and therefore favored monoalkylation of the complex \cite{popkov_TM}. 
Jenkins \textit{et al.} used the values of ellipticity at bond CPs to study the stability of various water clusters. 
They found a correlation between high ellipticity (moving away from singe bonds) and higher energy, less stable water clusters. 
They also calculated bond path lengths, number of cage CPs, and used the presence or absence of O--O bond CPs to study various water clusters.  
They concluded that the most stable clusters were actually those with the least number of hydrogen bonding interactions \cite{jenkins_water}.

Values of charge density, the Laplacian, directionality, and ellipticity at bond critical points in the charge density have been used to classify and understand bonding interactions. 
Since these values are based on the electron density rather than wave functions they can be determined both computationally and experimentally
Values at bond CPs have often been found to correlate to bond strengths and stability of molecules. 
In general, QTAIM gives insight into thermodynamic properties: stability of molecules and selectivity of reaction pathways.

\subsubsection{Bond Paths}

Bond paths are defined as the union of two gradient paths in $\rho(\bm{r})$ originating at a bond CP and terminating at neighboring nuclei \cite{Bader_rings_1977}.
Additional chemical bonding information can be gained as we move up in dimensionality from the study of zero-dimensional CPs to one-dimensional bond paths.
In highly symmetric systems, bond paths lie directly along the internuclear axis ,which is the straight line that can be drawn between two bonded nuclei. 
Bond paths can deviate from the internuclear axis to varying degrees, however, and this deviation often correlates to the stability of a bonding interaction. 
For example, hydrogen bond paths usually have at least a slight curvature to them, with weaker hydrogen bonds having more curved bond paths. 

In ring structures, it has been argued that bond paths curving outward from the ring indicate ring strain, while bond paths that curve in towards the center of the ring show stabilization \cite{Bader_rings_1977}. 
A simple example of this phenomenon can be seen in the C--C bond paths of cyclopropane in \ref{fig:cyclopropane}. 
The bond paths curve out from the internuclear axes, in line with the strong destabilization in this molecule due to ring strain.
In benzene, a resonance stabilized structure, the C--C bond paths curve in towards the central ring point.

\csmfigure{cyclopropane}{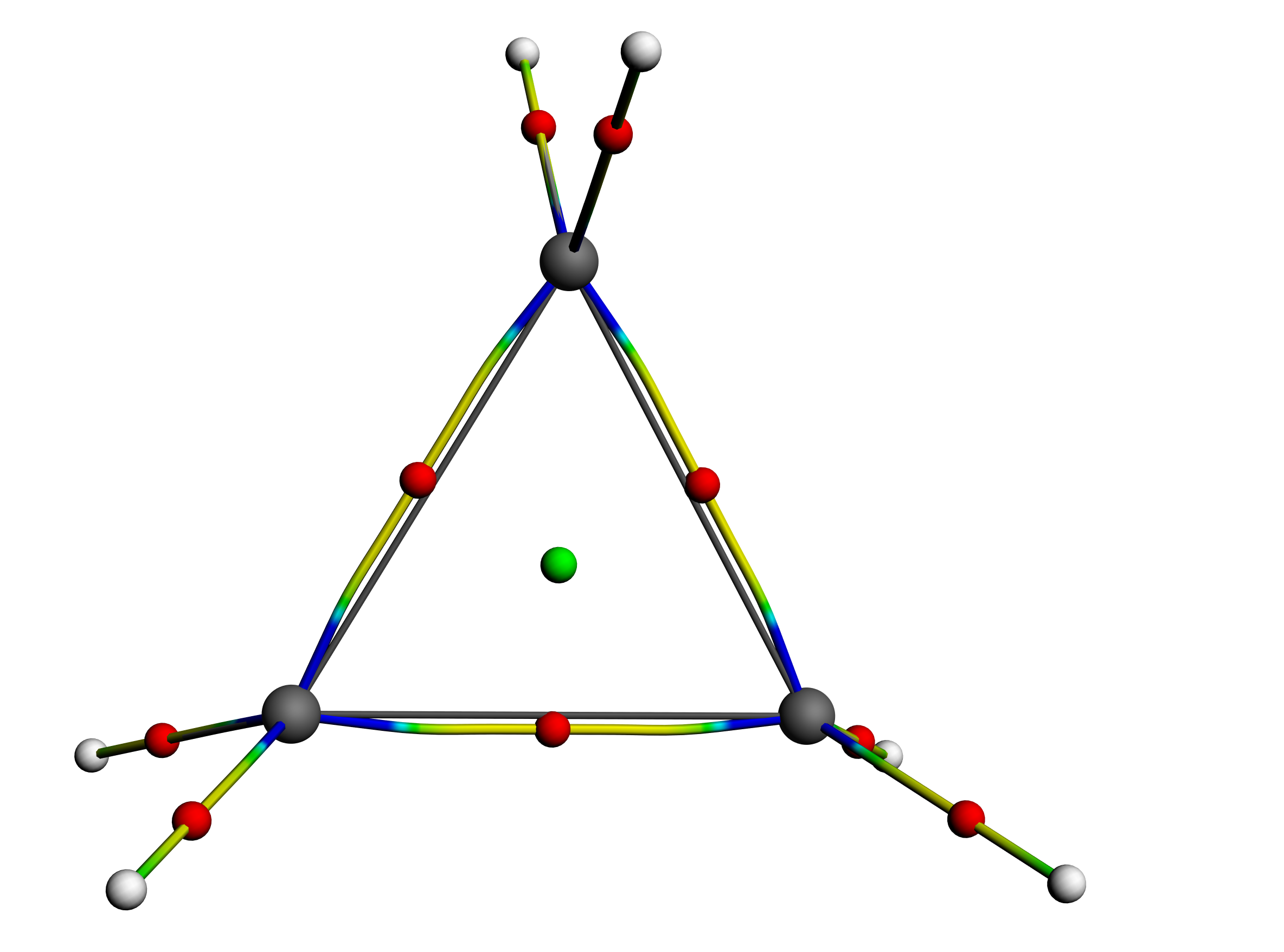}{.7\textwidth}{Bond paths in cyclopropane. The outward curved bond paths between the carbon atoms (colored lines) indicate ring strain in the molecule. The internuclear axes (grey lines) are shown for comparison.}

A bond path is an intuitive way to picture a chemical bond, as a ridge of charge density connecting nuclei. 
Some chemists have argued that the existence of a bond path is always indicative of a stabilizing bonding interaction \cite{Popelier_book, Bader_1985}. 
However, there are systems where bond paths are found, yet it has been argued that the interactions between the atoms that the bond path is connecting are repulsive in nature \cite{Bickelhaupt1, Miorelli, Haaland, Haaland1}. 
Additionally, there are instances where a bond path does not exist, yet most chemists would argue that a stabilizing bonding interaction exists \cite{Shahbazian_consistent}.
This disagreement over the interpretation of bond paths in $\rho(\bm{r})$ indicates a need to advance the current bonding model.

\subsubsection{Atomic Basins and The Zero-Flux Surface Condition}
\label{sec:atomicbasins}

While the original QTAIM only defines bonding interactions based on 1D lines, it does define physical boundaries for where 3D ``quantum atoms" exist within molecules \cite{baderbook}. 
An atom in a molecule (AIM), often referred to as a Bader atom or atomic basin, can be realized by picking any point in the charge density and following the gradient path of steepest ascent that will eventually terminate at a nuclear CP. 
All gradient paths that terminate at a single nuclear CP cover the space that defines an atomic volume. 
These regions are bounded by surfaces of  zero-flux in $\nabla \rho(\bm{r})$, meaning that no gradient path can cross these surfaces, and the zero-flux surface condition is satisfied,

\begin{equation}
\label{eqn:zeroflux}
\nabla  \rho(\bm{r}) \cdot \textup{n}(\bm{r}) = 0, \textup{for all } \bm{r} \in \partial \Omega
\end{equation}

where $\partial \Omega$ is the boundary of each mononuclear region and n$(\bm{r})$ is a normal vector. 
An atom in a molecule can therefore be recognized as the union of a nuclei and its attractive basin in the gradient vector field. 
This partitioning of space allows additive properties of atoms within molecules to be calculated, such as charge, energy, and volume. 
These properties can be found by integration over atomic basins.
\addsymbol{Boundary of mononuclear region}{$\partial \Omega$}
\addsymbol{Normal vector}{n$(\bm{r})$}

Atomic basins are classified as quantum atoms because they are defined by a quantum boundary condition in terms of a measurable expectation value, an observable (the charge density) \cite{Bader_FunctionalGroups}. 
The fact that these volumes have well-defined properties such as energy is due to the zero-flux surface condition, eqn (\ref{eqn:zeroflux}). 
An arbitrary volume in $\rho(\bm{r})$ does not have a well-defined kinetic energy since the local kinetic energy density at any point is ambiguous \cite{Hernandez_KE, Cohen_KE}. 
The total kinetic energy of a molecule can equivalently be calculated by integration of the Schr\"odinger kinetic energy,

\begin{equation}
\label{eqn:SchKE}
\textup{K}(\Omega) = - \frac{\hbar^2}{4m} N \int_\Omega dr \int d \tau'[\Psi \nabla^2 \Psi^* + \Psi^* \nabla^2 \Psi]
\end{equation}

\noindent or the gradient kinetic energy,

\begin{equation}
\label{eqn:graKE}
\textup{G}(\Omega) = \frac{\hbar^2}{2m} N \int_\Omega dr \int d \tau' \nabla_i \Psi^* \cdot \nabla_i \Psi
\end{equation}

\addsymbol{Schr\"odingier kinetic energy}{K}
\addsymbol{Gradient kinetic energy}{G}

\noindent or through any linear combination of these two equations \cite{Hernandez_KE}.

In order for the kinetic energy to be well-defined, K$(\Omega)$ must be equal to G$(\Omega)$. 
Locally, eqns (\ref{eqn:SchKE}) and (\ref{eqn:graKE}) differ by a term that is proportional to the laplacian,

\begin{equation}
\label{eqn:KEdifference}
\textup{K}(\bm{r}) = \textup{G}(\bm{r}) -   \frac{\hbar^2}{4m} \nabla^2 \rho(\bm{r}).
\end{equation}

\noindent To compute the difference of the average kinetic energy in a volume, $\Omega$, one can integrate eqn (\ref{eqn:KEdifference}). 
We use Gauss's theorem to write this difference in terms of a surface integral,

\begin{equation}
\label{eqn:gauss}
\textup{K}(\Omega) = \textup{G}(\Omega) -  \frac{\hbar^2}{4m} N \int dS(\Omega, \bm{r}) \nabla \rho(\bm{r}) \cdot \textup{n}(\bm{r}).
\end{equation}

\noindent The surface integral will be zero for any volume bounded by a zero-flux surface in the gradient of the charge density. This means that any volume in the charge density bounded by a ZFS has a well-defined kinetic energy.

Furthermore, the virial theorem gives the total energy of a system based on the kinetic energy, and can be written in terms of density functional theory (DFT) as \cite{Parr_Virial}

\begin{equation}
\label{eqn:virial}
E = -T[\rho]-\sum_A X_A \frac{\partial E}{\partial X_A}
\end{equation}

\noindent where $T$ is the total kinetic energy functional and $X_A$ are the nuclear coordinates. 
The total energy is also given by
\addsymbol{Total kinetic energy}{$T$}
\addsymbol{Total potential energy}{$V$}
\addsymbol{Total energy}{$E$}
\addsymbol{Nuclear coordinates}{$X_A$}
\addabbreviation{Density functional theory}{DFT}

\begin{equation}
\label{eqn:virial2}
E = T[\rho] + V[\rho]
\end{equation}

\noindent where $V$ is the total potential energy functional. At a stationary point (i.e. an equilbrium geometry), the force term goes to zero and the total energy can be calculated as

\begin{equation}
\label{eqn:virialfunctional}
E = -T[\rho].
\end{equation}

\noindent Therefore, in regions over which $T$ is well-defined, the kinetic energy alone can be used to calculate the total energy of a volume in $\rho(\bm{r})$.

\subsubsection{Delocalization Index}
\label{sec:DI}

In general, atomic basins are used to study atomic properties. 
Atomic properties are calculated using one-electron operators and integrating over an atomic basin using

\begin{equation}
\label{eqn:atomicoperator}
O(\Omega) = \langle \hat{O} \rangle_\Omega = \frac{N}{2} \int_{\Omega} d\bm{r} \int d \tau' [\Psi^* \hat{O} \Psi + (\hat{O} \Psi)^* \Psi]
\end{equation}

where $\hat{O}$ is any one-electron operator. Common values calculated using eqn (\ref{eqn:atomicoperator}) include atomic populations (used to determine atomic charge, referred to as the Bader charge), energies, electrostatic moments, and atomic volumes.
\addsymbol{One-electron operator}{$\hat{O}$}

Of particular interest when discussing bonding in terms of QTAIM is the electron delocalization index (DI) between atomic basins. Atomic basins can be used to calculate a bond order (defined in this  case as the number of shared electrons pairs between atoms) by integrating the exchange density over two bonded atomic basins. The DI between two atoms is defined as
\addabbreviation{Delocalization Index}{DI}

\begin{equation}
\label{eqn:delocalization}
\delta(A,B) = 2 |F^{\alpha}(A,B)| + 2|F^{\beta} (A,B)|
\end{equation}

\noindent in which F is the Fermi correlation defined as
\addsymbol{Fermi correlation}{F}
\addsymbol{Delocalization index}{$\delta$}

\begin{equation}
\label{eqn:fermi}
F^{\sigma}(A,B) = -\sum\limits_i \sum\limits_j S_{ij}(A)S_{ji}(B)
\end{equation}

\noindent where $S_{ij}(\Omega)$ is the overlap integral between two spin orbitals over an atomic region \cite{Mattabook}.
\addsymbol{Overlap integral}{$S$}

Matta \textit{et al.} extended the idea of DI between atoms and calculated the delocalization between two phenyl rings in a biphenyl system to study hydrogen-hydrogen bonding. It was found that the delocalization of electrons between the two rings was actually maximized for a planar configuration rather than a twisted (lower energy) configuration. This is used as part of an argument that the controversial hydrogen-hydrogen bond path in biphenyl is a stabilizing interaction \cite{Matta_delocalization}.

\subsection{Topological Models of Kinetic Energy Density}
\label{sec:KEmodels}

The topology of scalar fields other than $\rho(\bm{r})$ have also been used to create chemical bonding models. 
Many of these functions strive to define delocalization of electrons and are based on the kinetic energy density. 
Because these functions focus on kinetic energy, they are often interpreted as providing information on the motion of the electrons. 
Therefore, delocalization functions generally give insight into how the electrons will rearrange and move, making them useful chemical bonding models for predicting reactivity. 
Here, I review three of the most prominent delocalization functions: the electron localization function (ELF), the electron localizability indicator (ELI), and the localized orbital locator (LOL) method.
\addabbreviation{Electron localization function}{ELF}
\addabbreviation{Electron localizability indicator}{ELI}
\addabbreviation{Localized orbital locator}{LOL}

ELF was developed with the goal of defining a more rigorous means of bond classification based on local quantum-mechanical functions that depend on the Pauli exclusion principle. 
Becke and Edgecombe originally defined the function using spin densities \cite{ELF_1990}, but Savin \textit{et al.} generalized ELF for the total electron density  as \cite{ELF_Savin}

\begin{equation}
\label{eqtn:ELF}
\text{ELF} = \frac{1}{1+(\frac{D}{D_0})^2}
\end{equation}

which is mapped onto a finite range of $0 \leq ELF \leq 1$, where

\begin{equation}
\label{eqn:ELFterms}
\begin{split}
D &= \tau - \frac{1}{4} \frac{(\nabla \rho )^2}{\rho}, \\
D_0 &= \frac{3}{5} (6 \pi^2)^{2/3} \rho^{5/3}  \text{, and}\\
\tau &= \sum_i |\nabla \psi_i|^2.\\
\end{split}
\end{equation}

\noindent $D$ is the difference between the positive kinetic energy, $\tau$, and the von Weizs\"acker kinetic energy for non-interacting bosons. 
$D_0$ is the kinetic energy of a uniform electron gas. ELF is often interpreted as showing the amount of excess kinetic energy there is due to Pauli repulsion  \cite{ELF_1992, poater_AIM_ELF}.
\addsymbol{Uniform electron gas kinetic energy}{$D_0$}
\addsymbol{Positive kinetic energy}{$\tau$}

A topological analysis of ELF can be used to define bonding and non-bonding regions in molecules.  
As opposed to having nuclear attractors in a topological analysis of the charge density, there are three types of attractors in ELF: core, bonding, and non-bonding, each with its own attractive basin surrounding an electron pair \cite{ELF_1990}. 
Between each attractive basin there are surfaces of zero-flux in $\nabla ELF$ (see \ref{fig:Ge_ELF}). 
Saddle points of the (3, $-1$) variety are deemed bifurcation points, and the value of ELF at these points is interpreted as the amount of interaction between adjacent basins, i.e. a measure of delocalization. 

\csmfigure{Ge_ELF}{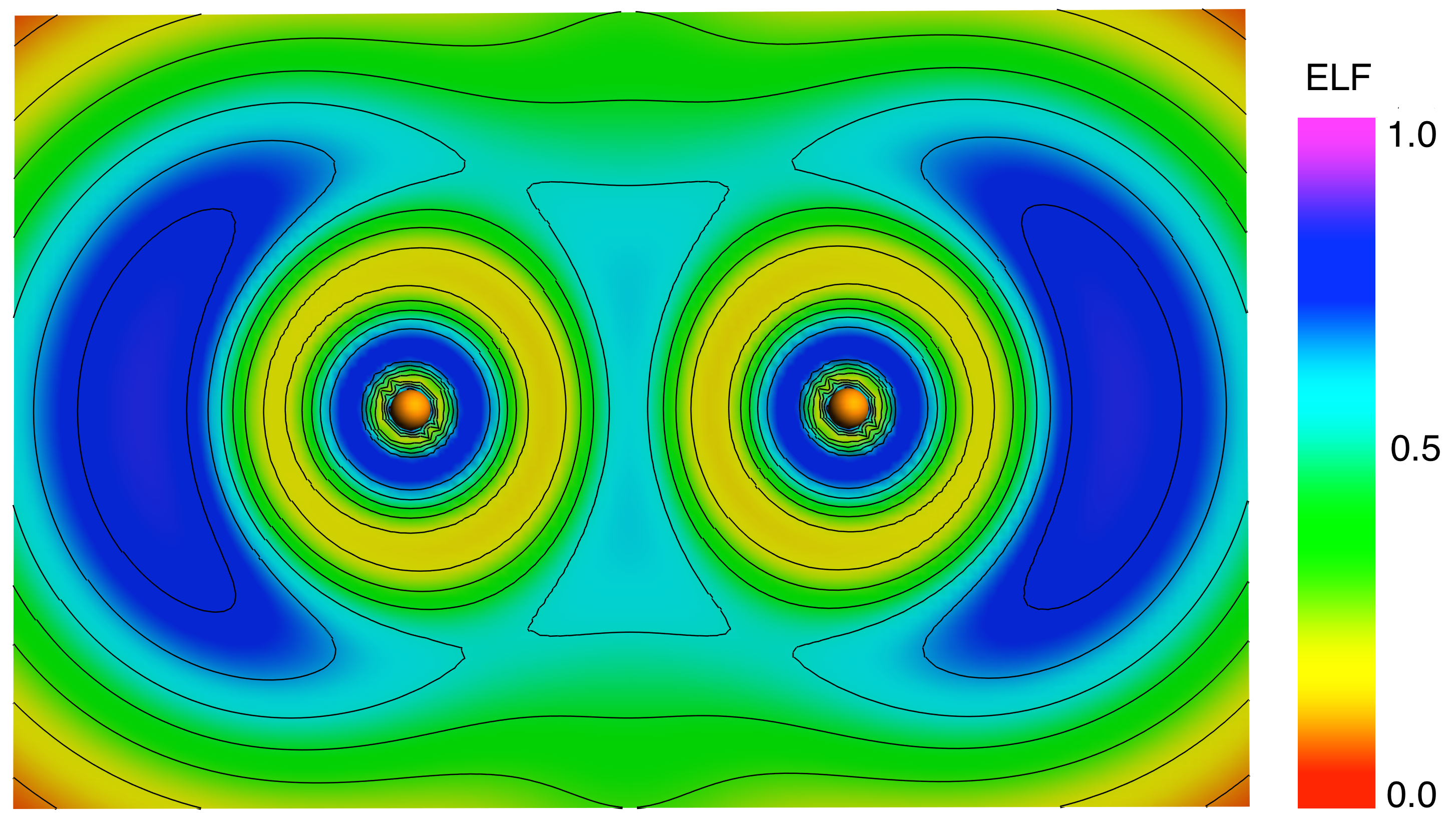}{.8\textwidth}{A contour plot of ELF for a Ge$_2$ molecule showing topological separation of core, valence, and bonding regions into separate regions bounded by ZFSs.}

One of the main applications of ELF has been to aromatic compounds. 
Poater \textit{et al.} recently published a detailed review comparing the use of QTAIM and ELF methods for analyzing aromatic bonding \cite{poater_AIM_ELF}.  
It has been shown that it is possible to characterize aromaticity in some molecules by separating ELF into $\pi$ and $\sigma$ orbital contributions \cite{Santos_ELF}. 
While the bifurcation values in the total ELF were not able to distinguish between aromatic, anti-aromatic, and non-aromatic systems, $\pi$ ELF values were. 
Santos \textit{et al.} used $\pi$ ELF values to classify a variety of molecules including benzene, a proposed B$_6$(CO)$_6$ molecule, and cyclohexatriene, into each of these categories of aromaticity. 
Fuster \textit{et al.} performed a study looking at ortho/para versus meta orienting substituents on benzene rings. 
They were able to distinguish the preferential site of attack (ortho/para or meta) for a second substituent for a set of substituted rings based on isosurfaces of ELF.

It has been argued that ELF is able to predict site reactivity because this function quantifies the importance of Pauli repulsion at a certain point in a molecule. 
When electrons are unpaired, or in pairs of antiparallel spin, Pauli repulsion is not a large contributor to the total energy. 
There is not excess kinetic energy in this region compared to a uniform electron gas. 
At the boundaries between paired electrons though, Pauli repulsion is extremely important causing an increase in the kinetic energy of electrons and low values of ELF. 


A recent variation to ELF is the electron localization index D (ELI-D) which analyzes the average number of electrons per fixed fraction of a same-spin electron pair. 
ELI-D holds the number of same-spin electron pairs fixed for a variable volume cell allowing for different values of charge, Q$_i$. 
Therefore, ELI-D can be interpreted as being proportional to the average number of electrons required to form a fixed fraction of the same-spin electrons. 
The index has been defined by Kohout and coworkers \cite{Kohout_ELI} as
\addabbreviation{Electron localization index D}{ELI-D}
\addsymbol{Charge in ELI-D cell}{$Q_i$}

\begin{equation}
\label{eqn:ELID}
\text{ELI-D} = \sum\limits_i^{occ} \rho_{\sigma,i} \Bigg[\frac{12}{\rho_{\sigma} (\tau - \frac{1}{4} \frac{\left(\nabla \rho_{\sigma}\right)^2}{\rho_{\sigma}})}\Bigg]^{3/8}
\end{equation}

\noindent where $\tau$ is the same positive kinetic energy used in ELF (eqn (\ref{eqn:ELFterms})), and $\sigma$ indicates spin.

While this index is topologically identical to ELF, there are some distinctions in the interpretation and calculation. 
Mainly, ELI-D does not require the use of a uniform electron gas as a reference state. 
The partial ELI-D (pELI-D) further decomposes this index into individual contributions from each orbital \cite{ELID_Pi}. 
The pELI-D provides a measure of which orbitals contribute to a portion of Q$_i$. 
Using pELI-D, Grin \textit{et al.} were able to gain additional insight into transition metal bonding, enabling them to quantify the contribution of specific orbitals to bonding in Sc$_2^+$ and TiF$_4$ \cite{Grin_ELI_2007}.
\addabbreviation{Partial electron localization index D}{pELI-D}


One of the newest topological methods for analyzing bonding interactions is the localized orbital locator (LOL) originally proposed by Becke and Schmider \cite{Becke_LOL}. 
LOL leaves out the term for the kinetic energy of non-interacting bosons, and only uses the positive form of the kinetic energy \cite{jacobsen_LOL,ayers_KE,  Hernandez_KE}. 
As in ELF, LOL, $\nu(\bm{r})$, is mapped onto a finite range, $ 0 \leq \nu(\bm{r}) \leq 1$ using the formula

\begin{equation}
\label{eqn:LOL}
\nu(\bm{r}) = \frac{1}{1+\frac{\tau}{D_0}}.
\end{equation}

\noindent LOL is the easiest to compute of the three delocalization formulas presented here, and it has been argued that it is also the easiest to interpret \cite{ELID_Pi}. 
\addsymbol{Localized orbital locator}{$\nu(\bm{r})$}

Rather than focusing on electron delocalization, Jacobsen emphasizes an interpretation of LOL based on the velocity of electrons. 
A value of $\nu(\bm{r})  = \frac{1}{2}$ means that electrons at $\bm{r}$ are moving at the same speed as they would be if they were a uniform electron gas. 
$\nu(\bm{r}) < \frac{1}{2}$ indicates high kinetic energy and thus faster electrons. 
Finally, $\nu(\bm{r}) > \frac{1}{2}$ corresponds to slow moving electrons. 
Faster electrons are generally found near the nuclei and are thus core electrons. 
Areas with slower moving electrons indicate valence electrons. 
LOL was specifically designed to analyze regions of covalent bonding based on the idea that the driving force for covalent bonding is the lowering of kinetic energy \cite{Ruedenberg_KE}. 

In the original paper introducing the LOL method, covalent bonds and lone pair regions were accurately predicted for simple reactions. 
This was accomplished by calculating contour plots of $\nu(\bm{r})$ for reactants, transition states, and products \cite{Becke_LOL}. 
The authors observed that the shape of regions with $\nu(\bm{r}) >\frac{1}{2}$ can be correlated with $\sigma$ bonding, $\pi$ bonding, or lone pair regions. 
It has been shown that LOL generally does a better job of distinguishing between weak and enhanced $\pi$-delocalization in aromatic compounds than ELF does, such as in benzene and borazine \cite{ELID_Pi}. 
Additionally, Yang performed a study using a multitude of bonding metrics to characterize various pthalocyanides. 
He found that LOL distinguished between core, bonding, and non-bonding regions in a much clearer fashion than ELF \cite{Yang_aromaticity}.

While LOL is still a very new function, there is potential for it to become a powerful tool in predicting reactivity as it focuses on the motion of electrons in molecules. 
This idea will be developed further when discussing the decomposition of atomic basins into gradient bundles in Chapter \ref{cha:intrinsic}.


\subsection{Bond Bundles}

Although the only volumes recognized in the original QTAIM are atomic basins, they are not the only meaningful volumes in the charge density bounded by ZFSs and containing well-defined properties.
Pend\'{a}s \textit{et al.} defined a primary bundle as a set of all gradient paths in the charge density that start at one minimum (cage CP) and end at the same maximum (nuclear CP) \cite{pendas1}. 
Primary bundles are bounded by ZFSs and have one nuclear CP, one cage CP, $n$ ring CPs, and $n$ bond CPs on the surface of each bundle. 
Primary bundles take the shape of polyhedra, with 2$n$+2 vertices and 2$n$ faces. 
The union of all the primary bundles sharing a single nuclear CP is the atomic basin of attraction discovered by Bader. 
Pend\'{a}s \textit{et al.} then went on to define a basin of repulsion as the union of all primary bundles sharing a single cage point. 

The elegance of the definitions of atomic and repulsive basins is that there is only one ZFS bounding a nuclear or cage CP that does not pass through the point itself. 
This makes it relatively easy computationally to locate attractive and repulsive basins in the charge density. 
Saddle points in $\rho(\bm{r})$ have an in an infinite number of ZFSs bounding them. 
A 2-dimensional cut plane of some of the ZFSs surrounding the C--C bond CP in ethene is shown in \ref{fig:ZFS}. 

\csmfigure{ZFS}{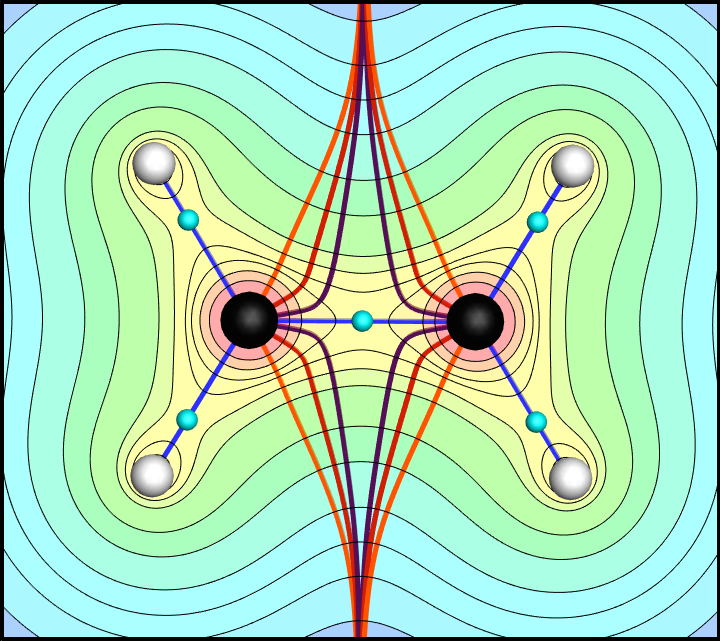}{.5\textwidth}{2D cut plane of the charge density of ethene in the plane of the molecule. Thick colored lines show a sampling of ZFSs bounding the C--C bond CP.}

It was noted by Eberhart  that while there are an infinite number of ZFSs bounding bond CPs, some of these ZFSs are special \cite{eberhart_IB} as they are also contained in a 2-dimensional relative critical set of $\rho(\bm{r})$. 
Relative critical sets are a higher dimensional extension of critical points applicable to lines and surfaces for a function in $\mathbb{R}^3$ (see \cite{Miller, criticalsets}). 
Relative critical points, which are the elements of a relative critical set, can be defined \cite{Miller} similarly to the elements of the Eberly height-ridge \cite{Eberly}, which is one example of a relative critical set. 
Let $I=\{i_1,\cdots,i_{3-d} \} \subset \{ 1,2,3 \}$. 
A point $\bm{r}_0$ is a d-dimensional relative $I$-critical point of $\rho$ if the eigenvalues 
of the Hessian, $H(\rho)$, are distinct from each other and the eigenvectors, $e_i$, for $i=1,2,3$ of $H(\rho)$ satisfy the condition

\begin{equation}
\label{eqn:ridge}
\nabla \rho(\bm{r}_0) \cdot e_i = 0, \quad \text{for } 1\leq i \leq 3-d.
\end{equation}

\noindent The relative $I$-critical set of $\rho$ is the set of all such relative $I$-critical points. 
Note that the number of entries in $I$, and thus the number of orthogonality conditions required in eqn (\ref{eqn:ridge}), determines the dimensionality of the critical set. Critical points, lines and surfaces require three, two, and one orthogonality conditions, respectively.
\addsymbol{Eigenvectors of the Hessian}{$e_i$}

A bond bundles is therefore defined as a volume bounded by ZFSs that satisfy the definition of 2-dimensional relative critical surfaces, and containing a single bond CP. 
A BB has well-defined, additive properties just like an atomic basin, due to the ZFS condition. 
Eberhart and Jones have shown that bond bundles recover properties consistent with the standard picture of a bond such as electron count. For example, when the valence charge density is integrated over the bond bundle between bonded carbons for ethane, benzene, ethene, and ethyne; two, three, four, and six electrons are recovered, respectively \cite{Jones_KE}. 
This is in agreement with the standard model of these bonds being single, one and a half, double, and triple bonds. 
Bond bundles have been used to uncover structure--property relationships using two different approaches \cite{Jones_functionality}. 
The first is to investigate the geometry of ZFSs bounding a BB, and the second is through integration of properties over BB volumes.

As an example of the first approach, Jones showed that the shape and size of bond bundles can be used to predict the tendency of molecules to undergo nucleophilic substitution or addition \cite{Jones_nucl}. 
Phosphirane (a 3-membered ring) more readily undergoes ring-opening reactions than phosphetane (a 4-membered ring). 
While the 3-membered ring is more strained than phosphetane, the strain energy alone does not account for the large discrepancy in activation energies for these two reactions \cite{ringstrain}. 
The C--P BB cleaved during the reaction in phosphirane has closed surfaces creating a bond bundle with finite volume. 
The C--P BB in phosphetane is larger and open on one end. 
Bond bundles with smaller volumes generally contain fewer valence electrons, and when a bond breaks, a bond bundle must collapse.
Therefore, Jones argued that closed BBs will more readily undergo nucleophilic attack, as these bonding interactions are energetically easier to disrupt.

An example of the second approach to the application of bond bundles demonstrates a relationship between electron count and material toughness. 
The number of valence electrons in BBs was found to correlate with the work of separation, $W_{\infty}$, in a computational study of alloying elements for high strength steels \cite{Jones_alloys}. 
$W_{\infty}$  is defined as the minimum amount of energy required to separate an interface into two free surfaces. 
Finding alloying elements that increase $W_{\infty}$ allow for the creation of tougher materials. 
This study found that the number of valence electrons in second-neighbor Fe--Fe BBs correlated linearly with $W_{\infty}$ for an iron-ceramic interface common in dispersion-strengthened steels. 
This correlation was then used to explain why tougher materials can be created using various transition metals as alloying elements. 
There is very little $p_z$ character in the second-neighbor Fe--Fe BBs. 
By substituting Ni (which possesses low-lying $p$ character) for some of the Fe sites, the number of electrons in the second-neighbor Fe--Fe BBs is reduced and $W_{\infty}$ increases. 
This prediction is consistent with the known properties of Ni alloyed steels such as BlastAlloy, which is remarkably tough \cite{BlastAlloy}. 
This is an important example of where a 3-dimensional, physical bond must be studied to understand a property. When the bond CPs between various bonding interactions in the system alone were examined, no correlation was found with the work of separation. 
\addsymbol{Work of separation}{$W_{\infty}$}

As these two examples show, a model that presents the chemical bond as both a topological connection between atoms and as a region with well-characterized properties can potentially broaden our understanding of the relationships between chemical structure and properties.
The investigation of bond bundles adds a new dimension to ways the charge density can be studied.
Rather than looking only at the topology of $\rho(\bm{r})$, the study of bond bundles involves investigating the geometry of $\rho(\bm{r})$, specifically by observing the shape of meaningful topological connections. 
By expanding on this model it may be possible to extract additional useful information from the structure of the charge density. 
Gradient bundle analysis is being developed with the goal of creating an even higher resolution picture of chemical bonding based on the topology and geometry of $\rho(\bm{r})$. 


\chapter{In Search of An Intrinsic Chemical Bond}
\label{cha:intrinsic}

\begin{center}
\noindent Modified from a paper published in \textit{Computational and Theoretical Chemistry}.\\
Amanda Morgenstern\footnote{Primary researcher and author}, Tim Wilson\footnote{Developed code}, Jonathan Miorelli\footnote{Performed distribution statistics and some calculations}, Travis Jones\footnote{Provided advice on computations and document editing}, and M. E. Eberhart\footnote{Corresponding author}. \textit{Comput. Theor. Chem.} 1053: 31--37, 2015.
\end{center}

\subsection{Abstract}
The chemical bond, as a link between atoms, is an intrinsic  property of the charge density.  
However, bond energy, which is commonly seen as the energy difference between a molecular state and an arbitrary dissociated state,  depends extrinsically on the charge density.  
The view of a bond as a natural link possessing properties that are externally determined often leads to contradictory interpretations as to the origins of the structure and properties of molecules and solids.   
Ideally, one would like to uncover an intrinsic property of the chemical bond that gives similar information content as that provided by bond energy.   
To this end, we report on our ongoing work exploring the intrinsic geometry imposed on the charge density by mapping it onto volumes bounded by zero-flux surfaces in the gradient of the charge density, deemed gradient bundles.   
These natural volume elements of QTAIM have well defined properties.
Hence, this mapping produces a set of property distributions with a quantifiable  geometric structure that varies from molecule to molecule.  
Here, we examine the intrinsic geometry of the kinetic energy distribution in gradient bundles for a series of homonuclear diatomic molecules.
We find that the structure given by gradient bundles reproduces standard valence shell electron pair repulsion diagrams for the set of molecules, H$_2$, N$_2$, O$_2$, and F$_2$.
Furthermore, there is evidence that the geometric properties of these distributions correlate with bond energies.  

\subsection{Introduction}

Chemistry is the study of atoms and bonds.  
QTAIM  provides a rigorous and well studied representation of an atom in terms of the intrinsic topology and geometry of the charge density \cite{baderbook}.  
While QTAIM revealed that a three-dimensional atom with well-defined properties can be unambiguously identified using the topology of the electron charge density, this same theory provides only a one-dimensional representation of the chemical bond, as a bond path.
Atomic energies can be calculated by integrating energy values over atomic basins. 
Since bond paths are only one-dimensional, bond energies can not be obtained in this same manner. 
Bond energy is generally calculated as the energy difference between the equilibrium conformation of a molecule and some  \textit{arbitrary} dissociated state of the same molecule. 
The bond as a topological linkage is an intrinsic property of the  charge density, while its most notable property, the bond energy,  is an extrinsic property that depends on the coordinates of the dissociated state.  
A QTAIM appropriate representation of the chemical bond and its properties must be based in the intrinsic geometry of the charge density.  
Here we propose one approach toward developing such a representation using gradient bundles in the charge density.  

Just as there are charge density ridges (bond paths), there are also one-dimensional critical sets that are formerly termed valley paths \cite{Eberly}, which went unnoticed in the original formulation of QTAIM.  
Because valleys and ridges differ only by the sign of the curvatures along the path, both are often referred to as ``ridges'', which is the nomenclature we shall adopt here for any relative critical set. 
By taking into account all the various types of ridges in the charge density a more complete topological structure of molecules and solids can be constructed in a similar fashion as the primary bundles defined by Pendas \textit{et al.} \cite{pendas1}.

For extended systems  there will always be four kinds of charge density CPs (0-ridges), six kinds of 1-ridges, and four kinds of 2-ridges \cite{bondbundle2,Jones_KE, eberhart_IB}.  
The 1-ridges pairwise connect the four types of critical points and the 2-ridges are surfaces containing three pairwise connected CPs. 
This ridge structure forms a set of space filling volumes homeomorphic to tetrahedra.  
Coincident with the four vertices of each tetrahedron is a nuclear, bond, ring, and cage CP. 
(Though this is the typical connectivity, tetrahedra can also be formed with a nuclear, cage and two bond or two ring CPs.)
The six edges of each tetrahedron are 1-ridges, and the four faces are 2-ridges. 
In an open system---a molecule or a solid with a free surface---all four types of CPs need not be present, in which case some of the 1-ridges will be of infinite length and some of the 2-ridges will have infinite area \cite{bondbundle2,Jones_KE}.  
In such systems the 2-ridges form a set of open tetrahedra \cite{Jones_functionality, Jones_nucl}.

In either an open or closed system the tetrahedra are simplices, which means that they may be ``glued'' together to form a simplicial complex that is homeomorphic to the charge density topology of any molecular system \cite{eberhart_cauchy,foundations}.  Accordingly, these simplices have been designated irreducible bundles, IBs, where bundle is used to evoke an image of a bundle of gradient paths (see \ref{fig:ethene}). In open systems some of the IBs must be homeomorphic to open tetrahedra and are referred to as open IBs and are said to be open in the direction normal to the tetrahedral face, 2-ridge, of infinite area.  
\addabbreviation{Irreducible bundle}{IB}

\csmfigure{ethene}{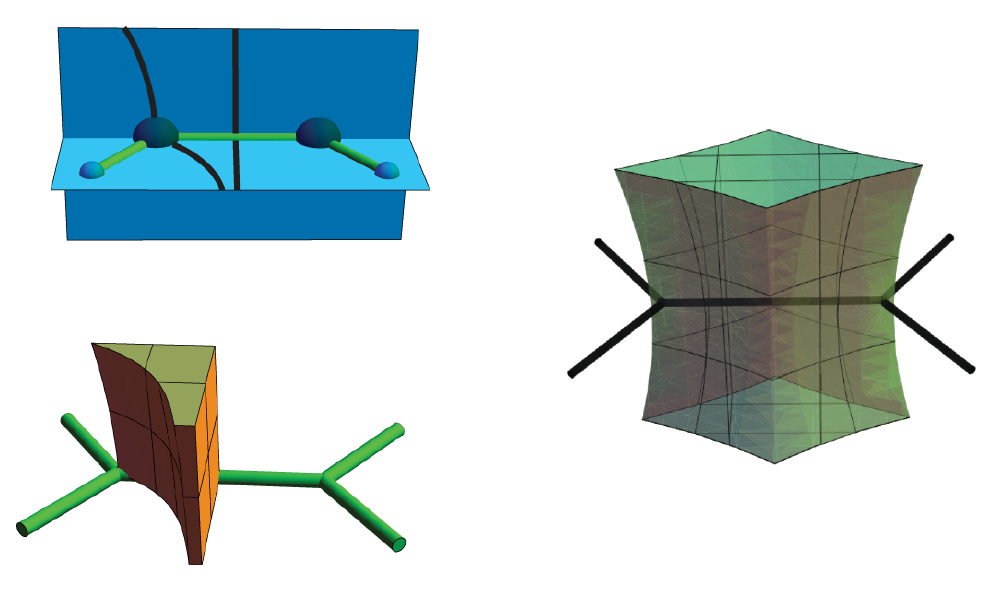}{.8\textwidth}{A C-C bond bundle in ethene built up from its irreducible bundles. The black lines shown in the top left are the 1-ridges that form the edges of one IB that is a part of the C-C bond bundle in ethene. The 3D IB bounded by ZFSs is shown in the bottom left, and the full bond bundle composed of the eight symmetry equivalent IBs is on the right.}

In addition to the extended simplicial complexes, local structures can be generated by gluing together a finite number of IBs. The most basic of these are given through the union of IBs sharing a single CP.  The union of all IBs sharing the same nuclear CP generates atomic basins.  The union of all IBs sharing the same cage CP yields the repulsive basin first noted by Pendas \textit{et al.} \cite{pendas1}.  In addition, one can construct the union of all IBs sharing the same ring point.  Finally, there is the union of all IBs sharing the same bond CP.  This volume contains a single bond critical point and its associated bond path and recovers the bond bundle volume first identified by Eberhart \textit{et al.} \cite{bondbundle2}. \ref{fig:ethene} shows the construction of the C--C bond bundle of ethene. 

\subsection{Gradient Bundles}

While the IB is the basic structural element capturing the connections between critical points, it too can be built from smaller volumes bounded by ZFSs \cite{foundations}. 
We define a gradient bundle as \textit{any} volume in the charge density bounded by a zero-flux surface in $\nabla \rho(\bm{r})$. 
To help with visualization, begin by noting that sufficiently close to a nuclear CP all charge density gradient vectors are radial and of the same magnitude. 
As a result, it is always possible to find a spherical isosurface centered on each nuclear CP. 

\ref{fig:SphereGB} illustrates a method for constructing gradient bundles from a set of neighboring gradient paths.
We can triangulate the surface of the spherical isosurface of $\rho(\bm{r})$. 
The set of gradient paths passing through the points interior to a triangular element on the sphere form a gradient bundle.
Atomic basins, basins of repulsion, and bond bundles can all be decomposed into sets of neighboring gradient bundles, and are in fact gradient bundles themselves. 
Any property that can be found by integration over one of these larger volumes can equivalently be calculated by integrating the property over any set of gradient bundles within the larger basin or bond bundle.  
Gradient bundles can be made arbitrarily small to analyze the charge density to varying degrees of locality.

\csmfigure{SphereGB}{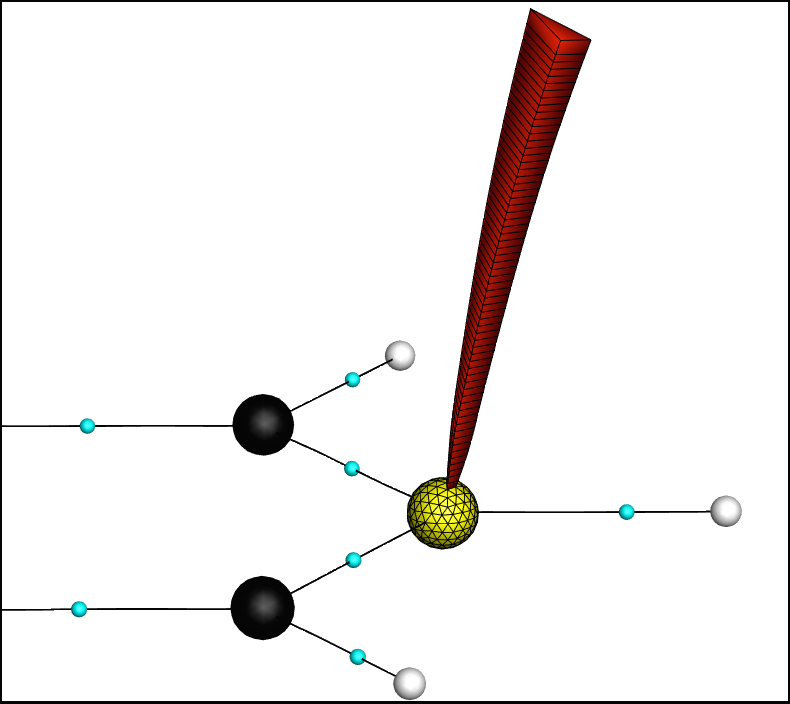}{.6\textwidth}{A gradient bundle created from the triangulation of a sphere (shown in yellow) around a carbon nucleus in benzene. Carbon nuclei are shown in black, hydrogen nuclei in white, bond critical points in cyan, and bond paths are black.}

As motivation for what follows, consider the property distributions of  ``van der Waals bonded'' Ne$_2$  and  ``covalent'' N$_2$.  
\ref{fig:gradpaths} shows a contour plot of the computed charge densities and the associated gradient fields for each molecule (see Section~\ref{sec:methods_intrinsic} for computational parameters).  
The gradient fields are visually different, with the  Ne$_2$ gradient paths being atomic-like (radial) except in the immediate neighborhood of the 2-ridge that is the ZFS of the Ne Bader atoms, i.e., the midplane between the nuclei. 
In contrast, for N$_2$ the gradient path curvature is distributed along a greater length of the path, and particularly,  closer to the nucleus. 
With such distinctive gradient fields, it seems reasonable that their property distributions will also be distinct.  

\csmfigure{gradpaths}{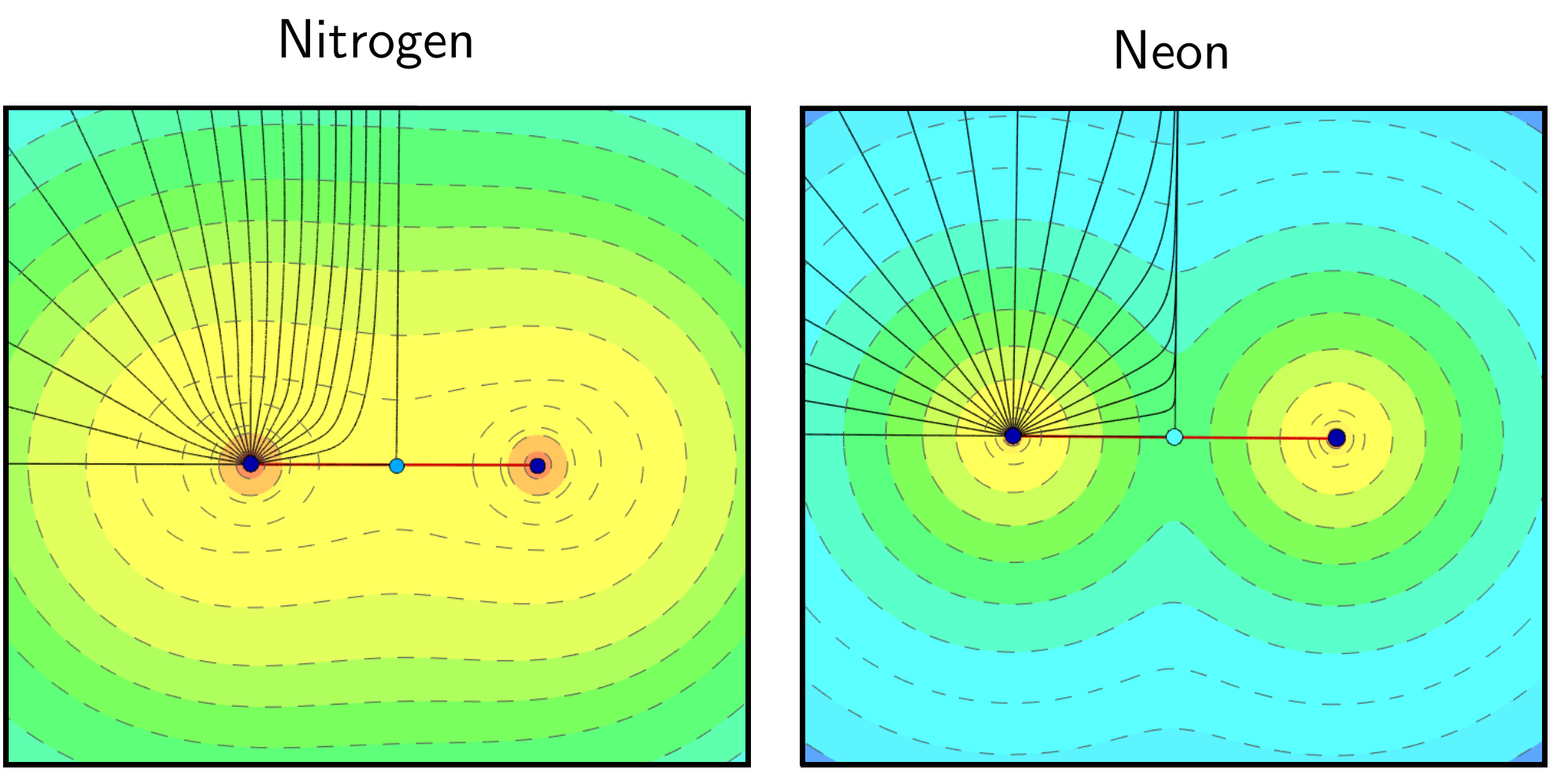}{\textwidth}{Gradient paths seeded every 10$^{\circ}$ in the charge density for N$_2$ and Ne$_2$ to show the general gradient path behavior. Dashed lines represent contour lines of $\rho(\bm{r})$, solid lines are gradient paths, blue circles are nuclear CPs, cyan circles are bond CPs, and red lines are bond paths.}

To examine these differences in more detail we explore the intrinsic properties of the chemical bond through a comparison of the charge density and kinetic energy distributions of H$_2$, N$_2$, O$_2$, F$_2$, and Ne$_2$.  
This series was chosen for the initial study of intrinsic bonding properties based on their symmetry, allowing for easier gradient bundle construction, and their small size, resulting in more expedient calculations.

\subsection{Methods}
\label{sec:methods_intrinsic}

The reported results were obtained using the Amsterdam Density Functional Package, ADF,  version 2013.01 \cite{ADF, ADF2}. 
All calculations were performed spin-restricted with a non-relativistic all-electron triple $\zeta$ singly polarized basis set and the Perdew-Burke-Ernzerhof (PBE) parametrization of the generalized gradient approximation (GGA) functional {\cite{Perdew, PBE, PW91}}.  
Bond distances for H$_2$, N$_2$, O$_2$, and F$_2$ were fully optimized using ADF and an internuclear distance of 2.98 \AA\ was used for Ne$_2$. 
Bader atomic properties were calculated using a grid based approach implemented in ADF using default parameters which includes using the exact density for the $V_{xc}$ \cite{ayers_ADF}.
\addabbreviation{Amsterdam density functional package}{ADF}
\addabbreviation{Perdew-Burke-Ernzerhof}{PBE}
\addabbreviation{Generalized gradient approximation}{GGA}

In an effort to ensure that our results were not highly dependent on the choice of functional, we also investigated the local density approximation (LDA), Becke-3-Lee-Yang-Parr with 20\% hartree-fock exchange (B3LYP), Meta-GGA-type Minnesota 06 local (M06L), and Becke-Lee-Yang-Parr (BLYP) functionals \cite{VWN, B3LYP, MO6L, correlation}.  
Our principal criteria for functional dependence was visual inspection of the gradient fields and qualitative comparisons of property distributions. 
For the purposes of comparing properties integrated over gradient bundles, our preliminary data shows that functional choice does not appear to greatly affect the results for this set of molecules (see Appendix~\ref{app:functionals}). 

\addabbreviation{Becke-3-Lee-Yang-Parr}{B3LYP}
\addabbreviation{Minnesota 06 local}{M06L}
\addabbreviation{Becke-Lee-Yang-Parr}{BLYP}
\addabbreviation{Local density approximation}{LDA}

We have chosen to investigate only the singlet O$_2$ molecule to avoid the additional complexity associated with the spin polarized charge density, though it can be treated with QTAIM \cite{spintopo}. 
We have enforced a ``closed-shell'' configuration by performing a spin-restricted calculation to approximate the excited singlet O$_2$ state \cite{Carbogno}.

The  geometry of the charge density was further analyzed using the Bondalyzer add-on package in Tecplot \cite{Tecplot}.  
For each molecule, 73 gradient paths were seeded every 2.5$^{\circ}$ around a semi circle of radius 0.05 \AA\ centered on a nuclear CP and lying on a plane through the molecule (\ref{fig:tubes}). 
Due to the D$_{\infty h}$ symmetry of the molecules, the rotation of these gradient paths through any angle about their internuclear axis (by default we used a rotation angle of 360$^{\circ}$) will create ZFSs that bound gradient bundles as shown in \ref{fig:gradient_bundles}.

\csmfigure{tubes}{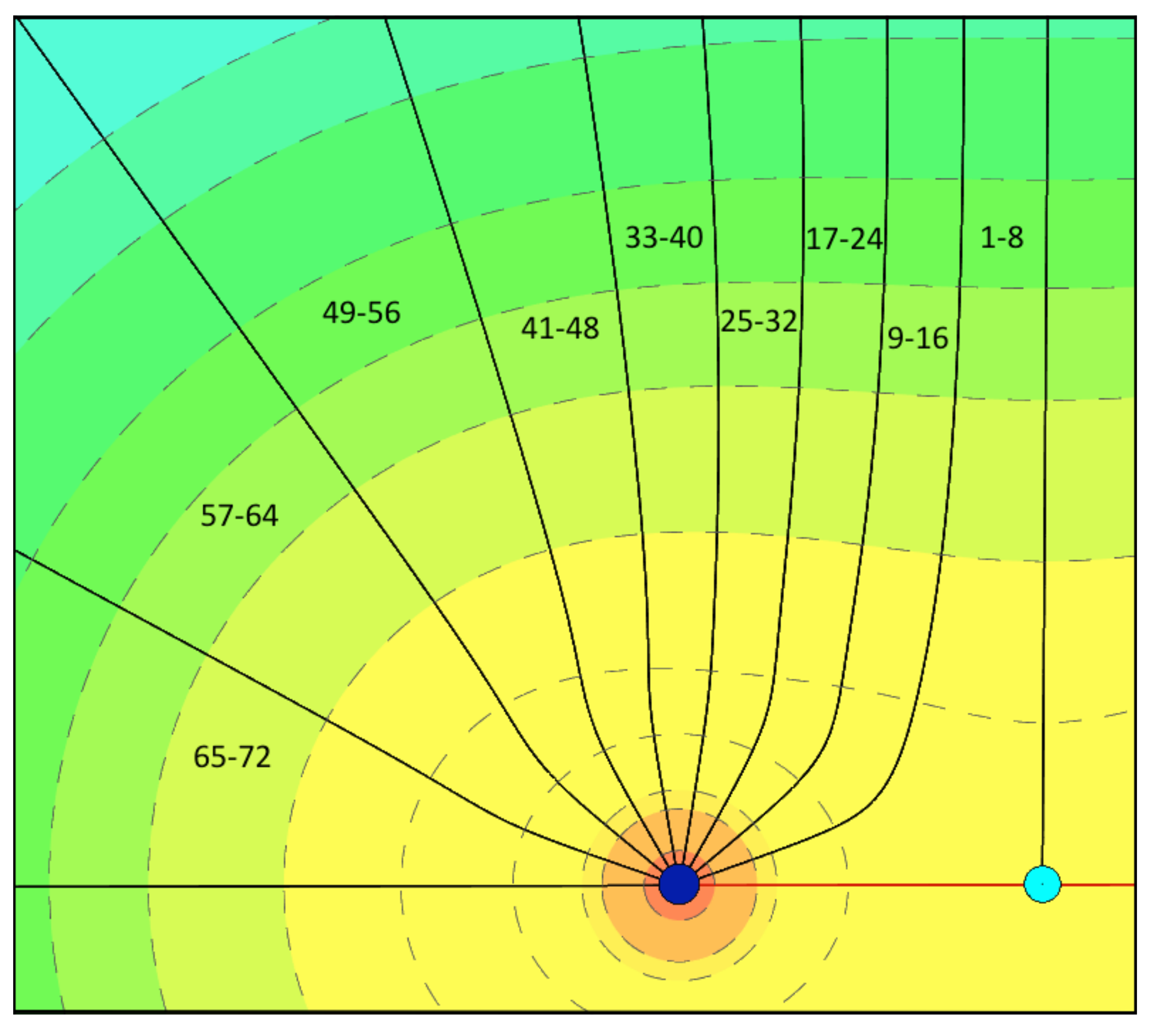}{.5\textwidth}{In all plots the gradient bundles are numbered in a counterclockwise manner with the first GB containing the bond path and the $72^{nd}$ GB containing the ridge from the nuclear CP in the direction opposite the bond path.  Shown for clarity are gradient paths used to create ZFSs every 20$^{\circ}$.}

The (intensive) properties of rotational gradient bundles are a function only of the azimuthal angle, $\theta$, effectively, a 1D property distribution. 
Here we define $\theta$ as the angle between each gradient path at its initial seed point and the bond path.  
Property distributions were found by integrating the calculated values of each property over gradient bundles using a second-order trapezoidal method \cite{Tecplot} and a  mesh size of 0.006 \AA\  for N$_2$, O$_2$, and F$_2$, 0.005 \AA\ for H$_2$, and 0.004 \AA\ for Ne$_2$. 
Since the behavior of the gradient paths in hydrogen and neon vary significantly from the other dimers tested (see Section~\ref{results_intrinsic}), a finer mesh was used for these molecules after an initial testing at 0.006 \AA\ to ensure the results were not caused by integration errors near the nucleus.

\csmfigure{gradient_bundles}{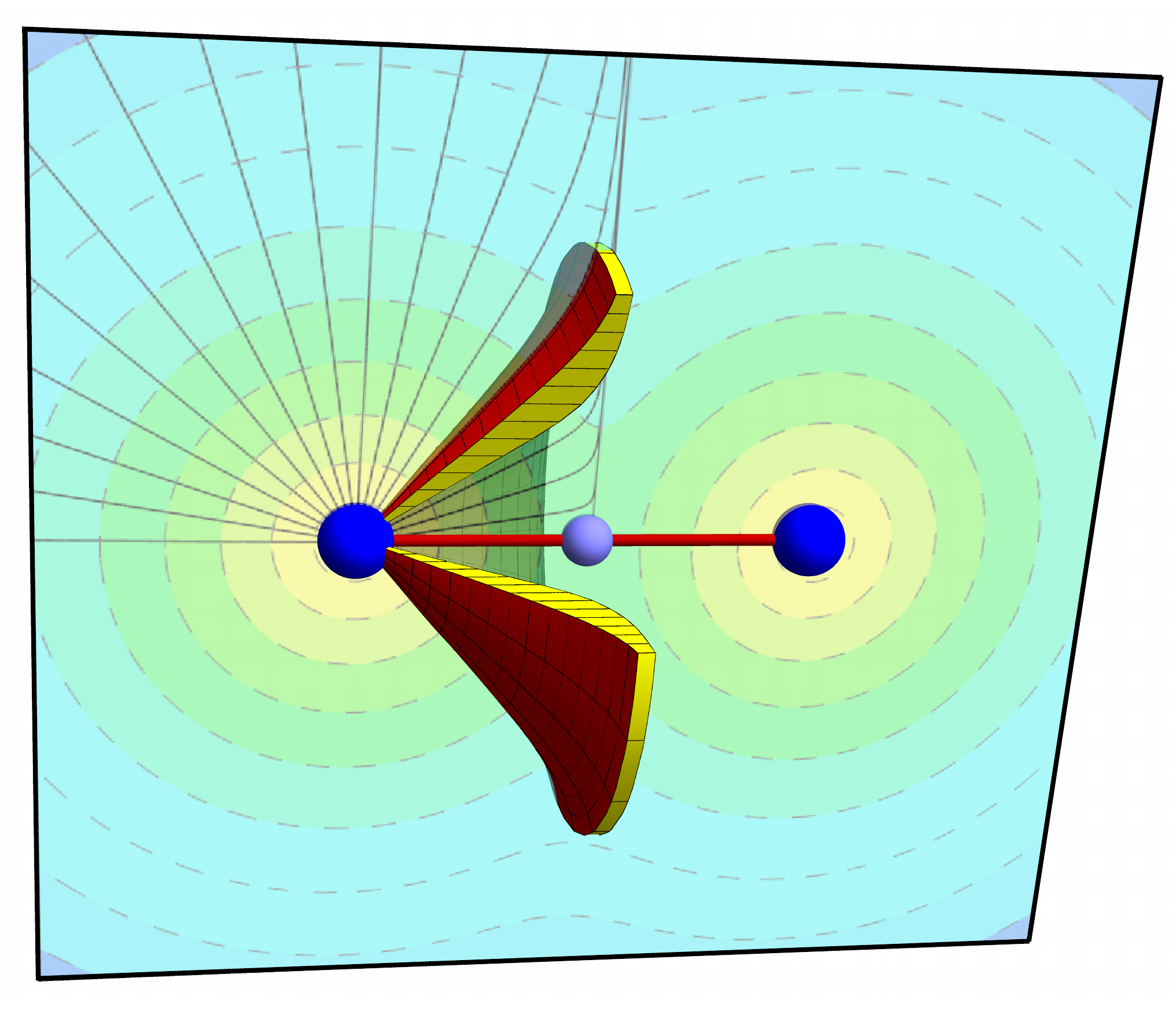}{.5\textwidth}{A representation of a gradient bundle in Ne$_2$ bounded by ZFSs at $\theta$ = 40$^{\circ}$ and $\theta$ = 50$^{\circ}$ that has been rotated around the internuclear axis by 280$^{\circ}$. The image has been tilted to emphasize the structure of a gradient bundle which takes the shape of an umbrella.  Note that the gradient bundles in this system will actually extend out to infinity; the GB shown here has been truncated in this image for clarity.}

The integration limits extended from  0.05 \AA\ from the nuclear CP to the charge density isosurface of 0.001 $electrons/bohr^3$ for each dimer. 
In open systems, bader atomic volumes are generally cut off at 0.001 a.u. as this encloses more than 99\% of the electron density \cite{baderbook}. 
To avoid numerical integration errors near the nuclei, the properties from the spherical region within the 0.05 \AA\ radius around the nuclear CP were found separately and proportioned based on volume among the gradient bundles. 
This is an adequate method of analysis due to the radial behavior of the gradient paths near the nuclei. 
With this procedure, the integrals were well converged, with electron count and kinetic energy to within 1\%\ of the ADF numerical integration results.

\subsection{Results and Discussion}
\label{results_intrinsic}

The first column of \ref{fig:GBplots} plots the integrated electron counts in the rotational gradient bundles for  H$_2$, N$_2$, O$_2$, F$_2$, and Ne$_2$.    
N$_2$ and H$_2$  are the only molecules with local maxima in the gradient volumes containing the bond path. 
The other three molecules have greater electron accumulation near the middle of each plot, i.e, $ \theta  \approx 90^{\circ}$, which is due, in part, to the larger volumes of these gradient bundles. 

The most obvious variation through the series is the shift of the mode of the distributions (the maximum) to smaller values of  $\theta$ through the series  N$_2$, O$_2$, F$_2$, and Ne$_2$.  
This shift points to an accumulation of electrons in the gradient bundles in the internuclear region, but not necessarily in those along the bond path. 
\ref{tab:data} presents the skewness of each plot, which is a useful metric for quantifying the asymmetry of a distribution. 
Because  electron count distributions are essential unimodal (N$_2$ is slightly bimodal) the skewness simply indicates on which side of the plot there is a greater number of electrons.
Negative values indicate that the distribution is weighted toward smaller values of $\theta$, and positive values toward larger values of $\theta$.  

\csmfigure{GBplots}{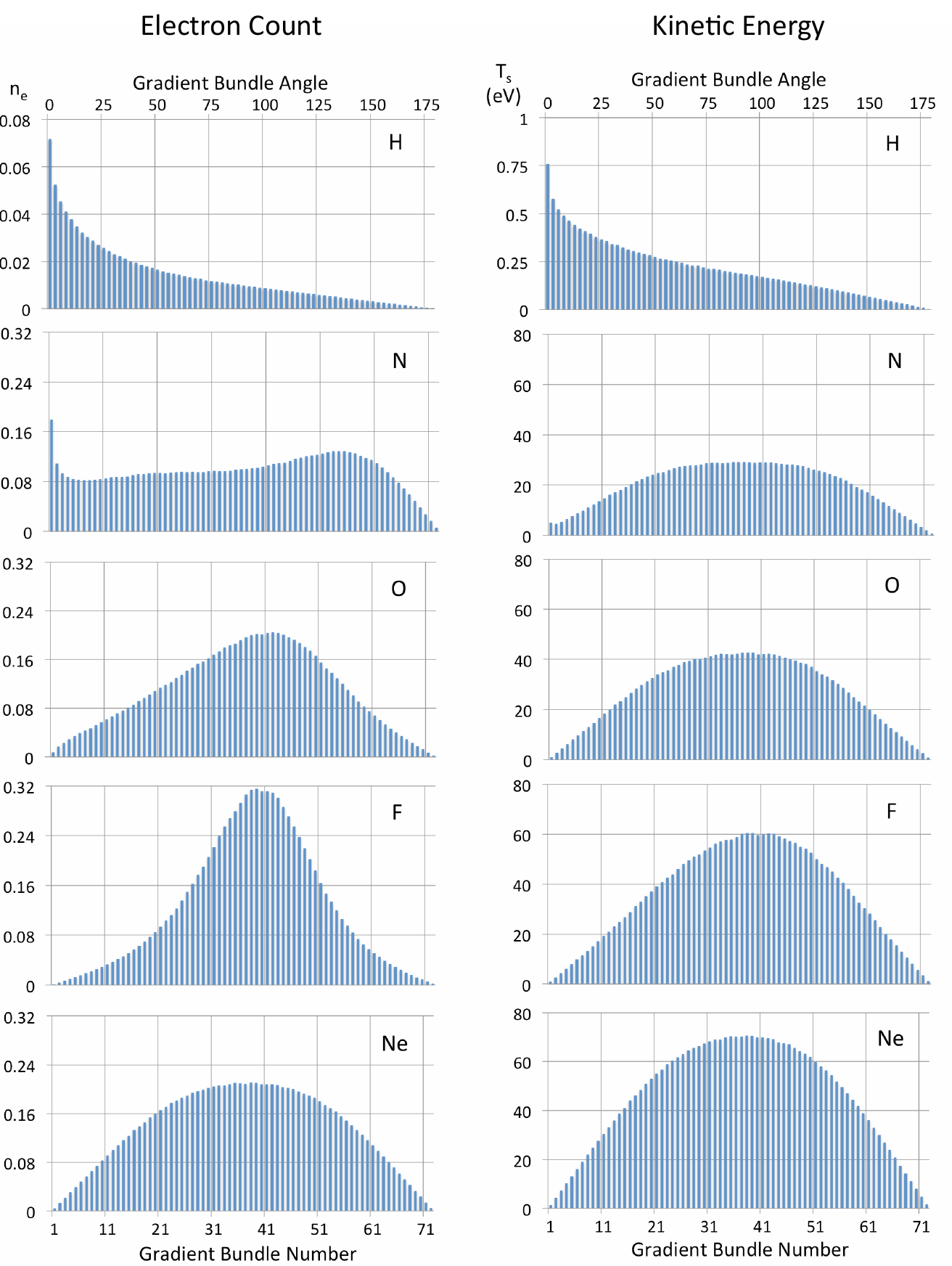}{.9\textwidth}{Electron and kinetic energy integrations over gradient bundles for a set of homonuclear diatomic molecules. Left: The electron count in each gradient bundle. Right: Kinetic energy density in each gradient bundle in electron volts (eV). Note that the scales on the plots for hydrogen vary from the rest of the dimers.}

\csmfigure{GBplots2}{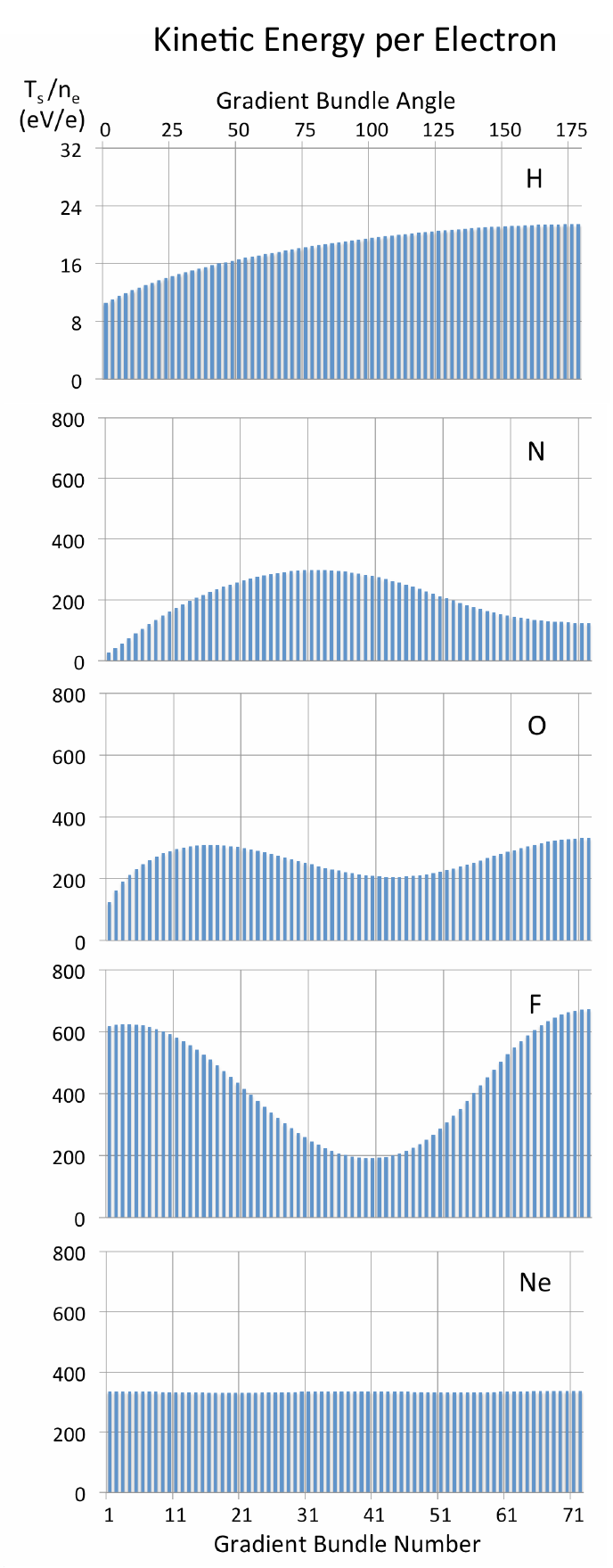}{.5\textwidth}{Average kinetic energy per electron in each gradient bundle. Note that the scales on the plot for hydrogen vary from the rest of the dimers.}

Each rotational gradient bundle is bounded by a ZFS and thus satisfies the virial theorem \cite{baderbook}.  
Accordingly, at the equilibrium internuclear distance  $\langle E \rangle = - \langle T \rangle $ and $-\frac{\langle V \rangle}{\langle T \rangle} = 2$, where $\langle E \rangle$, $\langle T \rangle$, and $\langle V \rangle$  are the average values over any volume bounded by a ZFS of  the total, kinetic, and  potential energies, respectively. 
Within Kohn-Sham DFT, the kinetic energy is separated into a non-interacting and a correlation kinetic energy,  $T_s$ and  $T_c$, respectively. 
While the virial relationship is only satisfied for the total kinetic energy,  because a molecule's $T_c$ is generally a small component of the total kinetic energy it has often been ignored \cite{ayers_ADF}. 
\addsymbol{Non-interacting kinetic energy}{$T_s$}
\addsymbol{Correlation kinetic energy}{$T_c$}
\addsymbol{Number of electrons}{$n_e$}

\begin{table}
\caption{\label{tab:data}Bond energies and distribution statistics for gradient bundles in homonuclear diatomic molecules.}
\vspace{2mm}
\begin{center}
\begin{tabular}{| c |  c  c | c  c  c |}
\hline
&&&& Skewness &\\
 Molecule & TBE.$^a$ & EBE$^b$ & n$_e$  & T$^c_s$ & T$_s$/n$_e$ \\ [.8ex]
 \hline
 H$_2$ & 6.75 &  4.51& 1.93 & 1.043 &  -0.809 \\[.8ex] 
 N$_2$ & 16.6 & 9.80  & -0.852  & -0.632 & -0.357 \\[.8ex]
 O$_2$  & 8.59  & 6.14$^d$  & -0.0122   & -0.488 & -0.395 \\[.8ex]
 F$_2$  & 3.51 & 1.64 & 0.572   & -0.293  & -0.0377  \\[.8ex]
Ne$_2$ & 0.0052 & $10^{-7}$ & -0.549 & -0.538  & 0.262 \\[.8ex]
 \hline
 \end{tabular}
 \end{center}
\footnotesize{$^a$Theoretical bond energies (TBE) are from the ADF calculations described in Methods using the PBE functional (Theoretical energies regardless of computational method are known to differ substantially from measured values.). $^b$Experimental bond energies (EBE) are reported at 0K and determined thermochemically \cite{BDE}.  $^c$All energy values are reported in eV. $^d$0.98 eV has been added to the ground-state triplet oxygen EBE to obtain the singlet oxygen EBE \cite{Oxygen}.}
\end{table}
\addabbreviation{Theoretical bond energy}{TBE}
\addabbreviation{Experimental bond energy}{EBE}

We verified that the virial theorem is satisfied for the Bader atoms in each molecule using the standard Bader analysis package in ADF where $-\frac{\langle V \rangle}{\langle T \rangle}$ = 2.0319, 2.0061, 2.0033, 2.0027, and 2.0022 for H$_2$, N$_2$, O$_2$, F$_2$, and Ne$_2$, respectively. 
Furthermore, the correlation energy ($T_c$) for each dimer was found to be less than 1\% of the total kinetic energy. 
It is then tempting to suggest that the average value of $T_c$ over a gradient bundle 
will also represent a small contribution to  the total kinetic energy over the same volume.   
Such an assumption would allow us to infer a fairly accurate value of the total energy of a gradient bundle from only a knowledge of the easily calculated non-interacting kinetic energy density, $T_s$.    
However, while $T_c$ is a small contribution to the total kinetic energy, it is not evenly distributed throughout a molecule due to the freedom in the choice of local kinetic energy \cite{ayers_ADF, ayers_KE,Cohen_KE,Cohen2}. 
In fact, atomic basins have been shown to sometimes have negative or large contributions of $T_c$ to the total kinetic energy \cite{ayers_ADF}. 
It has also been shown that the bond path, and consequently the gradient bundle around the bond path, is a ``privileged  channel''   for exchange and correlation \cite{Pendas_bondpath}.  
Therefore, it is likely that the contributions of $T_c$ to $\langle T \rangle$ are much greater in some gradient bundles than others.  
Nonetheless, we have calculated the distributions of $T_s$ for the five dimers and the results are presented in the right column of \ref{fig:GBplots}.

Since $T_s$ is much greater for core than valence electrons, the distribution for H$_2$ is significantly different than that of the other dimers. 
H$_2$ lacks any core electrons, creating a unique kinetic energy distribution based purely on valence electrons. 
For the remaining molecules with core and valence electrons, the plots are dominated by the high kinetic energy of radially distributed core electrons.
This artifact is especially noticeable for the rotational gradient bundles near 90$^\circ$, which contain a greater number of core electrons simply because these gradient bundles have larger volumes.

Inspection of \ref{tab:data} reveals that the kinetic energy density in the H$_2$ gradient bundles has a positive skewness while the other dimers are charactered by a negatively skewed distribution. In the case of H$_2$, this is simply a consequence of the fact that the maximum in the kinetic energy density occurs on the bond path. As a result, the tail of the distribution lies entirely at larger angles. In the other dimers the negative skew shows that the the kinetic energy distribution is skewed towards the bond path. 
For the most part however, the subtle distribution of the valence electrons, and hence bonding, are masked against the background of the core electrons as has been seen in previous studies of kinetic energy densities \cite{jacobsen_LOL}.

This situation can be remedied by calculating the average kinetic energy per electron, $T_s/n_e$,  by simply dividing the integrated $T_s$ of each gradient bundle by the total electron count in the same bundle.   
Since each core electron will be characterized by the same large kinetic energy (compared to valence electrons) \cite{jacobsen_LOL}, the regions where $T_s/n_e$ is minimal occur in the gradient bundles with the greatest valence density. 
The values of $T_s/n_e$ are plotted in \ref{fig:GBplots2}. 

Consistent with this interpretation, there are prominent minima for  N$_2$, O$_2$,  and  F$_2$  near 180$^\circ$, 110$^\circ$, and 100$^\circ$, respectively.  
These minima mark the locations of the ``lone pairs'' as predicted by valence shell electron pair repulsion (VSEPR) diagrams. 
The single lone pair of N$_2$ is accommodated as far as possible from the electrons in the gradient bundle nearest the bond path.  
The two lone pair of O$_2$, however, are forced through correlation effects into the larger volumes of the rotational gradient bundles near  110$^\circ$.  
Finally, the three lone pairs of electrons in F$_2$ are located in the deep minimum and larger gradient bundles 100$^\circ$ from the bond path.  

While the geometry of the $T_s/n_e$ manifold at larger angles provides information about the ``non-bonding'' electrons, it is the   geometry of this manifold  near the bond path bundles that may correlate with bond energies (see \ref{tab:data}).  
The most obvious correlation is that bound dimers are characterized by a minimum at $\theta = 0^{\circ}$.   
The three molecules with the greatest bond energies (N$_2$, O$_2$,  H$_2$)  have global minima in the gradient bundles containing the bond path. 
F$_2$ possesses a shallow local minimum in the first gradient bundle and this bond is substantially less energetically stable than the other bound dimers. 
The $T_s/n_e$ distribution across the gradient bundles in Ne$_2$ appears flat, indicating an extremely weak bonding interaction.  

\subsection{Conclusions}

While much work remains to be done, our initial results clearly suggest that it is possible to characterize bonding interactions in terms of intrinsic properties of the charge density, and in a fashion consistent with the principles of QTAIM.  
Central to this approach is the decomposition of a molecular system into gradient bundles.  
Gradient bundles are the natural volume elements of QTAIM, as their properties are well defined and additive.  
Using this approach, we were able to account for variations in the charge density and kinetic energy distribution that are typically assigned to less rigorous decompositions of the charge density, such as its separation into core and valence electrons and the further separation into lone pairs and bonding electrons.    
This study opens the door to the decomposition of more complex molecules into gradient bundles and the determination of the property distributions of these systems.  
We fully expect that these distributions will provide a much needed intrinsic representation of the chemical bond.


\chapter{Bond Dissociation  Energies From Gradient Bundle Analysis}
\label{cha:BDE}

\begin{center}
\noindent Modified from a paper published in \textit{Physica Scripta}.\\
Amanda Morgenstern\footnote{Primary researcher and author} and M. E. Eberhart\footnote{Corresponding author}. \textit{Phys. Scr.} 91: 023012,  2016.
\end{center}

\subsection{Abstract}
New and more robust models of chemical bonding are necessary to further our understanding of chemical phenomena.  
Among these are gradient bundle methods, which analyze bonding interactions in terms of property distributions over geometrically defined volumes in the charge density.  
Gradient bundle analysis provides a systematic framework from which to search for structure--property relationships.
Specifically, the kinetic energy density in gradient bundles can be used to recover the valence electron structure of molecules, and has been demonstrated to qualitatively recover bonding properties in small molecules.
Here we present new findings that quantitatively relate the lowering of kinetic energy in bonding regions to bond energies for a set of diatomic molecules. 
We find a linear correlation between experimental bond dissociation energies and the number of valence electrons in bonding regions defined by kinetic energy density in gradient bundles

\subsection{Introduction}

Chemical bonding has traditionally been explained by a lowering of potential energy ($V$) as atoms approach their equilibrium distance. 
However, many chemists argue that covalent bonding is actually due to a lowering of kinetic energy ($T$). 
This concept may at first appear to be impossible, since the only negative contribution to the energy of a stationary state of a bonded molecule in the Schr\"odinger equation is due to the attraction between the nucleus and electrons ($V$). 
However, as atoms go from a non-bonded to a bonded state, they will not be in a stationary state at all times \cite{Ruedenberg_KE}. 
In order to better understand the mechanism of chemical bonding one must look at more than the observable states given by eigenvalues from the Hamiltonian at the beginning and end of chemical bond formation.
Chemical bonding has been explained by a lowering of $T$ using three different approaches: 1) observing a reaction coordinate using the virial theorem, 2) energy decomposition methods, and 3) regional distributions of kinetic energy density.

The first of these approaches is due to Slater, who proposed that covalent bonding was due to a lowering of kinetic energy in his early work on the virial theorem \cite{Slater_virial}. 
When a molecule is not in an equilibrium state there is a force term on the nuclei, and the virial theorem can be written by taking the force term into account as an addition to the $T$ and $V$ terms, or as part of the kinetic and potential energy.
Using the latter method one gets

\begin{equation}
\label{eqn:virial3}
\begin{split}
T &= -E-r \left( \frac{dE}{dr} \right) , \textup{and}\\
V &= 2E +r \left(\frac{dE}{dr}\right).
\end{split}
\end{equation}

\noindent \ref{fig:slater_plot} shows that close to the equilibrium distance, $r_{eq}$, there is a decrease in $V$ as two non-bonded atoms approach one another to form a diatomic molecule.
Further from this point along the reaction coordinate (when the two atoms are far away) there is actually a large decrease in $T$ and an increase in $V$, before $V$ decreases near $r_{eq}$.
This is due to the large force term in eqn~\ref{eqn:virial3} as $r$ is increased.
\addsymbol{Equilibrium distance}{$r_{eq}$}

\csmfigure{slater_plot}{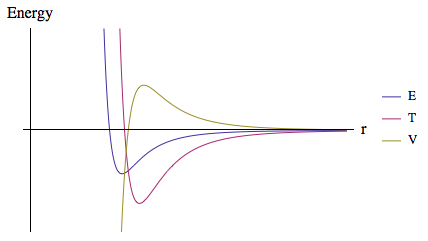}{.7\textwidth}{A plot of the relationship between total energy, kinetic energy, and potential energy in a homonuclear diatomic molecule using the definitions from eqn \ref{eqn:virial3}. The total energy was calculated using a pair potential of the form $-a/r^5 + b/r^9$.}

Klaus Ruedenberg has also long been a proponent that covalent bonding is caused by a lowering of kinetic energy.
He makes this argument based on energy decomposition methods \cite{Ruedenberg_bond_1962, Ruedenberg_KE}.
The energy change when two atoms or fragments approach to form a molecule can be decomposed into promotional, interference, sharing penetration, and quasi-classical interactions. 
The interference energy due to $T$ is greatly negative. 
While the total V decreases and the total $T$ increases upon bond formation, Ruedenberg argues that the lowering of the interference kinetic energy is what determines whether a covalent bond will form between two atoms as well as determining the total energy change of the system.
 
Finally, functions such as ELF and LOL define bonding interactions based on a lowering of T in spatial regions \cite{ELF_1992, ELF_Savin, Becke_LOL, jacobsen_LOL}. 
Mapping the kinetic energy density onto a normalized field (see Section \ref{sec:KEmodels}) shows that bonding regions between nuclei and lone pair regions have locally depleted $T$.  
While these methods define volumes corresponding to ``bonding regions," electron integrations in these volumes is rarely performed, as it is difficult to interpret any conceptual meaning tied to these volumes or electron counts.
This is because volumes based on the topology of ELF and LOL do not have well-defined energies that can be calculated.
ELF and LOL  bonding basins are not bounded by ZFSs in $\nabla \rho(\bm{r})$, but by ZFSs in the gradient of functions based on the kinetic energy density.

In Chapter \ref{cha:intrinsic}, we found a qualitative relationship between spatial regions with depleted kinetic energy and experimental BDE. 
Here we present a quantitative relationship between $T$ and BDE, that is based on well-defined energies in regions bounded by ZFSs in  $\nabla \rho(\bm{r})$.
We determine the volumes of gradient bundles in regions with a lowered kinetic energy density for a variety of diatomic molecules.
Then, we present a correlation between the number of valence electrons in these regions and experimentally determined BDEs.
Gradient bundles thus provide an avenue to further coalesce the ideas of a chemical bond as an energetically stabilizing interaction and a physical topological connection in $\rho(\bm{r})$.

\addabbreviation{Bond dissociation energy}{BDE}

\subsection{Methods}
\label{methods:BDE}

Charge densities and energies were obtained using the Amsterdam Density Functional Package, ADF,  version 2013.01 \cite{ADF, ADF2}. All molecules were optimized using spin-unpolarized calculations with a non-relativistic all-electron triple $\zeta$ singly polarized basis set, no frozen core, and the Perdew-Burke-Ernzerhof (PBE) parametrization of the generalized gradient approximation (GGA) functional {\cite{Perdew, PBE, PW91}}.

Values of $\rho(\vec{\bm{r}})$ and $T_s$ were imported into Tecplot \cite{Tecplot} on a grid with 0.015 \AA\ spacing for gradient bundle analysis. All gradient paths were seeded on the surface of a nuclear-centered sphere such that 75 $\pm$ 5\% of the core electrons were contained within the sphere. This ensures a radial distribution of electrons at the seed point while maintaining enough distance from the nuclei that the nuclear cusp does not disturb the gradient path trajectory. Kinetic energies and electron counts inside of this sphere were distributed amongst the rotational gradient bundles as described in Chapter \ref{cha:intrinsic}.

\subsection{Results and Discussion}
\label{gb}

With an infinite number of ZFSs in $\nabla \rho(\vec{\bm{r}})$,  there are an infinite number of ways to partition a molecule into regions with well-defined energies.
Gradient bundle analysis takes advantage of this fact by creating property distributions within larger atomic basins or bond bundles. 
Additionally, GBA requires as inputs only the charge density and values that can, in principle, be calculated as a functional of the charge density, e.g., the non-interacting kinetic energy density \cite{Parrbook}.
This allows for the determination of chemical properties either theoretically or by using high resolution data from x-ray diffraction experiments.

It has been shown \cite{Morgenstern} that valence shell electron pair repulsion (VSEPR) diagrams can be recovered by integrating the number of electrons ($n_e$) and the non-interacting kinetic energy ($T_s$) within gradient bundles. 
Minima in the average kinetic energy per electron (KE/e) correspond to bonding and lone pair regions containing valence electrons. 
\ref{fig:CO_GBA} shows the KE/e distribution for carbon monoxide. 
Gradient paths were seeded in a plane every 2.5$^{\circ}$ around the nuclei and rotated around the internuclear axis to create rotational gradient bundles. 
$T_s$ and $n_e$ are integrated within each GB producing an atomic basin property distribution. 
Minima in KE/e occur at 0$^{\circ}$ for both the carbon and oxygen atoms, which corresponds to the rotational gradient bundles bounded by the interatomic surface (IAS) and the first ZFS moving away from the bond path. 
A second minimum occurs around each atom at 180$^{\circ}$, which corresponds to the rotational gradient bundle in the direction opposite the bond path. 
These minima denote the bonding and lone pair regions in CO, respectively. 
Carbon monoxide is generally described as having a triple bond and single lone pair on both the carbon and oxygen atoms, consistent with the GBA picture. 
\addabbreviation{Kinetic energy}{KE}
\addabbreviation{Electron}{e}
\addabbreviation{Interatomic surface}{IAS}

\csmfigure{CO_GBA}{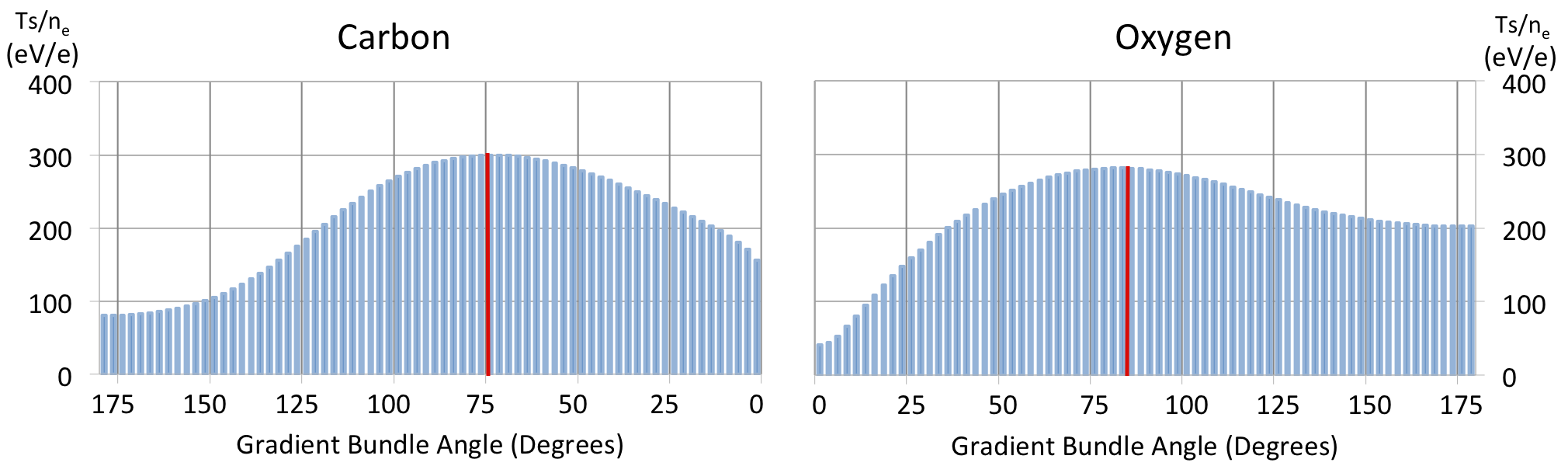}{\textwidth}{Integration of the average KE/e in each 2.5$^{\circ}$ gradient bundle in CO. The plots are oriented so that they line up geometrically with \ref{fig:CO_GP2}: the bonding region is in the center of the image (0$^{\circ}$) and the lone pair regions are on the outside edges of the plots (180$^{\circ}$). }

It was noted \cite{Morgenstern} that period 2 homonuclear diatomic molecules with higher bond dissociation energy have a deeper KE/e minimum in the bonding region. 
Specifically, H$_2$, N$_2$, and O$_2$ have pronounced minima in their KE/e plots at 0$^{\circ}$, and are characterized as strongly bound.
In contrast, F$_2$ has a shallow KE/e minimum in the bonding region and a much lower BDE. 
These results prompted the search for a meaningful relationship between the bonding region KE/e minima in gradient bundles and BDE, a property that has not previously been correlated with the molecular charge density.

As in the LOL and ELF methods,  maxima in gradient bundle KE/e plots can be used to define boundaries between bonding and lone pair regions. 
Kinetic energy bonding regions (KE BRs) are volumes in the charge density bounded by zero-flux surfaces. 
KE BRs are found by searching for the gradient bundle around each nuclei with the highest average kinetic energy per electron, then using the ZFS bounding this gradient bundle to define the bonding region. 
The KE BR for CO is highlighted in \ref{fig:CO_GP2}.
\addabbreviation{Kinetic energy bonding region}{KE BR}

\csmfigure{CO_GP2}{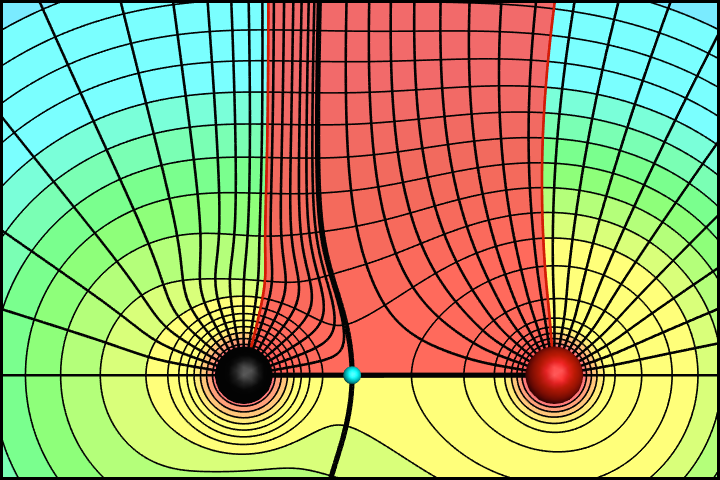}{.65\textwidth}{A cut plane of the kinetic energy bonding region in carbon monoxide. Contour lines of $\rho(\bm{r})$ are drawn on a logarithmic scale from 10$^{-3}-10$ e/bohr$^3$. Thick black lines show bond paths and interatomic surfaces. Thin black lines show gradient paths seeded every 10$^{\circ}$ around each nuclei. The red shaded region is a 2D cut plane of half of the KE BR.}

\ref{table:electrons} shows the total number of electrons, number of valence electrons, and ``normalized" electron counts in KE BRs for a set of diatomic molecules. 
In the spirit of using only the total charge density (and values directly calculable from $\rho(\vec{\bm{r}})$) we have used the following method to approximate the number of valence electrons in each bonding region. 

\begin{table}
\begin{center}
 \caption{\label{table:electrons}Total number of electrons, number of valence electrons, and normalized electron counts in kinetic energy bonding regions in a set of diatomic molecules.}
\begin{tabular}{| c | c c c |}
\hline
Molecule & Total e  & Valence e & Normalized e \\[.8ex]
 \hline
 HF & 1.31 & 1.11 & 0.56 \\[.8ex]
 HCl & 3.01 & 1.70 & 0.17\\[.8ex]
 HBr & 4.06 & 2.43 & 0.14 \\[.8ex]
 HI & 4.57 & 2.16 & 0.06 \\[.8ex]
 N$_2$ & 6.00 & 4.43 & 1.11\\[.8ex]
 O$_2$ & 1.59 & 1.17 & 0.29\\[.8ex]
 F$_2$ & 0.04 & 0.03 & 0.01\\[.8ex]
 CO & 6.01 & 4.35 & 1.09\\[.8ex]
 S$_2$ & 4.16 & 1.82 & 0.09 \\[.8ex]
 Cl$_2$ & 1.52 & 0.58 & 0.03 \\[.8ex]
  \hline
 \end{tabular}
 \end{center}
\end{table}

First, the total charge density is integrated over the KE BR to obtain the number of electrons in this region. 
Next, the angle between the bond path and ZFS bounding the KE BR is used to define a spherical sector around the nucleus. 
Since core electrons maintain a nearly spherical distribution around nuclei, the percentage of a sphere that the spherical sector occupies determines the percentage of core electrons contained in the KE BR. 
The approximate number of valence electrons in the KE BR is determined by subtracting the calculated core electrons from the total electron count. 
Finally, the normalized electron count is determined by dividing the number of valence electrons in the KE BR by the total number of core electrons in the molecule. 

\csmfigure{electronplot}{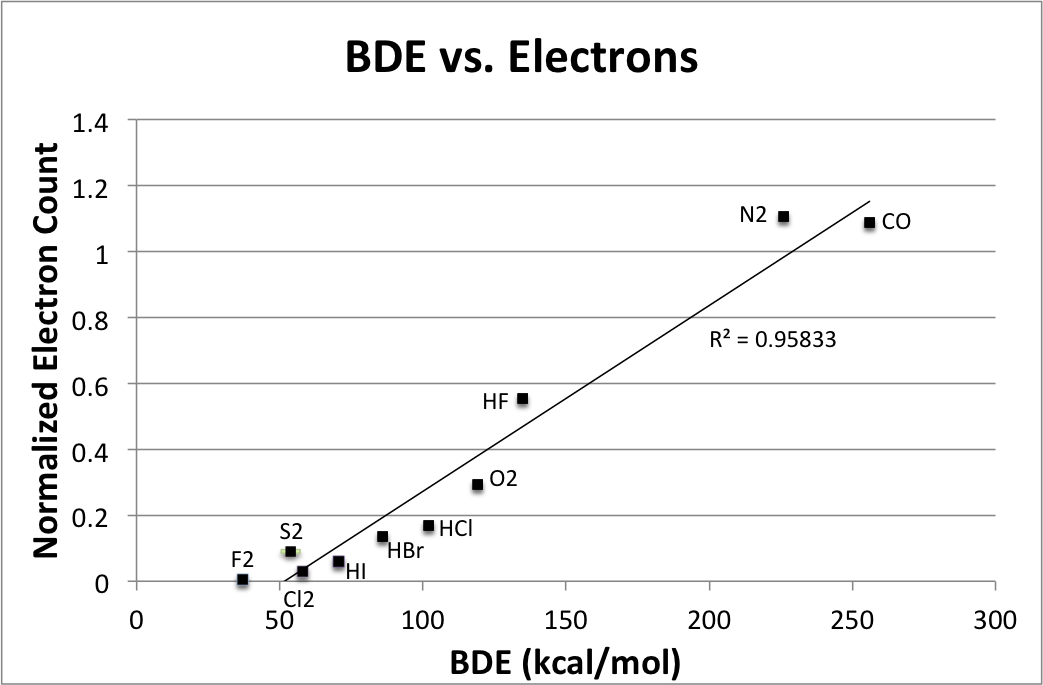}{.85\textwidth}{Relationship between the normalized electron count in KE BRs and BDE. Experimental BDEs were obtained from \cite{BDE}}

The relationship between the normalized electron count of the kinetic energy bonding regions and experimentally determined  bond dissociation energies is provided in \ref{fig:electronplot} and shows a linear trend with an R$^2$ value of 0.958. 
This observation suggests that there is a correlation between the number of valence electrons contained in bonding regions--defined by the kinetic energy density--and BDE. 
Generally, there will be more electrons in a KE BR if the kinetic energy is depleted over a larger region. 
Therefore, a greater lowering of kinetic energy in the bonding region corresponds to a stronger chemical bonding interaction.

\subsection{Conclusions}

The gradient bundle model of chemical bonding utilizes the zero-flux surface condition to define bonding regions with well-defined energies, similar to atomic basins. 
Gradient bundle analysis retains the topological picture of a chemical bond given by QTAIM and bond bundle methods.  
New and existing models should be evaluated based on their ability to  rationalize and predict chemical and physical properties.
With the extension of gradient bundle analysis, we have uncovered a quantitative relationship between the distribution of kinetic energy in small molecules and their bond dissociation energies, and further rationalized bonding interactions as originating from a lowering of kinetic energy.

By combining the representations of a bond as a stabilizing energetic interaction and as a topological connection in the charge density, we hope GBA will continue to provide useful chemical insights, which will be enhanced through further developments to the gradient bundle model. 
Specifically, we are searching for structure--property relationships involving concepts such as electronegativity, reactivity, and chemical hardness. 
Further investigation of the BDE relationship to GBs will involve using energy decomposition methods to investigate the lowering of KE in relation the degree of covalency attributed to specific bonding interactions. 
Also of interest is whether or not KE BRs correspond to bond bundle volumes. 
However, bond bundles are computationally difficult to find with our existing algorithms. This limitation is being addressed through our in-house software development program and we hope to soon be able to compute BBs and GBs for complex low symmetry molecules.


\chapter{The Influence of Zero-Flux Surface Motion on Chemical Reactivity}
\label{cha:ZFS}

\begin{center}
Modified from a paper published in \textit{Physical Chemistry Chemical Physics}.\\
Amanda Morgenstern\footnote{Primary researcher and author}, Charles Morgenstern\footnote{Provided mathematical guidance on derivation}, Jonathan Miorelli\footnote{Provided guidance on DFT concepts and usage}, Tim Wilson\footnote{Developed code}, and M. E. Eberhart\footnote{Corresponding author}. \textit{Phys Chem Chem Phys}. 91: 5638--5646, 2016.
\end{center}

\subsection{Abstract}
Visualizing and predicting the response of the electron density, $\rho(\bm{r})$, to an external perturbation provides a portion of the insight necessary to understand chemical reactivity.
One strategy used to portray electron response is the electron pushing formalism commonly utilized in organic chemistry, where electrons are 
pictured as flowing between atoms and bonds.  Electron pushing is a  powerful tool, but does not give a complete picture of electron response. 
We propose using the motion of zero-flux surfaces (ZFSs) in the gradient of the charge density, $\nabla \rho(\bm{r})$, as an adjunct to electron pushing.
Here we derive an equation rooted in conceptual density functional theory showing that the movement of ZFSs contributes to energetic changes in a molecule undergoing a chemical reaction.
Using a substituted acetylene, 1-iodo-2-fluoroethyne, as an example, we show the importance of both the boundary motion and the change in electron counts within the atomic basins of the quantum theory of atoms in molecules for chemical reactivity.
This method can be extended to study the  ZFS motion between smaller gradient bundles in $\rho(\bm{r})$ in addition to larger atomic basins. 
Finally, we show that the behavior of $\nabla \rho(\bm{r})$ within atomic basins contains information about electron response and can be used to predict chemical reactivity. 

\subsection{Introduction}

A central goal of chemistry is to explain and predict chemical reactions.   
For the better part of the last century our approach toward achieving this goal has been to seek a better understanding of electron response to a changing chemical potential. 
This endeavor has produced two complementary views of electron response.   
The first is an orbital perspective, where we picture electrons moving between occupied and unoccupied orbitals as the nuclear coordinates of reacting molecules change.  
This powerful perspective forms the foundation of frontier orbital theory \cite{frontier} and finds productive applications across all branches of chemistry, particularly in the study of coordination compounds.  
The second approach is density based and views electrons as flowing between and within molecules.  
The prototypical representation for such flow is provided by electron pushing formalisms where electron flow is represented with curved arrows \cite{e_pushing}.  
This representation of electron response provides the foundation for much of organic chemistry.

While the orbital perspective has profited from advances in computational methods, the density-based approaches have not benefited to the same extent.  
Although it is now a routine matter to calculate molecular charge densities, discerning the charge flow from these calculations is rarely as apparent as the electron pushing formalism portrays.  

Researchers have sought to reconcile electron pushing using calculated and experimentally obtained electron densities \cite{Jones_nucl} with the goal of predicting chemical reactivity.   
As part of this attempt, alternative ways to picture electron density response have been proposed.  
Rather than as electron flow, Bader presented chemical reactions as resulting from the movement of critical points in $\rho(\bm{r})$ to produce topological catastrophes \cite{baderbook, bader_catastrophe}.
Ayers added to this notion with the electron preceding picture (EPP). 
In this model, electron density around a critical point is assumed to respond more readily in the most compliant directions, where compliance has been shown to often correlate with the magnitude of the eigenvectors of the Hessian of $\rho(\bm{r})$ \cite{ayers_predict, ayers_2009, ayers_EPP2014}.   
From this vantage point, electron density responds more like an inhomogeneous substance subject to deformation than as a flowing fluid.  
\addabbreviation{Electron preceding picture}{EPP}

Though the EPP comes closer to capturing the essence of electron response than electron pushing arrows, this approach does not lend itself to visualization, as   electron flow, a vector quantity, has been replaced with the rank two compliance tensor.
Additionally, this picture is still imperfect, as critical points are but a single and incomplete characteristic of the charge density's gradient field, $\nabla \rho(\bm{r})$ \cite{Miorelli}.
Accordingly, here we explore the response of the charge density's full gradient field to a perturbation.
Our hope is that the understanding resulting from this line of investigation will allow for a closer marriage between the density based electron pushing schemes and the fundamental properties of electron response that can now be investigated using both theoretical and experimental techniques. 

Our approach is to use gradient bundle analysis, which reduces the gradient field to a set of regions bounded by zero-flux surfaces (ZFSs) and hence having well defined properties---including energies.  
While all molecules can be partitioned into sets of gradient bundles, for computational simplicity we have chosen to focus on a linear molecule in this study, 1-iodo-2-fluoroethyne (ICCF). 
\ref{fig:3DGB} shows an example of a rotational gradient bundle in a neutral ICCF molecule.
We then exploit  conceptual DFT formalisms to describe the energetic changes of these regions during a chemical reaction. 
In this paper we show that the motion of ZFSs (i.e. the gradient field), between and within atomic basins, plays an important role in energetic changes in molecules and understanding chemical reactivity.
Our long-term objective is to improve our ability to predict chemical reactivity using the topology and geometry of the charge density. 
Here we show that valuable insight into reactivity can be gained by viewing the rearrangement of the charge density alone, without performing (often costly) high level energy calculations.

\csmfigure{3DGB}{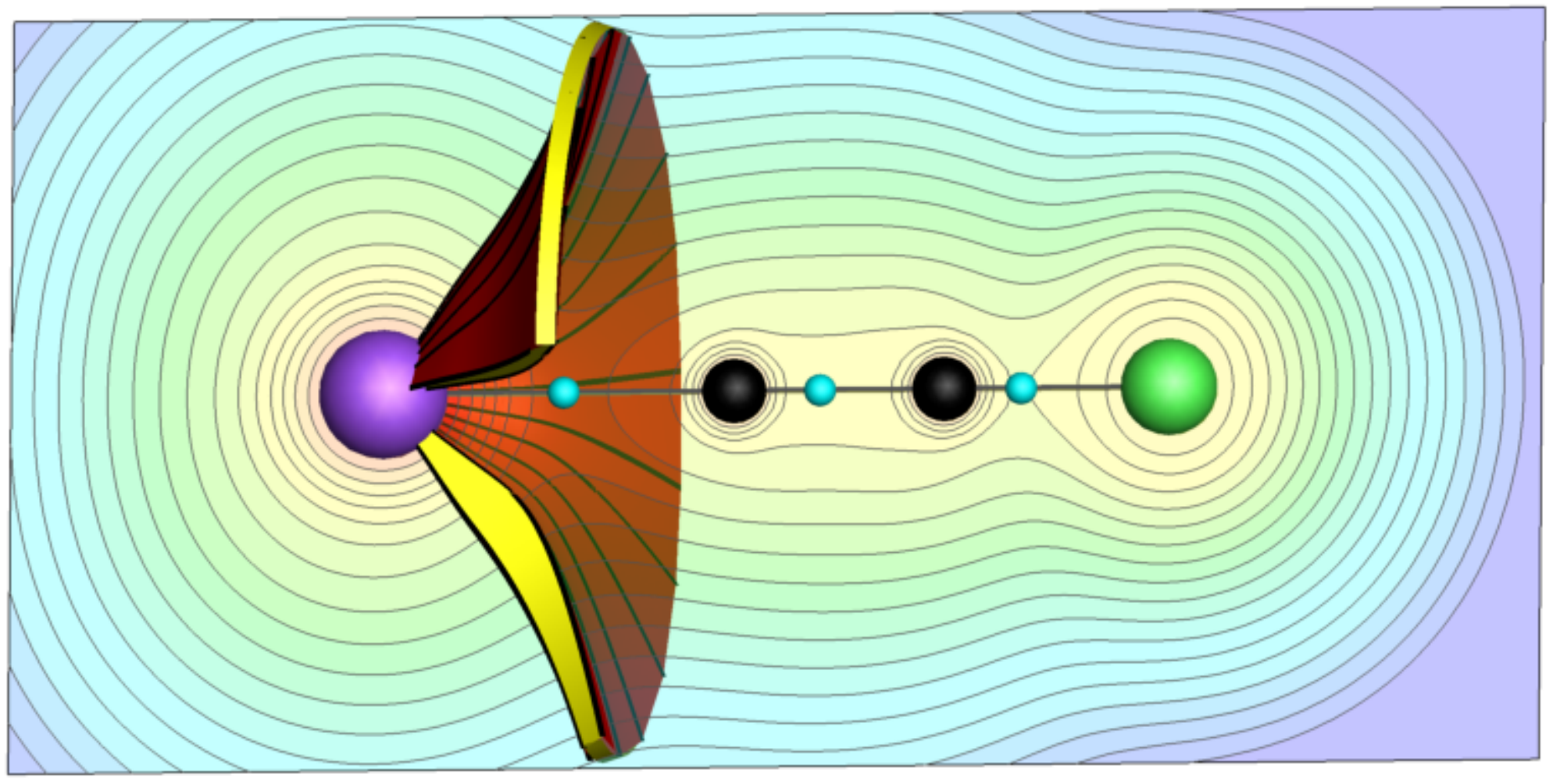}{.9\textwidth}{A rotational gradient bundle contained in the iodine atomic basin of an ICCF molecule. Contours on a cut plane of $\rho(\bm{r})$ through the molecule are drawn on a logarithmic scale from $10^{-4}-10$ e/bohr$^3$. Spheres are colored as: I-purple, C-black, F-green, bond CP-cyan; bond paths are grey, and a sampling of gradient paths on the ZFSs are shown in black. This coloring is used in all figures in this chapter.}

\subsection{Background}

Conceptual density functional theory (CDFT) provides density based definitions for chemical concepts such as chemical potential, electronegativity, and chemical hardness and softness.  
These definitions begin with the fundamental equation of density functional theory expressing the energy of a chemical system as a functional of the density.  
The minimum energy state of a system can be found variationally, 

\begin{equation}
\label{eqn:functional}
\delta (E - \mu \rho(\bm{r})) = 0
\end{equation}

where $\mu$ is a Lagrange multiplier and has the form of the system's electronic chemical potential \cite{Parr_EN}, i.e.,

\begin{equation}
\label{eqn:mu}
\mu = \left( \frac{\partial E}{\partial N} \right)_{\nu(\bm{r})}
\end{equation}

\noindent where $E$ is the energy of the molecule, $N$ is the number of its electrons, and $\nu(\bm{r})$ is the portion of the molecular potential due solely to the atomic nuclei---the so called ``external potential.'' 
CDFT associates electronegativity, $\chi$, with $-\mu$. 
In general, $\mu$  is approximated using a finite difference, where one electron is either added  or removed from a system.  
This approach yields two approximations to the chemical potential,  $\mu^+$ and $\mu^-$, which find useful applications in the study of  nucleophilic and electrophilic attack. 
\addsymbol{Chemical potential}{$\mu$}
\addsymbol{Chemical hardness}{$\eta$}
\addsymbol{Electrons count}{$N$}
\addsymbol{Electronegativity}{$\chi$}

CDFT also recognizes chemical hardness as the linear response of the chemical potential to electron count, which is equivalent to the second derivative of the energy with respect to electron count \cite{Parr_HSAB}

\begin{equation}
\label{eqn:hardness}
\eta = \left( \frac{\partial \mu}{\partial N} \right)_{\nu(\bm{r})} =\left( \frac{\partial^2 E}{\partial N^2} \right)_{\nu(\bm{r})} .
\end{equation}

Eqns (\ref{eqn:mu}) and (\ref{eqn:hardness}) give global values for chemical potential and hardness. There is no agreed upon definition of atomic  analogues of these properties, though several have been proposed \cite{Parr_ENequal, Politzer_EN, Tognetti_atomicEN, ayers_local_hardness1, ayers_local_hardness2, Ghosh_local_hardness1, Ghosh_local_hardness2}. 
A common definition for atomic chemical potential that we will use here is that due to Politzer et al. \cite{Politzer_EN}

\begin{equation}
\label{eqn:atomicmu}
\mu_i =\left(  \frac{\partial E}{\partial N_i} \right)_{\substack{N_j,\ \nu(\bm{r}) \\ i\neq j \hphantom{\ \nu(\bm{r}) }}}
\end{equation}

\noindent where $N_i$ is the number of electrons associated with atom $i$, and $j$ represents all other atoms in the molecule. 
This definition can be used in conjunction with a variety of methods that partition charge between atoms---Mulliken populations, Hirshfeld populations, and the  populations in the atomic basins of QTAIM can be used to determine $N_i$. 
This definition is also consistent with the electronegativity equalization principle requiring  $\chi$ and $\mu$ to be constant throughout a molecule at equilibrium \cite{SandersonEN, Politzer_EN}.

In  a similar way we define atomic hardness as

\begin{equation}
\label{eqn:atomiceta}
\eta_i =\left(  \frac{\partial^2 E}{\partial N^2_i} \right)_{\substack{N_j,\ \nu(\bm{r}) \\ i\neq j \hphantom{\ \nu(\bm{r}) }}}
 = \left(  \frac{\partial \mu}{\partial N_i} \right)_{\substack{N_j,\ \nu(\bm{r}) \\ i\neq j \hphantom{\ \nu(\bm{r}) }}} .
\end{equation}

\noindent During a chemical reaction---when the system's atoms are no longer in chemical equilibrium and there may be charge transfer---chemical potentials may change.  When the charge transfer is small, the new values of atomic chemical potentials are accurately approximated by \cite{Chermette_CDFT}

\begin{equation}
\label{eqn:mu_change}
\mu_i \approx \mu^{\circ}_i + \eta^{\circ}_i \Delta N_i
\end{equation}

\noindent where $\mu^{\circ}_i$ and $\eta^{\circ}_i$ are the chemical potential and hardness of atom $i$ before electron transfer, and $\Delta N_i$ is the number of electrons transferred to atom $i$. 

Changes in atomic population, as required by eqn ({\ref{eqn:mu_change}), can be calculated using condensed Fukui functions. The total Fukui function for a molecule describes the change in electron density resulting from a change in electron count and is given by \cite{Parr_fukui}

\begin{equation}
\label{eqn:fukui}
f(\bm{r},N) = \left( \frac{\partial \rho (\bm{r}, N)}{\partial N} \right)_{\nu(\bm{r})}
\end{equation}

\noindent which can again be approximated with a finite difference, yielding $f^{+/-}$. 
Eqn \ref{eqn:fukui} is general, but care must be taken in how the densities are calculated. If degenerate states exist for the $N$ or $N \pm 1$ molecule, the Fukui matrix must be calculated using degenerate perturbation theory rather than simply taking an average of the degenerate density functions \cite{Bultinck_fukui_2011, Bultinck_fukui_2012, Bultinck_fukui_2014}.
Integrating the Fukui function over atomic basins defined by QTAIM (or multiplying the Fukui function by atomic weight factors for other atomic population methods, see Ayers et al. \cite{Ayers_condensed}) yields a condensed Fukui function and provides the change in atomic electron count. 
\addsymbol{Fukui function}{$f(\bm{r})$}

However,  calculating $\Delta N_i$ in this way does introduce a degree of ambiguity.  
Specifically, one needs to choose whether to use the response of molecular fragments (RMF) or fragment of molecular response (FMR) approach \cite{Ayers_condensed, bultinck_fukui}.
For QTAIM atomic basin populations, RMF and FMR methods can give drastically different results \cite{bultinck_fukui, cioslowski_fukui}. 
Essentially, the FMR method calculates atomic populations using only the atomic basin from the neutral molecule. 
For the removal of an electron, the change in electron count for atom $i$ is calculated by
\addabbreviation{Response of molecular fragments}{RMF}
\addabbreviation{Fragments of molecular response}{FMR}

\begin{equation}
\label{eqn:FMR}
f^-_i = \int_{\Omega_i} \rho(\bm{r},N)dV - \int_{\Omega_i} \rho(\bm{r}, N-1)dV
\end{equation}
\addsymbol{Atom condensed Fukui function}{$f_i$}
\addsymbol{Atomic Basin}{$\Omega$}

\noindent where $\Omega_i$ is the original atomic basin of atom $i$ in a neutral molecule.

The RMF method uses both the atomic basin from the neutral molecule and the new atomic basin after an electron is removed from the molecule 

\begin{equation}
\label{eqn:RMF}
f^-_i = \int_{\Omega_i} \rho(\bm{r},N)dV - \int_{\Omega^-_i} \rho(\bm{r}, N-1)dV
\end{equation}

\noindent where $\Omega^-_i$ is the atomic basin of atom $i$ in the charged molecule. 
In the case of QTAIM, this method takes into account the change in the boundary of each atomic basin, i.e., the motion of their bounding ZFSs. 
The RMF condensed Fukui function gives the difference in the Bader charge of the atom before and after an electron is removed from the system. 
Advantages and disadvantages of each method have been discussed \cite{bultinck_fukui, contreras_fukui, arulmozhiraja_fukui, deproft_fukui}.
We believe that the RMF method is more meaningful for studying chemical reactivity, as atomic surface motion plays a role in mediating the energetics of a reaction. 
In fact, we show that electron response in general is more completely represented by studying the ZFS motion of all gradient bundles as well as those bounding atomic basins.

\subsection{Derivation of ZFS Influence on Chemical Reactivity}

Here we combine the atomic basin definition provided by QTAIM with CDFT to express the work of atomic surface motion as a contribution to a molecule's change in energy due to a change in electron count in the system. 
The motivation for this derivation is in line with our long-term objective, showing that the geometry of the charge density, which can be expressed in terms of the size and shape of its gradient bundles, may be an important tool for understanding chemical reactivity.

While a molecule is undergoing a chemical reaction, the electrons in its atomic basins need not be in an equilibrium state.  Electron transfer between atoms is driven by gradients in the chemical potential until a new chemical equilibrium is established with the chemical potential of the atomic basins once again being equal. 
This could be due to combining two aqueous chemicals, deposition of a gas on a solid surface, or allowing a molecule to interact with an electron reservoir of constant chemical potential. 
Regardless of the specifics of the process, we can capture its essence by considering an isolated molecule composed of two atoms (1 and 2), which are initially at equilibrium  and therefore $\mu_1 = \mu_2$, and the ZFS boundary between the two atomic basins is static. 
We then allow the electron count to vary and determine the change in energy due to this variation.  As this energy change is independent of the path, most generally $\mu_1 = \mu_2$ only at the beginning and end of the charge transfer.  The interatomic distance, and therefore the external potential, is held constant to satisfy the CDFT definition of chemical potential.

For our model diatomic system, energy is a function of the number of electrons associated with atom 1 and atom 2, and the total number of electrons in the molecule is given by $N = N_1 + N_2$, such that

\begin{equation}
\label{eqn:sumN}
\Delta N = \Delta N_1 + \Delta N_2 .
\end{equation}

\noindent During a chemical reaction, the chemical potentials of atoms 1 and 2, which are functions of $N_1$ and $N_2$ respectively, will change, as will the number of electrons associated with each atom. 
The change in energy of the molecule is given by the Taylor series expansion

\begin{equation}
\Delta E = \mu_1 \Delta N_1 + \mu_2 \Delta N_2 + \frac{\partial^2 E}{\partial N_1^2} \frac{\Delta N_1^2}{2} + \frac{\partial^2 E}{\partial N_1 \partial N_2} \Delta N_1 \Delta N_2 + \frac{\partial^2 E}{\partial N_2^2} \frac{\Delta N_2^2}{2} + \cdots.
\end{equation}

\noindent Assuming $\Delta N_1$ and $\Delta N_2$ are small, initially we adopt a linear approximation

\begin{equation}
\label{eqn:DeltaE}
\Delta E \approx \mu_1 \Delta N_1 + \mu_2 \Delta N_2 .
\end{equation}

The domain constituting the atomic basin volume of atom 1 is $\Omega_1$.  Accordingly, the number of electrons in an atomic domain $\Omega_i$ is given by

\begin{equation}
N_i = \int_{\Omega_i} \rho(\bm{r},N) dV.
\end{equation}

When the number of electrons in the system is changed, $\rho(r,N)$ will vary as will the boundary separating domains $\Omega_1$ and $\Omega_2$.  We denote the new domain of atom 1 as $\widetilde{\Omega}_1$. Using standard set notation \cite{setnotation}, the volume that is common to both the original and new basin is defined as $\Omega_1 \cap \widetilde{\Omega}_1 = \Omega_1^{com}$. The addition of any volume to the original $\Omega_1$ is $\Omega_1^+ = \widetilde{\Omega}_1 \setminus \Omega_1^{com}$ and any volume that is removed from the original basin upon changing the number of electrons in the system is $\Omega_1^- = \Omega_1 \setminus \Omega_1^{com}$. The new number of electrons in $\widetilde{\Omega}_1$ can be approximated by

\begin{align}
\label{deltan2}
N_1 + \Delta N_1 &\approx \int_{\widetilde{\Omega}_1} (\rho(\bm{r}) + \Delta N f(\bm{r})) dV \nonumber \\
  &= \int_{\Omega_1} \rho(\bm{r}) dV + \int_{\Omega_1} \Delta N f(\bm{r}) dV + \int_{\Omega_1^+} \rho(\bm{r}) dV \nonumber \\
  &+ \int_{\Omega_1^+} \Delta N f(\bm{r}) dV - \int_{\Omega_1^-} \rho(\bm{r}) dV - \int_{\Omega_1^-} \Delta N f(\bm{r}) dV
\end{align}

\noindent where $f(\bm{r})$ again indicates the Fukui function, and both $f(\bm{r})$ and $\rho(\bm{r})$ are for $N$ number of electrons for the remainder of the derivation.
Eqn (\ref{deltan2}) is exact if the Fukui function is calculated using a finite difference with the same $\Delta N$ that is used here.
The change in the number of electrons in $\Omega_1$ is

\begin{equation}
\Delta N_1 \approx \int_{\Omega_1} \Delta N f(\bm{r}) dV + \int_{\Omega_1^+} \rho(\bm{r}) dV + \int_{\Omega_1^+} \Delta N f(\bm{r}) dV - \int_{\Omega_1^-} \rho(\bm{r}) dV - \int_{\Omega_1^-} \Delta N f(\bm{r}) dV.
\end{equation}

\noindent Substituting this into eqn (\ref{eqn:DeltaE}) and combining with eqn (\ref{eqn:sumN}) gives the change of energy in the system as

\begin{equation}
\label{eqn:deltae}
\begin{split}
\Delta E &\approx \mu_2 \Delta N + (\mu_1-\mu_2) \Big( \int_{\Omega_1} \Delta N f(\bm{r}) dV + \int_{\Omega_1^+} \rho(\bm{r}) dV + \int_{\Omega_1^+} \Delta N f(\bm{r}) dV \\
 &- \int_{\Omega_1^-} \rho(\bm{r}) dV - \int_{\Omega_1^-} \Delta N f(\bm{r}) dV \Big)
\end{split}
\end{equation}

\noindent where $\mu_1 - \mu_2$ is the  driving force for charge transfer during a reaction. Noting from eqn (\ref{eqn:mu_change}) that

\begin{equation}
\label{eqn:change_hardness}
\mu_1 - \mu_2 \approx \eta^{\circ}_1 \Delta N_1 - \eta^{\circ}_2 \Delta N_2 
\end{equation}

\noindent we can combine eqns (\ref{eqn:mu_change}),  (\ref{eqn:deltae}), and (\ref{eqn:change_hardness}) to approximate the total change in energy as the sum of three terms

\begin{equation}
\label{eqn:3terms}
\Delta E  \approx \Delta E_{\Omega_2} + \Delta E_{\Omega_1} + \Delta E_{ZFS} 
\end{equation}

\noindent where

\begin{align}
\label{terms}
\Delta E_{\Omega_2} =& (\mu^{\circ}+\eta_2^{\circ} \Delta N_2) \Delta N, \nonumber \\
\Delta E_{\Omega_1} =& (\eta^{\circ}_1 \Delta N_1 - \eta^{\circ}_2 \Delta N_2) \left(\int_{\Omega_1} \Delta N f(\bm{r}) dV \right) \textup{, and } \nonumber \\
\Delta E_{ZFS} =& (\eta^{\circ}_1 \Delta N_1 - \eta^{\circ}_2 \Delta N_2) \Big( \int_{\Omega_1^+} \rho(\bm{r}) dV + \int_{\Omega_1^+} \Delta N f(\bm{r}) dV \nonumber \\
&- \int_{\Omega_1^-} \rho(\bm{r}) dV - \int_{\Omega_1^-} \Delta N f(\bm{r}) dV \Big).
\end{align}

As a particularly convenient way to associate meaning with the above equation, imagine the reaction occurring in three steps.
First, electron density from a nucleophile (electrophile) is transferred into (from)  $\Omega_2$, changing its chemical potential.  
The energy change associated with this process is $\Delta E_{\Omega_2}$. 
Next, as captured in $\Delta E_{\Omega_1}$, the chemical potential difference resulting from the original charge transfer will drive electron flow between $\Omega_1$ and $\Omega_2$. 
Finally, as part of the charge flow process, the atomic surface between atoms 1 and 2 may move as required by $\Delta E_{ZFS}$, which gives the work of boundary motion. 

We now rearrange eqn (\ref{eqn:deltae}) by combining terms  that depend only on atom 1 and those that depend only on atom 2 giving,

\begin{equation}
\label{eqn:equalterms}
\begin{split}
\Delta E &\approx  \mu_1 \left(\int_{\Omega_1} \Delta N f(\bm{r}) dV \right) + \mu_2 \left(\int_{\Omega_2} \Delta N f(\bm{r}) dV \right)\\
&+\mu_1 \left( \int_{\Omega^+_1} (\rho(\bm{r}) + \Delta N f(\bm{r})) dV -\int_{\Omega^-_1} (\rho(\bm{r}) + \Delta N f(\bm{r})) dV \right)\\
&+\mu_2 \left( \int_{\Omega^+_2} (\rho(\bm{r}) + \Delta N f(\bm{r})) dV -\int_{\Omega^-_2} (\rho(\bm{r}) + \Delta N f(\bm{r})) dV \right).
\end{split}
\end{equation}

\noindent The first two terms give the change in energy of the molecule due to the change in the number of electrons in the original atomic basins, modulated by the chemical potentials of atoms 1 and 2, respectively. 
The last two terms provide the energetic change due to the number of electrons that are in the regions of the atomic basins that have changed due to $\Delta N$, modulated by the same chemical potentials.   
From eqn (\ref{eqn:equalterms}), we can extend the linear approximation of  $\Delta E$ to a molecule with any number of atomic basins bounded by ZFSs as

\begin{equation}
\label{eqn:simplified}
\Delta E \approx \sum_i \mu_i (\Delta N_{\Omega_i} + N_{\Omega^{+/-}_i})
\end{equation}

\noindent where
\begin{equation}
\label{eqn:term1}
\Delta N_{\Omega_i} = \int_{\Omega_i} \Delta N f(\bm{r}) dV
\end{equation}

\noindent and

\begin{equation}
\label{eqn:term2}
N_{\Omega^{+/-}_i} = \int_{\Omega^+_i} (\rho(\bm{r}) + \Delta N f(\bm{r})) dV -\int_{\Omega^-_i} (\rho(\bm{r}) + \Delta N f(\bm{r})) dV.
\end{equation}

\noindent Noting that $\Delta N_{\Omega_i} + N_{\Omega^{+/-}_i}$ gives the total change in electron count in an atomic basin (using the RMF method), eqn (\ref{eqn:simplified}) can also be written as the sum of atomic energies within a molecule

\begin{equation}
\label{eqn:atomsum}
\Delta E = \sum_i \Delta E_i .
\end{equation}

In order to assess the magnitude of the terms in eqn (\ref{eqn:simplified}), we consider the substituted acetylene, 1-iodo-2-fluoroethyne.  
This molecule was chosen due to the difference in chemical hardness and electronegativity of iodine and fluorine.
We hypothesize that the mobility of ZFSs will vary depending on the hardness of the region. 
Our approach is to use a finite difference to compute the Fukui function and perform the integrations in eqns (\ref{eqn:term1}) and (\ref{eqn:term2}) for both [ICCF]$^-$ and [ICCF]$^+$.

\subsection{Methods}
\label{sec:ZFSmethods}

All reported charge densities were obtained within the Amsterdam Density Functional Theory package \cite{ADF, ADF2} utilizing the M06-2X functional with a QZ4P basis set, no frozen core electrons, and a scalar ZORA approximation to account for relativistic effects \cite{ADFbasis, zora1, zora2, zora3}. 
The standard Bader analysis package was run on the computed charge densities to obtain critical point locations and atomic charges \cite{ADFbader1, ADFbader2}. 
$N\pm1$ calculations were performed unrestricted, with spin $=1$, and charge $=\pm1$. 
Zero-flux surfaces were calculated using the Bondalyzer add-on package in Tecplot \cite{Tecplot}.
A grid spacing of 0.004 \AA\ was used for all molecules and electron counts were calculated using the standard integration package in Tecplot following the methods described in Chapter \ref{cha:intrinsic}.
\addabbreviation{Quadruple zeta 4 polarized}{QZ4P}
\addabbreviation{Triple zeta polarized}{TZP}
\addabbreviation{Double zeta polarized}{DZP}
\addabbreviation{Zeroth order regular approximation}{ZORA}

\subsection{Results and Discussion}

\ref{tbl:electrons} reports the integration values of the charge density for the terms in eqn (\ref{eqn:simplified}) for [ICCF]$^+$ and [ICCF]$^-$.
The integrals in eqn (\ref{eqn:term2}) were determined by finding the volumes between the neutral and $N\pm1$ atomic boundaries and integrating $\rho({\bm{r}},N\pm1)$ over these regions. 
Eqn (\ref{eqn:term1}) is simply the FMR condensed Fukui function, giving the change in electron count in the original atomic basin.  
\ref{fig:ZFS_motion} shows the atomic surfaces  for [ICCF]$^-$ and [ICCF]$^+$ compared to the neutral molecule while holding the nuclear coordinates fixed.

\begin{table}[ht]
\centering
\vspace{2mm}
 \caption{Integration of electrons for the integrals in eqn (\ref{eqn:simplified}) for [I-C$_1$-C$_2$-F] $\pm$ 1 electron.}
  \label{tbl:electrons}
\begin{tabular}{| c | c  c  c  c |}
\hline
$N-1$ & I & C$_1$ & C$_2$ & F  \\ [.8ex]
\hline
$\Delta N_{\Omega_i}$ & -0.576 & -0.076 & -0.223 & -0.108\\[1ex]
$N_{\Omega^{+}_i}$ & 0 & 0.018 & 0.037 & 0.019\\[1ex]
$N_{\Omega^{-}_i}$ & 0.016 & 0.035 & 0.022 & 0.002\\[1ex]
$\% \Delta N_{ZFS}$ & 2.7 & 41.2 & 20.8 & 16.5 \\[1ex]
$\Delta N_i$ & -0.592 & -0.092 &-0.208 & -0.090\\[1ex]
\hline
$N+1$ & I & C$_1$ & C$_2$ & F  \\ [.8ex]
\hline
$\Delta N_{\Omega_i}$ & 0.678 & 0.068 & 0.135 & 0.073\\[1ex]
$N_{\Omega^{+}_i}$ & 0.012 & 0 & 0.112 & 0.001\\[1ex]
$N_{\Omega^{-}_i}$ & 0 & 0.107 & 0.001 & 0.017\\[1ex]
$\% \Delta N_{ZFS}$ & 1.8 & 61.3 & 45.7 & 20.0 \\[1ex]
$\Delta N_i$ & 0.691 & -0.039 &0.246 & 0.057\\[1ex]
 \hline
 \end{tabular}
\end{table}

\csmfigure{ZFS_motion}{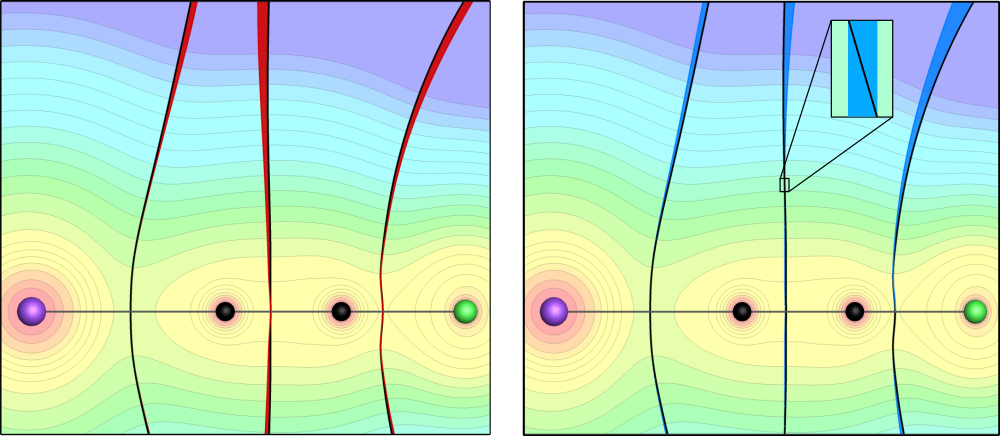}{\textwidth}{Atomic surface motion for ICCF $\pm$ 1 electron. Left: ICCF anion interatomic surface motion. Right: ICCF cation interatomic surface motion. Red (blue) regions indicate the motion of atomic boundaries for the N+1 (N--1) molecule,  black lines show the neutral molecule interatomic surfaces. Contours of $\rho(\bm{r})$ on the cut plane are drawn for the neutral molecule.}

The FMR condensed Fukui function does not fully capture the electron response of a molecule. 
There are at least two other factors that play a role in electron response in the electron density picture. 
Firstly, contributions to the change in electronic populations in the atomic basins from the surface motion are not negligible. 
Secondly, electron rearrangement in the original basin can occur without producing a dramatic effect on the integrated electron count within the volume. 

We consider the percent contribution of surface motion to the total electron response in the system using the following equation
\begin{equation}
\label{eqn:percent}
\%\Delta N_{ZFS}= 100 \left( \frac{N_{\Omega^+_i} + N_{\Omega^-_i}}{|\Delta N_{\Omega_i}| + N_{\Omega^+_i} + N_{\Omega^-_i}} \right) .
\end{equation}
\addsymbol{Percent contribution of ZFS motion to electron response}{$\%\Delta N_{ZFS}$}

\noindent The numerator provides the number of electrons that are both lost and gained in each atomic basin due to surface motion.
The denominator gives the magnitude of the change in electron count within the original basin (the FMR condensed Fukui function) as well as the separate contributions of the atomic basin gaining and losing electrons due to boundary motion.
The values of $\% \Delta N_{ZFS}$ for ICCF range from $1.8-61.3\%$. 
For C$_1$ in the [ICCF]$^-$ molecule, the motion of its ZFS contributes more to this atom's changing electron count than the  change in the original basin. 

The total change in electron population for atom $i$  is given by

\begin{equation}
\label{eqn:totaldeltaN}
\Delta N_i = \Delta N_{\Omega_i} + N_{\Omega^+_i} - N_{\Omega^-_i}
\end{equation}

\noindent which is the RMF condensed Fukui function---the change in Bader charge of the atoms in a molecule resulting from a change in electron count. Substituting the $\Delta N_i$ values into eqn (\ref{eqn:simplified}) for ICCF $\rightarrow$ [ICCF]$^+$ gives 

\begin{equation}
\label{eqn:E_plus}
\Delta E_{N-1} \approx \mu_I (-0.592) + \mu_{C_1} (-0.092) + \mu_{C_2} (-0.208) + \mu_F (-0.090)
\end{equation}

\noindent and for ICCF $\rightarrow$ [ICCF]$^-$

\begin{equation}
\label{eqn:E_minus}
\Delta E_{N+1} \approx \mu_I (0.691) + \mu_{C_1} (-0.039) + \mu_{C_2} (0.246) + \mu_F (0.057) .
\end{equation}

\noindent As would be expected, the chemically softer halogen, iodine, has a larger change in electron population compared to the harder fluorine atom for the addition or removal of an electron. 

It is important to keep in mind that eqns (\ref{eqn:E_plus}) and (\ref{eqn:E_minus}) are the linear approximations of $\Delta E$, and not valid for large values of $\Delta N_i$. 
Our preliminary results suggest a higher order response including chemical hardness and further derivatives would be required to accurately calculate $\Delta E$.
Incrementally adding fractions of electrons to the system and calculating the change in chemical potential over time, based on atomic hardness, gives an even better approximation of the change in energy. 
In this initial study however, we are interested in relative electron count changes of regions bounded by ZFSs in a molecule, which would ultimately contribute to the energy change of the molecule based on the chemical potential and higher order terms.

Interestingly, the motion of the C$_1$--C$_2$ bond critical point in ICCF $\rightarrow$ [ICCF]$^+$ does not fully represent the atomic surface motion. 
The inset in \ref{fig:ZFS_motion} shows how in some regions the boundary motion between the C$_1$ and C$_2$  atomic basin has added volume to C$_1$, while in other regions volume has been added to C$_2$. 
The bond critical point between the two carbon atoms, and therefore the atomic surface near the bond path, has moved towards C$_1$.
But away from the bond path, the atomic ZFS has actually moved into the original C$_2$  basin. 
This is further evidence that observing critical point motion alone does not fully represent the density picture of electron response.

While it is clear that the motion of atomic surfaces plays a role in determining the energetics of a reaction, atomic ZFS motion is also not the only factor in understanding and predicting chemical reactivity. 
\ref{fig:ZFS_motion} shows that atomic surfaces do not always move by a large amount. 
In fact, the atomic surface of the softer iodine atom moves less than the harder fluorine atom ZFS, especially for the $N-1$ case, contrary to what one might expect. 
For both the addition and removal of an electron, the electron density in $\Omega^{+/-}$ contributes less than 3\% of the total change in electron population for iodine.  
The movement of interatomic surfaces alone does not correctly depict the electron response.

Significant chemistry is also contained in the rearrangement of electron density within atomic basins. 
The large $\Delta N_i$ values for iodine are due to a change in electron count within the original iodine basin. 
However, electron density does not necessarily add to the original atomic basins in an equally distributed manner.
The electron response within atomic basins can be seen more clearly in a cut plane of ZFSs that form gradient bundles for ICCF before and after the change in electron count (\ref{fig:ICCF_GPs}).

\begin{figure}
	\begin{center}
		\subfigure{
			\resizebox{\textwidth}{!}{\includegraphics{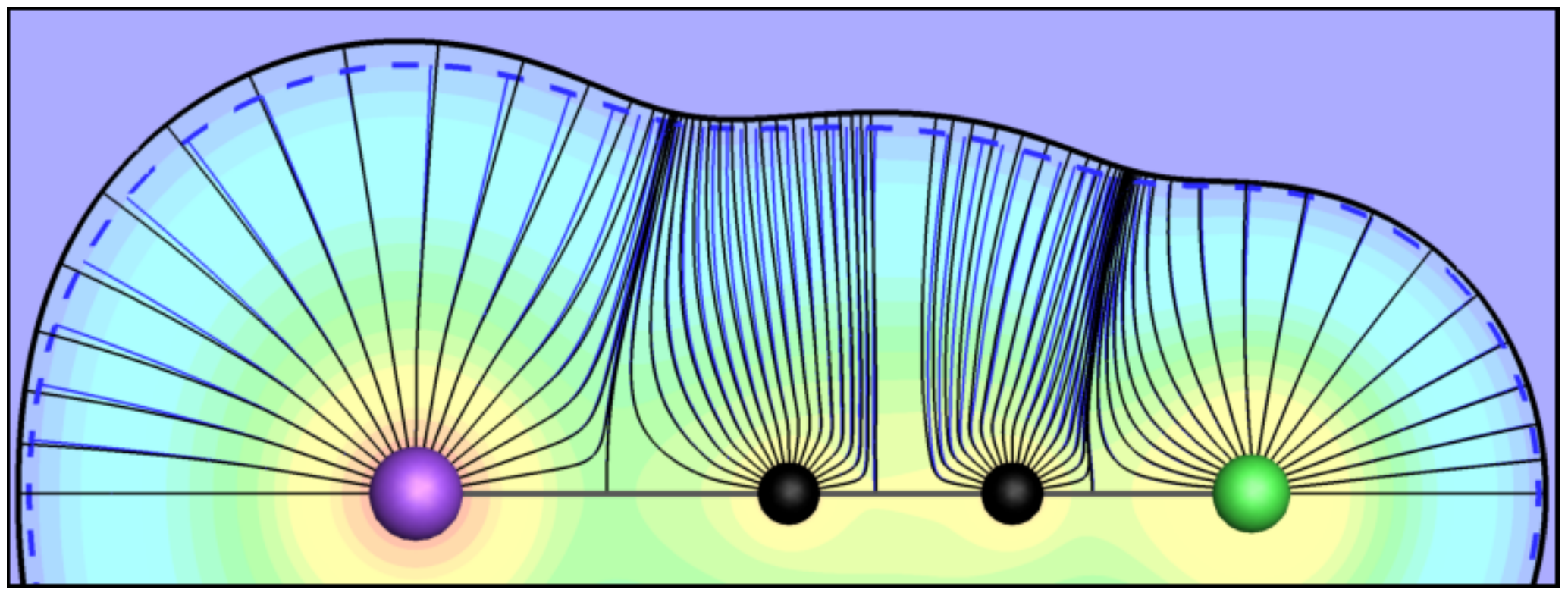}}
		} \\
		\subfigure{
			\resizebox{\textwidth}{!}{\includegraphics{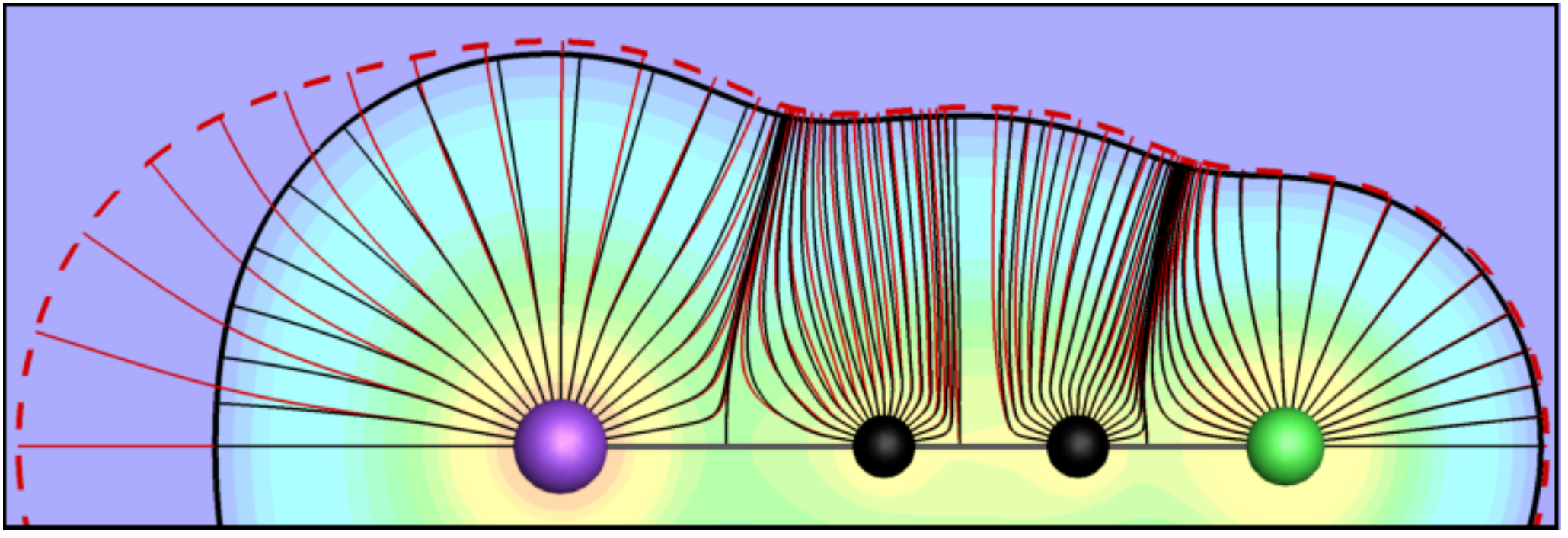}}
		}
		\caption{\label{fig:ICCF_GPs} Overlays of a cut plane of ZFSs for the top half of a neutral ICCF and [ICCF]$^+$ (top) and [ICCF]$^-$ (bottom) molecules. Blue (red) lines represent ZFSs for the $N-1$ ($N+1$) molecule and black ZFSs are for the neutral molecule. Contour lines at 10$^{-3}$ e/bohr$^3$ are shown in solid black for the neutral molecule and dashed blue (red) for the cation (anion). Contour flooding of $\rho(\bm{r})$ on the cut plane is for the neutral molecule.}
	\end{center}
\end{figure}

For the $N+1$ system, there is a large amount of electron rearrangement within the original iodine basin. 
This is not the case for the $N-1$ molecule. 
This result can be rationalized by returning to the frontier orbital view. 
Adding an electron to ICCF should put electron density into the LUMO while removing an electron should cause electron density to be lost from the HOMO of the neutral molecule. 
\ref{fig:HOMO_LUMO} shows that the HOMO of ICCF is concentrated most around the iodine atom, then the carbon atoms, and has the smallest contribution near the fluorine atom.
Around each nuclei, the density is fairly equally distributed causing little redistribution of the electron density within each atomic basin. 
Density is removed in a relatively uniform fashion around each nuclei.

\begin{figure}
	\begin{center}
		\subfigure{
			\resizebox{.48\textwidth}{!}{\includegraphics{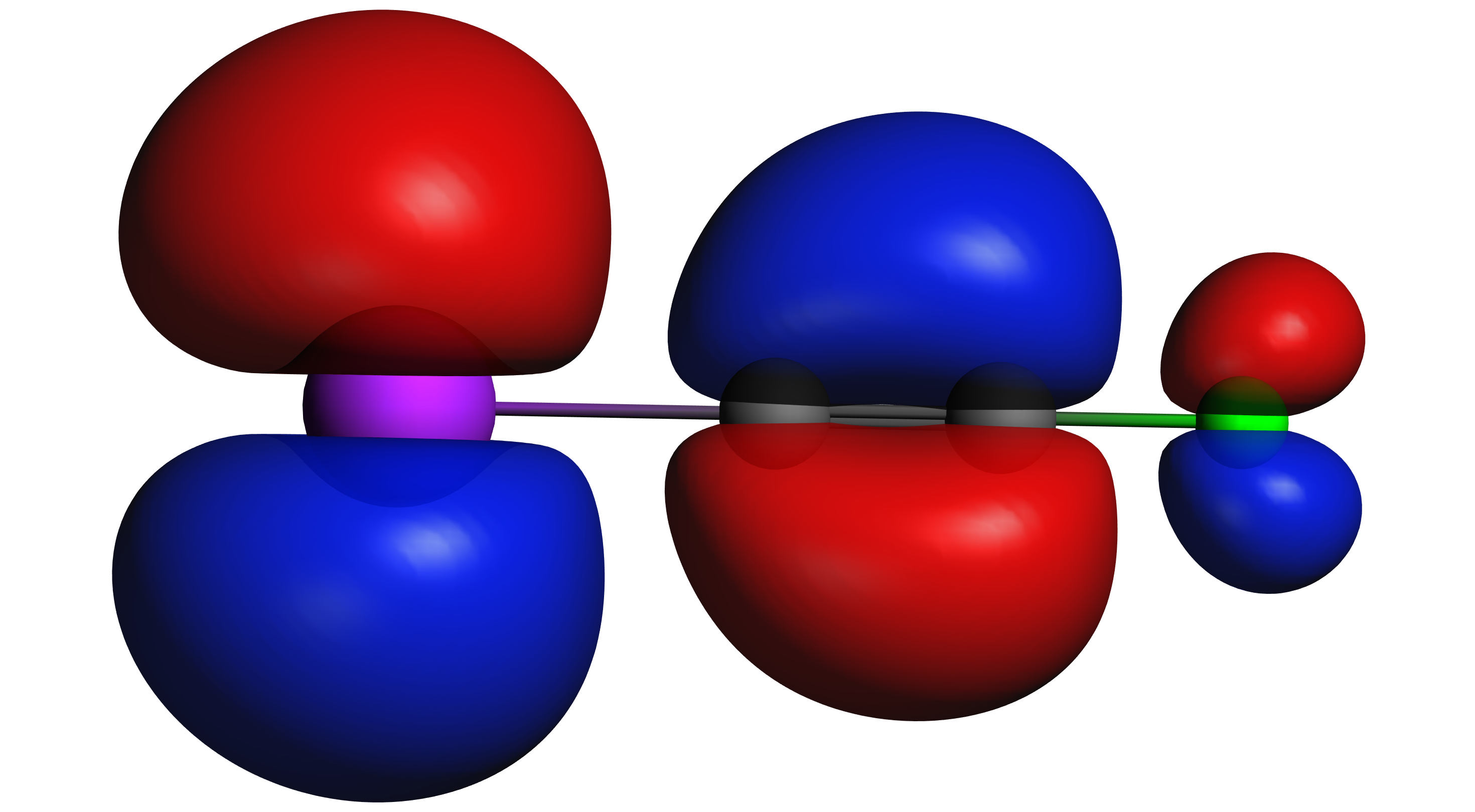}}
		} 
		\subfigure{
			\resizebox{.48\textwidth}{!}{\includegraphics{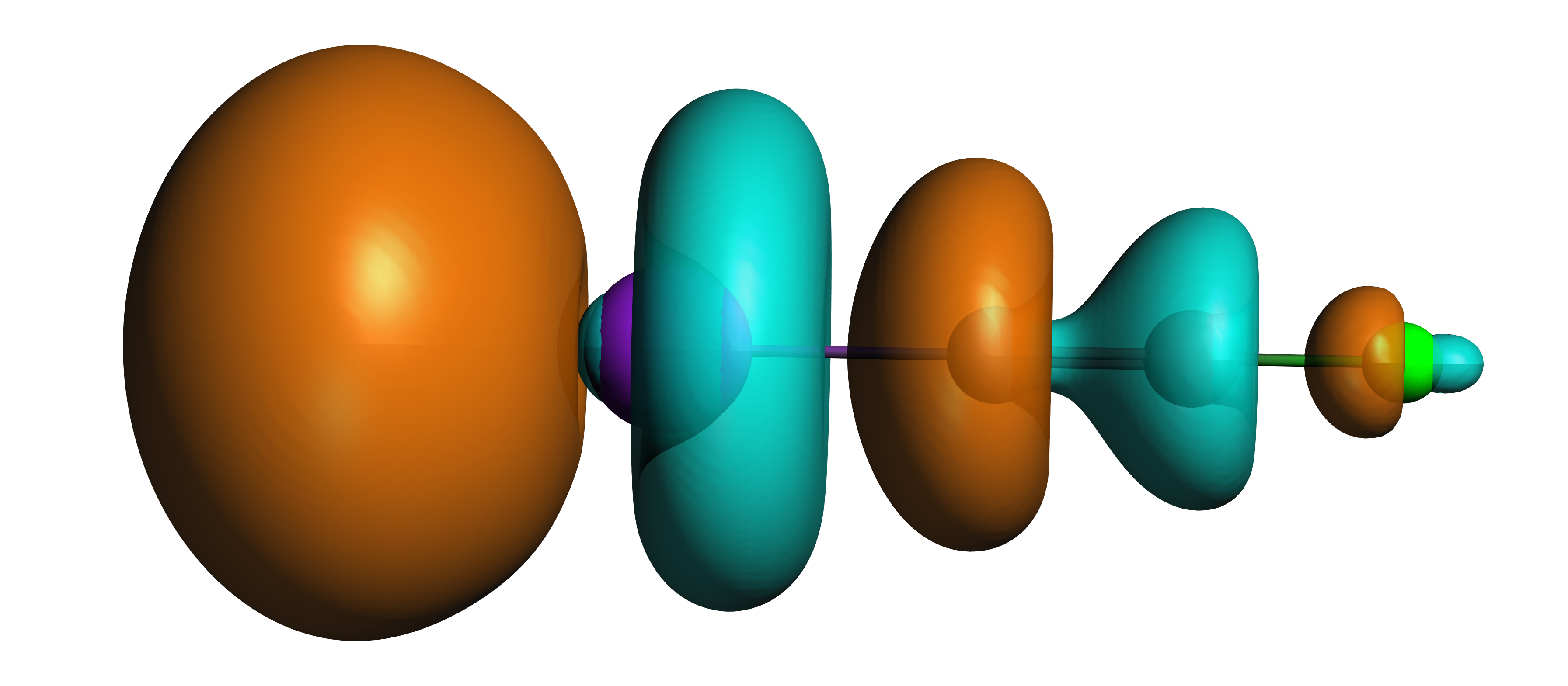}}
		}
		\caption{\label{fig:HOMO_LUMO} The HOMO (left) and LUMO (right) for a neutral ICCF molecule.}
	\end{center}
\end{figure}

The concentration of the LUMO for ICCF, however, varies throughout the molecule. 
The majority of electron density is added to a portion of the iodine basin, in the region pointing away from the rest of the molecule, as this is where the LUMO is concentrated. 
The addition of an electron causes a large rearrangement of the electron density within the iodine basin. 
The electron response can be seen by observing the curvature of the zero-flux surfaces within atomic basins.
When ICCF $\rightarrow$ [ICCF]$^-$, the curvature of the ZFSs in the iodine atom switch from positive to negative in the region pointing away from the I--C$_1$ bond path. 

Thus far, we have shown that eqn (\ref{eqn:simplified}) can be used to represent energetic changes in atomic basins.
However, there is no reason to limit energy calculations to atomic regions.
This equation can be used for any volume with well-defined energies, i.e., any gradient bundle.
In order to gain a highly localized picture of electron response, eqn (\ref{eqn:simplified}) can be used to study the energetic changes in gradient bundles.
Preliminary results show that the motion of ZFSs bounding gradient bundles can be used to understand and predict electron response, obtaining similar results to frontier orbital theory.

\subsection{Conclusions}

Electron response can be viewed from an orbital or an electron density perspective. 
A common way to rationalize and predict chemical reactivity from the density viewpoint is to use electron pushing arrows.
The electron pushing formalism is a powerful tool for the prediction of chemical reactions, but it does not give an entirely accurate picture of electron rearrangement.
Individual electrons do not flow between atoms and chemical bonds as chemical reactions occur.
Therefore, we propose viewing electron response  as the motion of zero-flux surfaces rather than electrons flowing between and within molecules.

Using concepts from QTAIM and conceptual DFT, we have shown that the motion of atomic boundaries plays a role in determining energetic changes of molecules when undergoing reactions. 
This method of rationalizing electron response is in line with the response of molecular fragment approach to calculating changes in atomic charges---the motion of atomic boundaries is included in the calculation.
Since ZFSs exist around any gradient bundle, not just atomic basins, studying the movement of the entire gradient field of $\rho(\bm{r})$ provides a more detailed picture of electron rearrangement. 
The motion of ZFSs between and within atomic basins shows where electron density changes when a chemical reaction occurs, providing similar results as frontier orbital theory.


\chapter{Applications to Enzyme Design}
\label{cha:enzymes}

Rational design of enzymes with novel and enhanced activity is an emerging field combining computational and experimental techniques. 
Enzyme design began as a purely experimental practice, using methods such as directed evolution and the creation of antibodies. 
Rational enzyme design involves making specific changes to existing enzymes in an attempt to catalyze reactions involving non-native substrates or to alter the catalytic rate.
In this chapter, I discuss applications of QTAIM and bond bundles to two rational design projects of metalloenzymes: (1) the redesign of carboxypeptidase A to catalyze a non-native substrate, and (2) the metal specificity of histone deacetylase.

Metalloenzymes catalyze vital reactions including oxidative metabolism of drugs (cytochrome P450) \cite{cytochrome}, hydrolysis of DNA and RNA (staphylococcal nuclease) \cite{nuclease}, and methane production and oxidation (methyl-coenzyme M reductase) \cite{MCR_2014}.
Unfortunately, metal ions present challenges for computationally modeling enzymes including determining correct metal coordination and spin states, the need for large basis sets in QM calculations, and parameterization of force fields for the metal in molecular dynamics (MD) simulations.
Furthermore, noncovalent interactions in the active site are know to play a key role in enzyme activity \cite{Mao_metalloenzyme_2013}, but our understanding of these interactions and ability to computationally determine their strength is not as advanced as with typical covalent bonds .
Hydrogen bonding is often important for properly orienting the substrate for reaction and stabilizing charges during the transition state \cite{KnowlesHbonds_2010}.
Other van der Waals interactions also play a role in enzyme catalysis, and the entire electrostatic environment of the active site has been proposed as being the key reason that enzymes are such efficient catalysts \cite{Warshel_1998, Warshel_2003}.
This phenomenon is known as electrostatic preorganization and will be discussed in more detail in regards to an ongoing project in Chapter \ref{cha:conclusion}.
\addabbreviation{Molecular dynamics}{MD}

QTAIM provides a method for examining all of these complex bonding features in the active sites of enzymes.
However, the use of QTAIM in understanding the mechanisms of enzymes is very recent and the application of topological studies to the enzyme design process is even more rare. 
Most often, when QTAIM is used for enzyme studies, it is used to predict inhibitors for drug design by evaluating properties at bond critical points \cite{Vega-Hissi_2015, Tosso_2013, Tosso_2014}.
To the best of my knowledge, the only direct application of QTAIM to enzyme design was a study of potential hydrogen bonding interactions in a mutant of \textit{Bacillus subtilis} lipase performed by Lin \textit{et al.} \cite{Lin_design}.
Here I present the application of QTAIM to two metalloenzyme design projects, as well as the first ever application of bond bundle analysis to biological systems.
While QTAIM is a useful tool for modeling and designing enzymes, it does not always fully recover the bonding character in enzyme active sites, necessitating the development of more refined chemical bonding models such as gradient bundle analysis.

\subsection{QM/DMD Methods}

Both of the following DFT studies use structures initially obtained from QM/DMD simulations, where DMD is short for discrete molecular dynamics \cite{QMDMD1, QMDMD2,QMDMD3,QMDMD4,QMDMD5}.
DMD is a simplified method of molecular dynamics using square well potentials to describe interactions.
This allows for faster sampling and enables one to efficiently capture dynamics on the order of tens of nanoseconds, in conjunction with the quantum mechanical description of the active site. 
It has been shown to perform exceptionally well for recapitulating and recovery of metalloenzyme structures, including subtle structural details in active sites \cite{QMDMD1,QMDMD_catechol,Valdez_Lactamase,Valdez_ARD}. 
QM/DMD is especially useful for metalloenzymes compared to other methods since the metal is treated purely quantum-mechanically, meaning there is no need to parameterize the force field for the metal.

\addabbreviation{Quantum mechanics}{QM}
\addabbreviation{Discrete molecular dynamics}{DMD}

The enzymes are divided into three regions: QM-only, DMD-only, and a shared QM-DMD region. 
DMD is used to sample the entire protein except for the metal and its immediately coordinated shell, which is held frozen during the sampling. 
During the DMD phase, the protein undergoes a simulated annealing procedure followed by DMD sampling at a constant temperature. 
The DMD parameters for the QM-DMD region are adjusted on-the-fly, based on the QM calculations. 
Solvent effects are included implicitly in the force field of DMD.

After each DMD sampling step, the lowest energy structure proceeds to the QM step.
The QM region is geometrically optimized, allowing all atoms in the QM-only region to move.
Atoms at the QM-DMD boundary are capped with hydrogens and the boundary atoms along with the capping hydrogens are frozen.
DFT single point energy calculations are then performed on the optimized QM structure.
The DMD and QM steps are performed iteratively, until both energy and geometry (based on RMSD calculations) are converged. 
The nuclear positions from the lowest energy QM step are used for single point active site DFT calculations and the resulting charge density is used for topological analysis.

\addabbreviation{Root mean square deviation}{RMSD}

\subsection{Predictive Methods for Computational Metalloenzyme Redesign - A Test Case with Carboxypeptidase A}
\begin{center}
Modified from a manuscript submitted to \textit{Physical Chemistry Chemical Physics}.\\
Crystal E. Valdez\footnote{Primary researcher and author}, Amanda Morgenstern\footnote{Performed QTAIM analysis}, Mark E. Eberhart\footnote{Provided guidance on QTAIM analysis}, and Anastassia N. Alexandrova\footnote{Corresponding author}. 2016.
\end{center}

\subsubsection{Introduction}

Carboxypeptidase A (CPA) is an exopeptidase that preferentially cleaves C-terminal aliphatic and aromatic amino acids from dietary proteins. 
In this study we predict mutations to CPA that could allow the enzyme to be active toward  a non-native substrate, hippuryl-L-aspartate.
\ref{fig:cpa_structure} shows the active site of CPA, which consists of a Zn$^{2+}$ ion coordinated to two histidines (His69 and His196), a glutamate residue (Glu72), and one water molecule. 
The zinc ion acts as a Lewis acid, lowering the pKa of the bound water molecule to make it a more potent nucleophile, and polarizing the carbonyl oxygen of the substrate (in this case, hippuryl-L-phenylalanine (HPA)). 
Key components of an active mutant enzyme include proper positioning of the substrate with respect to the catalytic Zn and important residues in the active site, resulting in favorable energetics of the rate determining step. 

\addabbreviation{Carboxypeptidase A}{CPA}
\addabbreviation{Hippuryl-L-phenylalanine}{HPA}

\csmfigure{cpa_structure}{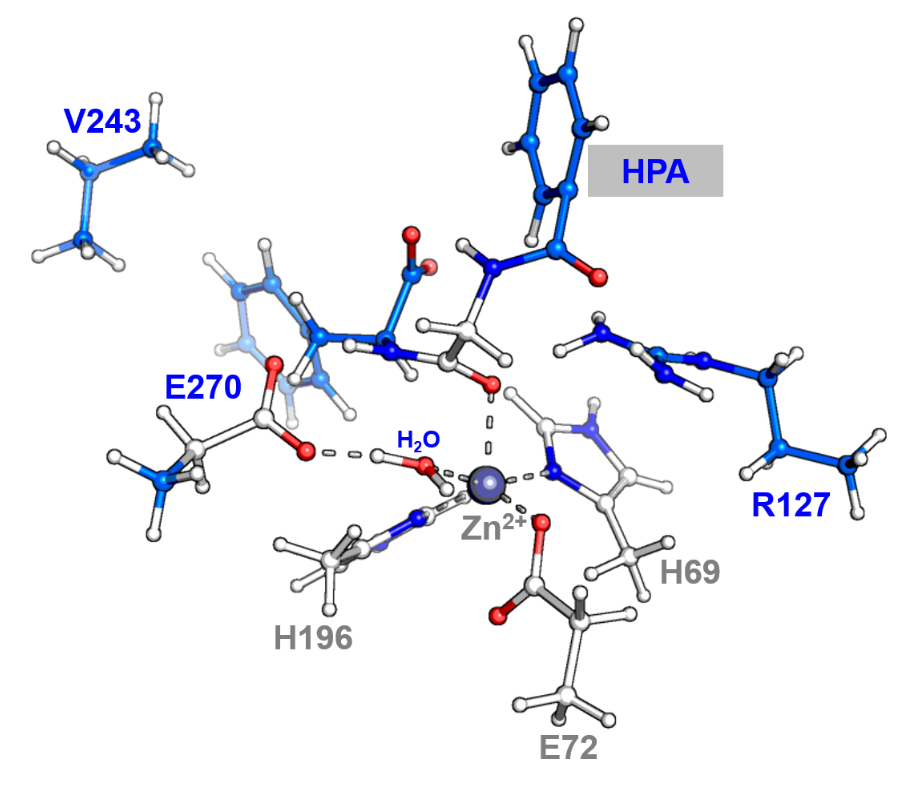}{.5\textwidth}{Active site of CPA showing partitioning for QM/DMD calculations: grey-QM/DMD region, blue-QM region. Single point DFT calculations were performed on the entire active site pictured here.}

While the exact mechanism of CPA is debated, a water-promoted mechanism (\ref{fig:cpa_mechanism}) is believed to be the most energetically favorable \cite{CPA_mechanism1, CPA_mechanism2}. 
The C-terminal peptide specificity of CPA comes from the highly organized hydrogen bond network within the binding pocket that includes Arg127. 
Additionally, Arg127 interacts not only with the terminal carboxylic acid, but also with the carbonyl oxygen of the peptide backbone coordinated to zinc, an interaction believed to contribute to catalysis and substrate binding \cite{CPA_2009}. 
During the course of the reaction, Arg127 stabilizes the oxyanion developed on the tetrahedral intermediate. 
Glu270 is also thought to play a vital role in catalysis by providing a hydrogen bond acceptor for the zinc-bound water. 
Our goal in the design process is to preserve these valuable interactions.

\csmfigure{cpa_mechanism}{CPA_mechanism}{\textwidth}{The proposed water-promoted mechanism of peptide hydrolysis in CPA (the peptide substrate, shown in blue, is truncated for clarity). The water molecule is shown in red.} 

A mutant enzyme (V243R) is computationally predicted to catalyze hydrolysis of the non-native substrate.
However, the mutant enzyme is predicted to be less catalytically efficient than the native enzyme based on a higher calculated activation energy barrier (13.3 $kcal/mol$ for the native enzyme vs. 21.6 $kcal/mol$ for the mutant). 
Here I present the charge density topology analysis of both native CPA and V243R, rationalizing why the native enzyme should be a more efficient catalyst.

\subsubsection{Methods}
The initial structure of CPA was obtained from the Protein Data Bank (code: 6CPA) \cite{CPA_PDB}.
The coordinates from this file were also used to create the V243R mutant enzyme.
The crystal structure contains a bound inhibitor, which served as a scaffold to build the substrates for both systems.
QM/DMD was performed using the QM and QM-DMD regions shown in \ref{fig:cpa_mechanism} for dynamic sampling for both the native and mutant enzyme. 
After annealing, protein sampling occurred at a constant temperature of 0.10 $kcal/(mol*k_B)$ (approximately 50K). 
QM/DMD convergence was reached after 100 iterations, corresponding to around 50 ns of dynamics.
The QM steps were performed with \textit{Turbomole} using a TPSS functional, double $\zeta$ quality basis set for all atoms except the metal which utilized a def2-TZVPP basis set \cite{Turbomole2, Perdew1, Perdew2, Gaussianbasis, basis2}.
The Conductor-like Screening Model (COSMO) continuum solvent model was applied with a dielectric constant of 20.0 \cite{COSMO}. 
A full mechanistic study using the converged active site geometry was performed in \textit{Turbomole}.
The rate determining step for the peptide hydrolysis mechanism shown in \ref{fig:cpa_mechanism} is predicted to have an activation barrier of 13.3 $kcal/mol$ for the native enzyme and 21.6 $kcal/mol$ in the mutated enzyme.
\addabbreviation{Conductor-like screening model}{COSMO}
\addabbreviation{Tao Perdew Staroverov Scuseria}{TPSS}

The electronic charge densities of the stationary points resulting from a QM mechanistic study for the native enzyme and V243R mutant with the bound substrates were calculated with the Amsterdam Density Functional Package (ADF) version 2014 \cite{ADF, ADF2, ADF2014}  using similar computational parameters as in \textit{Turbomole}. 
This includes the enzyme-substrate (ES), first transition state (TS1), and enzyme-intermediate (EI) structures.
A TPSS functional and COSMO solvent model with a dielectric of 20.0 were utilized \cite{TPSS_ADF, COSMO_ADF}.  
A double $\zeta$ quality basis set, DZP, was employed for all atoms except the metal, which was calculated using a triple $\zeta$ quality basis set, TZP \cite{ADFbasis}.
The Bondalyzer add-on package in Tecplot was used to analyze the calculated charge densities \cite{Tecplot}.\\
\addabbreviation{Enzyme-substrate}{ES}
\addabbreviation{Transition state}{TS}
\addabbreviation{Enzyme-intermediate}{EI}

\subsubsection{Results and Discussion}

Two key components to the catalytic activity of CPA are correct positioning of the water molecule for nucleophilic attack and stabilization of the partially charged substrate carbonyl. 
\ref{fig:CPAreac} shows bond paths of interest for the ES complexes. 
In both the native CPA and V243R mutant, the water molecule is positioned near the substrate carbonyl due to coordination with the Zn ion and hydrogen bond donation to E270. 
The native enzyme has additional hydrogen bonding from the second water hydrogen to E72. 
In V243R, E72 is doubly coordinated to the Zn ion, rather than acting as a hydrogen bond acceptor for the water molecule. 
The donation of the second hydrogen atom in the native enzyme results in a more negative Bader charge on the water oxygen (see \ref{tbl:badercharges}). 
This should make the native water molecule a better nucleophile as is indicated by the lower activation energy in the native enzyme as compared to V243R.

\csmfigure{CPAreac}{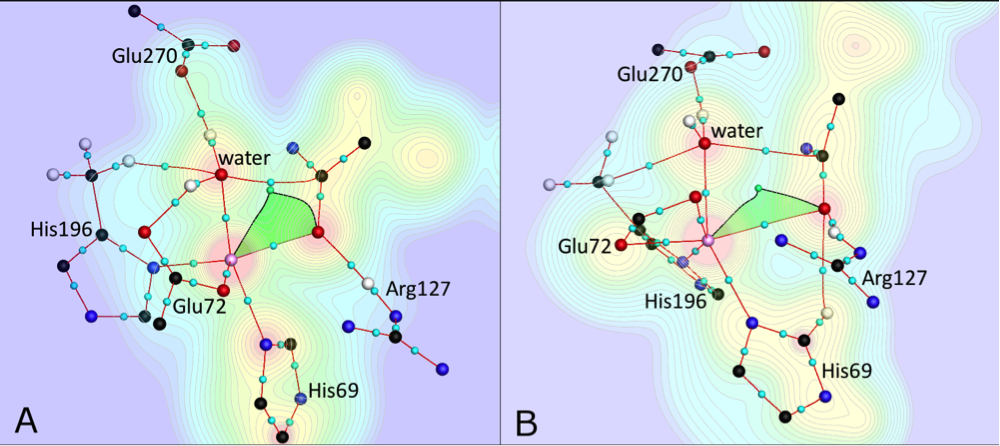}{\textwidth}{Bond paths of interest in the native CPA (A) and V243R mutant (B) ES complexes. Contours in $\rho(\bm{r})$ are drawn on a cut plane on a logarithmic scale from $10^{-3}-1$ $e/bohr^3$. Red lines indicate bond paths. The pictured portion of the Zn--O1 bond bundles are shaded green with black lines showing approximate edges. The following coloring scheme is used: Zn-purple, O-red, C-black, H-white, N-blue, bond CP-cyan, ring CP-green.}

\begin{table}[ht]
\centering
\vspace{2mm}
 \caption{Bader charges of atoms participating in the tetrahedral intermediate in the CPA hydrolysis mechanism. O$_{\textup{wat}}$ is the water oxygen and C1 and O1 are the carbon and oxygen atoms on the substrate carbonyl.}
  \label{tbl:badercharges}
\begin{tabular}{| r | c  c | c  c | c  c |}
\hline
& \multicolumn{2}{c |}{ES} & \multicolumn{2}{c |}{TS1} & \multicolumn{2}{c|}{EI}  \\ [.8ex]
& Native & Mutant & Native & Mutant & Native & Mutant\\[1ex]
\hline
O$_{\textup{wat}}$ &-1.181 & -1.135 & -1.014 & -1.038 & -1.003 & -1.003\\[1ex]
C1 & 1.120 & 1.162 & 1.100 & 1.080 & 1.117 & 1.073\\[1ex]
O1 & -1.087 & -1.103 & -1.102 & -1.102 & -1.089 & -1.085\\[1ex]
Zn & 1.314 & 1.308 & 1.297 & 1.304 & 1.250 & 1.262\\[1ex]
 \hline
 \end{tabular}
\end{table}

The partially negative carbonyl oxygen on the substrate is stabilized through hydrogen bonding to R127 and coordination to the Zn center in both enzymes. 
The hydrogen bond from R127 to O1 in the reactants is not as strong in V243R as compared to the native enzyme. 
This is indicated by the charge density at the H$_{\textup{R127--O1}}$ bond CP, which has decreased from 0.0487 in the native enzyme to 0.0128 $e/bohr^3$ in the mutant (\ref{tbl:rhoCPs}).
Additionally, the O1--H$_{\textup{R127}}$--N$_{\textup{R127}}$ angle has decreased from almost linear in CPA, 175$^{\circ}$, to 147$^{\circ}$ in V243R. 
However, there is an additional weak hydrogen bonding interaction between the carbonyl oxygen and H69 in the mutant enzyme.

\begin{table}[ht]
\centering
\vspace{2mm}
 \caption{Charge density at bond and ring critical points pictured in \ref{fig:CPAreac} for the rate-determining step of peptide hydrolysis in native CPA and the V243R mutant.}
  \label{tbl:rhoCPs}
\begin{tabular}{| r | c  c | c  c | c  c |}
\hline
& \multicolumn{2}{c |}{ES} & \multicolumn{2}{c |}{TS1} & \multicolumn{2}{c |}{EI}  \\ [.8ex]
& Native & Mutant & Native & Mutant & Native & Mutant\\[1ex]
\hline
C1--O$_{\textup{wat}}$ &-0.0166 & 0.0155 & 0.0953 & 0.0817 & 0.217 & 0.204\\[1ex]
C1--O1 & 0.352 & 0.368&0.338&0.342&0.301&0.298\\[1ex]
Zn--O1 & 0.0575 & 0.0183 & 0.0699 & 0.0599 & 0.0762 & 0.0746\\[1ex]
Zn--O$_{\textup{wat}}$ & 0.0558 & 0.0596&0.0327&0.0395& n/a & n/a\\[1ex]
H$_{\textup{R127}}$--O1 &0.0487&0.0128&0.0596&0.0205&0.0412&0.0287\\[1ex]
Ring CP & 0.0166&0.0129&0.0262&0.0269&n/a&n/a\\[1ex]
H$_{\textup{H69}}$--O1 &n/a&0.0157&n/a&0.0160&n/a&0.0127\\[1ex]

 \hline
 \end{tabular}
\end{table}

Similarly, the native enzyme has greater stabilization of the carbonyl oxygen from the Zn ion as compared to V243R. 
This can be seen in the amount of charge density at the Zn--O1 bond CP and also in the size of the Zn--O1 bond bundle. 
The charge density decreases from 0.0575 in the native enzyme to 0.0183 $e/bohr^3$ at the bond CP in V243R. 
A cut plane of a portion of the bond bundle in the plane of the Zn--O1--ring CP is shown in \ref{fig:CPAreac}. 
These points define three vertices of the Zn--O1 bond bundle. 
The pictured portion of the bond bundle in the native enzyme is significantly larger than in V243R, indicating a more stabilizing bonding interaction. 

In both native CPA and V243R mutant, the Zn ion and hydrogen bond network are able to stabilize the partial charge on the substrate carbonyl. 
Throughout this step of the reaction, the charge on the oxygen varies by only 0.178 electrons and 0.132 electrons for the native and mutated enzyme, respectively. 
Overall, the hydrogen bonding networks in both the native and mutated enzyme promote the reaction. 
The higher activation energy in V243R may be partially due to less stabilizing interactions from both hydrogen bonding and the coordination of the substrate carbonyl to the Zn ion. 

In the rate-determining step of the reaction, the Zn--O$_{\textup{wat}}$ bonding interaction is broken, opening the tetrahedral ring pictured in \ref{fig:CPAreac}. 
Topologically, the ring opening is achieved by the ring CP in the tetrahedral intermediate moving towards the Zn--O$_{\textup{wat}}$ bond CP in the TS (see \ref{fig:CPATS}) until it ultimately annihilates the bond CP in the EI complex. 
While the ring CP in V243R is geometrically closer to the Zn--O$_{\textup{wat}}$ bond CP in the ES complex, the charge density region between the ring and bond CP in the native enzyme is ``flatter". 
The change in $\rho(\bm{r})$ between the ring-bond CPs is only 0.0392 $e/bohr^3$ in the native enzyme while it is 0.0467 $e/bohr^3$ in V243R. 
In accordance with the electron-preceding picture \cite{ayers_predict, ayers_2009, ayers_EPP2014,Jones_functionality}, it should therefore be less energy intensive for the native enzyme ring CP to annihilate the bond CP than in the mutant enzyme, which agrees with our mechanistic predictions.

\csmfigure{CPATS}{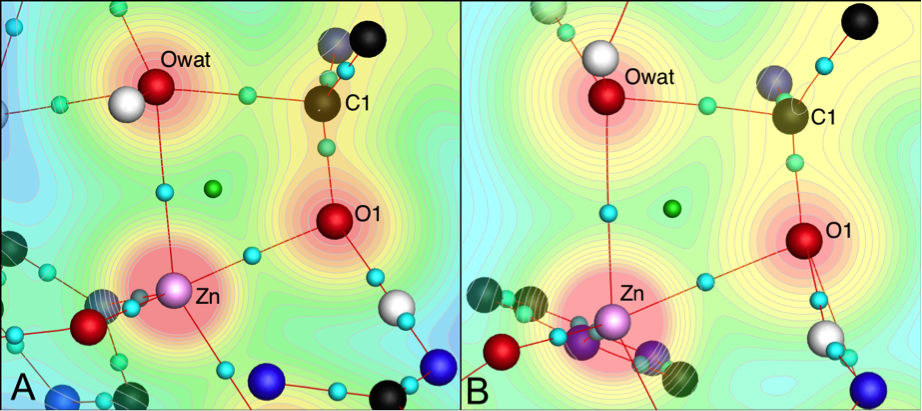}{\textwidth}{Bond paths and critical points for the both the native CPA (A) and V243R mutant (B) TS. The ring CP (green) has moved towards the Zn--O$_{\textup{wat}}$ bond CP in both enzymes. After the transition state, the ring CP converges with the Zn--O$_{\textup{wat}}$ bond CP, causing the two critical points to be topologically annihilated and the Zn--O$_{\textup{wat}}$ bonding interaction to break. }

The observed correspondence between values of $\rho(\bm{r})$ at bond and ring CPs in the reacting region combined with bond bundle analysis and the energetic barrier to the reaction outlines a novel strategy for predicting enzyme reactivity. 
Differences in $\rho(\bm{r})$ in the reactant states of related enzymes, for example designed computationally to catalyze the same reaction, could serve as a probe of relative catalytic activity prior to the mechanistic study. 
We believe this probe is very sensitive and non-empirical, responding to the embedding in the protein charges, i.e. the electrostatic pre-organization. 
Importantly, analysis can be performed on the reactant state alone, prior to performing more costly and often tedious transition state calculations during the mechanistic study. 
Further in-depth investigation of the utility of this parameter for computational enzymology is discussed in section \ref{sec:preorganization}.

\subsection{Metal Specificity of Histone Deacetylase}
\label{sec:HDAC}

\begin{center}
Modified from a paper published in \textit{Journal of Physical Chemistry B}\footnote{Reproduced in part with permission from  \textit{Michael R. Nechay, Nathan M. Gallup, Amanda Morgenstern, Quentin A. Smith, Mark E. Eberhart, and Anastassia N. Alexandrova. J Phys. Chem. B. Article ASAP, DOI: 10.1021/acs.jpcb.6b00997.} Copyright 2016 American Chemical Society.}.\\
M. R. Nechay\footnote{Primary researcher and author}, N. M. Gallup\footnote{Performed calculations and analysis for active site models}, A. Morgenstern\footnote{Performed QTAIM analysis}, Q. A. Smith\footnote{Performed preliminary QM/DMD study}, M. E. Eberhart\footnote{Provided guidance on QTAIM analysis}, A. N. Alexandrova\footnote{Corresponding author}. 2016. DOI: 10.1021/acs.jpcb.6b00997.
\end{center}

\subsubsection{Introduction}

Histone deacetylases (HDACs) are responsible for the removal of acetyl groups from histones, which results in gene silencing (21). 
Overexpression of HDACs is associated with cancer, and their inhibitors are of particular interest as chemotherapeutics \cite{HDAC_Haberland_2009, HDAC_Bolden_2006, HDAC_Marks_2007, HDAC_Bradner_2010, HDAC_Paris_2008,HDAC_Suzuki_2009}. 
HDAC8 is the most studied HDAC, and has traditionally been considered to be a Zn-dependent enzyme. 
However, recent experimental assays have challenged this assumption and shown that HDAC8 is catalytically active with a variety of different metals, and that it may be an Fe-dependent enzyme \textit{in vivo} \cite{HDAC_M_Gantt_2006}. 
Here we study the mechanism of HDAC8 utilizing a series of divalent metal ions in physiological abundance (Zn$^{2+}$, Fe$^{2+}$, Co$^{2+}$, Mn$^{2+}$, Ni$^{2+}$, and Mg$^{2+}$). 
Two important factors for evaluating the metal dependency of HDAC8 are (1) activation energies of the rate-limiting step with each metal cation, and (2) the binding affinity of the protein for each metal.
In this section I present the topological analysis of the charge density in an ongoing effort to explain the calculated and experimentally observed energetics in HDAC8 with each metal ion.

\addabbreviation{Histone deacetylase}{HDAC}

The active site of HDAC8 contains a metal-binding center coordinated to one histidine (H180) and two aspartate (D178 and D267) residues (see \ref{fig:HDACstructure}). 
A water molecule is also present in the crystal structure and interacts with the metal center and two additional histidine residues, H142 and H143. 
These histidines form dyads with the aspartates D176 and D183, respectively. 
In the crystal structure of a Y306F mutant (PDB code 2V5W), the acetyl-lysine substrate appears coordinated to the zinc, with a carbonyl oxygen. 

\csmfigure{HDACstructure}{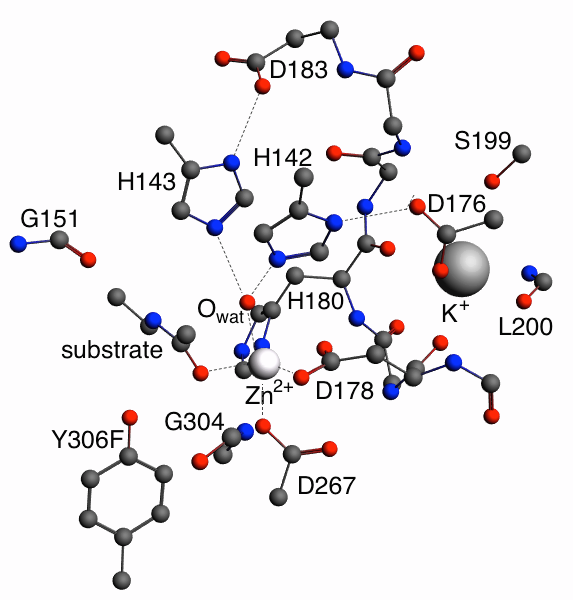}{.65\textwidth}{The active site of HDAC8 extracted from the crystal structure, highlighting the most critical residues relevant to catalysis within the active site. In this work, F306 was mutated to tyrosine to facilitate catalysis. This full structure was used as the QM region in QM/DMD calculations and for the QM mechanistic study and charge density calculations.}

We found a proposed proton shuttle mechanism \cite{HDAC_mech_Wu_2010} to be the most viable through detailed QM/DMD calculations \cite{HDAC_Nechay}. 
In the proton shuttle mechanism (\ref{fig:HDACrxnmech}), H142 and H143 are initially singly protonated. 
H143 alternates between acting as a base and acting as an acid, first abstracting a proton from water and subsequently transferring it to the substrate nitrogen.
This results in two transition steps during the process of cleaving the C-N bond in the substrate, which we refer to as ts1 and ts2. 
The reaction coordinate for the proton shuttle mechanism with all six metal ions was calculated using the methods described in Section \ref{HDACmethods}.

\csmfigure{HDACrxnmech}{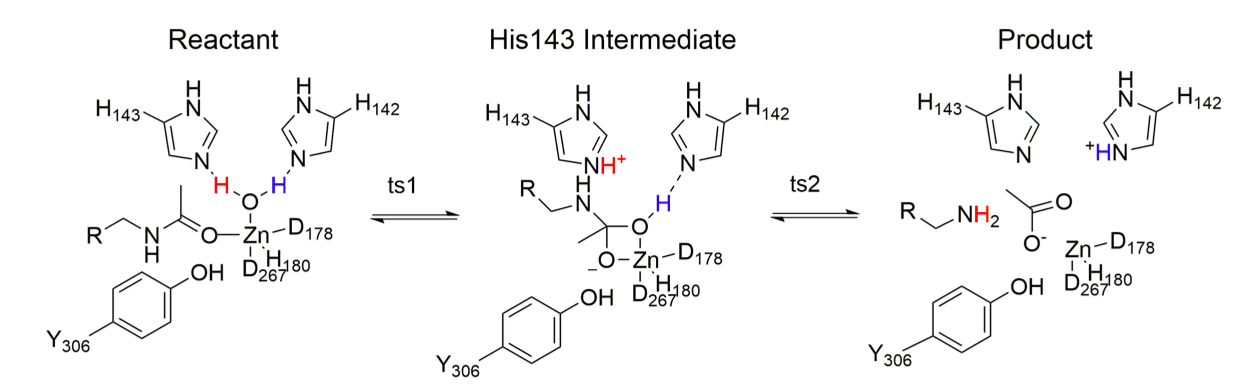}{\textwidth}{Proposed proton shuttle mechanism for HDAC8.}

For the second part of this study, we explored how metal binding affinity may affect the overall reactivity of HDAC8. Gantt and coworkers have shown that several of the metal ions in this investigation were inactive \textit{in vitro} \cite{HDAC_M_Gantt_2006}. 
However, many of our computed rate-limiting transition state energies are well under the ceiling of catalytic inactivity. 
It is possible that the binding affinity of each metal towards HDAC8 could have an effect on \textit{in vitro} catalytic activity, and may play an important role for \textit{in vivo} metal selection. 
Here, we attempt to elucidate the significance of this effect using metal swapping QM calculations.

The combination of $\Delta \Delta G$ data for metal swapping with the catalytic rate constants from mechanistic studies is required to reproduce the experimentally observed trend in metal-dependent performance. 
We predict Co$^{2+}$ and Zn$^{2+}$ to be the most active metals in HDAC8, followed by Fe$^{2+}$, and Mn$^{2+}$ and Mg$^{2+}$ to be the least active.
Ni$^{2+}$ proves to be an interesting outlier for which we recommend further study.
We show that results from the metal swapping portion of this study can be quantitatively rationalized using QTAIM analysis on the charge densities for each metal-substituted active site.
For the activation energy study, however, QTAIM is only able to provide a qualitative explanation for energetic results, presenting a clear case where higher resolution bonding models, such as gradient bundle analysis, may be required.

\addsymbol{Gibbs free energy}{$G$}

\subsubsection{Methods}
\label{HDACmethods}

The initial structure of HDAC8 was obtained from the Protein Data Bank crystal structure of the Y306F mutant (code: 2V5W) \cite{HDAC8_xtal_Vannini_2007}. Residue 306 was transformed from phenylalanine to tyrosine using UCSF Chimera. The protonation states of all residues were chosen in accord with their pKa values at neutral pH, including singly protonated His180 on the $\delta$ site, and singly protonated H142 and H143 on the $\epsilon$ site.

An appropriate choice of QM and DMD subsystem sizes was critical for QM/DMD simulations. 
A QM region of 165 atoms was found to be necessary in order to produce logical results (such as binding of the substrate) and is pictured in \ref{fig:HDACstructure}. 
QM calculations were performed using DFT as implemented in \textit{Turbomole} v6.5 \cite{Turbomole2}.
Solvent effects were included implicitly in the force field of DMD. 
In the QM region, solvent was modeled using COSMO \cite{COSMO} with a dielectric constant of 4.0 (in consideration of the buried active site).
Separate calculations were also performed using non-polarizable point charges (AMBER force field), but no significant changes in the mechanism were observed.
(The results of these calculations are reported in the supplementary information of \cite{HDAC_Nechay}).
All reported charge densities and energies are therefore from the COSMO calculations.
DFT single point energy calculations were carried out using the TPSS functional \cite{Perdew1,Perdew2}. 
The def2-TZVPP basis set \cite{basis2} was used for the metal atoms, while all other atoms were described by def2-SVP \cite{basis2}. The Grimme dispersion correction \cite{Grimme_2004} was included for all QM calculations. 

The lowest energy structures from QM/DMD proceeded to the QM mechanistic study.
All calculations of the stationary points on the reaction profile were optimized with \textit{Turbomole}  using the TPSS functional and Def2-TZVPP basis set for all metals, and Def2-SVP basis sets for non-metal atoms. 
Vibrational frequency analysis was done using the same level of theory, and the nature of all transition states was determined by the presence of a single imaginary frequency aligned with the reaction coordinate. 
Single point energy calculations were carried out on all stationary points using the larger Def2-TZVPP basis set for all atoms. 
The full reaction coordinate for all metals, including zero point energy and thermal and entropic corrections, is shown in \ref{fig:HDACrxncoord}.

\csmfigure{HDACrxncoord}{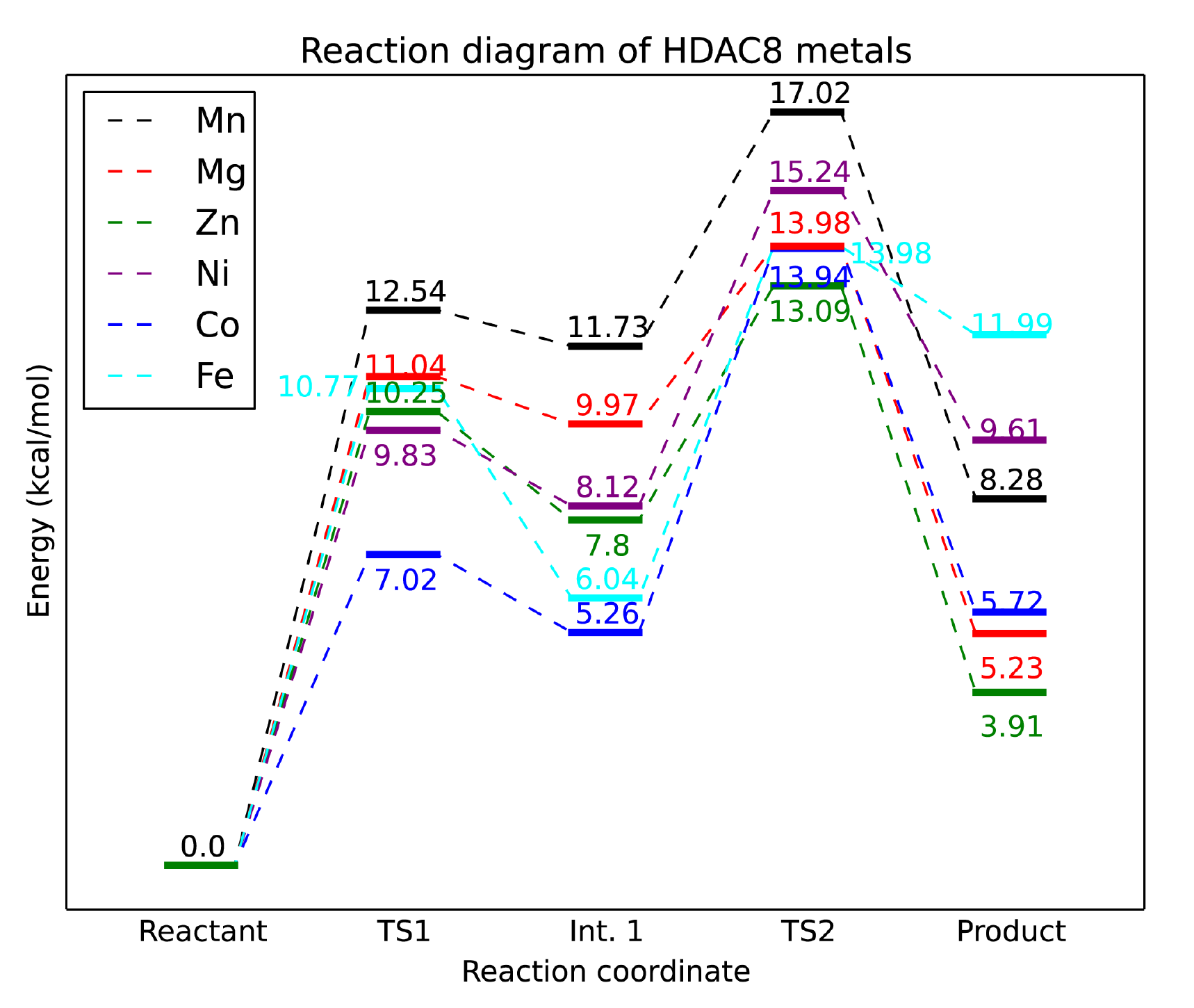}{.8\textwidth}{Calculated reaction pathway for HDAC8 with a variety of divalent metal ions.}

Born-Oppenheimer molecular dynamics (BOMD) simulations were done for the full DFT sampling of the complexes of the studied metals with organic chelators in order to assess the relative metal binding affinities to the HDAC8 protein.
To calculate the $\Delta \Delta G$ for metal swapping, we first compared the accuracy of DFT to experiment using the experimentally known stability constants of the complexes of the studied metals with GEDTA and DTPA \cite{Menergies}.
From these stability constants, the free energies of reactions (1-4) in \ref{fig:Mswap} are known. 
Further, from $\Delta \Delta G$ of reactions (1) and (2), one can close the thermodynamic cycle for DPTA on the upper left in \ref{fig:Mswap}. 
Analogously, one can close the cycle for GEDTA on the upper right in \ref{fig:Mswap}, using $\Delta \Delta G$ of reactions (3) and (4). 
Through combination of these two cycles, one can then calculate the $\Delta G$ of metal swapping between the two chelators,
DTPA--M$_a^{2-}$ + GEDTA--M$_b^{2-}$ $\rightarrow$ DTPA--M$_b^{2-}$ + GEDTA--M$_a^{2-}$,
bypassing the complicated calculation of the solvated metal ions, M$_a^{2+}$ (aq) and M$_b^{2+}$ (aq). 

The $\Delta G$ values of metal swapping were calculated both from the experimental data, and with DFT (See Appendix \ref{app:Mswap} for results). 
The structures of the metal--GEDTA and metal--DTPA complexes were geometry-optimized and also subjected to 5 ps of BOMD, to verify the ground state geometries. 
To avoid additional desolvation complications, polydentate ligation of the chelating agent was selected such that the bound metal had no solvent access, preventing explicit water coordination, and allowing for the exclusive use of COSMO. 
The computed $\Delta G$ values were found to be in good agreement with experiment, having an average error of $0.5 \pm 1.5$ \textit{kcal/mol}.
\addabbreviation{Born-Oppenheimer molecular dynamics}{BOMD}

\csmfigure{Mswap}{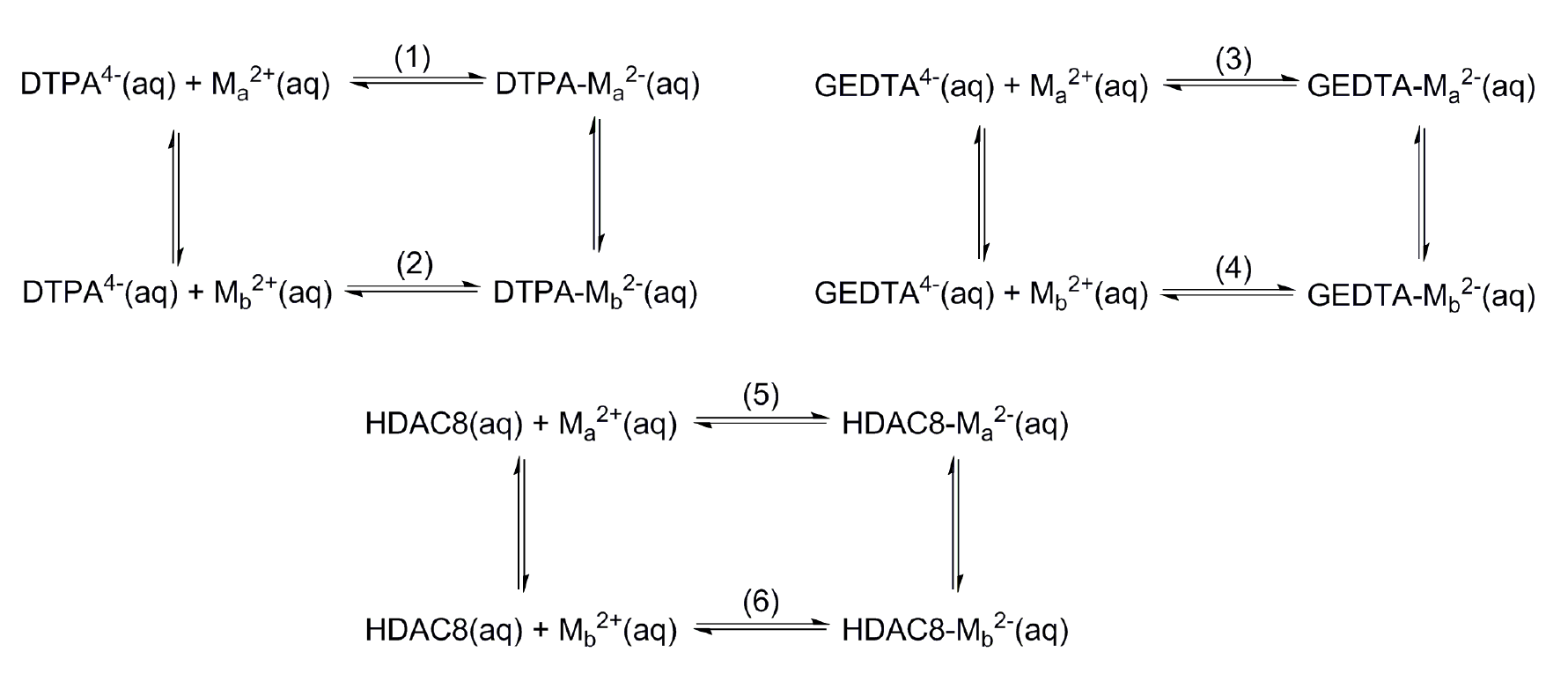}{\textwidth}{Schematic of all considered thermodynamic cycles exploited for relative metal binding affinities in HDAC8.}

This methodology was then applied to a theoretical metal swap between DTPA and HDAC8:
DTPA--M$_b^{2-}$ + HDAC8--Co$^{2-}$ $\rightarrow$ DTPA--Co$^{2-}$ + HDAC8--M$_b^{2-}$,
using $ \Delta G$ of reactions (1) and (2), and (5) and (6) in \ref{fig:Mswap}, and with M$_a$ set to Co$^{2+}$ as a reference. 
The above reaction can be paired with the experimental $\Delta G$ of reactions to yield $\Delta G$ of the following process:
M$_b^{2+}$ (aq) + HDAC8--Co$^{2-}$ $\rightarrow$ Co$^{2+}$(aq) + HDAC8--M$_b^{2-}$.
This gives the desired relative affinities of the different metals to HDAC8, collected in \ref{tbl:Mswap}.

\begin{table}[ht]
\centering
\vspace{2mm}
 \caption{$\Delta \Delta G$ of binding between metal ions and HDAC8, relative to Co$^{2+}$.  Energies are in \textit{kcal/mol}.}
  \label{tbl:Mswap}
\begin{tabular}{| c | c |}
\hline
Metal & $\Delta \Delta G$ binding\\[1ex]
\hline
Co$^{2+}$ & 0.00\\[1ex]
Zn$^{2+}$ & 1.99\\[1ex]
Fe$^{2+}$ & 3.52\\[1ex]
Mn$^{2+}$ & 6.28\\[1ex]
Mg$^{2+}$ & 16.75\\[1ex]
Ni$^{2+}$ & -4.33\\[1ex]
\hline
\end{tabular}
\end{table}

The charge densities for the reactant geometries of HDAC8 with all 6 metal ions were calculated with the Amsterdam Density Functional Package (ADF) version 2014.01 \cite{ADF, ADF2, ADF2014} using similar computational parameters as in the QM mechanistic study. 
A TPSS functional \cite{Perdew2, Perdew3} and COSMO solvent model with a dielectric constant of 4.0 were utilized. 
A double $\zeta$ quality basis set, DZP, was employed for all atoms except the metal, which was calculated using a triple $\zeta$ quality basis set, TZP \cite{ADFbasis}. 
Spin-unrestricted calculations were performed on the Zn and Mg active sites, while all other metals were calculated in their high-spin states. The Bondalyzer add-on package in Tecplot \cite{Tecplot} was used to analyze the resultant charge densities.

\subsubsection{Results and Discussion}

The reaction coordinate in \ref{fig:HDACrxncoord} shows the second transition state (ts2) to be rate limiting in all cases. 
This result marks a pronounced deviation from previously published models, which claimed the first transition state, ts1, to be rate-limiting \cite{HDAC_mech_Wu_2010, HDAC_mech_Corminboeuf_2006} for Zn$^{2+}$. 
This may be an artifact of not using a large enough QM region in the earlier studies. 
We further find that Zn$^{2+}$ facilitates the fastest catalytic path, followed by Co$^{2+}$, Fe$^{2+}$ and Mg$^{2+}$, then Ni$^{2+}$, and finally Mn$^{2+}$. 
This is in contrast to experimental assays published by Gantt and coworkers \cite{HDAC_M_Gantt_2006} who found Co$^{2+}$ and Fe$^{2+}$ to be the most active, followed by Zn$^{2+}$, and little reactivity for Ni$^{2+}$ and Mn$^{2+}$.

The main role of the metals in the reaction comes in the first step, where the carbonyl of the substrate is activated for nucleophilic attack. 
This determines both the activation barrier and the energy of the resultant intermediate, thereby shaping the entire reaction profile. 
We examine the reactant states using QTAIM, to shed light on the electronic effects leading to the metal-dependent performance. 
The lower energy barriers of ts1 for Zn$^{2+}$, Ni$^{2+}$, Fe$^{2+}$, and Co$^{2+}$ may be in part due to the higher number of occupied d orbitals than for Mn$^{2+}$ (Mg$^{2+}$ obviously lacking any).

The greater number of d-electrons implies that orbitals higher up in the enzyme's molecular orbital manifold will be filled. 
In general, orbitals higher in the manifold are characterized by a greater number of antibonding orbitals and hence more interatomic nodal planes. 
It is orbitals of this type that stabilize topological rings and cages by contributing density along bond paths though not along nodal planes. 
This combination leads to deeper rings and greater curvature at bond CPs. 
In the plane of a ring of nuclei as shown in \ref{fig:HDACbondpaths}, the ring point is a minimum in the charge density. 
In order for a ring point to exist in this region, a bond path must form between the water molecule and substrate carbonyl.

\csmfigure{HDACbondpaths}{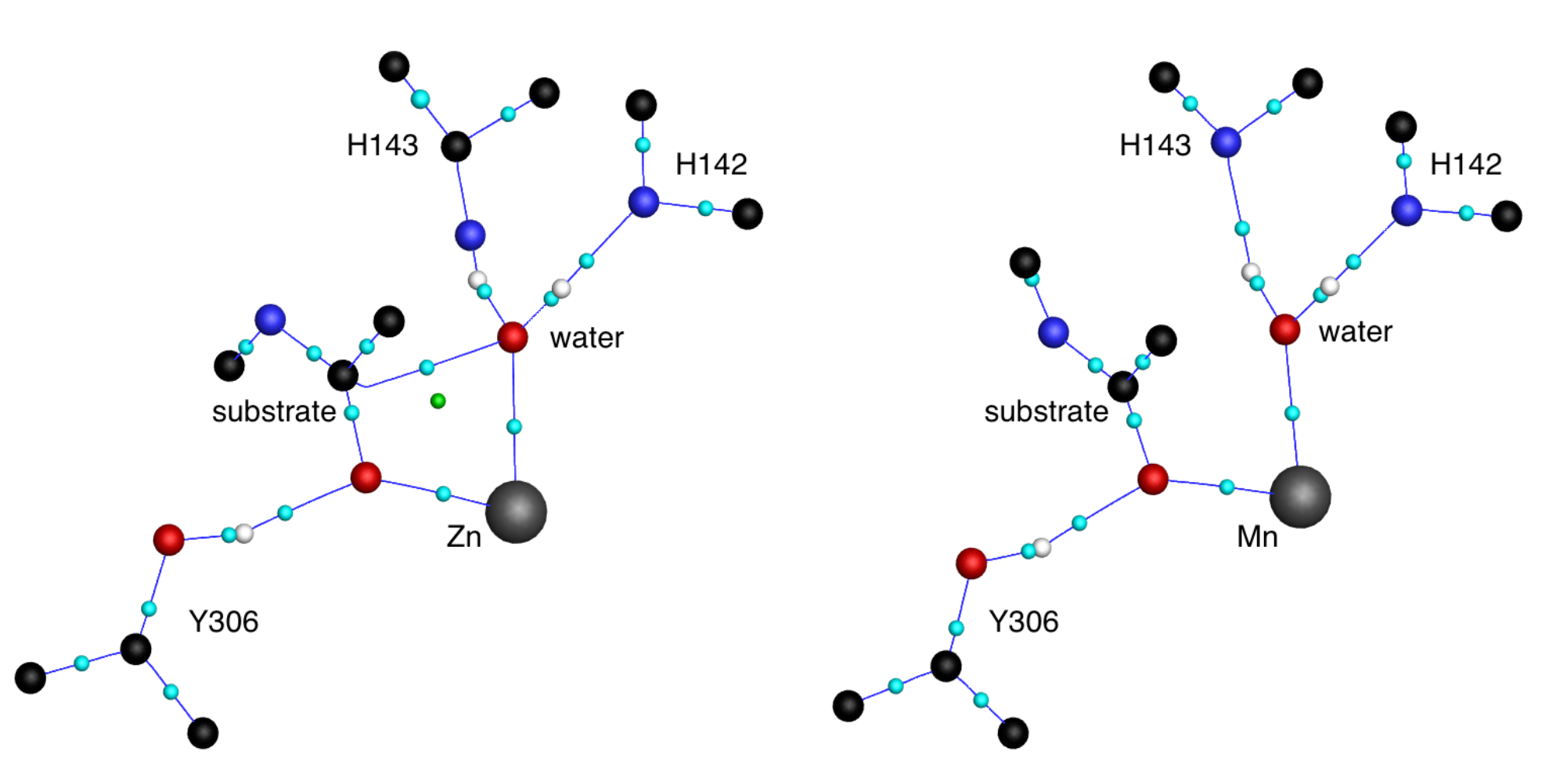}{\textwidth}{Critical points and bond paths of interest in the active site of HDAC8 with Zn (left) and Mn (right). When Zn is present in the active site a bond path forms between the water oxygen and the carbon atom from the substrate carbonyl. This topologically necessitates a ring critical point to exist in the active site as well. When Mn is present, no ring CP is found as there is not a bond path between the water and substrate carbonyl. Ni, Fe, and Co give the same topology as Zn, while Mg has the same topology as Mn. Sphere coloring is as follows: C-black, N-blue, O-red, Metal-grey, bond CP-cyan, ring CP-green.}

A bond path is present in the reactant state between the substrate carbonyl and water molecule when Zn$^{2+}$, Ni$^{2+}$, Fe$^{2+}$, or Co$^{2+}$ is placed in the active site of HDAC8. 
This bond path is not present for Mg$^{2+}$ or Mn$^{2+}$.
While this bond is not indicative of a strong interaction (based on low values of charge density at the bond CP as well as the curved nature of the bond path), it still indicates a stabilizing interaction between the water molecule and substrate. 
It can be seen in the proton shuttle mechanism (\ref{fig:HDACrxnmech}) that when a proton is transferred from the water molecule to H143, a bond forms between the carbonyl carbon and water oxygen. 
This bond has already begun to form in the reactant state for Zn$^{2+}$, Ni$^{2+}$, Fe$^{2+}$, and Co$^{2+}$, which facilitates the removal of a water hydrogen and ultimately lowers the activation energy of this reaction step. 

However, the correlation of this topological effect with the barrier height only provides a qualitative rational for activation energies.
The ts1 activation energies for  Zn$^{2+}$, Ni$^{2+}$, Fe$^{2+}$, and Mg$^{2+}$ are all within 1.2 $kcal/mol$, generally considered the margin of error for DFT energy calculations. 
The presence/absence of the bond path does not provide information to predict that Co$^{2+}$ would have a lower ts1 activation barrier by almost 3 $kcal/mol$ compared to any other metal tested, or that Mn$^{2+}$ would have a higher ts1 barrier of 1.5 $kcal/mol$.
Values of $\rho(\bm{r})$ and $\nabla^2 \rho(\bm{r})$ at bond and ring CPs also did not correlate to the activation energies.
While the existence of a  bond path sheds some light on the abilities of various metals to catalyze deacetylation, the lack of quantitativeness of this analysis highlights a need for advancements in analyzing the topology of the charge density, which will be discussed in more detail in sections \ref{sec:advancements} and \ref{sec:preorganization}.

The metal binding affinities in HDAC8, however, can be more fully explained using QTAIM.
According to the $\Delta \Delta G$ calculations described in section \ref{HDACmethods}, Co$^{2+}$ is predicted to bind to HDAC8 very strongly. 
It is followed by Zn$^{2+}$ and Fe$^{2+}$. 
Mn$^{2+}$ and Mg$^{2+}$ have considerably smaller affinities for HDAC8. 
Ni$^{2+}$ is calculated to have the highest binding affinity towards HDAC8 among all the studied metals, in contrast to experimental results. 
This ordering can be related to the active site geometries and corresponding electronic effects. 
A particularly direct interaction exists between the metal and H180, which is apparently highly important for metal binding. 
The relationship between binding affinity and the amount of charge density at the H$_{\textup{H180}}$--metal bond CP is linear (\ref{fig:Mplot}) with an R$^2$ value of 0.976. 
The charge density at the bond CP is a consequence of $\sigma$-bonding and $\pi$-back-donation with the available d-atomic orbitals (AOs) on the given metal. 
Mg, lacking any occupied d-AOs, has a significantly lower charge density at the bond CP and lower binding affinity than any of the transition metals, despite having a relatively short bond length (which is simply due to its smaller size). 
Thus, the geometric and electronic parameters of this H$_{\textup{H180}}$--metal bond can serve as predictors of the metal binding affinity in this case.
\addabbreviation{Atomic orbital}{AO}

\csmfigure{Mplot}{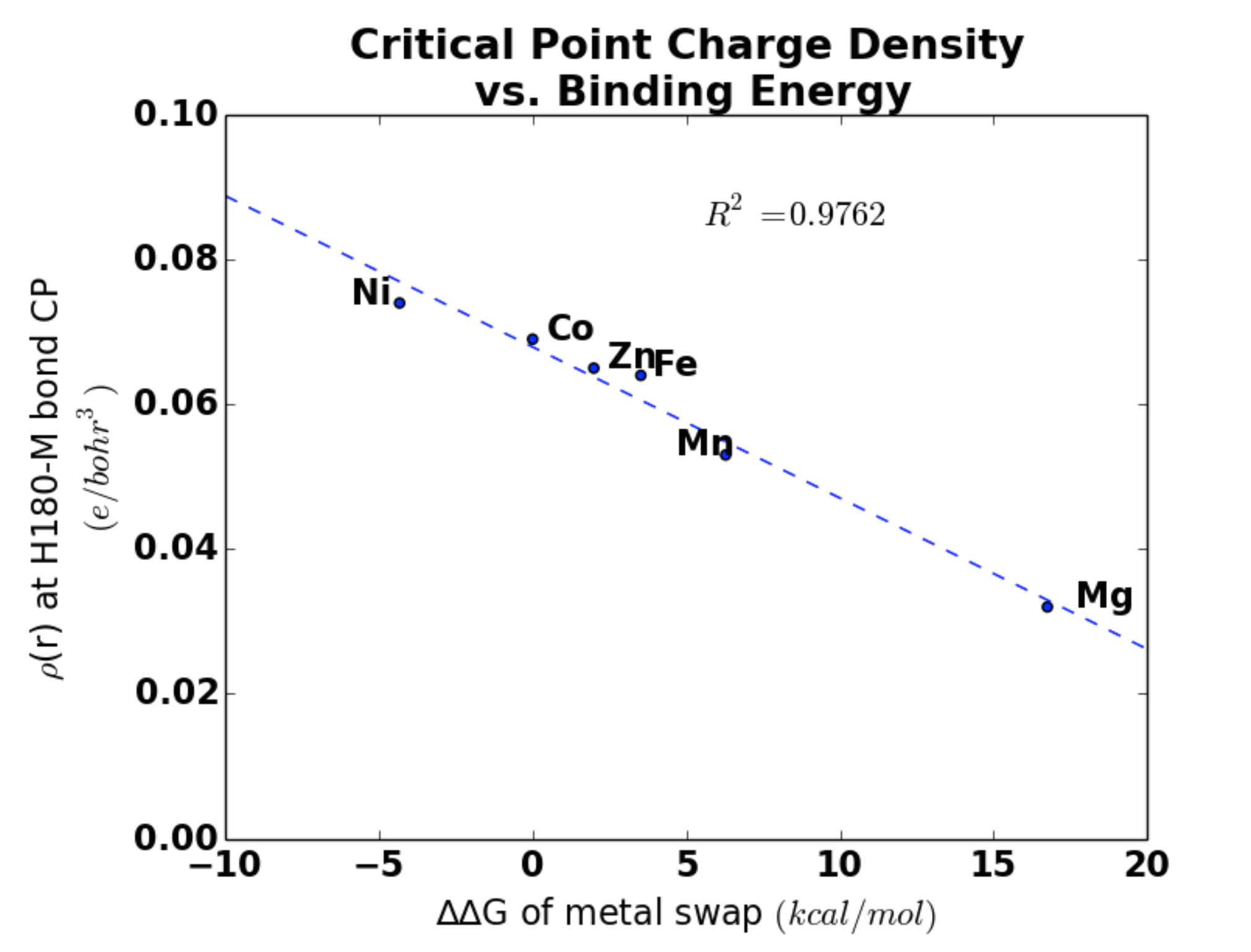}{.75\textwidth}{Relationship between $\rho(\bm{r})$ at H$_{\textup{H180}}$--metal bond CPs and $\Delta \Delta G$ of metal swapping.}

The difference in metal binding affinity may play a key role in the overall catalytic activity of enzymes such as HDAC8 capable of utilizing different metals, and elucidate its preference for any one particular ion. 
We evaluate the ability for the protein to retain a metal and subsequently use it for catalysis. 
Our simple probabilistic model captures the combined effects of metal-binding equilibration and subsequent catalysis relative to any particular ion. We use the equation,

\begin{equation}
k_{rel}= k_{cat} * k_{binding}
\end{equation}

 where $k_{rel}$ stands for the relative total catalytic activity of a particular protein-metal complex, $k_{cat}$ represents the catalytic activity as determined by QM mechanistic analysis, and $k_{binding}$ embodies the relative equilibrium binding constant between a metal and an enzyme. These values can be represented as Boltzmann distributions utilizing the above computed quantities in catalysis and binding affinity to yield

\begin{equation}
k_{rel}=exp\Big( -\frac{\Delta G^{\ddagger}}{RT}\Big) * exp\big( - \frac{\Delta \Delta G_{binding}}{RT}\Big)
\end{equation}

\noindent whose values are summarized in \ref{tbl:krel}.

\begin{table}[ht]
\centering
\vspace{2mm}
 \caption{Predicted and experimental \cite{HDAC_M_Gantt_2006} total catalytic activities for different metal ions in HDAC8. Binding affinity energies were taken relative to Co$^{2+}$. Normalized values are relative to the predicted most active metal ion-protein complex.}
  \label{tbl:krel}
\begin{tabular}{| l | c c c |}
\hline
Metal & $k_{rel}$ & $k_{rel}$ & K$_{cat}$ (s$^{-1}$) \\[1ex]
 & Predicted & Normalized & Experimental  \\[1ex]
\hline
Ni$^{2+}$ & 1.89E-08&1.00 & n/a\\[1ex]
Co$^{2+}$ & 7.64E-11& 4.03E-03 & 1.2\\[1ex]
Zn$^{2+}$ & 1.27E-11&6.70E-04 & 0.90\\[1ex]
Fe$^{2+}$ & 3.11E-14&1.64E-06 & 0.48\\[1ex]
Mn$^{2+}$ & 1.37E-17&7.11E-10 & n/a\\[1ex]
Mg$^{2+}$ & 1.46E-23&7.71E-16 & n/a\\[1ex]
\hline
\end{tabular}
\end{table}

This model qualitatively predicts the trend seen in catalytic activity for these metals in experiment \cite{HDAC_M_Gantt_2006}, though the trend is slightly perturbed, either due to experimental errors, or inaccuracies of our model. 
Co$^{2+}$ is predicted as the most active, followed by Zn$^{2+}$ and Fe$^{2+}$. 
Additionally, the experimentally non-catalytic metal Mn$^{2+}$ is predicted as being four orders of magnitude less reactive than Fe$^{2+}$. 
This is due to both manganese's poor binding affinity to the active site of HDAC8 and higher $\Delta G^{\ddagger}$. Zn$^{2+}$ is predicted to be less catalytically active than Co$^{2+}$, despite its low reaction barrier, due to its poor affinity, while Co$^{2+}$ exhibits high catalytic activity in spite of a higher reaction barrier. 
Fe$^{2+}$ competes with Co$^{2+}$ and Zn$^{2+}$, even though it has a higher reaction barrier than both of these cations. 
\addsymbol{Activation energy}{$\Delta G^{\ddagger}$}

Ni$^{2+}$, which was experimentally shown to be inactive, is a clear outlier, and predicted to be exceptionally catalytic, due to high binding affinity. 
This may be an artifact of the simple model, and requires further investigation.  
Another possibility is that the coordination of Ni$^{2+}$ in HDAC8 may be different from other metals tested here. 
Mg$^{2+}$ has not been tested experimentally, but we found that it has reasonable reaction energetics.
However, its affinity to HDAC8 is very low, and Mg$^{2+}$ is ultimately predicted by our calculations to have a very low catalytic activity. 
Thus, it seems that the d-AO structure is required in HDAC8, despite the simple Lewis acid catalysis performed by the metal in this enzyme.

\subsection{Conclusions}

Using QTAIM and bond bundle analysis, we were able to justify the differences in activation energies for native CPA and a V243R mutant for a non-native substrate. 
The hydrogen bonding network in the native enzyme was seen to be more favorable than the mutant enzyme due to a greater number of hydrogen bond paths with a higher value of $\rho(\bm{r})$ at bond CPs.
The Zn ion was also shown to interact and activate the substrate carbonyl in the native CPA more than in V243R.
This analysis was based on values of charge density as well as by the size of the Zn--O$_1$ bond bundle.
Finally, the charge density between the ring CP in the tetrahedral intermediate and the Zn--O$_{\textup{wat}}$ bond CP was flatter in the native enzyme, leading to a more energetically facile annihilation of the bond and ring CPs, which is topologically required for the reaction to occur.

In our second enzyme study, we were able to understand the metal binding affinities to HDAC based on the amount of $\rho(\bm{r})$ at bond critical points.
However, we were only able to qualitatively rationalize the predicted activation energies of the reaction based on the existence of a bond path (and therefore ring CP) between the water oxygen and substrate carbonyl.
This study highlights a need for a higher resolution picture of the charge density to better understand how and why chemical reactions occur.
We are currently working to extend gradient bundle analysis techniques to be applicable to non symmetric systems such as enzymes. 
Details of these improvements to GBA are discussed in the next chapter.


\chapter{Future Work and Conclusions}
\label{cha:conclusion}

Thus far, we have shown that gradient bundles can be used to understand valence electron structure, recover bond dissociation energies, and visualize chemical reactivity in small molecules.
The next task in developing GBA is to search for structure--property relationships that provide chemical insight into bonding interactions that can not be explained using existing chemical bonding models.
One instance where this advancement is needed was highlighted in the previous chapter on metalloenzyme design.
While the existence of a bond path correlated to the initial activation energies for deacetylation catalyzed by HDAC with most metals, the iron cation did not follow the predicted trend.
In order to further analyze the bonding interactions between the substrate carbonyl and water molecule in HDAC, GBA methods must be adapted to be applicable to more complex systems. 
We hypothesize that the full gradient field in the active site of HDAC does hold the information necessary to predict which metal cations would make the most efficient catalysis, we just need the methods to extract this information from the charge density.

To this end, the following work is being done to advance gradient bundle analysis. 
We are quantifying the ZFS picture of reactivity by calculating gradient bundle condensed Fukui functions, $f_{GB}$.
In addition to viewing the movement of ZFSs, this will enable a change in electron count to be calculated within each gradient bundle.
Observing the motion of ZFSs and $f_{GB}$ highlights a method for predicting which gradient bundles will be most energetically favorable to change their electron count, based only on the differential geometry of gradient paths and ZFSs. 
Our goal with this project is to answer the almost century old question originally posed by Slater, ``where are the HOMO and LUMO in the ground state charge density?".

Work is also being done to expand GBA to nonlinear systems.
We are currently developing a volume integration algorithm to allow 2D plots of electron count and energy to be created around bond paths in Tecplot.
This project requires finding meaningful ways to interpret gradient bundle analysis results based on using stereographic projections of data. 
Once these extensions to GBA are in place we will be able to complete projects studying the gradient bundles in aromatic and anti-aromatic compounds, as well as in solid state systems, such as the ongoing work attempting to elucidate the unique properties of Ir. 
Finally, we are using QTAIM and GBA to determine the size of active site cluster models required to properly model enzyme mechanisms and kinetics, as well as determining if the charge density in the active site alone can be used to capture electrostatic preorganization in enzymes.

\subsection{Gradient Bundle Condensed Fukui Functions}

The ability to qualitatively analyze reactivity using the movement of ZFSs bounding gradient bundles was demonstrated in chapter \ref{cha:ZFS}. 
This analysis can be made quantitative by additionally calculating the change in electron density in gradient bundles as the electron count in a system changes.
The result of this calculation is a gradient bundle condensed Fukui function, $f_{GB}$, which provides a more detailed picture of site reactivity than atom condensed Fukui functions.
GB condensed Fukui functions also provide evidence of, and explanation for, negative Fukui functions, which is a topic of debate \cite{Ayers_NegFukui_2006, Ayers_NegFukui_2007, Ayers_NegFukui_2012, Bulat_NegFukui_2004, Fuentealba_NegFukui_2000, Roy_NegFukui_1999}.
To demonstrate this novel method, here I present $f_{GB}$ calculations for F$_2$ and N$_2$. 
\addsymbol{Gradient bundle condensed Fukui function}{$f_{GB}$}

The computational methods for all of the following data was the same as in section \ref{sec:ZFSmethods}, except that a triple $\zeta$ basis set was used and no relativistic corrections were made. 
The RMF method for the condensed Fukui function was integrated over gradient bundles using the equation,

\begin{equation}
\label{eqn:RMF_minus}
f^-_{GB} = \int_{\omega_i} \rho(\bm{r},N)dV - \int_{\omega^-_i} \rho(\bm{r}, N-1)dV
\end{equation}

for the removal of an electron and with

\begin{equation}
\label{eqn:RMF_plus}
f^+_{GB} =  \int_{\omega^-_i} \rho(\bm{r}, N+1)dV - \int_{\omega_i} \rho(\bm{r},N)dV
\end{equation}

\noindent for the addition of an electron.
$\omega_i$ indicates the integration is performed over any gradient bundle, not just an atomic basin.
Note that both equations are set up such that the Fukui function will be positive in the region most likely to undergo nucleophilic attack ($f^+_{GB}$) or electrophilic attack ($f^-_{GB}$).
\addsymbol{Gradient bundle}{$\omega$}

The gradient bundle condensed Fukui functions for F$_2$ are shown in the left of \ref{fig:F_GBCF_cation} and \ref{fig:F_GBCF_anion}. 
The plots on the right side of the figures are the percent change in the number of valence electrons in each gradient bundle using

\begin{align}
\label{eqn:RMF_percent_minus}
f^-_{GB,\%} &= 100 \left(  \frac{\int_{\omega_i} \rho_V(\bm{r},N)dV - \int_{\omega^-_i} \rho_V(\bm{r}, N-1)dV}{ \int_{\omega_i}\rho_V(\bm{r},N)dV} \right) \textup{, and} \\
f^+_{GB,\%} &= 100 \left( \frac{  \int_{\omega^-_i} \rho_V(\bm{r}, N+1)dV - \int_{\omega_i} \rho_V(\bm{r},N)dV}{ \int_{\omega^-_i} \rho_V(\bm{r}, N+1)dV} \right).
\end{align}

\noindent where $\rho_V$ indicates valence electron density which is calculated using the methods described in chapter \ref{cha:BDE}. 
The percent change in electron count may be a more meaningful number than $f_{GB}$ when using rotational gradient bundles, as it accounts for the fact that gradient bundles near the internuclear axis have a much smaller volume and will therefore always have a smaller change in electron count than the gradient bundles near the center of each atomic basin.
\addsymbol{Valence electron density}{$\rho_V$}

\begin{figure}
	\begin{center}
		\subfigure{
			\resizebox{.45\textwidth}{!}{\includegraphics{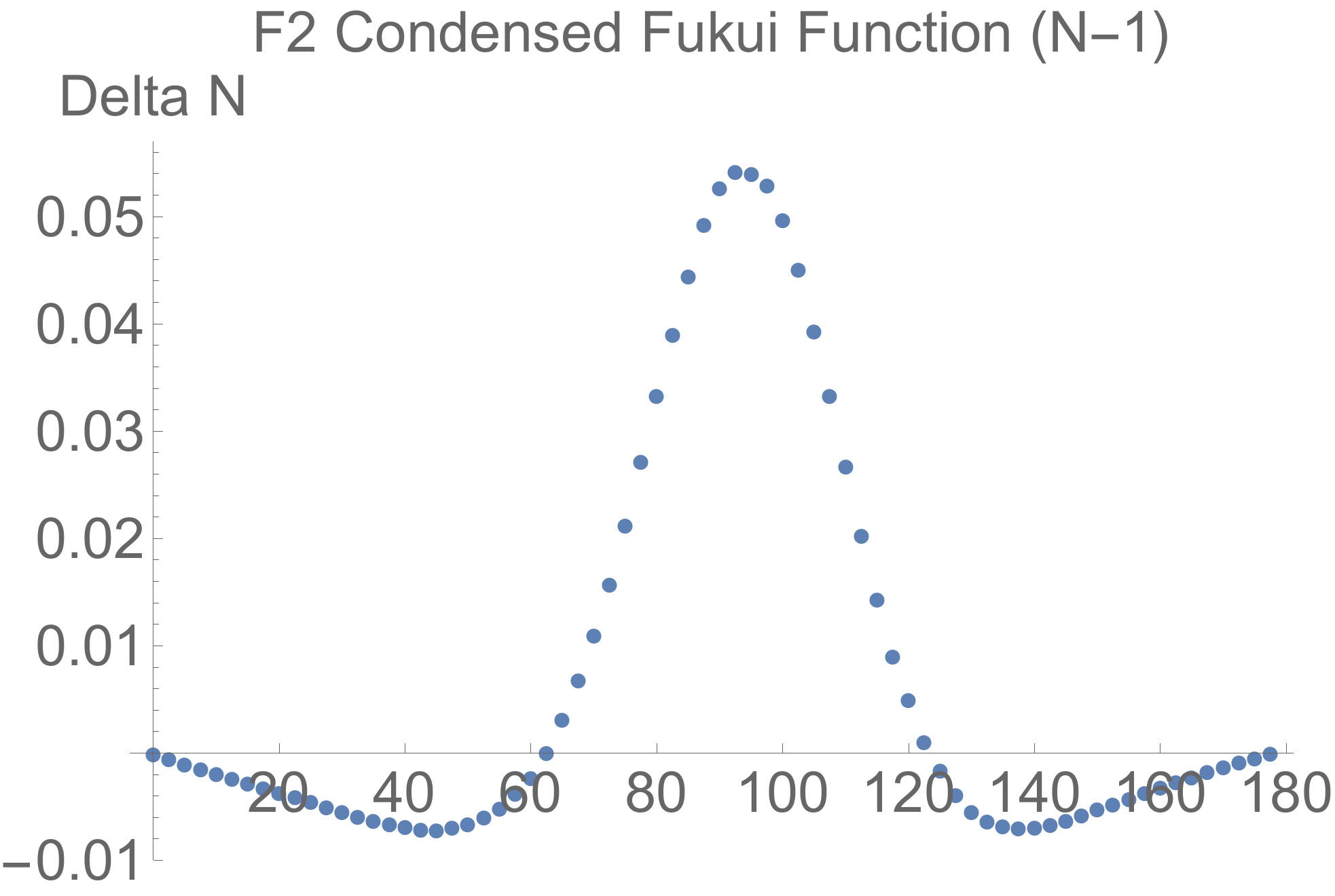}}
		} 
		\subfigure{
			\resizebox{.45\textwidth}{!}{\includegraphics{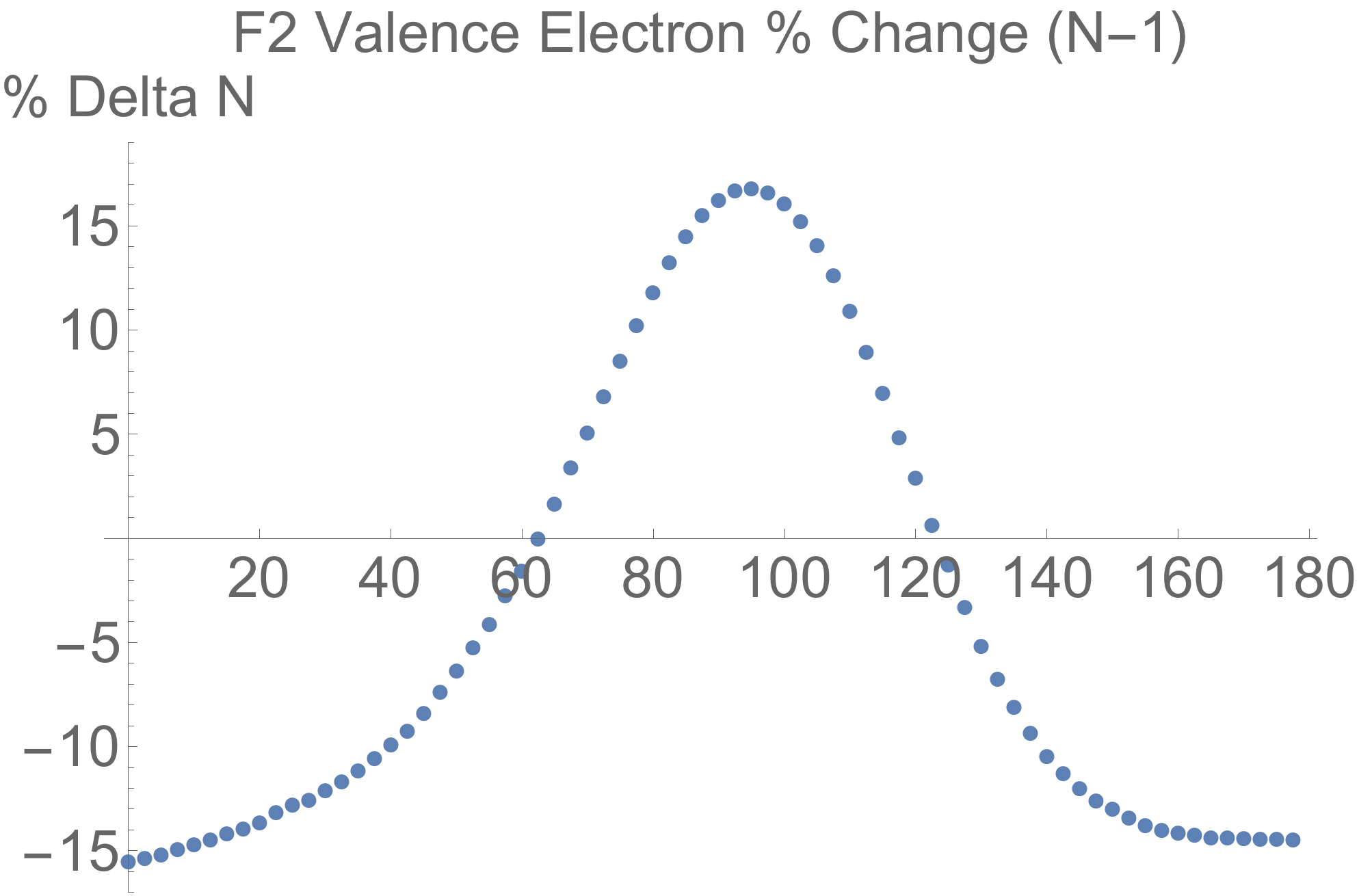}}
		}
		\caption{\label{fig:F_GBCF_cation} Plots of $f^-_{GB}$ (left) and $f^-_{GB,\%}$ (right) around one fluorine atom in F$_2$. Gradient bundle angles along the x-axes are measured with respect to the bond path, such that the GB at 0$^{\circ}$ contains the bond path and is bounded by the interatomic surface. }
	\end{center}
\end{figure}

 \begin{figure}
	\begin{center}
		\subfigure{
			\resizebox{.45\textwidth}{!}{\includegraphics{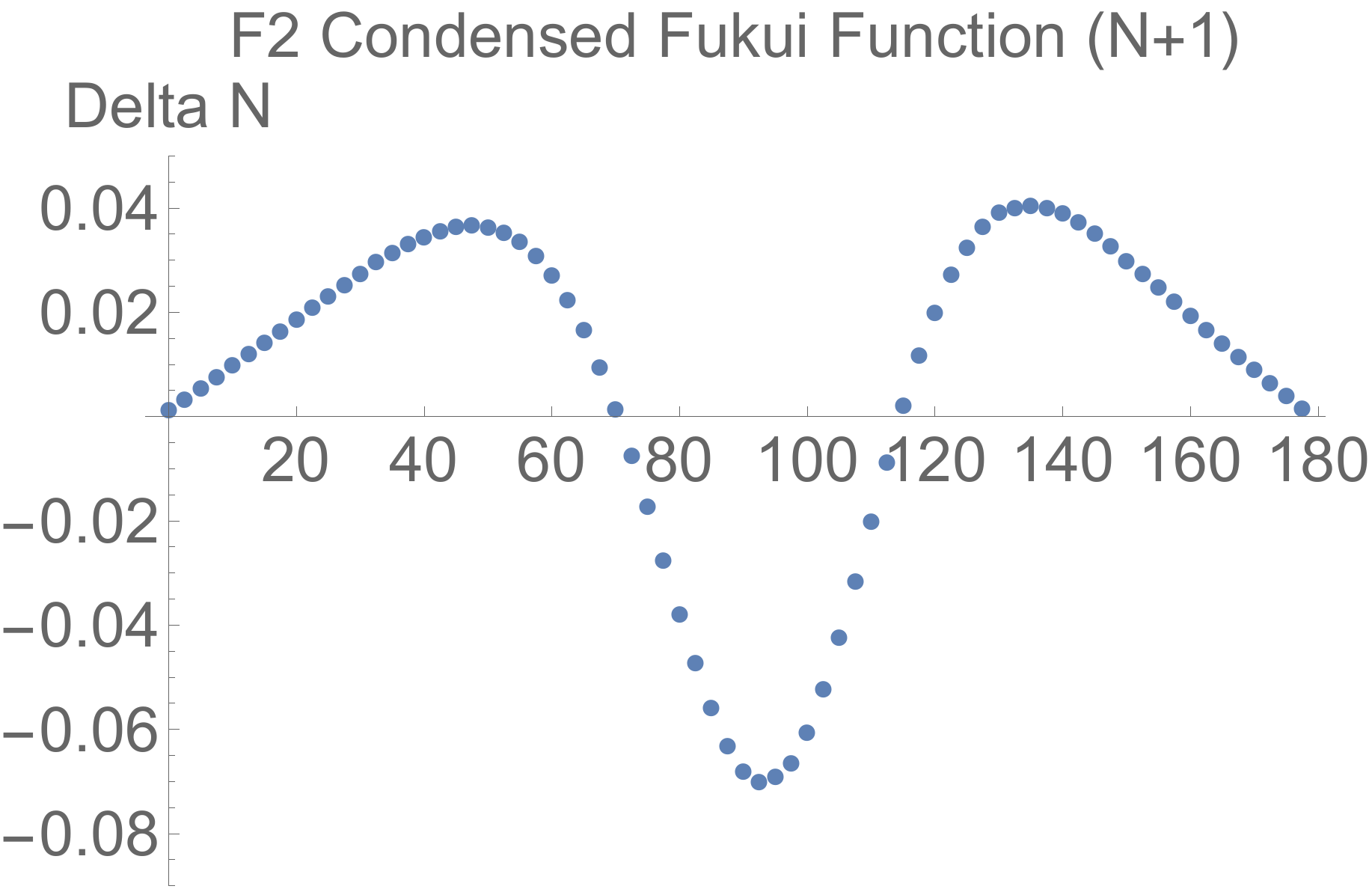}}
		} 
		\subfigure{
			\resizebox{.45\textwidth}{!}{\includegraphics{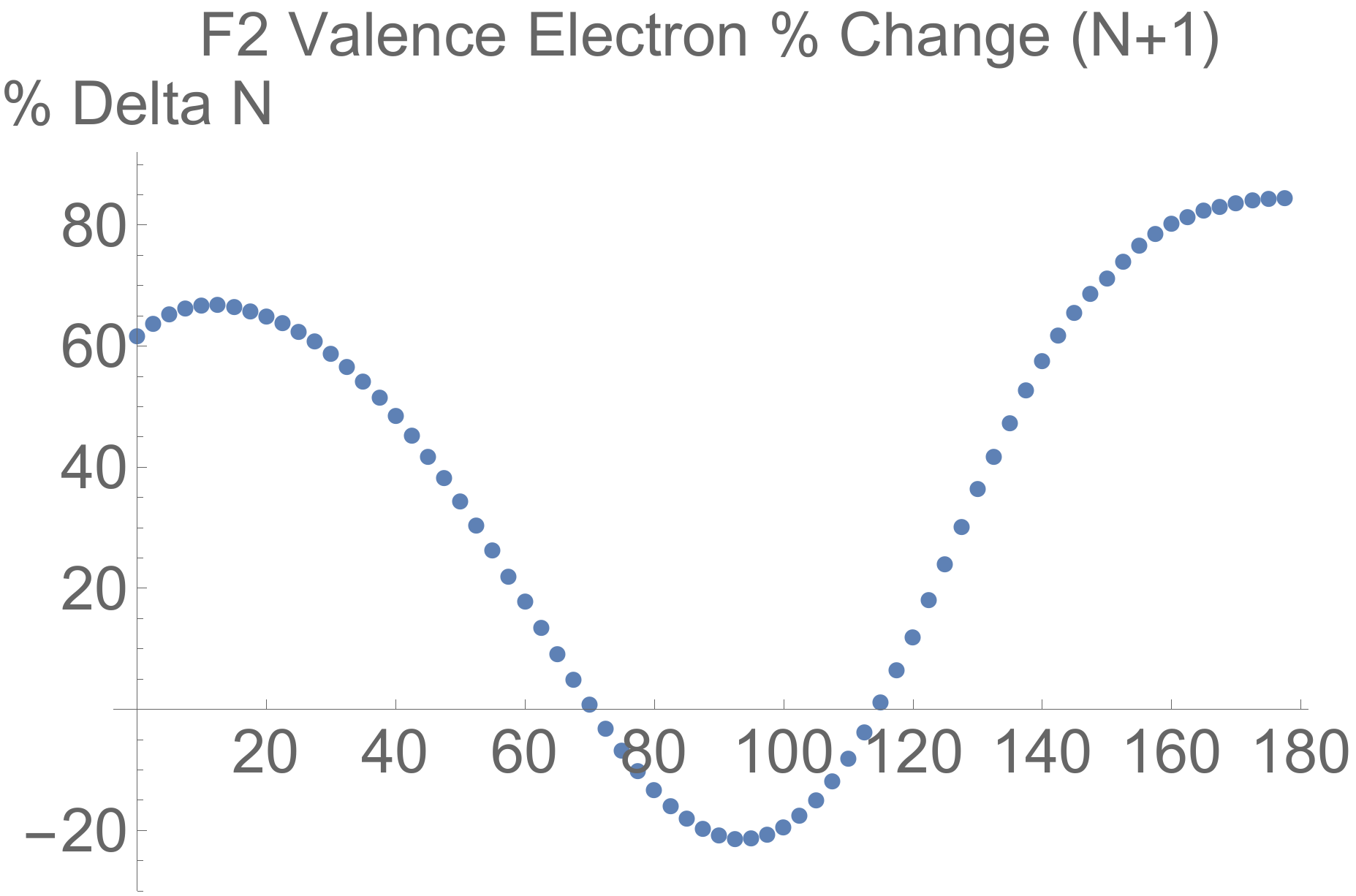}}
		}
		\caption{\label{fig:F_GBCF_anion}  Plots of $f^+_{GB}$ (left) and $f^+_{GB,\%}$ (right) for one fluorine atom in F$_2$. }
	\end{center}
\end{figure}

 \begin{figure}
	\begin{center}
		\subfigure{
			\resizebox{.45\textwidth}{!}{\includegraphics{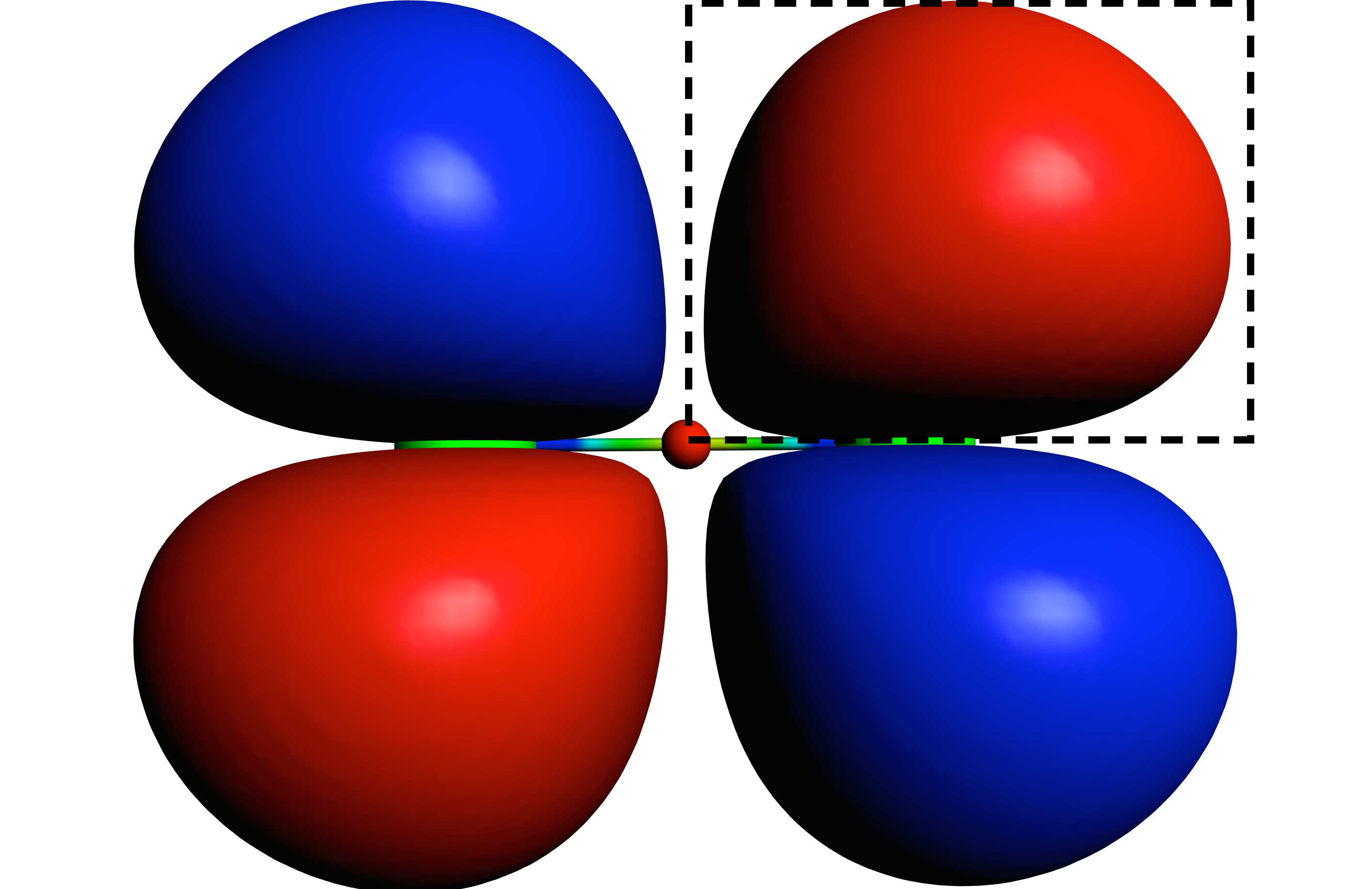}}
		} 
		\subfigure{
			\resizebox{.45\textwidth}{!}{\includegraphics{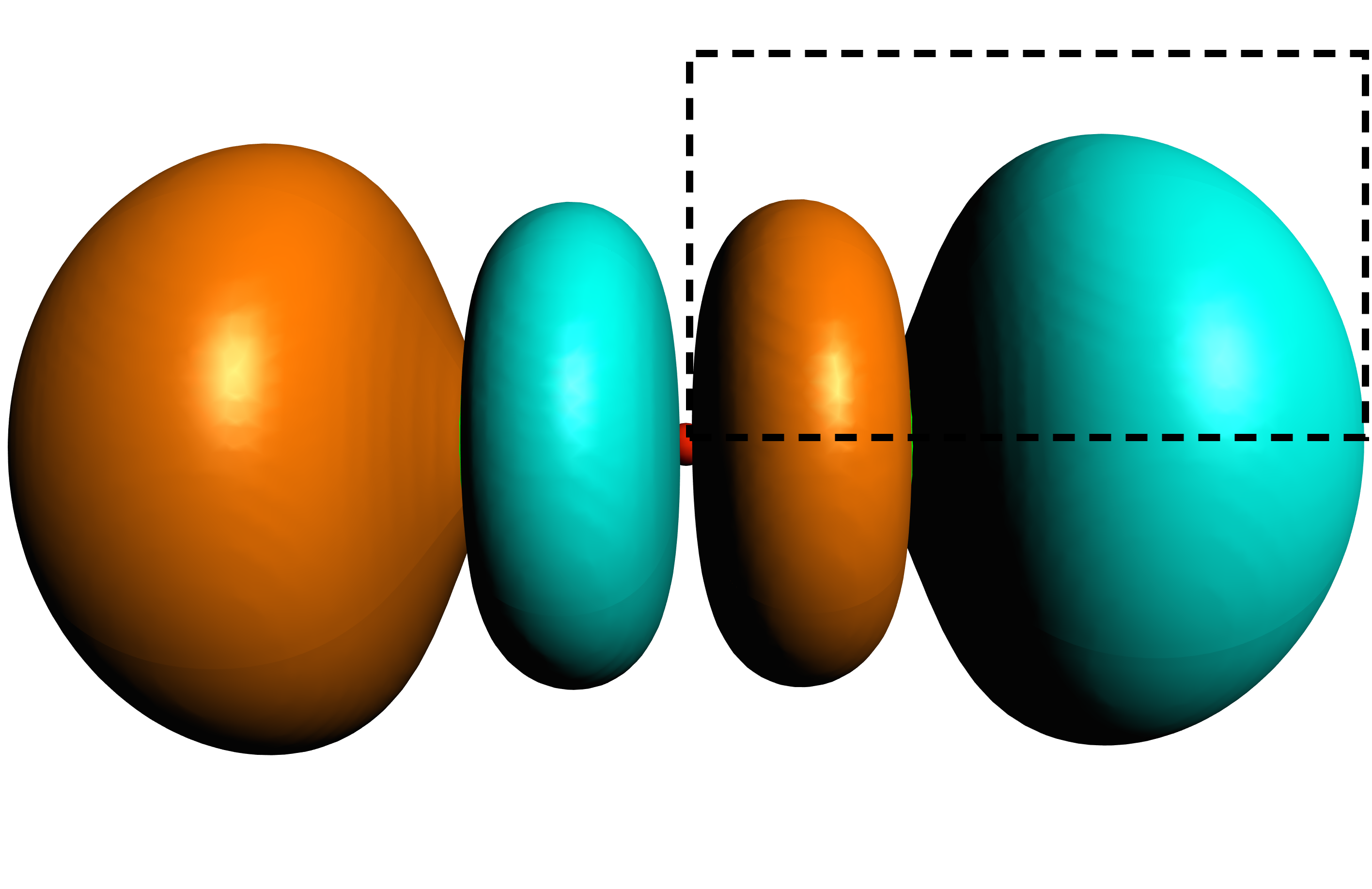}}
		}
		\caption{\label{fig:F_MO} The HOMO (left) and LUMO (right) for a neutral fluorine molecule. Black dashed boxes indicate the regions of the orbital plots that geometrically match the Fukui function plots in \ref{fig:F_GBCF_cation} and \ref{fig:F_GBCF_anion}.}
	\end{center}
\end{figure}

Maxima in $f^-_{GB,\%}$ for molecular fluorine match the frontier orbital picture of reactivity. 
Electrophilic attack on F$_2$ is predicted to occur 90$^{\circ}$ from the bond path, near the center of each atomic basin, while the Fukui function is negative near the bond path and interatomic surface (0$^{\circ}$) and pointing away from the bond path (180$^{\circ}$).
This indicates that if an electron is removed from F$_2$,  charge density will actually be added to the regions with $-f_{GB}$.
The HOMO of F$_2$ is shown in the left side of \ref{fig:F_MO} and has a lobe in the center of each fluorine atomic basin.
There is a nodal plane through the center of the molecule that contains the IAS and bond path.
Orbital lobes are present where maxima in $f^-_{GB,\%}$ occur and nodal planes match the locations of $-f^-_{GB,\%}$.

The LUMO of fluorine is pictured on the right side of \ref{fig:F_MO} and has two lobes around each fluorine atom, and a nodal plane through the center of the atomic basins, normal to the bond path. 
This is also recovered in the plot of $f^+_{GB,\%}$ in \ref{fig:F_GBCF_anion}, which shows that electron density will be added to both sides of the atomic basin, and there will be a depletion of charge in the center of the basin, 90$^{\circ}$ from the bond path.
The majority of the charge density will add to the gradient bundles away from the bond path, closer to 180$^{\circ}$, which matches where the largest part of the LUMO is located.
The nodal plane through the interatomic surface is also recovered as $f^+_{GB,\%}$ has local minimum at 0$^{\circ}$.

\ref{fig:N_GBCF} shows the percent change in valence electron count in gradient bundles for molecular nitrogen. 
Again, the most reactive sites for electrophilic and nucleophilic attack on nitrogen as predicted by maxima in $f_{GB,\%}$ match the HOMO and LUMO, respectively (\ref{fig:N_MO}).
$f^-_{GB,\%}$ has two maxima, one along the interatomic surface (0$^{\circ}$), and another at 180$^{\circ}$, matching the locations of the two lobes of the HOMO.
The Fukui function is negative through the center of the atom, matching the nodal plane through the nuclei.
There is a single maximum in $f^+_{GB,\%}$, 120$^{\circ}$ from the bond path, which is in line with the single LUMO lobe in each atomic basin. 
There are two regions with $-f_{GB}$, which match the two orthogonal nodal planes through the bond path and interatomic surface.

 \begin{figure}
	\begin{center}
		\subfigure{
			\resizebox{.48\textwidth}{!}{\includegraphics{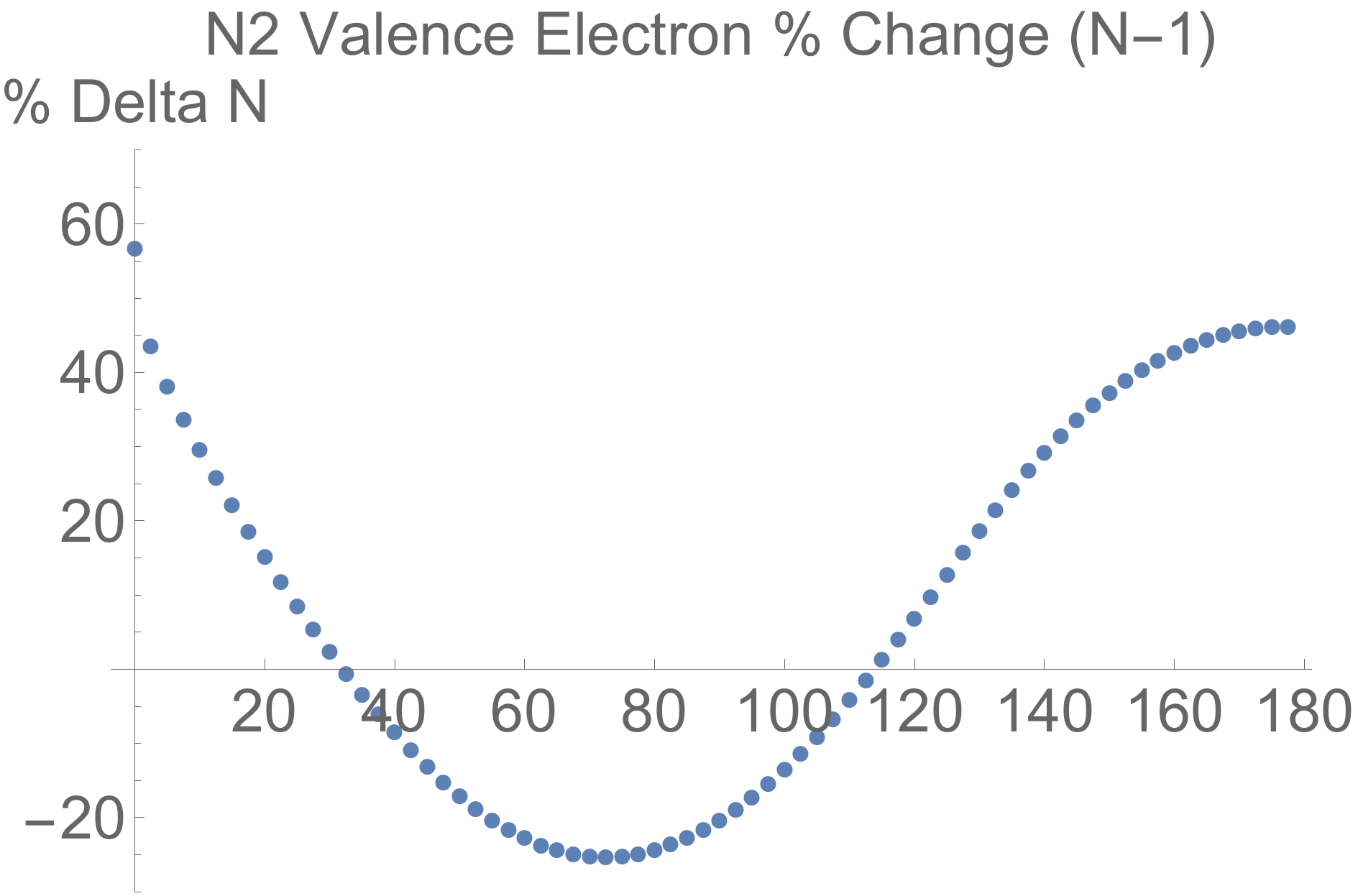}}
		} 
		\subfigure{
			\resizebox{.48\textwidth}{!}{\includegraphics{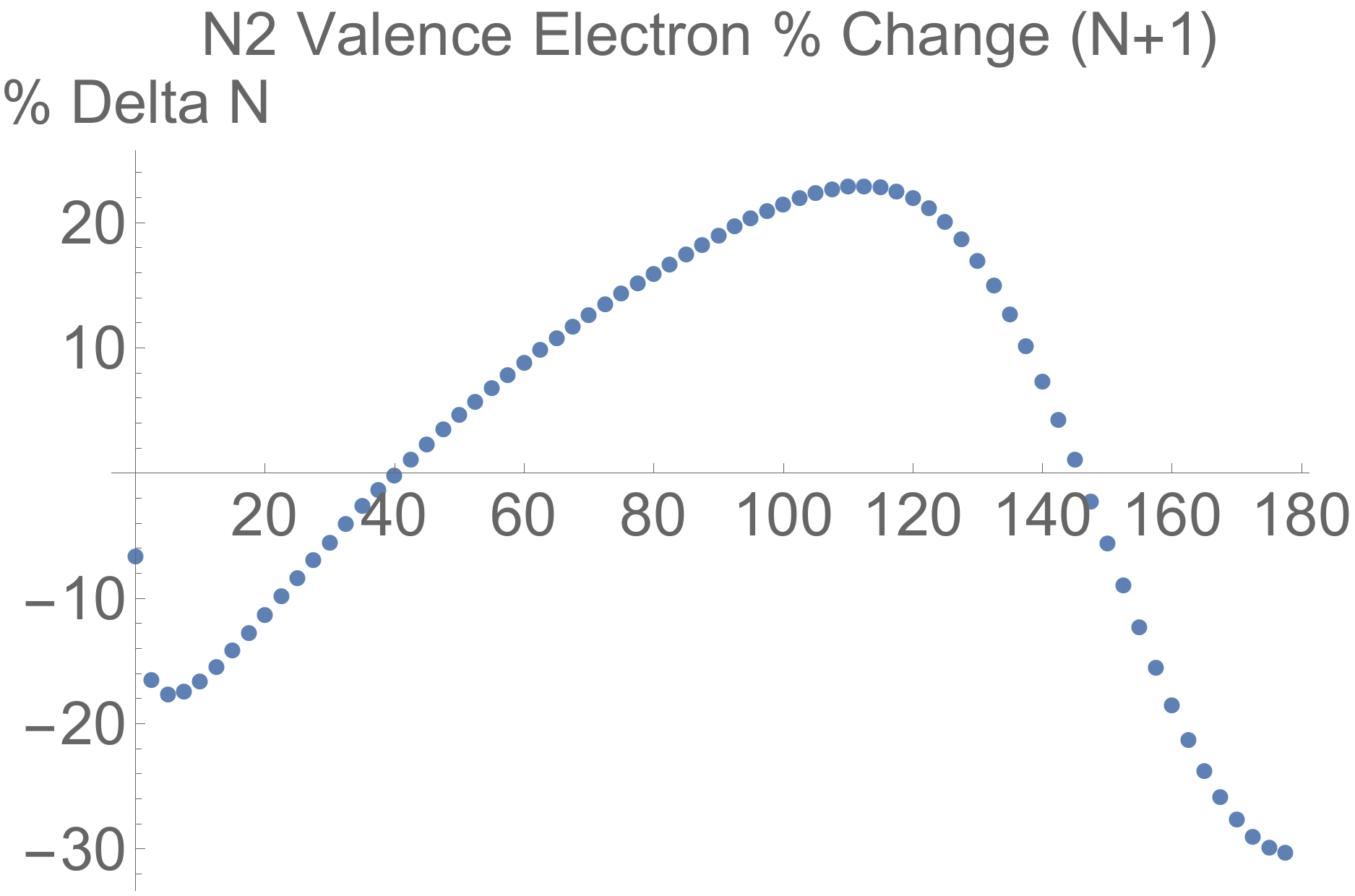}}
		}
		\caption{\label{fig:N_GBCF} Plots of $f^-_{GB,\%}$ (left) and $f^+_{GB,\%}$ (right) around one nitrogen atom in N$_2$. }
			\end{center}
\end{figure}

\begin{figure}
	\begin{center}
		\subfigure{
			\resizebox{.45\textwidth}{!}{\includegraphics{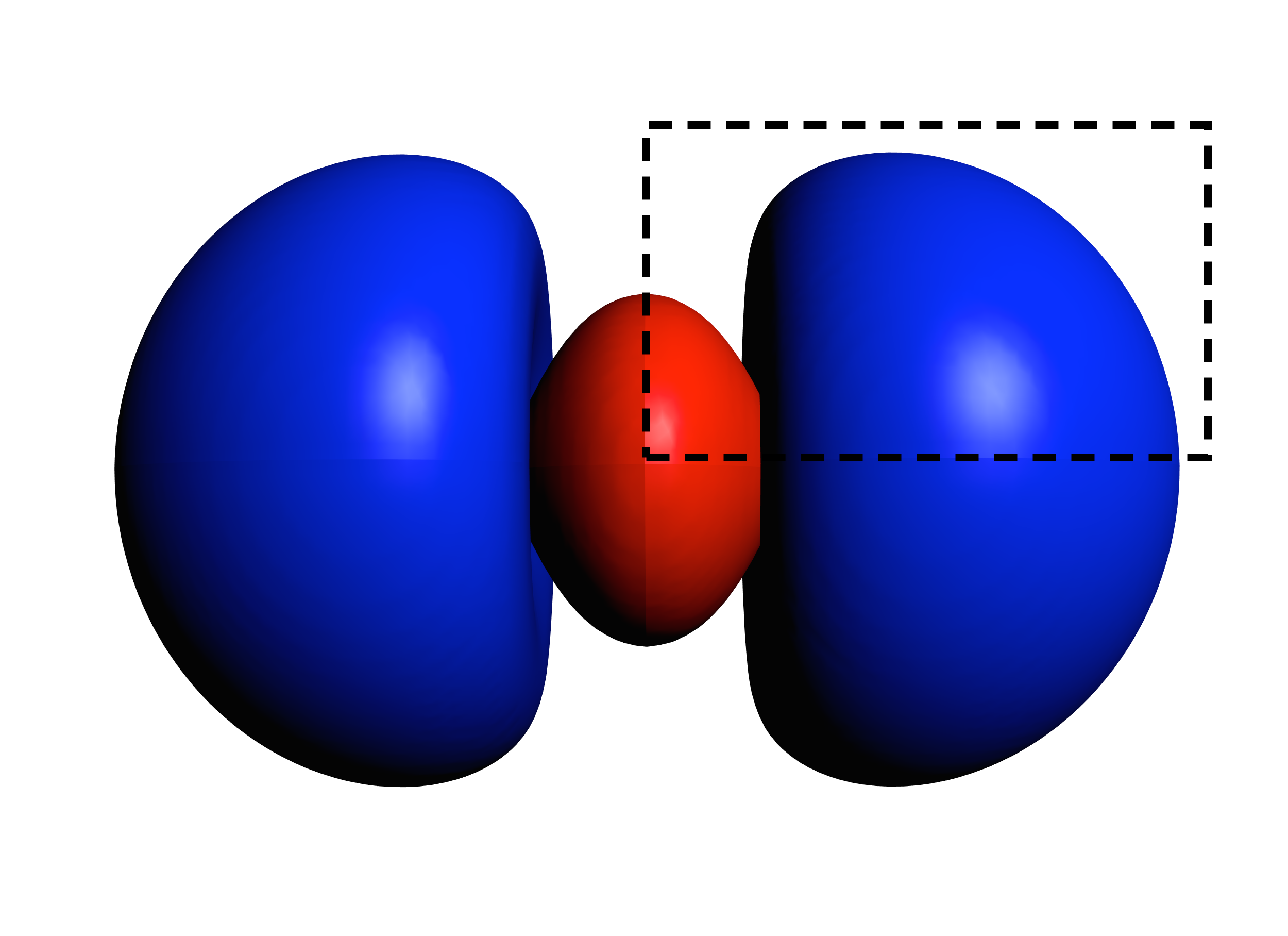}}
		} 
		\subfigure{
			\resizebox{.45\textwidth}{!}{\includegraphics{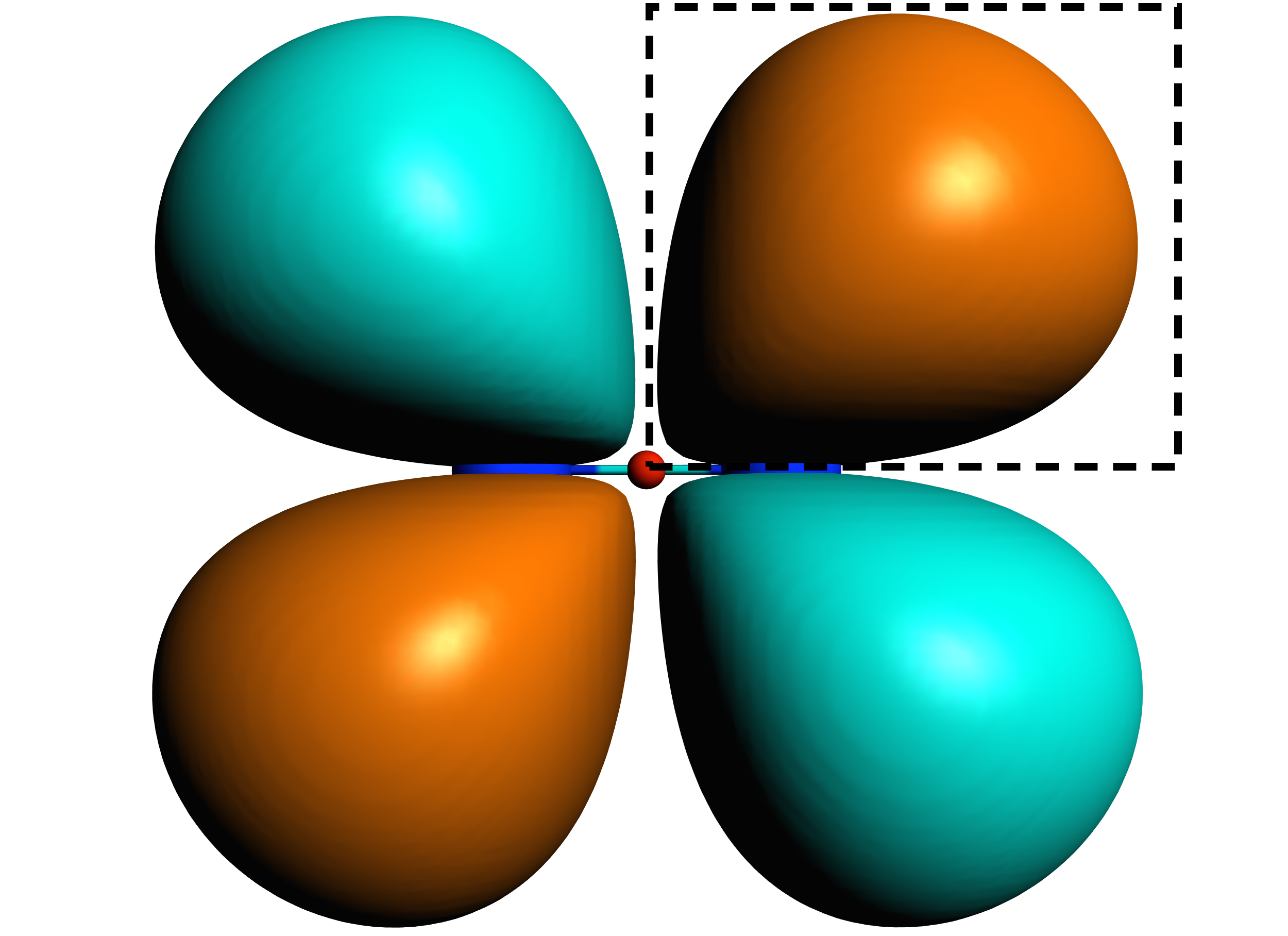}}
		}
		\caption{\label{fig:N_MO} The HOMO (left) and LUMO (right) for a neutral nitrogen molecule. Black dashed boxes indicate the regions of the orbital plots that geometrically match the Fukui function plots in \ref{fig:N_GBCF}.}
	\end{center}
\end{figure}

Condensed Fukui functions may be particularly useful for locating regions where the Fukui function is negative. 
It was originally argued that that electron density could not be removed from any regions in a molecule when electrons were added to the system, and vice versa, and therefore negative Fukui functions were not meaningful, but were the result of inaccurate calculation methods \cite{Bulat_NegFukui_2004, Fuentealba_NegFukui_2000, Roy_NegFukui_1999}. 
In the frozen orbital approximation, this is indeed impossible, as the HOMO and LUMO only show regions where electrons will be lost or gained, respectively. 
There is no ``negative" HOMO or LUMO.

Recently,  it has not only been accepted that negative Fukui functions exist, but it is speculated that they may be useful in finding redox induced electron transfer (RIET) molecules \cite{Ayers_NegFukui_2006, Ayers_NegFukui_2007, Ayers_NegFukui_2012}.
A RIET molecule has the property that when the molecule as a whole is oxidized (removal of electrons), one of the atoms or functional groups in the molecule is reduced (gains electrons). 
Very few  molecules have been shown to have the RIET property, since it is rare for the integration of $\rho(\bm{r})$ over an entire atomic region to change in the direction opposite of the entire molecule.
The same principle works for gradient bundles however, rather than just for atomic basins, and is much more common.
\addabbreviation{Redox induced electron transfer}{RIET}

Regions of negative Fukui functions have been rationalized by orbital relaxation. 
When an electron is removed from a molecule, and the orbitals are allowed to relax, the orbitals that are still occupied contract toward the nucleus \cite{Ayers_relaxation_2002}.
This causes electron density to actually increase in the nodal surfaces of the HOMO.
The volume over which the Fukui function is negative varies from molecule to molecule. 
It has been postulated that molecules whose chemistry is not highly controlled by its frontier molecular orbitals will have more prominent negative Fukui function regions \cite{Ayers_NegFukui_2012}.

While the gradient bundle condensed Fukui functions calculated here match the frontier orbital picture, this may not necessarily always be true. 
The HOMO and LUMO neglect orbital relaxation and are only an approximation to where electron density will change most. 
A more accurate orbital picture of reactivity can be gained by calculating the frontier orbitals using the Fukui matrix.
The Fukui matrix is defined as the derivative of the density matrix \cite{Bultinck_fukui_2011}. 
The density matrix can be expressed in terms of its basis functions, $\varphi$, as

\begin{equation}
\label{densitymatrix}
P_{ij} = \langle \varphi_i(\bm{r}) | \rho(\bm{r,r'}) | \varphi_j(\bm{r'}) \rangle
\end{equation}
\addsymbol{Density matrix}{$P$}
\addsymbol{Fukui matrix}{$\pmb{\bm{\textup{f}}}$}

\noindent giving rise to the Fukui matrix

\begin{equation}
\label{fukuimatrix}
\pmb{\bm{\textup{f}}} = \left[ f_{ij} = \left( \frac{\partial P_{ij}}{\partial N} \right)_{\nu(\bm{r})}  \right]
\end{equation}

\noindent where the Fukui function is given by the diagonal of $\pmb{\bm{\textup{f}}}$. 
In general, the Fukui matrix is calculated using a finite difference just as the Fukui function is.
The Fukui matrix can be diagonalized and decomposed as 

\begin{equation}
\pmb{\bm{\textup{f}}}^{\pm} = \sum_i \phi_i \bm{F}^{\pm}_{ii} \phi^T_i
\end{equation}

\noindent where $\bm{F}_{ii}$ are known as the Fukui eigenvalues and $\phi_i$ are the Fukui orbitals.
\addsymbol{Fukui eigenvalues}{$\bm{F}_{ii}$}
\addsymbol{Fukui orbitals}{$\phi_i$}
\addsymbol{Basis functions}{$\varphi$}

For instances when the the frozen orbital approximation is completely accurate, there will be a single Fukui eigenvalue equal to one that completely represents the frontier molecular orbital. 
In general, one eigenvalue will be slightly less than one, and there will be other eigenfunctions that contribute to the actual regions where electron density is added or lost. 
Furthermore, Fukui eigenvalues can be negative, indicating orbitals that correspond to regions where negative Fukui functions are observed.
We hypothesize that results from gradient bundle condensed Fukui functions will match the Fukui orbitals well, including negative regions. 
We plan on also extending this study to nonlinear molecules using the method advancements described in section \ref{sec:advancements}.

\subsection{Predicting Site Reactivity from the Ground State Charge Density}

Calculating gradient bundle condensed Fukui functions has proven to be an indicator of chemical reactivity, but this method is not always computationally advantageous. 
When the electron count in a neutral system changes, electrons become unpaired and unrestricted calculations must be performed.
These calculations are not always straightforward and often difficult to converge.
Furthermore, from a theoretical standpoint they should not be necessary. 
The first theorem of DFT states that all properties are obtainable from the ground state charge density, without the need for a second calculation for the charged molecule.
To this end, we are currently exploring ways of finding the reactive regions of a molecule from the ground state charge density alone.

Our method is to use differential geometry to observe the curvature and torsion of neighboring gradient bundles around each nuclei in a molecule. 
A related study was performed calculating the curvature of atomic basin surfaces, in an attempt to associate the curvature of interatomic surfaces with properties such as electronegativity \cite{Popelier_difgeom_1996}.
Any surface in 3D space can be characterized by its Gaussian curvature, which is the product of the two principle curvatures at a point on the surface.
The total curvature of a surface is given by integration of the Gaussian curvature over the entire surface. 
Rather than directly calculating the curvature of ZFSs bounding gradient bundles (though this is also a path we plan to explore in future work), we are currently observing the effect of the differential geometry of the gradient vector field of $\rho(\bm{r})$ that manifests in the size of gradient bundles.
Specifically, we are calculating the cross-sectional area of gradient bundles at an isovalue of $\rho(\bm{r})$.

For linear molecules, this procedure is greatly simplified and can be approximated by simply measuring the width of rotational gradient bundles. 
Previous work plotting the Fukui function to predict reactivity has used the van der Waals radii of atoms as an isosurface cut-off with good results \cite{Ayers_fukui_2005}.
\ref{fig:N_divergence} shows the width of gradient bundles around a nitrogen atom in N$_2$ using the van der Waals radii defined at $\rho(\bm{r})$ = 0.002 $e/bohr^3$.
We find that electron density is most likely to be removed from the widest gradient bundles, i.e., the bundles with the largest surface area.

\begin{figure}
	\begin{center}
		\subfigure{
			\resizebox{.55\textwidth}{!}{\includegraphics{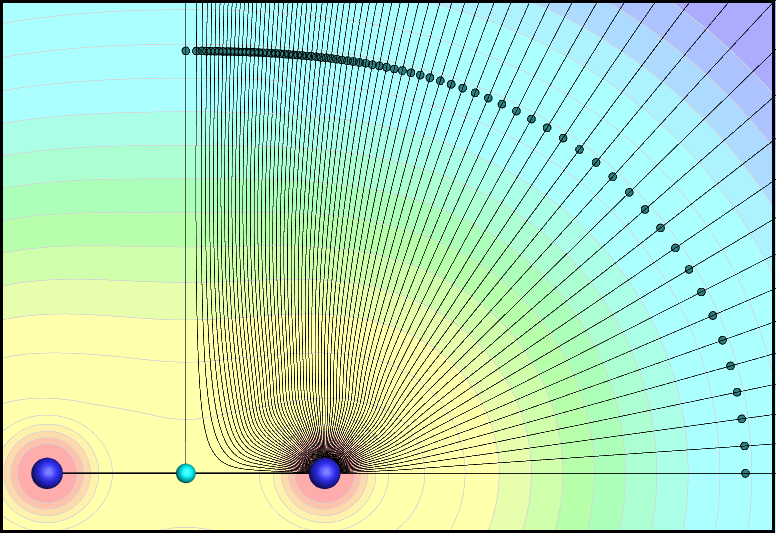}}
		} 
		\subfigure{
			\resizebox{.45\textwidth}{!}{\includegraphics{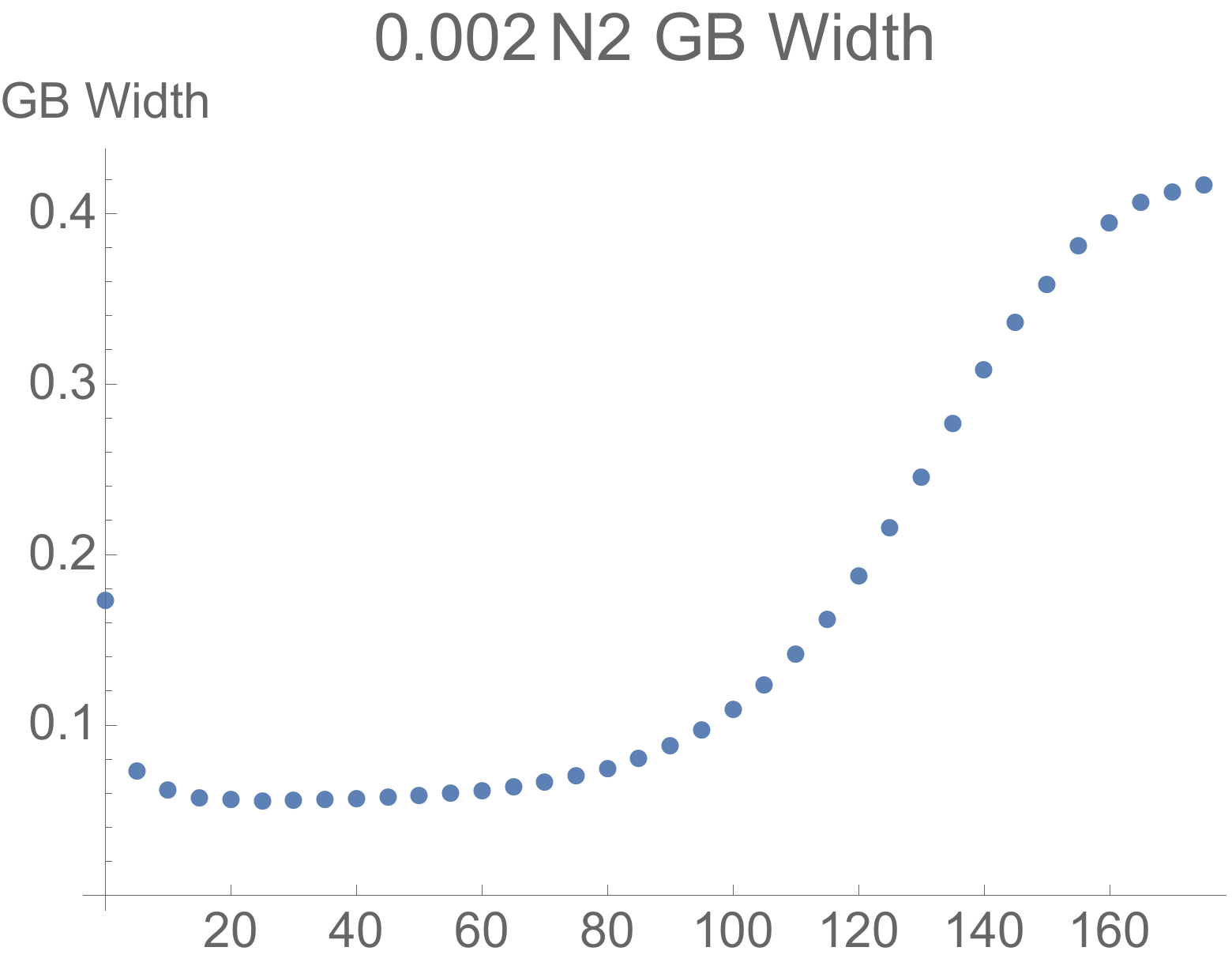}}
		}
		\caption{\label{fig:N_divergence} Left: A cutplane of rotational gradient bundles in N$_2$. Black circles indicate points on the bounding zero-flux surfaces at an isovalue of $\rho(\bm{r}) = 0.002$. Right: The distance between GB points on the left, which we refer to as the width of the gradient bundles. The x-axis displays the gradient bundle angle, where 0$^{\circ}$ is the gradient bundle bounded by the interatomic surface and containing the bond path.}
	\end{center}
\end{figure}

The rational for this method is based on lower kinetic energy being an indicator of valence electrons as discussed in chapters \ref{cha:intrinsic} and \ref{cha:BDE}.
Electrons that are less confined (have lower density) will have lower kinetic energy. 
This idea can be seen in the simple Thomas-Fermi approximation of the kinetic energy, which is proportional to $\rho(\bm{r})^{5/3}$ \cite{TF_KE}. 
Gradient bundles have increased cross-sectional area when the charge density is decreasing rapidly from the center of the gradient bundle in a direction normal to the gradient paths in the bundle (i.e. have high Gaussian curvature).
This is similar to calculating the sign of the Laplacian (which provides the direction the charge density is changing in the fastest), but goes beyond a local density approximation where only $\rho(\bm{r})$ at a point is used. 
Instead, we are observing the more global property of the curvature of the charge density over surfaces which affects the volume of gradient bundles.
Since electrons in larger gradient bundles have more volume to occupy, they will move slower and have a lower average kinetic energy.
Wider rotational gradient bundles indicate regions with low kinetic energy, and therefore can be used to locate valence electrons.

The widest gradient bundles in N$_2$ are pointing in the direction opposite the bond path, 180$^{\circ}$, and there is a second maximum near the interatomic surface at 0$^{\circ}$.
This matches the two maxima in the gradient bundle condensed Fukui function and the HOMO for N$_2$, as well as the location of valence electrons provided by analysis of the kinetic energy presented in chapter \ref{cha:intrinsic}.
\ref{fig:F_divergence} shows the same analysis for F$_2$. 
In this molecule there is a single maximum indicating the most energetically favorable region to remove electron density near 100$^{\circ}$, exactly the same region that valence electrons were predicted to occupy using the kinetic energy density.
Again, this matches the single lobe of the HOMO on each fluorine atom and the single maximum in $f^-_{GB}$ in F$_2$.
The widest gradient bundles (at an isosurface of $\rho(\bm{r})$ at the van der Waals radius) indicate the gradient bundles that will lose the most electron density when an electron is removed from the molecule as a whole.

\begin{figure}
	\begin{center}
		\subfigure{
			\resizebox{.53\textwidth}{!}{\includegraphics{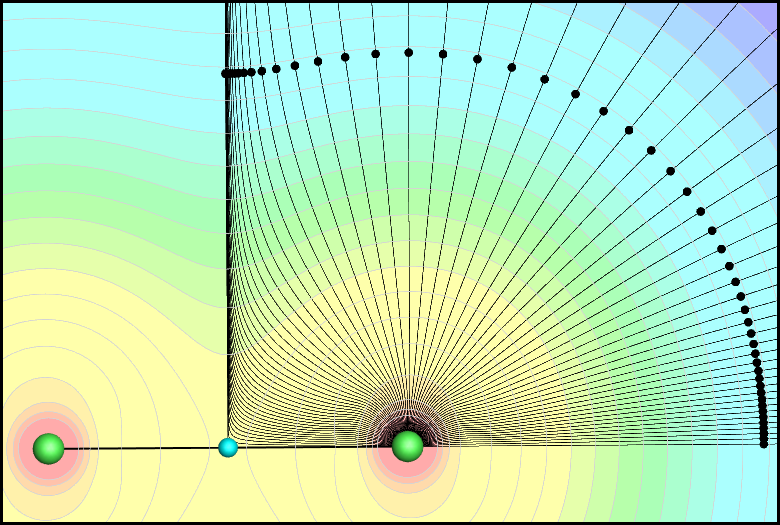}}
		} 
		\subfigure{
			\resizebox{.47\textwidth}{!}{\includegraphics{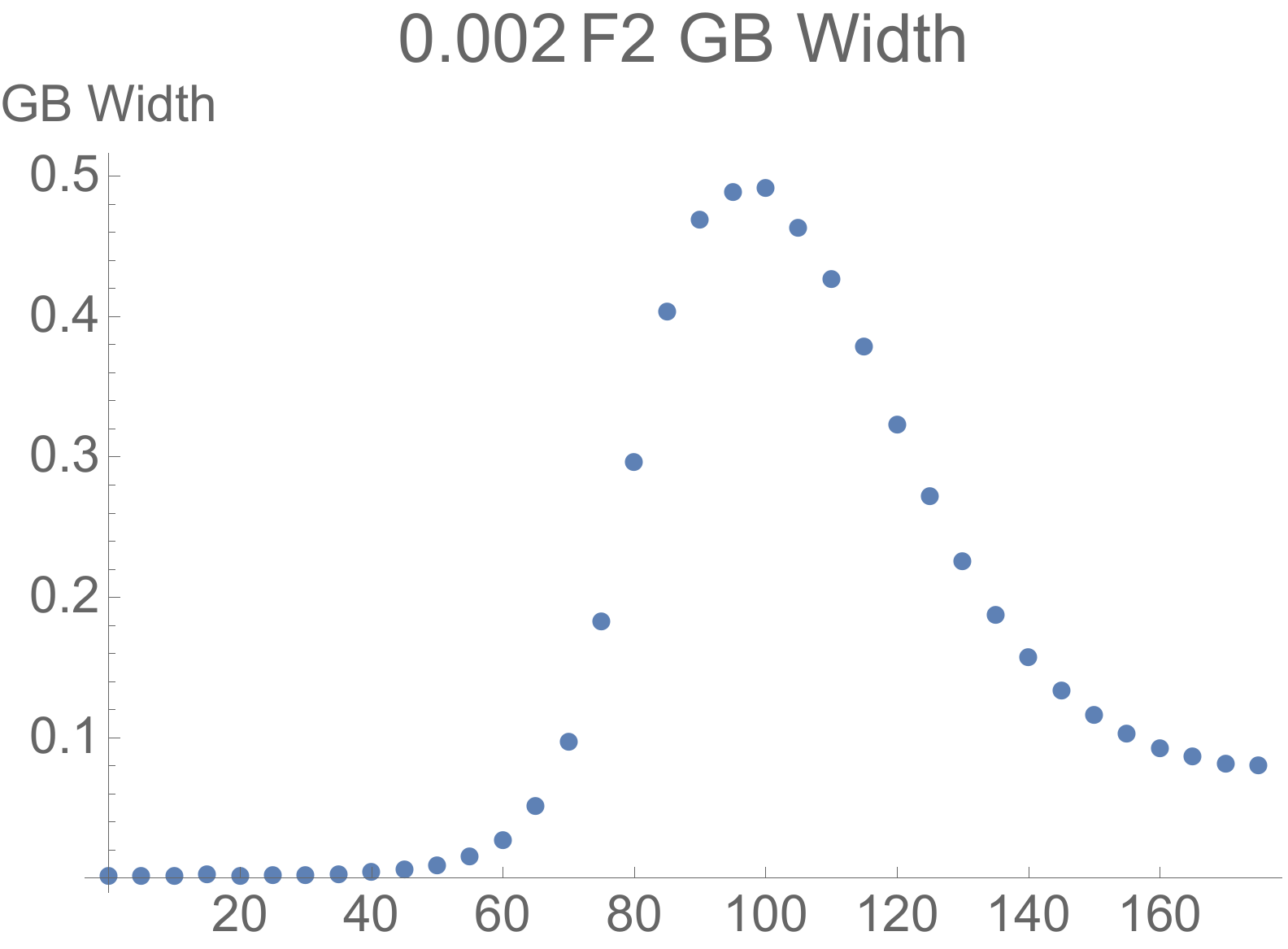}}
		}
		\caption{\label{fig:F_divergence} Left: A cutplane of rotational gradient bundles in F$_2$. Black circles indicate points on the bounding zero-flux surfaces at an isovalue of $\rho(\bm{r}) = 0.002$. Right: The width of the gradient bundles measured at the points marked in the left plot.}
	\end{center}
\end{figure}

We are currently comparing the condensed Fukui function and gradient bundle sizes for 1-iodo-2-fluoro-ethyne, the molecule studied in chapter \ref{cha:ZFS}.
ICCF has a more complex gradient bundle structure and bonding interactions than F$_2$ or N$_2$ and raises additional questions.
\ref{fig:ICCF_divergence} shows the width of rotational gradient bundles at the van der Waals radii for each atom, as well as $f^-_{GB,\%}$.
For the most part, maxima in the plots match up, showing that the size of gradient bundles is a good predictor of where electron density will be removed. 
However, the gradient bundle at 0$^{\circ}$ in C2 (along the C1--C2 interatomic axis) does not fit this pattern. 
This gradient bundle is extremely large, yet its electron count does not decrease upon creating the cation of ICCF.
In fact, this gradient bundle has a negative condensed Fukui function.

\begin{figure}
	\begin{center}
		\subfigure{
			\resizebox{\textwidth}{!}{\includegraphics{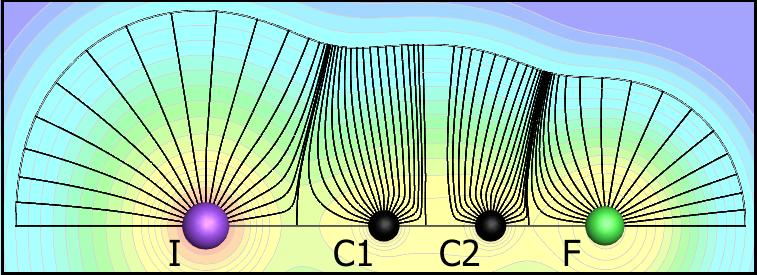}}
		} \\
		\subfigure{
			\resizebox{\textwidth}{!}{\includegraphics{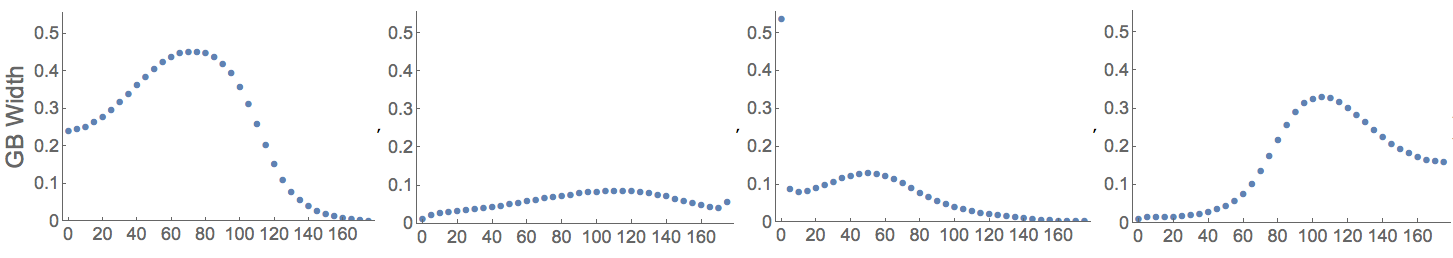}}
		} \\
		\subfigure{
			\resizebox{\textwidth}{!}{\includegraphics{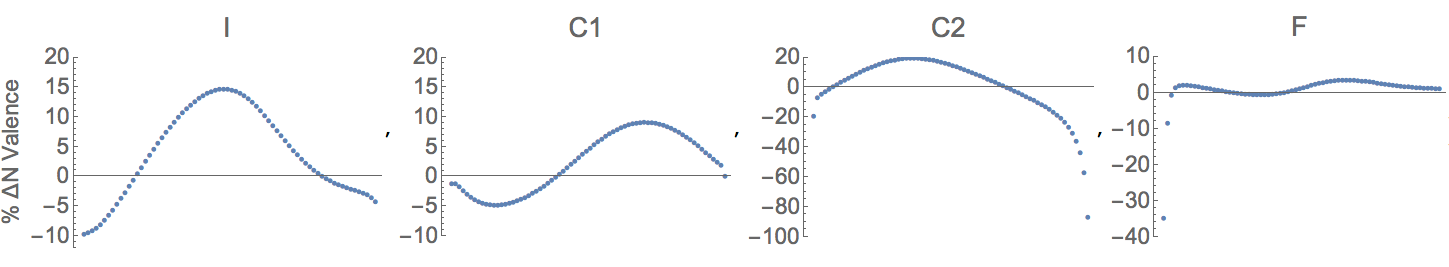}}
		}
		\caption{\label{fig:ICCF_divergence} Top: A cut plane of ZFSs bounding rotational gradient bundles in ICCF. The isosurface of $\rho(\bm{r})$ = 0.002 is traced with a black line to show where the width of gradient bundles is measured. 
Middle: The width of each GB at the van der Waals radius. 
Bottom: $f^-_{GB,\%}$ for each atom in ICCF.} 
	\end{center}
\end{figure}

Further investigation of this gradient bundle in the C2 atomic basin is underway. 
The large difference in electronegativity between the fluorine and iodine atoms on the molecule may be what is causing the unusual gradient bundle to exist, since we do not observe the same feature in the gradient of the charge density for HCCH, HCCI or HCCF.

We are also working on the more difficult task of predicting which gradient bundles will increase their electron counts when an electron is added to a molecule. 
How to find the LUMO, so to say, in the ground state charge density is an open question in the field of CDFT. 
The ground state charge density is calculated as the sum of all occupied orbitals in the molecule, which includes the HOMO.
The LUMO is unoccupied, making it difficult to understand where this orbital would be visible in the ground state charge density.
However, $\rho(\bm{r})$ contains all properties of a molecule, and therefore regions where it is most energetically favorable to add electrons must be identifiable using only $\rho(\bm{r})$. 
It is just a matter of discerning a method for capturing this property within the charge density structure.

We are currently investigating if a relationship exists between the LUMO and the rate at which gradient bundles are changing size as we move around each nuclei.
Preliminary results show that electron density is added to one of two areas.
One, the gradient bundles with the smallest cross-sectional areas match the LUMO.
Or two, the regions with the steepest slope in the gradient bundle size plots, i.e. the gradient bundles that are changing size most quickly as we move radially around each nuclei.
A full study of the differential geometry of the gradient bundles, including their integrated curvatures, may be necessary.

\subsection{Gradient Bundle Analysis Method Improvements} 
\label{sec:advancements}

It has been argued that studying the differential geometry of the full gradient vector field of $\rho(\bm{r})$ would provide useful chemical information \cite{Popelier_book}.
Bader stated that ``further study of the gradient vector field of the electron density leads to a  complete theory of structure..." \cite{Bader_FunctionalGroups}.
While we have shown that studying the full gradient vector field provides additional information to standard QTAIM analysis, it is also apparent that this is a highly complex task from both a mathematical and computational standpoint.
This is why our methods have only been applied to linear molecules. 
In order to apply GBA  methods to more complex systems such as enzymes and metals, software and theoretical developments must be made.

The main obstacle in using gradient bundle analysis methods on non-linear systems is performing volume integrations.
The program that we currently use to perform GBA, Tecplot \cite{Tecplot}, does not have a volume meshing algorithm.
We are in the process of writing add-on packages for Tecplot that will enable volume integrations, rather than using our current ``trick" of multiplying properties in a molecular plane by $2 \pi r$, where r is the distance from the internuclear axis (see section \ref{sec:methods_intrinsic}).

There are some recently developed programs that have algorithms for calculating atomic basin properties (see \cite{Popelier_integration_2011, Popelier_integration_2007} and references therein), but volume integration is still a major challenge in the entire field of QTAIM. 
Most of the available integration routines take advantage of the fact the the charge density and energy both decrease quickly as you move away from the nuclear CP, so most of the properties of interest are contained within the $\beta$ sphere.
The $\beta$ sphere is defined as the largest sphere centered on a nuclear CP that does not extend past an interatomic surface in any direction. 
Therefore, it will be completely contained within the atomic basin but will generally not include parts of the basin near the bond critical points themselves.
Since gradient bundles are not centered on nuclei, $\beta$ sphere algorithms can not be utilized.
Instead, a full volume meshing and quadrature integration method must be implemented.

Once volume integration of gradient bundles is possible, we will also be able to extend the work predicting site reactivity based on the size of gradient bundles. 
Rather than simply calculating the width of gradient bundles, we can compare the surface area of gradient bundles defined by the triangulation of a sphere around nuclear CPs as was shown in \ref{fig:SphereGB}. 
We plan to have one gradient bundle that contains the bond path of interest and then create a stereographic projection of gradient bundle properties (such as electron count and kinetic energy) centered around the bond path gradient bundle. 
This will enable us to see how gradient bundle properties vary in three-dimensions around a bonding interaction.
Finally, we would like to test all of the gradient bundle methods described in this thesis using non-DFT methods.
When these advancements to gradient bundle analysis are complete, we will be able to complete the projects discussed in the remaining sections of this chapter.

\subsection{Applications to Aromaticity}
\label{aromaticity}

The concept of aromaticity is often linked to electron delocalization.
Electron delocalization is not an observable property however, meaning there is not a unique definition of aromaticity \cite{poater_AIM_ELF}.
Multiple theories try to describe and quantify delocalization of electrons based on the weighting of resonance structures \cite{Parrondo_resonance_1995}, energy calculations from natural bonding orbitals \cite{Giuffreda_NBO_delocalization_2004}, loge theory \cite{Bader_loge_1974}, and the Laplacian of the electron charge density \cite{Shahbazian_aromaticity_2011}, to name a few examples.
There are a variety of aromaticity indicators based on these principles and others, such as the geometry-based harmonic oscillator model of aromaticity (HOMA) \cite{Krygowski_HOMA_1993} and the magnetic-based nucleus-independent chemical shift (NICS) \cite{Schleyer_NICS_1996}.

The use of aromaticity indicators based on QTAIM and ELF principles has been reviewed by Poater \textit{et al.} \cite{poater_AIM_ELF}.
The delocalization index (DI) (see section \ref{sec:DI}) between atoms in aromatic compounds \cite{Bader_DI_fermi_1996}, as well as the value of $\rho(\bm{r})$ and $\nabla \rho(\bm{r})$ at bond and ring CPs, has been found to often correlate to other aromaticity measures such as NICS and HOMA \cite{Krygowski_aromaticity_ringCP_1997}. 
Values at ($3,-1$) CPs in the gradient field of ELF have similarly been used to quantify delocalization and hence aromaticity \cite{Savin_ELF_delocalization_1996}.
Santos \textit{et al.} was able to better distinguish between many aromatic and anti-aromatic compounds using the ELF$_{\pi}$ and ELF$_{\sigma}$ values at CPs \cite{Santos_ELF, Santos_ELF_aromaticity_2005}.
However, when aromaticity values are compared using multiple indicators for the same set of molecules, it is not uncommon to have a different trend emerge depending on which indicators are used \cite{Krygowski_aromaticity_substituents_2004, Sola_aromaticity_discrepency_2004, Schleyer_aromaticity_unique_2002}.
\addabbreviation{Nucleus-independent chemical shift}{NICS}
\addabbreviation{Harmonic oscillator model of aromaticity}{HOMA}

Poater \textit{et al.} conclude their review on aromaticity by stating that, ``the lack of universal nature of all indices available until now reinforces the stated belief of multidimensional character of aromaticity..." \cite{poater_AIM_ELF}.
Studying the full gradient field of $\rho(\bm{r})$, as opposed to looking at DI between basins and values at critical points, may provide a method for capturing the multidimensional character ascribed to aromaticity.
We are currently conducting a study comparing the gradient bundles of typical aromatic and anti-aromatic compounds including benzene, furan, cyclobuta-1,3-diene, and the anion and cation of cyclopenta-1,3-diene. 
We predict that the curvature of gradient paths terminating at ring CPs in cyclic aromatic compounds will be distinct between molecules that are aromatic and anti-aromatic in character.

\subsection{Gradient Bundle Analysis in Closed Systems}
\label{sec:closed}

The brittle behavior of face-centered cubic (FCC) iridium remains an unexplained property since it was first observed decades ago.
Most FCC metals are highly ductile, whereas single crystal Ir is known to undergo brittle failure at room temperature and polycrystalline Ir undergoes intergranular fracture around 1000$^{\circ}$C \cite{Hecker_Ir_brittle_1978}.
Many contradictory theories for the underlying cause of brittleness in Ir have been proposed including segregation of impurities at grain boundaries \cite{Fortes_Ir_1967}, bond directionality \cite{Gornostyrev_Ir_2000,  Gong_Ir_2011}, and ELF values \cite{Zhao_Ir_2008, Cerny_Ir_2010}.
The search for an answer to this interesting behavior remains an area of active investigation \cite{Gong_Ir_brittle_2012}.
\addabbreviation{Face-centered cubic}{FCC}

The metallurgical community believes that the brittle behavior of iridium can be explained using the electron charge density.
We have studied the values of $\rho(\bm{r})$, $\nabla \rho(\bm{r})$, and directionality at bond and ring CPs in FCC Ir and Pt.
As in studies of ELF values at these points, the charge density values between the two metals are similar, yet their observed properties (including Cauchy pressure) are distinct.
Additionally, the full topology of the charge density, including all the edges of irreducible bundles, are identical between Ir and Pt.
However, we believe there may be some distinguishing features in the full gradient field of $\rho(\bm{r})$.

To test this hypothesis, we are performing full gradient bundle analysis on Pt and Ir, using both cluster calculations performed within ADF and full periodic structures.
We are currently qualitatively comparing the gradient fields in planes that should be important to determination of the shear moduli and hence Cauchy pressure in these materials. 
Once the advancements to GBA proposed in section \ref{sec:advancements} are complete, we will also quantitatively compare 3D gradient paths and volumes of gradient bundles in Ir and Pt, with the goal of discerning the unique properties of these materials despite their apparent physical similarities in $\rho(\bm{r})$.

\subsection{Electrostatic Preorganization and Active Site Model Size}
\label{sec:preorganization} 

The origin of the catalytic power of enzymes is a matter of debate.
A popular theory is that the active site of the enzyme binds strongly to the transition state of the substrate molecule, which lowers the activation energy of the reaction.
It is also often stated that the catalytic groups in the active site are perfectly oriented to facilitate the reaction.
Some chemists attribute the catalytic efficiency of enzymes to transition state stabilization (TSS) \cite{Pauling_TSS_1946} while others attempt to explain the catalytic rates based on reactant state destabilization (RSD) \cite{Jencks_RSD_1986}.
All of these proposals are contingent on the method used to define the reference state used for comparison.
\addabbreviation{Transition state stabilization}{TSS}
\addabbreviation{Reactant state destabilization}{RSD}

Warshel has argued that electrostatic preorganization is the main contributor to the high catalytic ability of enzymes \cite{Warshel_1998, Warshel_2006_preorganization}. 
When a reaction is catalyzed in a polar solution such as water, the solvent molecules around the solute are randomly oriented with dipole vectors pointing in various directions. 
As the reaction proceeds through the transition state, the water dipoles must orient themselves in such a way as to facilitate the lowering of the activation barrier.
This rearrangement of the electron density costs energy (due to water--water interactions being broken in order for hydrogen bonds to the substrate to be formed that stabilize the TS) and is contained in the activation energy of the transition state.
In an enzyme, the active site is preorganized, meaning that the electrostatic environment is already oriented to stabilize the transition state. 
Dipole vectors are properly arranged through the positioning of charged and polar residues as well as the orientation of any water molecules in the active site.
There is little or zero energy cost for organizing the electron density in the active site, since it is already in the most favorable position. 

Capturing the electrostatic preorganization in a computer simulation requires taking into account the entire protein and solvent environment. 
However, this does not necessarily mean one must calculate the charge density of the entire protein to a high degree of accuracy. 
The electron density has been described as near-sighted, meaning that the density in one region is not highly dependent on the charge density further away \cite{Kohn_nearsightedness_2005}.
The electron density based on a calculation including only the atoms in the active site should closely match the density calculated using the full enzyme, as long as the nuclear positions are provided from a full enzyme calculation. 
Therefore, it should be possible to capture the electrostatic preorganization in the active site of an enzyme just by performing a single point QM calculation on the active site, keeping in mind that  the nuclear coordinates must first be calculated using full protein simulations.

We are currently designing a computational experiment to see if this hypothesis is valid. 
There are two goals of this study.
The first is to see if we do indeed capture the electrostatic preorganization of the active site just by running QM calculations on the nuclear coordinates from a full protein simulation (without any electrostatic embedding).
The second task is to see how large of an active site model needs to be modeled in order to capture the correct electron charge density distribution.
De Proft recently performed a study \cite{DeProft_convergence_2015} showing that the Hirshfeld charges on a neutral residue in a protein converge very quickly as the active site model size changes (from a 3 to 11 $\AA$ radii around a central atom). 
However, the atomic charges did not converge as well when the residue was charged. 
This study only looked at one system (human 2-cysteine peroxiredoxin thioredoxin peroxidase B) and reported the convergence of atomic charges using the Hirshfeld, Hirshfeld-I, and natural population analysis (NPA) charges.

The nuclear coordinates from QM/DMD calculations for both the carboxypeptidase A (CPA) and histone deacetylase 8 (HDAC8) enzymes presented in chapter \ref{cha:enzymes} will be used in this study.
Following the methods from \cite{DeProft_convergence_2015}, active site models for QM single point calculations are being created using a 3, 5, 7, 9, and 11 $\AA$ radius around the metal cations.
Any amino acid residue that has at least one atom within this cut-off will be completely retained, with dangling residues being capped with hydrogen atoms.
Additionally, each of the five size models for each enzyme will be run as a gas-phase calculation, using COSMO, and with point charges to simulate the electrostatic environment from the full protein, for a total of thirty calculations.
The charge densities will be analyzed using QTAIM and gradient bundle analysis.
The bader charges of the metal ions, substrate carbonyls, and water molecule (which is a key component of the reaction mechanism in both of the enzymes tested) will be compared for each calculation.
The full gradient field of the charge density will be compared by overlaying gradient paths and comparing the stereographic projections of the amount of charge density in gradient bundles around key bond paths as described in section \ref{sec:advancements}.

The results of this study may help to determine the level of computational cost required in order to properly describe the electrostatic environment in the active site of a metalloenzyme.
The size of the active site proved to be an important factor in correctly calculating the energetics of reaction in HDAC8, so it will be interesting to see if this is also important for determining the electron density distribution, or if the theory of the near-sightedness of $\rho(\bm{r})$ holds.
Initial testing of electrostatic embedding (using point charges) did not qualitatively change the reaction profile in HDAC8, but it did change the exact energies at stationary points.
Here we will see if electrostatic embedding or implicit solvation affects the charge density calculated in the active site of enzymes compared to purely gas phase calculations.
Following this study, we will also focus on further quantifying the way changes in $\rho(\bm{r})$ equate to changes in activation energies and reaction rates.

\subsection{Conclusions}

In this thesis I have demonstrated that gradient bundle analysis is a useful and meaningful method for understanding and predicting chemical properties.
Partitioning the electron charge density into gradient bundles allows for distributions of well-defined properties to be calculated within atomic basins and bond bundles.
The structure of the valence electrons in molecules--bonding and lone pair regions--can be determined using gradient bundles, highlighting regions that are most likely to undergo reaction.
This structure can be used to define kinetic energy bonding regions, whose valence electron counts correlate to experimentally determined bond dissociation energies in diatomic molecules.
The motion of zero-flux surfaces bounding gradient bundles can be used to explain chemical reactivity, reproducing results obtained using standard frontier orbital theory, but without the need to decompose the charge density into orbital contributions.
We are currently working on enhancing the predictive capabilities of GBA, allowing for the most reactive sites on a molecule to be located using only the ground state charge density.

Advancements to gradient bundle analysis highlighted in this chapter will enable quantitative chemical bonding analysis of non-symmetric molecules.
The motivation for this work is to improve our ability to rationally design materials and enzymes with novel properties.
Current bonding models can not always fully explain the relationships between the structure of molecules and solids with their observed properties, such as our lack of understanding the brittle behavior of iridium described in section \ref{sec:closed}.
Rational enzyme design methods could also benefit from the use of gradient bundle analysis, as we showed in section \ref{sec:HDAC} that more traditional methods for analyzing the charge density, such as QTAIM, can not always connect the topology of $\rho(\bm{r})$ in active sites to predicted and experimental activation energies.
Using gradient bundles to analyze the full topology and geometry of the electron charge density will enable a higher resolution model of chemical bonding to be created, allowing for improved prediction of the reactivity in designed enzymes, materials, and molecules.

\backmatter


\bibliography{GB_refs}

\begin{thebibliography}{204}
\providecommand{\natexlab}[1]{#1}
\providecommand{\url}[1]{\texttt{#1}}
\expandafter\ifx\csname urlstyle\endcsname\relax
  \providecommand{\doi}[1]{doi: #1}\else
  \providecommand{\doi}{doi: \begingroup \urlstyle{rm}\Url}\fi

\bibitem[Bader(1990)]{baderbook}
R.~F.~W. Bader.
\newblock \emph{Atoms in Molecules: A quantum theory}.
\newblock Clarendon Press, Oxford, UK, 1990.

\bibitem[Geerlings et~al.(2003)Geerlings, Proft, and Langenaeker]{CDFT}
P.~Geerlings, F.~De Proft, and W.~Langenaeker.
\newblock Conceptual density functional theory.
\newblock \emph{Chemical Reviews}, 103\penalty0 (5):\penalty0 1793--1874, 2003.

\bibitem[Privett et~al.(2012)Privett, Kiss, Lee, Blomberg, Chica, Thomas,
  Hilvert, Houk, and Mayo]{Mayo_review}
Heidi~K. Privett, Gert Kiss, Toni~M. Lee, Rebecca Blomberg, Roberto~A. Chica,
  Leonard~M. Thomas, Donald Hilvert, Kendall~N. Houk, and Stephen~L. Mayo.
\newblock Iterative approach to computational enzyme design.
\newblock \emph{Proceedings of the National Academy of Sciences}, 109\penalty0
  (10):\penalty0 3790--3795, 2012.

\bibitem[Kiss et~al.(2013)Kiss, {\c C}elebi-{\"O}l{\c c}{\"u}m, Moretti, Baker,
  and Houk]{Houk_enzyme_review_2013}
Gert Kiss, Nihan {\c C}elebi-{\"O}l{\c c}{\"u}m, Rocco Moretti, David Baker,
  and K.~N. Houk.
\newblock Computational enzyme design.
\newblock \emph{Angewandte Chemie International Edition}, 52\penalty0
  (22):\penalty0 5700--5725, 2013.

\bibitem[Frushicheva et~al.(2014)Frushicheva, Mills, Schopf, Singh, Prasad, and
  Warshel]{Warshel_design_review_2014}
Maria~P. Frushicheva, Matthew J.~L. Mills, Patrick Schopf, Manoj~K. Singh,
  Ram~B. Prasad, and Arieh Warshel.
\newblock Computer aided enzyme design and catalytic concepts.
\newblock \emph{Current Opinion in Chemical Biology}, 21:\penalty0 56--62,
  2014.

\bibitem[Zanghellini(2014)]{Zanghellini_design_review_2014}
Alexandre Zanghellini.
\newblock de novo computational enzyme design.
\newblock \emph{Current Opinion in Biotechnology}, 29:\penalty0 132--138, 2014.

\bibitem[Wiberg et~al.(1987)Wiberg, Bader, and Lau]{Bader_hydrocarbons}
Kenneth~B. Wiberg, Richard F.~W. Bader, and Clement D.~H. Lau.
\newblock Theoretical analysis of hydrocarbon properties. 2. additivity of
  group properties and the origin of strain energy.
\newblock \emph{Journal of the American Chemical Society}, 109\penalty0
  (4):\penalty0 1001--1012, 1987.

\bibitem[Popelier(1999)]{Popelier_similarity}
P.~L.~A. Popelier.
\newblock Quantum molecular similarity. 1. bcp space.
\newblock \emph{The Journal of Physical Chemistry A}, 103\penalty0
  (15):\penalty0 2883--2890, 1999.

\bibitem[Matta et~al.(2006)Matta, Castillo, and Boyd]{Matta_AA}
Ch{\'e}rif~F. Matta, Norberto Castillo, and Russell~J. Boyd.
\newblock Extended weak bonding interactions in dna:  $\pi$-stacking
  (base-base), base-backbone, and backbone-backbone interactions.
\newblock \emph{The Journal of Physical Chemistry B}, 110\penalty0
  (1):\penalty0 563--578, 2006.

\bibitem[Matta and Bader(2003)]{Matta_protein}
C.~F. Matta and R.~F.~W. Bader.
\newblock Atoms-in-molecules study of the genetically encoded amino acids.
  {III}. {B}ond and atomic properties and their correlations with experiment
  including mutation-induced changes in protein stability and genetic coding.
\newblock \emph{Proteins}, 52\penalty0 (3):\penalty0 360--399, 2003.

\bibitem[Chaudry et~al.(2004)Chaudry, , and Popelier]{Popelier_pka}
U.~A. Chaudry, , and P.~L.~A. Popelier.
\newblock Estimation of pka using quantum topological molecular similarity
  descriptors:  {A}pplication to carboxylic acids, anilines and phenols.
\newblock \emph{The Journal of Organic Chemistry}, 69\penalty0 (2):\penalty0
  233--241, 2004.

\bibitem[Zou and Bader(1994)]{wigner-seitz}
P.~F. Zou and R.~F.~W. Bader.
\newblock A topological definition of a wigner{--}seitz cell and the atomic
  scattering factor.
\newblock \emph{Acta Crystallographica Section A}, 50\penalty0 (6):\penalty0
  714--725, Nov 1994.

\bibitem[Mart\'{\i}n~Pend\'as et~al.(1997)Mart\'{\i}n~Pend\'as, Costales, and
  Lua\~na]{pendas1}
A.~Mart\'{\i}n~Pend\'as, Aurora Costales, and V\'{\i}ctor Lua\~na.
\newblock Ions in crystals: {T}he topology of the electron density in ionic
  materials. i. {F}undamentals.
\newblock \emph{Physical Review B}, 55:\penalty0 4275--4284, Feb 1997.

\bibitem[Eberhart(2001{\natexlab{a}})]{eberhart_shear}
Mark Eberhart.
\newblock Charge-density-shear-moduli relationships in aluminum-lithium alloys.
\newblock \emph{Physical Review Letters}, 87:\penalty0 205503, Oct
  2001{\natexlab{a}}.

\bibitem[Bader(1987)]{personal}
Richard F.~W. Bader.
\newblock Private communication, April 1987.

\bibitem[Shahbazian and Zahedi(2006)]{fuzzybond}
Shant Shahbazian and Mansour Zahedi.
\newblock Letter to the editor: {T}he concept of chemical bond - some like it
  fuzzy but others concrete.
\newblock \emph{Foundations of Chemistry}, 9\penalty0 (1):\penalty0 85--95,
  2006.

\bibitem[Poater et~al.(2006)Poater, Sol\`{a}, and Bickelhaupt]{Bickelhaupt1}
Jordi Poater, Miquel Sol\`{a}, and F.~Matthias Bickelhaupt.
\newblock A model of the chemical bond must be rooted in quantum mechanics,
  provide insight, and possess predictive power.
\newblock \emph{Chemistry -- A European Journal}, 12\penalty0 (10):\penalty0
  2902--2905, 2006.

\bibitem[Dyall and Jr.(1993)]{Dyall199327}
Kenneth~G. Dyall and Knut~F{\ae}gri Jr.
\newblock Finite nucleus effects on relativistic energy corrections.
\newblock \emph{Chemical Physics Letters}, 201\penalty0 (1--4):\penalty0
  27--32, 1993.

\bibitem[Andrae(2000)]{andrae}
Dirk Andrae.
\newblock Finite nuclear charge density distributions in electronic structure
  calculations for atoms and molecules.
\newblock \emph{Physics Reports}, 336\penalty0 (6):\penalty0 413--525, 2000.

\bibitem[te~Velde et~al.(2001)te~Velde, Bickelhaupt, Baerends, Fonseca~Guerra,
  van Gisbergen, Snijders, and Ziegler]{ADF}
G.~te~Velde, F.~M. Bickelhaupt, E.~J. Baerends, C.~Fonseca~Guerra, S.~J.~A. van
  Gisbergen, J.~G. Snijders, and T.~Ziegler.
\newblock Chemistry with adf.
\newblock \emph{Journal of Computational Chemistry}, 22\penalty0 (9):\penalty0
  931--967, 2001.

\bibitem[Popelier(1998)]{popelier_dihydrogen}
P.~L.~A. Popelier.
\newblock Characterization of a dihydrogen bond on the basis of the electron
  density.
\newblock \emph{The Journal of Physical Chemistry A}, 102\penalty0
  (10):\penalty0 1873--1878, 1998.

\bibitem[Matta and Boyd(2007)]{Mattabook}
C.~F. Matta and R.~J. Boyd.
\newblock \emph{The Quantum Theory of Atoms in Molecules: From Solid State to
  DNA and Drug Design}.
\newblock Wiley-VCH, Verlag GmbH and Co. KGaA, Weinheim, 2007.

\bibitem[Grabowski(2001)]{grabowski_hbond}
S{\l}awomir~Janusz Grabowski.
\newblock Ab initio calculations on conventional and unconventional hydrogen
  bonds--study of the hydrogen bond strength.
\newblock \emph{The Journal of Physical Chemistry A}, 105\penalty0
  (47):\penalty0 10739--10746, 2001.

\bibitem[Ford(2014)]{ford_li}
Thomas~A. Ford.
\newblock An ab initio study of the properties of some lithium-bonded complexes
  - comparison with their hydrogen-bonded analogues. 2. {N}atural bond orbital
  and quantum theory of atoms in molecules analysis.
\newblock \emph{Computational and Theoretical Chemistry}, 1042:\penalty0
  63--68, 2014.

\bibitem[Eberhart(1996)]{Eberhart:1996}
M.E. Eberhart.
\newblock The metallic bond: {E}lastic properties.
\newblock \emph{Acta Materialia}, 44\penalty0 (6):\penalty0 2495--2504, 1996.

\bibitem[Eberhart and Jones(2012)]{eberhart_cauchy}
Mark~E. Eberhart and Travis~E. Jones.
\newblock Cauchy pressure and the generalized bonding model for nonmagnetic bcc
  transition metals.
\newblock \emph{Physical Review B}, 86:\penalty0 134106, Oct 2012.

\bibitem[Eberhart and Giamei(1998)]{Eberhart:Giamei}
M.E. Eberhart and A.F. Giamei.
\newblock The visualization and use of electronic structure for metallurgical
  applications.
\newblock \emph{Materials Science and Engineering: A}, 248\penalty0
  (1--2):\penalty0 287--295, 1998.

\bibitem[Wang et~al.(2014)Wang, Jones, Wu, Lu, Halas, Durakiewicz, and
  Eberhart]{Jones_brittle}
X.~F. Wang, T.~E. Jones, Y.~Wu, Z.~P. Lu, S.~Halas, T.~Durakiewicz, and M.~E.
  Eberhart.
\newblock An electronic criterion for assessing intrinsic brittleness of
  metallic glasses.
\newblock \emph{The Journal of Chemical Physics}, 141\penalty0 (2):\penalty0
  024503, 2014.

\bibitem[Popkov and Breza(2010)]{popkov_TM}
Alexander Popkov and Martin Breza.
\newblock Why is monoalkylation versus bis-alkylation of the {N}i({II}) complex
  of the schiff base of
  ({S})-{N}-(2-benzoylphenyl)-1-benzylpyrrolidine-2-carboxamide and glycine so
  selective? {MP2} modelling and topological {QTAIM} analysis of chiral
  metallocomplex synthons of $\alpha$-amino acids used for the preparation of
  radiopharmaceuticals for positron emission tomography.
\newblock \emph{Journal of Radioanalytical and Nuclear Chemistry}, 286\penalty0
  (3):\penalty0 829--833, 2010.

\bibitem[Jenkins et~al.(2011)Jenkins, Restrepo, David, Yin, and
  Kirk]{jenkins_water}
Samantha Jenkins, Albeiro Restrepo, Jorge David, Dulin Yin, and Steven~R. Kirk.
\newblock Spanning {QTAIM} topology phase diagrams of water isomers w$_4$,
  w$_5$, and w$_6$.
\newblock \emph{Physical Chemistry Chemical Physics}, 13:\penalty0
  11644--11656, 2011.

\bibitem[Runtz et~al.(1977)Runtz, Bader, and Messer]{Bader_rings_1977}
G.~R. Runtz, R.~F.~W. Bader, and R.~R. Messer.
\newblock Definition of bond paths and bond directions in terms of the
  molecular charge distribution.
\newblock \emph{Canadian Journal of Chemistry}, 55\penalty0 (16):\penalty0
  3040--3045, 1977.

\bibitem[Popelier(2000)]{Popelier_book}
Paul L.~A. Popelier.
\newblock \emph{Atoms in Molecules: {A}n introduction}.
\newblock Pearson Education Limited, Essex, England, 2000.

\bibitem[Bader(1985)]{Bader_1985}
R.~F.~W. Bader.
\newblock Atoms in molecules.
\newblock \emph{Accounts of Chemical Research}, 18\penalty0 (1):\penalty0
  9--15, 1985.

\bibitem[Miorelli et~al.(2015)Miorelli, Wilson, Morgenstern, Jones, and
  Eberhart]{Miorelli}
Jonathan Miorelli, Tim Wilson, Amanda Morgenstern, Travis Jones, and Mark~E.
  Eberhart.
\newblock A full topological analysis of unstable and metastable bond critical
  points.
\newblock \emph{ChemPhysChem}, 16\penalty0 (1):\penalty0 152--159, 2015.

\bibitem[Haaland et~al.(2004{\natexlab{a}})Haaland, Shorokhov, and
  Tverdova]{Haaland}
Arne Haaland, Dimitry~J. Shorokhov, and Natalya~V. Tverdova.
\newblock Topological analysis of electron densities: {I}s the presence of an
  atomic interaction line in an equilibrium geometry a sufficient condition for
  the existence of a chemical bond?
\newblock \emph{Chemistry -- A European Journal}, 10\penalty0 (18):\penalty0
  4416--4421, 2004{\natexlab{a}}.

\bibitem[Haaland et~al.(2004{\natexlab{b}})Haaland, Shorokhov, and
  Tverdova]{Haaland1}
Arne Haaland, Dimitry~J. Shorokhov, and Natalya~V. Tverdova.
\newblock Topological analysis of electron densities: Is the presence of an
  atomic interaction line in an equilibrium geometry a sufficient condition for
  the existence of a chemical bond?
\newblock \emph{Chemistry -- A European Journal}, 10\penalty0 (24):\penalty0
  6210--6210, 2004{\natexlab{b}}.

\bibitem[Foroutan-Nejad et~al.(2014)Foroutan-Nejad, Shahbazian, and
  Marek]{Shahbazian_consistent}
Cina Foroutan-Nejad, Shant Shahbazian, and Radek Marek.
\newblock Toward a consistent interpretation of the {QTAIM}: {T}ortuous link
  between chemical bonds, interactions, and bond/line paths.
\newblock \emph{Chemistry -- A European Journal}, 20\penalty0 (32):\penalty0
  10140--10152, 2014.

\bibitem[Bader et~al.(1994)Bader, Popelier, and Keith]{Bader_FunctionalGroups}
Richard Frederick~William Bader, Paul Lode~Albert Popelier, and Todd~Alan
  Keith.
\newblock Theoretical definition of a functional group and the molecular
  orbital paradigm.
\newblock \emph{Angewandte Chemie International Edition in English},
  33\penalty0 (6):\penalty0 620--631, 1994.

\bibitem[Anderson et~al.(2010)Anderson, Ayers, and Hernandez]{Hernandez_KE}
James S.~M. Anderson, Paul~W. Ayers, and Juan I.~Rodriguez Hernandez.
\newblock How ambiguous is the local kinetic energy?
\newblock \emph{The Journal of Physical Chemistry A}, 114\penalty0
  (33):\penalty0 8884--8895, 2010.

\bibitem[Cohen(1979)]{Cohen_KE}
Leon Cohen.
\newblock Local kinetic energy in quantum mechanics.
\newblock \emph{The Journal of Chemical Physics}, 70\penalty0 (2):\penalty0
  788--789, 1979.

\bibitem[Ghosh and Parr(1985)]{Parr_Virial}
Swapan~K. Ghosh and Robert~G. Parr.
\newblock Density-determined orthonormal orbital approach to atomic energy
  functionals.
\newblock \emph{The Journal of Chemical Physics}, 82\penalty0 (7):\penalty0
  3307--3315, 1985.

\bibitem[Matta and Hern\'andez-Trujillos(2003)]{Matta_delocalization}
Ch\'{e}rif~F. Matta and Jes\'us Hern\'andez-Trujillos.
\newblock Bonding in polycyclic aromatic hydrocarbons in terms of the electron
  density and of electron delocalization.
\newblock \emph{The Journal of Physical Chemistry A}, 107\penalty0
  (38):\penalty0 7496--7504, 2003.

\bibitem[Becke and Edgecombe(1990)]{ELF_1990}
A.~D. Becke and K.~E. Edgecombe.
\newblock A simple measure of electron localization in atomic and molecular
  systems.
\newblock \emph{The Journal of Chemical Physics}, 92\penalty0 (9):\penalty0
  5397--5403, 1990.

\bibitem[Savin et~al.(1997)Savin, Nesper, Wengert, and F\"{a}ssler]{ELF_Savin}
Andreas Savin, Reinhard Nesper, Steffen Wengert, and Thomas~F. F\"{a}ssler.
\newblock {ELF}: {T}he electron localization function.
\newblock \emph{Angewandte Chemie International Edition in English},
  36\penalty0 (17):\penalty0 1808--1832, 1997.

\bibitem[Savin et~al.(1992)Savin, Jepsen, Flad, Andersen, Preuss, and von
  Schnering]{ELF_1992}
Andreas Savin, Ove Jepsen, J\"{u}rgen Flad, Ole~Krogh Andersen, Heinzwerner
  Preuss, and Hans~Georg von Schnering.
\newblock Electron localization in solid-state structures of the elements: the
  diamond structure.
\newblock \emph{Angewandte Chemie International Edition in English},
  31\penalty0 (2):\penalty0 187--188, 1992.

\bibitem[Poater et~al.(2005)Poater, Duran, Sol\`{a}, and Silvi]{poater_AIM_ELF}
Jordi Poater, Miquel Duran, Miquel Sol\`{a}, and Bernard Silvi.
\newblock Theoretical evaluation of electron delocalization in aromatic
  molecules by means of atoms in molecules ({AIM}) and electron localization
  function ({ELF}) topological approaches.
\newblock \emph{Chemical Reviews}, 105\penalty0 (10):\penalty0 3911--3947,
  2005.

\bibitem[Santos et~al.(2004)Santos, Tiznado, Contreras, and
  Fuentealba]{Santos_ELF}
J.~C. Santos, W.~Tiznado, R.~Contreras, and P.~Fuentealba.
\newblock Sigma--pi separation of the electron localization function and
  aromaticity.
\newblock \emph{The Journal of Chemical Physics}, 120\penalty0 (4):\penalty0
  1670--1673, 2004.

\bibitem[Kohout(2004)]{Kohout_ELI}
M.~Kohout.
\newblock A measure of electron localizability.
\newblock \emph{International Journal of Quantum Chemistry}, 97\penalty0
  (1):\penalty0 651--658, 2004.

\bibitem[Steinmann et~al.(2011)Steinmann, Mo, and Corminboeuf]{ELID_Pi}
Stephan~N. Steinmann, Yirong Mo, and Clemence Corminboeuf.
\newblock How do electron localization functions describe $\pi$-electron
  delocalization.
\newblock \emph{Physical Chemistry Chemical Physics}, 13:\penalty0
  20584--20592, 2011.

\bibitem[Wagner et~al.(2007)Wagner, Bezugly, Kohout, and Grin]{Grin_ELI_2007}
Frank~R. Wagner, Viktor Bezugly, Miroslav Kohout, and Yuri Grin.
\newblock Charge decomposition analysis of the electron localizability
  indicator: A bridge between the orbital and direct space representation of
  the chemical bond.
\newblock \emph{Chemistry -- A European Journal}, 13\penalty0 (20):\penalty0
  5724--5741, 2007.

\bibitem[Schmider and Becke(2000)]{Becke_LOL}
H.~L. Schmider and A.~D. Becke.
\newblock Chemical content of the kinetic energy density.
\newblock \emph{Journal of Molecular Structure: THEOCHEM}, 527:\penalty0
  51--61, 2000.

\bibitem[Jacobsen(2013)]{jacobsen_LOL}
Heiko Jacobsen.
\newblock Topology maps of bond descriptors based on the kinetic energy density
  and the essence of chemical bonding.
\newblock \emph{Physical Chemistry Chemical Physics}, 15:\penalty0 5057--5066,
  2013.

\bibitem[Ayers et~al.(2002{\natexlab{a}})Ayers, Parr, and Nagy]{ayers_KE}
Paul~W. Ayers, Robert~G. Parr, and Agnes Nagy.
\newblock Local kinetic energy and local temperature in the density-functional
  theory of electronic structure.
\newblock \emph{International Journal of Quantum Chemistry}, 90:\penalty0
  309--326, 2002{\natexlab{a}}.

\bibitem[Bitter et~al.(2007)Bitter, Ruedenberg, and Schwarz]{Ruedenberg_KE}
T.~Bitter, K.~Ruedenberg, and W.~H.~E. Schwarz.
\newblock Toward a physical understanding of electron-sharing two-center bonds.
  {I}. {G}eneral aspects.
\newblock \emph{Journal of Computational Chemistry}, 28\penalty0 (1):\penalty0
  411--422, 2007.

\bibitem[Yang(2010)]{Yang_aromaticity}
Yang Yang.
\newblock Hexacoordinate bonding and aromaticity in silicon phthalocyanide.
\newblock \emph{Journal of Physical Chemistry A}, 114:\penalty0 13257--13267,
  2010.

\bibitem[Eberhart(2001{\natexlab{b}})]{eberhart_IB}
Mark Eberhart.
\newblock A quantum description of the chemical bond.
\newblock \emph{Philosophical Magazine Part B}, 81\penalty0 (8):\penalty0
  721--729, 2001{\natexlab{b}}.

\bibitem[Miller(1998)]{Miller}
Jason Miller.
\newblock \emph{Relative Critical Sets in $\mathbb{R}^n$ and Applications to
  Image Analysis}.
\newblock PhD thesis, Chapel Hill, 1998.

\bibitem[Damon(1998)]{criticalsets}
James Damon.
\newblock Generic structure of two-dimensional images under gaussian blurring.
\newblock \emph{SIAM Journal on Applied Mathematics}, 59\penalty0 (1):\penalty0
  97--138, 1998.

\bibitem[Eberly et~al.(1994)Eberly, Gardner, Morse, Pizer, and
  Scharlach]{Eberly}
D.~Eberly, R.~Gardner, B.~Morse, S.~Pizer, and C.~Scharlach.
\newblock Ridges for image analysis.
\newblock \emph{Journal of Mathematical Imaging and Vision}, 4:\penalty0
  353--373, 1994.

\bibitem[Jones and Eberhart(2009)]{Jones_KE}
Travis~E. Jones and Mark~E. Eberhart.
\newblock The irreducible bundle: {F}urther structure in the kinetic energy
  distribution.
\newblock \emph{Journal of Chemical Physics}, 130:\penalty0 204108, 2009.

\bibitem[Jones et~al.(2011)Jones, Eberhart, Imlay, and
  Mackey]{Jones_functionality}
Travis~E. Jones, Mark~E. Eberhart, Scott Imlay, and Craig Mackey.
\newblock Bond bundles and the origins of functionality.
\newblock \emph{Journal of Physical Chemistry A}, 115:\penalty0 12582--12585,
  2011.

\bibitem[Jones(2012)]{Jones_nucl}
Travis~E. Jones.
\newblock Nucleophic substitution: A charge density perspective.
\newblock \emph{Journal of Physical Chemistry A}, 116:\penalty0 4233--4237,
  2012.

\bibitem[Wolk et~al.(2001)Wolk, Hoz, Basch, and Hoz]{ringstrain}
Joel~L. Wolk, Tova Hoz, Harold Basch, and Shmaryahu Hoz.
\newblock Quantification of the various contributors to rate enhancement in
  nucleophilic strain releasing reactions.
\newblock \emph{Journal of Organic Chemistry}, 66:\penalty0 915--918, 2001.

\bibitem[Jones et~al.(2012)Jones, Eberhart, Imlay, Mackey, and
  Olson]{Jones_alloys}
Travis~E. Jones, Mark~E. Eberhart, Scott Imlay, Craig Mackey, and Greg~B.
  Olson.
\newblock Better alloys with quantum design.
\newblock \emph{Physical Review Letters}, 109:\penalty0 125506, 2012.

\bibitem[Latourte et~al.(2012)Latourte, Wei, Feinberg, de~Vaucorbeil, Tran,
  Olson, and Espinosa]{BlastAlloy}
Felix Latourte, Xiaoding Wei, Zechariah~D. Feinberg, Alban de~Vaucorbeil,
  Phuong Tran, Gregory~B. Olson, and Horacio~D. Espinosa.
\newblock Design and identification of high performance steel alloys for
  structures subjected to underwater impulsive loading.
\newblock \emph{International Journal of Solids and Structure}, 49\penalty0
  (13):\penalty0 1573--1587, 2012.

\bibitem[Jones and Eberhart(2010)]{bondbundle2}
T.E. Jones and M.E. Eberhart.
\newblock The bond bundle in open systems.
\newblock \emph{International Journal of Quantum Chemistry}, 110:\penalty0
  1500--1505, 2010.

\bibitem[Eberhart and Jones(2013)]{foundations}
Mark~E. Eberhart and Travis~E. Jones.
\newblock The two faces of chemistry: can they be reconciled?
\newblock \emph{Foundations of Chemistry}, 15\penalty0 (3):\penalty0 277--285,
  2013.

\bibitem[Fonseca~Guerra et~al.(1998)Fonseca~Guerra, Snijders, te~Velde, and
  Baerends]{ADF2}
C.~Fonseca~Guerra, J.~G. Snijders, G.~te~Velde, and E.~J. Baerends.
\newblock Towards an order-{N} {DFT} method.
\newblock \emph{Theoretical Chemistry Accounts}, 99\penalty0 (6):\penalty0
  391--403, 1998.

\bibitem[Perdew(1986)]{Perdew}
John~P. Perdew.
\newblock Density-functional approximation for the correlation energy of the
  inhomogeneous electron gas.
\newblock \emph{Physical Review B}, 33:\penalty0 8822--8824, 1986.

\bibitem[Perdew et~al.(1996)Perdew, Burke, and Ernzerhof]{PBE}
John~P. Perdew, Kieron Burke, and Matthias Ernzerhof.
\newblock Generalized gradient approximation made simple.
\newblock \emph{Physical Review Letters}, 77\penalty0 (18):\penalty0
  3865--3868, 1996.

\bibitem[Perdew et~al.(1992)Perdew, Chevary, Vosko, Jackson, Pederson, Singh,
  and Fiolhais]{PW91}
J.~P. Perdew, J.~A. Chevary, S.~H. Vosko, K.~A. Jackson, M.~R. Pederson, D.~J.
  Singh, and C.~Fiolhais.
\newblock Atoms, molecules, solids, and surfaces: {A}pplications of the
  generalized gradient approximation for exchange and correlation.
\newblock \emph{Physical Review B}, 46\penalty0 (11):\penalty0 6671--6687,
  1992.

\bibitem[Rodr\'iguez et~al.(2009)Rodr\'iguez, Ayers, G\"otz, and
  Castillo-Alvarado]{ayers_ADF}
Juan~I. Rodr\'iguez, Paul~W. Ayers, Andreas~W. G\"otz, and F.~L.
  Castillo-Alvarado.
\newblock Virial theorem in the kohn-sham density-functional theory formalism:
  Accurate calculation of the atomic quantum theory of atoms in molecules
  energies.
\newblock \emph{Journal of Chemical Physics}, 131:\penalty0 21101, 2009.

\bibitem[Vosko et~al.(1980)Vosko, Wilk, and Nusair]{VWN}
S.~H. Vosko, L.~Wilk, and M.~Nusair.
\newblock Accurate spin-dependent electron liquid correlation energies for
  local spin-density calculations- a critical analysis.
\newblock \emph{Canadian Journal of Physics}, 58\penalty0 (8):\penalty0
  1200--1211, 1980.

\bibitem[Stephens et~al.(1994)Stephens, Devlin, Chabalowski, and Frisch]{B3LYP}
P.~J. Stephens, F.J. Devlin, C.F. Chabalowski, and M.J. Frisch.
\newblock Ab initio calculation of vibrational absorption and circular
  dichroism spectra using density functional force fields.
\newblock \emph{The Journal of Physical Chemistry.}, 98\penalty0 (45):\penalty0
  11623--11627, 1994.

\bibitem[Zhao and Truhlar(2006)]{MO6L}
Y.~Zhao and D.G. Truhlar.
\newblock A new local density functional for main-group thermochemistry,
  transition metal bonding, thermochemical kinetics, and non covalent
  interactions.
\newblock \emph{Journal of Chemical Physics}, 125:\penalty0 194101, 2006.

\bibitem[Lee et~al.(1988)Lee, Yang, and Parr]{correlation}
Chengteh Lee, Weitao Yang, and Robert~G. Parr.
\newblock Development of the colle-salvetti correlation-energy formula into a
  functional of the electron density.
\newblock \emph{Physical Review B}, 37:\penalty0 785--789, Jan 1988.

\bibitem[Jones et~al.(2008)Jones, Eberhart, and Clougherty]{spintopo}
Travis~E. Jones, Mark~E. Eberhart, and Dennis~P. Clougherty.
\newblock Topology of the spin-polarized charge density in bcc and fcc iron.
\newblock \emph{Physical Review Letters}, 100\penalty0 (1):\penalty0 017208,
  Jan 2008.

\bibitem[Carbogno et~al.(2013)Carbogno, Gross, Meyer, and Reuter]{Carbogno}
Christian Carbogno, Axel Gross, J\"{o}rg Meyer, and Karsten Reuter.
\newblock O2 adsorption dynamics at metal surfaces: Non-adiabatic effects,
  dissociation and dissipation.
\newblock In Ricardo D\'{i}ez Mui\~{n}o and Heriberto~Fabio Busnengo, editors,
  \emph{Dynamics of Gas-Surface Interactions}, volume~50 of \emph{Springer
  Series in Surface Sciences}, pages 389--419. Springer Berlin Heidelberg,
  2013.

\bibitem[Tecplot(2014)]{Tecplot}
Tecplot.
\newblock http://www.tecplot.com/, 2014.

\bibitem[Stans and Branscomb(1970)]{BDE}
M.~H. Stans and Lewis~M. Branscomb.
\newblock \emph{Bond Dissociation Energies in Simple Molecules}.
\newblock NSRDS, http://www.nist.gov/data/nsrds/NSRDS-NBS31.pdf, 1970.

\bibitem[Schweitzer and Schmidt(2003)]{Oxygen}
C.~Schweitzer and R.~Schmidt.
\newblock Physical mechanisms of generation and deactivation of singlet oxygen.
\newblock \emph{Chemical Reviews}, 103:\penalty0 1685--1757, 2003.

\bibitem[Cohen(1984)]{Cohen2}
Leon Cohen.
\newblock Representable local kinetic energy.
\newblock \emph{The Journal of Chemical Physics}, 80\penalty0 (9):\penalty0
  4277--4279, 1984.

\bibitem[Pend{\'a}s et~al.(2007)Pend{\'a}s, Francisco, Blanco, and
  Gatti]{Pendas_bondpath}
A. Mart{\'\i}n Pend{\'a}s, Evelio Francisco, Miguel A. Blanco, and Carlo
  Gatti.
\newblock Bond paths as privileged exchange channels.
\newblock \emph{Chemistry -- A European Journal}, 13\penalty0 (33):\penalty0
  9362--9371, 2007.

\bibitem[Slater(1933)]{Slater_virial}
J.~C. Slater.
\newblock The virial and molecular structure.
\newblock \emph{Journal of Chemical Physics}, 1:\penalty0 687--691, 1933.

\bibitem[Ruedenberg(1962)]{Ruedenberg_bond_1962}
Klaus Ruedenberg.
\newblock The physical nature of the chemical bond.
\newblock \emph{Reviews of Moderns Physics}, 34\penalty0 (2):\penalty0
  326--376, 1962.

\bibitem[Parr and Yang(1989)]{Parrbook}
R.~G. Parr and W.~Yang.
\newblock \emph{Density-Functional Theory of Atoms and Molecules}.
\newblock Clarendon Press, Oxford, 1989.

\bibitem[Morgenstern et~al.(2015)Morgenstern, Wilson, Miorelli, Jones, and
  Eberhart]{Morgenstern}
Amanda Morgenstern, Tim Wilson, Jonathan Miorelli, Travis Jones, and M.~E.
  Eberhart.
\newblock In search of an intrinsic chemical bond.
\newblock \emph{Computational and Theoretical Chemistry}, 1053:\penalty0
  31--37, 2015.

\bibitem[Fukui et~al.(1952)Fukui, Yonezawa, and Shingu]{frontier}
Kenichi Fukui, Teijiro Yonezawa, and Haruo Shingu.
\newblock A molecular orbital theory of reactivity in aromatic hydrocarbons.
\newblock \emph{The Journal of Chemical Physics}, 20\penalty0 (4):\penalty0
  722--725, 1952.

\bibitem[Kermack and Robinson(1922)]{e_pushing}
William~Ogilvy Kermack and Robert Robinson.
\newblock An explanation of the property of induced polarity of atoms and an
  interpretation of the theory of partial valencies on an electronic basis.
\newblock \emph{Journal of Chemical Society{,} Transactions}, 121:\penalty0
  427, 1922.

\bibitem[Bader et~al.(1979)Bader, Nguyen‐Dang, and Tal]{bader_catastrophe}
Richard F.~W. Bader, T.~Tung Nguyen‐Dang, and Yoram Tal.
\newblock Quantum topology of molecular charge distributions. ii. molecular
  structure and its change.
\newblock \emph{The Journal of Chemical Physics}, 70\penalty0 (9):\penalty0
  4316--4329, 1979.

\bibitem[Guevara-Garc\'ia et~al.(2011)Guevara-Garc\'ia, Echegaray, Toro-Labbe,
  Jenkins, Kirk, and Ayers]{ayers_predict}
Alfredo Guevara-Garc\'ia, Eleonora Echegaray, Alejandro Toro-Labbe, Samantha
  Jenkins, Steven~R. Kirk, and Paul~W. Ayers.
\newblock Pointing the way to the products? {C}omparison of the stress tensor
  and the second-derivative tensor of the electron density.
\newblock \emph{Journal of Chemical Physics}, 134:\penalty0 234106, 2011.

\bibitem[Ayers and Jenkins(2009)]{ayers_2009}
Paul~W. Ayers and Samantha Jenkins.
\newblock An electron-preceding perspective on the deformation of materials.
\newblock \emph{Journal of Chemical Physics}, 130\penalty0 (15):\penalty0
  154104, 2009.

\bibitem[Guevara-Garc\'ia et~al.(2014)Guevara-Garc\'ia, Ayers, Jenkins, Kirk,
  Echegaray, and Toro-Labbe]{ayers_EPP2014}
A.~Guevara-Garc\'ia, P.~W. Ayers, S.~Jenkins, S.~R. Kirk, E.~Echegaray, and
  A.~Toro-Labbe.
\newblock Electronic stress as a guiding force for chemical bonding.
\newblock \emph{Topics in Current Chemistry}, 351:\penalty0 103--124, 2014.

\bibitem[Parr et~al.(1978)Parr, Donelly, Levy, and Palke]{Parr_EN}
Robert~G. Parr, Robert~A. Donelly, Mel Levy, and William~E. Palke.
\newblock Electronegativity: The density functional viewpoint.
\newblock \emph{Journal of Chemical Physics}, 68:\penalty0 3801--3807, 1978.

\bibitem[Parr and Pearson(1983)]{Parr_HSAB}
Robert~G. Parr and Ralph~G. Pearson.
\newblock Absolute hardness: Companion parameter to absolute electronegativity.
\newblock \emph{Journal of American Chemical Society}, 105:\penalty0
  7512--7516, 1983.

\bibitem[Ray et~al.(1979)Ray, Samuels, and Parr]{Parr_ENequal}
Naba~K. Ray, Leonard Samuels, and Robert~G. Parr.
\newblock Studies of electronegativity equalization.
\newblock \emph{Journal of Chemical Physics}, 70:\penalty0 3680--3684, 1979.

\bibitem[Politzer and Weinstein(1979)]{Politzer_EN}
Peter Politzer and Harel Weinstein.
\newblock Some relations between electronic distribution and electronegativity.
\newblock \emph{Journal of Chemical Physics}, 71:\penalty0 4218--4220, 1979.

\bibitem[Tognetti et~al.(2015)Tognetti, Morell, and Joubert]{Tognetti_atomicEN}
Vincent Tognetti, Christophe Morell, and Laurent Joubert.
\newblock Atomic electronegativities in molecules.
\newblock \emph{Chemical Physics Letters}, 635:\penalty0 111--115, 2015.

\bibitem[Ayers and Parr(2008)]{ayers_local_hardness1}
Paul~W. Ayers and Robert~G. Parr.
\newblock Local hardness equalization: Exploiting the ambiguity.
\newblock \emph{Journal of Chemical Physics}, 128:\penalty0 184108, 2008.

\bibitem[Ayers and Parr(2000)]{ayers_local_hardness2}
P.~W. Ayers and R.~G. Parr.
\newblock Variational principles for describing chemical reactions: {T}he
  {F}ukui function and chemical hardness revisited.
\newblock \emph{Journal of American Chemical Society}, 122:\penalty0 2010,
  2000.

\bibitem[Ghosh(1990)]{Ghosh_local_hardness1}
S.~K. Ghosh.
\newblock Energy derivatives in density-functional theory.
\newblock \emph{Chemical Physics Letters}, 172:\penalty0 77, 1990.

\bibitem[Ghosh and Berkowitz(1985)]{Ghosh_local_hardness2}
S.~K. Ghosh and M.~Berkowitz.
\newblock A classical fluid-like approach to density-functional formalism of
  many-electron systems.
\newblock \emph{Journal of Chemical Physics}, 83:\penalty0 2976, 1985.

\bibitem[Sanderson(1955)]{SandersonEN}
R.~T. Sanderson.
\newblock Partial charges on atoms in organic compounds.
\newblock \emph{Science}, 121:\penalty0 207--208, 1955.

\bibitem[Chermette(1999)]{Chermette_CDFT}
H.~Chermette.
\newblock Chemical reactivity indexes in density functional theory.
\newblock \emph{Journal of Computational Chemistry}, 20:\penalty0 129--154,
  1999.

\bibitem[Parr and Yang(1984)]{Parr_fukui}
Robert~G. Parr and Weitao Yang.
\newblock Density functional spproach to the frontier-electron theory of
  chemical reactivity.
\newblock \emph{Journal of American Chemical Society}, 106:\penalty0
  4049--4050, 1984.

\bibitem[Bultinck et~al.(2011)Bultinck, Clarisse, Ayers, and
  Carb\'{o}-Dorca]{Bultinck_fukui_2011}
Patrick Bultinck, Dorien Clarisse, Paul~W. Ayers, and Ramon Carb\'{o}-Dorca.
\newblock The {F}ukui matrix: a simple approach to the analysis of the {F}ukui
  function and its positive character.
\newblock \emph{Physical Chemistry Chemical Physics}, 13:\penalty0 6110--6115,
  2011.

\bibitem[Bultinck et~al.(2012)Bultinck, Neck, Acke, and
  Ayers]{Bultinck_fukui_2012}
Patrick Bultinck, Dimitri~Van Neck, Guillame Acke, and Paul~W. Ayers.
\newblock Influence of electron correlation and degeneracy on the {F}ukui
  matrix and extension of frontier molecular orbital theory to correlated
  quantum chemical methods.
\newblock \emph{Physical Chemistry Chemical Physics}, 14:\penalty0 2408--2416,
  2012.

\bibitem[Bultinck et~al.(2014)Bultinck, Cardenas, Fuentealba, Johnson, and
  Ayers]{Bultinck_fukui_2014}
Patrick Bultinck, Carlos Cardenas, Patricio Fuentealba, Paul~A. Johnson, and
  Paul~W. Ayers.
\newblock How to compute the {F}ukui matrix and function for systems with
  (quasi-) degenerate states.
\newblock \emph{Journal of Chemical Theory and Computation}, 10:\penalty0
  202--210, 2014.

\bibitem[Ayers et~al.(2002{\natexlab{b}})Ayers, Morrison, and
  Roy]{Ayers_condensed}
Paul~W. Ayers, Robert~C. Morrison, and Ram~K. Roy.
\newblock Variational principles for describing chemical reactions: {C}ondensed
  reactivity indices.
\newblock \emph{Journal of Chemical Physics}, 116:\penalty0 8731--8744,
  2002{\natexlab{b}}.

\bibitem[Bultinck et~al.(2007)Bultinck, Fias, Alsenoy, Ayers, and
  Carb\'{o}-Dorca]{bultinck_fukui}
Patrick Bultinck, Stijn Fias, Christian~Van Alsenoy, Paul~W. Ayers, and Ramon
  Carb\'{o}-Dorca.
\newblock Critical thoughts on computing atom condensed fukui functions.
\newblock \emph{Journal of Chemical Physics}, 127:\penalty0 034102, 2007.

\bibitem[Cioslowski et~al.(1993)Cioslowski, Martinov, and
  Mixon]{cioslowski_fukui}
Jerzy Cioslowski, Martin Martinov, and Stacey~T. Mixon.
\newblock Atomic fukui indices from the topological theory of atoms in
  molecules applied to {H}artree-{F}ock and correlated electron densities.
\newblock \emph{Journal of Physical Chemistry}, 97:\penalty0 10948--10951,
  1993.

\bibitem[Contreras et~al.(1999)Contreras, Fuentealba, Galv\'{a}n, and
  P\'{e}rez]{contreras_fukui}
Renato~R. Contreras, Patricio Fuentealba, Marcelo Galv\'{a}n, and Patricia
  P\'{e}rez.
\newblock A direct evaluation of regional {F}ukui functions in molecules.
\newblock \emph{Chemical Physics Letters}, 304:\penalty0 405--413, 1999.

\bibitem[Arulmozhiraja and Kolandaivel(1997)]{arulmozhiraja_fukui}
S.~Arulmozhiraja and P.~Kolandaivel.
\newblock Condensed {F}ukui function: dependency on atomic charges.
\newblock \emph{Molecular Physics}, 90\penalty0 (1):\penalty0 55--62, 1997.

\bibitem[Proft et~al.(2002)Proft, Alsenoy, Peeters, Langenaeker, and
  Geerlings]{deproft_fukui}
F.~De Proft, C.~Van Alsenoy, A.~Peeters, W.~Langenaeker, and P.~Geerlings.
\newblock Atomic charges, dipole moments, and {F}ukui functions using the
  hirshfeld partitioning of the electron density.
\newblock \emph{Journal of Computational Chemistry}, 23:\penalty0 1198--2002,
  2002.

\bibitem[12(2009)]{setnotation}
ISO/TC 12.
\newblock Iso 80000-s:2009(en) quantities and unites -- part 2: Mathematical
  signs and symbols to be used in the natural sciences and technology, 2009.

\bibitem[van Lenthe and Baerends(2003)]{ADFbasis}
E.~van Lenthe and E.J. Baerends.
\newblock Optimized slater-type basis sets for the elements 1-118.
\newblock \emph{Journal of Computational Chemistry}, 24:\penalty0 1142, 2003.

\bibitem[van Lenthe et~al.(1993)van Lenthe, Baerends, and Snijders]{zora1}
E.~van Lenthe, E.~J. Baerends, and J.~G. Snijders.
\newblock Relativistic regular two-component hamiltonians.
\newblock \emph{Journal of Chemical Physics}, 99:\penalty0 4597--4610, 1993.

\bibitem[van Lenthe et~al.(1994)van Lenthe, Baerends, and Snijders]{zora2}
E.~van Lenthe, E.~J. Baerends, and J.~G. Snijders.
\newblock Relativistic total energy using regular approximations.
\newblock \emph{Journal of Chemical Physics}, 101:\penalty0 9783--9792, 1994.

\bibitem[van Lenthe et~al.(1999)van Lenthe, Ehlers, and Baerends]{zora3}
E.~van Lenthe, A.~E. Ehlers, and E.~J. Baerends.
\newblock Geometry optimization in the zero order regular approximation for
  relativistic effects.
\newblock \emph{Journal of Chemical Physics}, 110:\penalty0 8943--8953, 1999.

\bibitem[Rodriguez et~al.(2009)Rodriguez, Bader, Ayers, Michel, G\"otz, and
  Bo]{ADFbader1}
J.I. Rodriguez, R.F.W. Bader, P.W. Ayers, C.~Michel, A.W. G\"otz, and C.~Bo.
\newblock A high performance grid-based algorithm for computing {QTAIM}
  properties.
\newblock \emph{Chemical Physics Letters}, 472:\penalty0 149, 2009.

\bibitem[Rodriguez(2013)]{ADFbader2}
J.I. Rodriguez.
\newblock An efficient method for computing the {QTAIM} topology of a scalar
  field: The electron density case.
\newblock \emph{Journal of Computational Chemistry}, 34:\penalty0 681, 2013.

\bibitem[Meunier et~al.(2004)Meunier, de~Visser, and Shaik]{cytochrome}
Bernard Meunier, Samu\"{e}l~P. de~Visser, and Sason Shaik.
\newblock Mechanism of oxidation reactions catalyzed by cytochrome p450
  enzymes.
\newblock \emph{Chemical Reviews}, 104\penalty0 (9):\penalty0 3947--3980, 2004.

\bibitem[Cotton et~al.(1979)Cotton, Hazen, and Legg]{nuclease}
F.~Albert Cotton, Edward~E. Hazen, and Margaret~J. Legg.
\newblock Staphylococcal nuclease: Proposed mechanism of action based on
  structure of enzyme---thymidine 3′,5′-bisphosphate---calcium ion complex
  at 1.5-{\aa} resolution.
\newblock \emph{Proceedings of the National Academy of Sciences}, 76\penalty0
  (6):\penalty0 2551--2555, 1979.

\bibitem[Ragsdale(2014)]{MCR_2014}
Stephen~W. Ragsdale.
\newblock \emph{The Metal-Driven Biogeochemistry of Gaseous Compounds in the
  Environment}, chapter Biochemistry of Methyl-Coenzyme M Reductase: The Nickel
  Metalloenzyme that Catalyzes the Final Step in Synthesis and the First Step
  in Anaerobic Oxidation of the Greenhouse Gas Methane, pages 125--145.
\newblock Springer Netherlands, Dordrecht, 2014.

\bibitem[Zhao et~al.(2013)Zhao, Wang, Ji, and Mao]{Mao_metalloenzyme_2013}
Meng Zhao, Hai-Bo Wang, Liang-Nian Ji, and Zong-Wan Mao.
\newblock Insights into metalloenzyme microenvironments: biomimetic metal
  complexes with a functional second coordination sphere.
\newblock \emph{Chemical Society Reviews}, 42:\penalty0 8360--8375, 2013.

\bibitem[Knowles and Jacobsen(2010)]{KnowlesHbonds_2010}
Robert~R. Knowles and Eric~N. Jacobsen.
\newblock Attractive noncovalent interactions in asymmetric catalysis: Links
  between enzymes and small molecule catalysts.
\newblock \emph{Proceedings of the National Academy of Sciences}, 107\penalty0
  (48):\penalty0 20678--20685, 2010.

\bibitem[Warshel(1998)]{Warshel_1998}
Arieh Warshel.
\newblock Electrostatic origin of the catalytic power of enzymes and the role
  of preorganized active sites.
\newblock \emph{Journal of Biological Chemistry}, 273\penalty0 (42):\penalty0
  27035--27038, 1998.

\bibitem[Warshel(2003)]{Warshel_2003}
Arieh Warshel.
\newblock Computer simulations of enzyme catalysis: Methods, progress, and
  insights.
\newblock \emph{Annual Review of Biophysics and Biomolecular Structure},
  32\penalty0 (1):\penalty0 425--443, 2003.

\bibitem[Vega-Hissi et~al.(2015)Vega-Hissi, Tosso, Enriz, and
  Gutierrez]{Vega-Hissi_2015}
Esteban~Gabriel Vega-Hissi, Rodrigo Tosso, Ricardo~Daniel Enriz, and Lucas~Joel
  Gutierrez.
\newblock Molecular insight into the interaction mechanisms of
  amino-2h-imidazole derivatives with bace1 protease: A qm/mm and qtaim study.
\newblock \emph{International Journal of Quantum Chemistry}, 115\penalty0
  (6):\penalty0 389--397, 2015.

\bibitem[Tosso et~al.(2013)Tosso, Andujar, Gutierrez, Angelina, Rodr{\'\i}guez,
  Nogueras, Baldoni, Suvire, Cobo, and Enriz]{Tosso_2013}
Rodrigo~D. Tosso, Sebastian~A. Andujar, Lucas Gutierrez, Emilio Angelina,
  Ricaurte Rodr{\'\i}guez, Manuel Nogueras, H{\'e}ctor Baldoni, Fernando~D.
  Suvire, Justo Cobo, and Ricardo~D. Enriz.
\newblock Molecular modeling study of dihydrofolate reductase inhibitors.
  molecular dynamics simulations, quantum mechanical calculations, and
  experimental corroboration.
\newblock \emph{Journal of Chemical Information and Modeling}, 53\penalty0
  (8):\penalty0 2018--2032, 2013.

\bibitem[Angelina et~al.(2014)Angelina, Andujar, Tosso, Enriz, and
  Peruchena]{Tosso_2014}
Emilio~L. Angelina, Sebasti{\'a}n~A. Andujar, Rodrigo~D. Tosso, Ricardo~D.
  Enriz, and N{\'e}lida~M. Peruchena.
\newblock Non-covalent interactions in receptor--ligand complexes. a study
  based on the electron charge density.
\newblock \emph{Journal of Physical Organic Chemistry}, 27\penalty0
  (2):\penalty0 128--134, 2014.

\bibitem[Ni et~al.(2013)Ni, Jin, Chen, and Lin]{Lin_design}
Zhong Ni, Rongzhong Jin, Huayou Chen, and Xianfu Lin.
\newblock Just an additional hydrogen bond can dramatically reduce the
  catalytic activity of bacillus subtilis lipase a {I12T} mutant: An
  integration of computational modeling and experimental analysis.
\newblock \emph{Computers in Biology and Medicine}, 43\penalty0 (11):\penalty0
  1882--1888, 2013.

\bibitem[Sparta et~al.(2012)Sparta, Shirvanyants, Ding, Dokholyan, and
  Alexandrova]{QMDMD1}
Manuel Sparta, David Shirvanyants, Feng Ding, Nikolay~V. Dokholyan, and
  Anastassia~N. Alexandrova.
\newblock Hybrid dynamics simulation engine for metalloproteins.
\newblock \emph{Biophysical Journal}, 103\penalty0 (4):\penalty0 767--776,
  2012.

\bibitem[Dokholyan et~al.(1998)Dokholyan, Buldyrev, Stanley, and
  Shakhnovich]{QMDMD2}
Nikolay~V. Dokholyan, Sergey~V. Buldyrev, H~Eugene Stanley, and Eugene~I.
  Shakhnovich.
\newblock Discrete molecular dynamics studies of the folding of a protein-like
  model.
\newblock \emph{Folding and Design}, 3\penalty0 (6):\penalty0 577--587, 11
  1998.

\bibitem[Dokholyan(2006)]{QMDMD3}
Nikolay~V Dokholyan.
\newblock Studies of folding and misfolding using simplified models.
\newblock \emph{Current Opinion in Structural Biology}, 16\penalty0
  (1):\penalty0 79--85, 2 2006.

\bibitem[Ding et~al.(2005)Ding, Guo, Dokholyan, Shakhnovich, and Shea]{QMDMD4}
Feng Ding, Weihua Guo, Nikolay~V. Dokholyan, Eugene~I. Shakhnovich, and
  Joan-Emma Shea.
\newblock Reconstruction of the src-sh3 protein domain transition state
  ensemble using multiscale molecular dynamics simulations.
\newblock \emph{Journal of Molecular Biology}, 350\penalty0 (5):\penalty0
  1035--1050, 7 2005.

\bibitem[Ding et~al.(2008)Ding, Tsao, Nie, and Dokholyan]{QMDMD5}
Feng Ding, Douglas Tsao, Huifen Nie, and Nikolay~V. Dokholyan.
\newblock Ab initio folding of proteins with all-atom discrete molecular
  dynamics.
\newblock \emph{Structure}, 16\penalty0 (7):\penalty0 1010--1018, 7 2008.

\bibitem[Sparta and Alexandrova(2012)]{QMDMD_catechol}
Manuel Sparta and Anastassia~N Alexandrova.
\newblock How metal substitution affects the enzymatic activity of
  catechol-o-methyltransferase.
\newblock \emph{PLoS ONE}, 7\penalty0 (10):\penalty0 e47172, 2012.

\bibitem[Valdez and Alexandrova(2012)]{Valdez_Lactamase}
Crystal~E. Valdez and Anastassia~N. Alexandrova.
\newblock Why urease is a di-nickel enzyme whereas the ccra β-lactamase is a
  di-zinc enzyme.
\newblock \emph{The Journal of Physical Chemistry B}, 116\penalty0
  (35):\penalty0 10649--10656, 09 2012.

\bibitem[Sparta et~al.(2013)Sparta, Valdez, and Alexandrova]{Valdez_ARD}
Manuel Sparta, Crystal~E. Valdez, and Anastassia~N. Alexandrova.
\newblock Metal-dependent activity of fe and ni acireductone dioxygenases: How
  two electrons reroute the catalytic pathway.
\newblock \emph{Journal of Molecular Biology}, 425\penalty0 (16):\penalty0
  3007--3018, 8 2013.

\bibitem[\'{A}lvarez Santos et~al.(1998)\'{A}lvarez Santos,
  Gonz\'{a}lez-Lafont, M.~Lluch, Oliva, and X.~Avile's]{CPA_mechanism1}
Silvia \'{A}lvarez Santos, \`{A}ngels Gonz\'{a}lez-Lafont, Jose' M.~Lluch,
  Baldomero Oliva, and Francesc X.~Avile's.
\newblock Theoretical study of the role of arginine 127 in the water-promoted
  mechanism of peptide cleavage by carboxypeptidase a.
\newblock \emph{New Journal of Chemistry}, 22:\penalty0 319--326, 1998.

\bibitem[Vardi-Kilshtain et~al.(2003)Vardi-Kilshtain, Shoham, and
  Goldblum]{CPA_mechanism2}
Alexandra Vardi-Kilshtain, Gil Shoham, and Amiram Goldblum.
\newblock Anhydride formation is not a valid mechanism for peptide cleavage by
  carboxypeptidase-a: a semiempirical reaction pathway study.
\newblock \emph{Molecular Physics}, 101\penalty0 (17):\penalty0 2715--2724,
  2003.

\bibitem[Szeto et~al.(2009)Szeto, Mujika, Zurek, Mulholland, and
  Harvey]{CPA_2009}
Michelle~W.Y. Szeto, Jon~I. Mujika, Jolanta Zurek, Adrian~J. Mulholland, and
  Jeremy~N. Harvey.
\newblock {QM/MM} study on the mechanism of peptide hydrolysis by
  carboxypeptidase {A}.
\newblock \emph{Journal of Molecular Structure:{THEOCHEM}}, 898\penalty0
  (1--3):\penalty0 106--114, 2009.

\bibitem[Kim and Lipscomb(1990)]{CPA_PDB}
Hidong Kim and William~N. Lipscomb.
\newblock Crystal structure of the complex of carboxypeptidase a with a
  strongly bound phosphonate in a new crystalline form: comparison with
  structures of other complexes.
\newblock \emph{Biochemistry}, 29\penalty0 (23):\penalty0 5546--5555, 1990.

\bibitem[Ahlrichs et~al.(1989)Ahlrichs, B{\"a}r, H{\"a}ser, Horn, and
  K{\"o}lmel]{Turbomole2}
Reinhart Ahlrichs, Michael B{\"a}r, Marco H{\"a}ser, Hans Horn, and Christoph
  K{\"o}lmel.
\newblock Electronic structure calculations on workstation computers: The
  program system turbomole.
\newblock \emph{Chemical Physics Letters}, 162\penalty0 (3):\penalty0 165--169,
  1989.

\bibitem[Perdew and Wang(1992)]{Perdew1}
John~P. Perdew and Yue Wang.
\newblock Accurate and simple analytic representation of the electron-gas
  correlation energy.
\newblock \emph{Physical Review B}, 45:\penalty0 13244--13249, Jun 1992.

\bibitem[Tao et~al.(2003)Tao, Perdew, Staroverov, and Scuseria]{Perdew2}
Jianmin Tao, John~P. Perdew, Viktor~N. Staroverov, and Gustavo~E. Scuseria.
\newblock Climbing the density functional ladder: Nonempirical meta-generalized
  gradient approximation designed for molecules and solids.
\newblock \emph{Physical Review Letters}, 91:\penalty0 146401, Sep 2003.

\bibitem[Sch{\"a}fer et~al.(1992)Sch{\"a}fer, Horn, and
  Ahlrichs]{Gaussianbasis}
Ansgar Sch{\"a}fer, Hans Horn, and Reinhart Ahlrichs.
\newblock Fully optimized contracted gaussian basis sets for atoms {L}i to
  {K}r.
\newblock \emph{The Journal of Chemical Physics}, 97\penalty0 (4):\penalty0
  2571--2577, 1992.

\bibitem[Weigend and Ahlrichs(2005)]{basis2}
Florian Weigend and Reinhart Ahlrichs.
\newblock Balanced basis sets of split valence, triple zeta valence and
  quadruple zeta valence quality for {H} to {R}n: Design and assessment of
  accuracy.
\newblock \emph{Physical Chemistry Chemical Physics}, 7:\penalty0 3297--3305,
  2005.

\bibitem[Klamt and Schuurmann(1993)]{COSMO}
A.~Klamt and G.~Schuurmann.
\newblock Cosmo: a new approach to dielectric screening in solvents with
  explicit expressions for the screening energy and its gradient.
\newblock \emph{Journal of Chemical Society{,} Perkin Transactions 2}, pages
  799--805, 1993.

\bibitem[SCM(2014)]{ADF2014}
\emph{ADF2014}.
\newblock SCM, Theoretical Chemistry, Vrije Universiteit, Amsterdam, The
  Netherlands, 2014.

\bibitem[Seth and Ziegler(2012)]{TPSS_ADF}
Michael Seth and Tom Ziegler.
\newblock Range-separated exchange functionals with slater-type functions.
\newblock \emph{Journal of Chemical Theory and Computation}, 8\penalty0
  (3):\penalty0 901--907, 2012.

\bibitem[Pye and Ziegler(1999)]{COSMO_ADF}
C.~Cory Pye and Tom Ziegler.
\newblock An implementation of the conductor-like screening model of solvation
  within the amsterdam density functional package.
\newblock \emph{Theoretical Chemistry Accounts}, 101\penalty0 (6):\penalty0
  396--408, 1999.

\bibitem[Haberland et~al.(2009)Haberland, Montgomery, and
  Olson]{HDAC_Haberland_2009}
Michael Haberland, Rusty~L. Montgomery, and Eric~N. Olson.
\newblock The many roles of histone deacetylases in development and physiology:
  implications for disease and therapy.
\newblock \emph{Nature Reviews Genetics}, 10\penalty0 (1):\penalty0 32--42, 01
  2009.

\bibitem[Bolden et~al.(2006)Bolden, Peart, and Johnstone]{HDAC_Bolden_2006}
Jessica~E. Bolden, Melissa~J. Peart, and Ricky~W. Johnstone.
\newblock Anticancer activities of histone deacetylase inhibitors.
\newblock \emph{Nature Reviews Drug Discovery}, 5\penalty0 (9):\penalty0
  769--784, 09 2006.

\bibitem[Marks and Breslow(2007)]{HDAC_Marks_2007}
Paul~A Marks and Ronald Breslow.
\newblock Dimethyl sulfoxide to vorinostat: development of this histone
  deacetylase inhibitor as an anticancer drug.
\newblock \emph{Nature Biotechnology}, 25\penalty0 (1):\penalty0 84--90, 01
  2007.

\bibitem[Bradner et~al.(2010)Bradner, West, Grachan, Greenberg, Haggarty,
  Warnow, and Mazitschek]{HDAC_Bradner_2010}
James~E Bradner, Nathan West, Melissa~L Grachan, Edward~F Greenberg, Stephen~J
  Haggarty, Tandy Warnow, and Ralph Mazitschek.
\newblock Chemical phylogenetics of histone deacetylases.
\newblock \emph{Nature Chemical Biology}, 6\penalty0 (3):\penalty0 238--243, 03
  2010.

\bibitem[Paris et~al.(2008)Paris, Porcelloni, Binaschi, and
  Fattori]{HDAC_Paris_2008}
Marielle Paris, Marina Porcelloni, Monica Binaschi, and Daniela Fattori.
\newblock Histone deacetylase inhibitors: From bench to clinic.
\newblock \emph{Journal of Medicinal Chemistry}, 51\penalty0 (6):\penalty0
  1505--1529, 2008.

\bibitem[Suzuki et~al.(2009)Suzuki, Suzuki, Ota, Nakano, Kurihara, Okuda,
  Yamori, Tsumoto, Nakagawa, and Miyata]{HDAC_Suzuki_2009}
Nobuaki Suzuki, Takayoshi Suzuki, Yosuke Ota, Tatsuya Nakano, Masaaki Kurihara,
  Haruhiro Okuda, Takao Yamori, Hiroki Tsumoto, Hidehiko Nakagawa, and Naoki
  Miyata.
\newblock Design, synthesis, and biological activity of boronic acid-based
  histone deacetylase inhibitors.
\newblock \emph{Journal of Medicinal Chemistry}, 52\penalty0 (9):\penalty0
  2909--2922, 2009.

\bibitem[Gantt et~al.(2006)Gantt, Gattis, , and Fierke]{HDAC_M_Gantt_2006}
Stephanie~L. Gantt, Samuel~G. Gattis, , and Carol~A. Fierke.
\newblock Catalytic activity and inhibition of human histone deacetylase 8 is
  dependent on the identity of the active site metal ion.
\newblock \emph{Biochemistry}, 45\penalty0 (19):\penalty0 6170--6178, 2006.

\bibitem[Wu et~al.(2010)Wu, Wang, Zhou, Cao, and Zhang]{HDAC_mech_Wu_2010}
Ruibo Wu, Shenglong Wang, Nengjie Zhou, Zexing Cao, and Yingkai Zhang.
\newblock A proton-shuttle reaction mechanism for histone deacetylase 8 and the
  catalytic role of metal ions.
\newblock \emph{Journal of the American Chemical Society}, 132\penalty0
  (27):\penalty0 9471--9479, 2010.

\bibitem[Nechay et~al.(2016)Nechay, Gallup, Morgenstern, Smith, Eberhart, and
  Alexandrova]{HDAC_Nechay}
Michael~R. Nechay, Nathan~M. Gallup, Amanda Morgenstern, Quentin~A. Smith,
  Mark~E. Eberhart, and Anastassia~N. Alexandrova.
\newblock Histone deacetylase 8: Characterization of physiological divalent
  metal catalysis.
\newblock \emph{Journal of Physical Chemistry B}, 2016.
\newblock DOI: 10.1021/acs.jpcb.6b00997.

\bibitem[Vannini et~al.(2007)Vannini, Volpari, Gallinari, Jones, Mattu,
  Carf{\'\i}, De~Francesco, Steink{\"u}hler, and
  Di~Marco]{HDAC8_xtal_Vannini_2007}
Alessandro Vannini, Cinzia Volpari, Paola Gallinari, Philip Jones, Marco Mattu,
  Andrea Carf{\'\i}, Raffaele De~Francesco, Christian Steink{\"u}hler, and
  Stefania Di~Marco.
\newblock Substrate binding to histone deacetylases as shown by the crystal
  structure of the hdac8{\textendash}substrate complex.
\newblock \emph{EMBO reports}, 8\penalty0 (9):\penalty0 879--884, 2007.

\bibitem[Grimme(2004)]{Grimme_2004}
Stefan Grimme.
\newblock Accurate description of van der waals complexes by density functional
  theory including empirical corrections.
\newblock \emph{Journal of Computational Chemistry}, 25\penalty0 (12):\penalty0
  1463--1473, 2004.

\bibitem[Analytical and Products(pp 252-253)]{Menergies}
Analytical and Biological Products.
\newblock \emph{Dojindo.com Metal Chelates}, pp 252-253.

\bibitem[Staroverov et~al.(2003)Staroverov, Scuseria, Tao, and Perdew]{Perdew3}
Viktor~N. Staroverov, Gustavo~E. Scuseria, Jianmin Tao, and John~P. Perdew.
\newblock Comparative assessment of a new nonempirical density functional:
  Molecules and hydrogen-bonded complexes.
\newblock \emph{The Journal of Chemical Physics}, 119\penalty0 (23):\penalty0
  12129--12137, 2003.

\bibitem[Corminboeuf et~al.(2006)Corminboeuf, Hu, Tuckerman, and
  Zhang]{HDAC_mech_Corminboeuf_2006}
Cl\'emence Corminboeuf, Po~Hu, Mark~E. Tuckerman, and Yingkai Zhang.
\newblock Unexpected deacetylation mechanism suggested by a density functional
  theory qm/mm study of histone-deacetylase-like protein.
\newblock \emph{Journal of the American Chemical Society}, 128\penalty0
  (14):\penalty0 4530--4531, 2006.

\bibitem[Ayers(2006)]{Ayers_NegFukui_2006}
Paul~W. Ayers.
\newblock Can one oxidize an atom by reducing the molecule that contains it?
\newblock \emph{Physical Chemistry Chemical Physics}, 8:\penalty0 3387--3390,
  2006.

\bibitem[Melin et~al.(2007)Melin, Ayers, and Ortiz]{Ayers_NegFukui_2007}
Junia Melin, Paul~W. Ayers, and Joseph~Vincent Ortiz.
\newblock Removing electrons can increase the electron density:  a
  computational study of negative fukui functions.
\newblock \emph{The Journal of Physical Chemistry A}, 111\penalty0
  (40):\penalty0 10017--10019, 2007.

\bibitem[Echegaray et~al.(2012)Echegaray, C{\'a}rdenas, Rabi, Rabi, Lee, Zadeh,
  Toro-Labbe, Anderson, and Ayers]{Ayers_NegFukui_2012}
Eleonora Echegaray, Carlos C{\'a}rdenas, Sandra Rabi, Nataly Rabi, Sungmin Lee,
  Farnaz~Heidar Zadeh, Alejandro Toro-Labbe, James S.~M. Anderson, and Paul~W.
  Ayers.
\newblock In pursuit of negative fukui functions: examples where the highest
  occupied molecular orbital fails to dominate the chemical reactivity.
\newblock \emph{Journal of Molecular Modeling}, 19\penalty0 (7):\penalty0
  2779--2783, 2012.

\bibitem[Bulat et~al.(2004)Bulat, Chamorro, Fuentealba, and
  Toro-Labbe]{Bulat_NegFukui_2004}
Felipe~A. Bulat, Eduardo Chamorro, Patricio Fuentealba, and Alejandro
  Toro-Labbe.
\newblock Condensation of frontier molecular orbital fukui functions.
\newblock \emph{The Journal of Physical Chemistry A}, 108\penalty0
  (2):\penalty0 342--349, 2004.

\bibitem[Fuentealba et~al.(2000)Fuentealba, P{\'e}rez, and
  Contreras]{Fuentealba_NegFukui_2000}
P.~Fuentealba, P.~P{\'e}rez, and R.~Contreras.
\newblock On the condensed fukui function.
\newblock \emph{The Journal of Chemical Physics}, 113\penalty0 (7):\penalty0
  2544--2551, 2000.

\bibitem[Roy et~al.(1999)Roy, Pal, and Hirao]{Roy_NegFukui_1999}
Ram~Kinkar Roy, Sourav Pal, and Kimihiko Hirao.
\newblock On non-negativity of fukui function indices.
\newblock \emph{The Journal of Chemical Physics}, 110\penalty0 (17):\penalty0
  8236--8245, 1999.

\bibitem[Ayers et~al.(2002{\natexlab{c}})Ayers, Morrison, and
  Roy]{Ayers_relaxation_2002}
Paul~W. Ayers, Robert~C. Morrison, and Ram~K. Roy.
\newblock Variational principles for describing chemical reactions: Condensed
  reactivity indices.
\newblock \emph{The Journal of Chemical Physics}, 116\penalty0 (20):\penalty0
  8731--8744, 2002{\natexlab{c}}.

\bibitem[Popelier(1996)]{Popelier_difgeom_1996}
Paul~L.A. Popelier.
\newblock On the differential geometry of interatomic surfaces.
\newblock \emph{Canadian Journal of Chemistry}, 74\penalty0 (6):\penalty0
  829--838, 1996.

\bibitem[Bartolotti and Ayers(2005)]{Ayers_fukui_2005}
Libero~J. Bartolotti and Paul~W. Ayers.
\newblock An example where orbital relaxation is an important contribution to
  the fukui function.
\newblock \emph{The Journal of Physical Chemistry A}, 109\penalty0
  (6):\penalty0 1146--1151, 2005.

\bibitem[March(1983)]{TF_KE}
N.H. March.
\newblock \emph{Thoery of the Inhomogeneous Electron Gas}.
\newblock Plenum Press, 1983.

\bibitem[Popelier(2011)]{Popelier_integration_2011}
Paul L.~A. Popelier.
\newblock Fully analytical integration over the 3d volume bounded by the β
  sphere in topological atoms.
\newblock \emph{The Journal of Physical Chemistry A}, 115\penalty0
  (45):\penalty0 13169--13179, 2011.

\bibitem[Rafat and Popelier(2007)]{Popelier_integration_2007}
M.~Rafat and P.~L.~A. Popelier.
\newblock Visualization and integration of quantum topological atoms by spatial
  discretization into finite elements.
\newblock \emph{Journal of Computational Chemistry}, 28\penalty0 (16):\penalty0
  2602--2617, 2007.

\bibitem[Parrondo et~al.(1995)Parrondo, Karafiloglou, Pappalardo, and
  Marcos]{Parrondo_resonance_1995}
Ramon~M. Parrondo, Padeleimon Karafiloglou, Rafael~R. Pappalardo, and
  Enrique~Sanchez Marcos.
\newblock Calculation of the weights of resonance structures of molecules in
  solution.
\newblock \emph{The Journal of Physical Chemistry}, 99\penalty0 (17):\penalty0
  6461--6467, 1995.

\bibitem[Giuffreda et~al.(2004)Giuffreda, Bruschi, and
  L{\"u}thi]{Giuffreda_NBO_delocalization_2004}
Maria~Grazia Giuffreda, Maurizio Bruschi, and Hans~Peter L{\"u}thi.
\newblock Electron delocalization in linearly $\pi$-conjugated systems: A
  concept for quantitative analysis.
\newblock \emph{Chemistry -- A European Journal}, 10\penalty0 (22):\penalty0
  5671--5680, 2004.

\bibitem[Daudel et~al.(1974)Daudel, Bader, Stephens, and
  Borrett]{Bader_loge_1974}
R.~Daudel, R.~F.~W. Bader, M.~E. Stephens, and D.~S. Borrett.
\newblock The electron pair in chemistry.
\newblock \emph{Canadian Journal of Chemistry}, 52\penalty0 (8):\penalty0
  1310--1320, 1974.

\bibitem[Foroutan-Nejad et~al.(2011)Foroutan-Nejad, Badri, Shahbazian, and
  Rashidi-Ranjbar]{Shahbazian_aromaticity_2011}
Cina Foroutan-Nejad, Zahra Badri, Shant Shahbazian, and Parviz Rashidi-Ranjbar.
\newblock The laplacian of electron density versus nics$_{zz}$ scan: Measuring
  magnetic aromaticity among molecules with different atom types.
\newblock \emph{The Journal of Physical Chemistry A}, 115\penalty0
  (45):\penalty0 12708--12714, 2011.

\bibitem[Krygowski(1993)]{Krygowski_HOMA_1993}
Tadeusz~Marek Krygowski.
\newblock Crystallographic studies of inter- and intramolecular interactions
  reflected in aromatic character of $\pi$-electron systems.
\newblock \emph{Journal of Chemical Information and Computer Sciences},
  33\penalty0 (1):\penalty0 70--78, 1993.

\bibitem[von Ragu{\'e}~Schleyer et~al.(1996)von Ragu{\'e}~Schleyer, Maerker,
  Dransfeld, Jiao, and van Eikema~Hommes]{Schleyer_NICS_1996}
Paul von Ragu{\'e}~Schleyer, Christoph Maerker, Alk Dransfeld, Haijun Jiao, and
  Nicolaas J.~R. van Eikema~Hommes.
\newblock Nucleus-independent chemical shifts:  a simple and efficient
  aromaticity probe.
\newblock \emph{Journal of the American Chemical Society}, 118\penalty0
  (26):\penalty0 6317--6318, 1996.

\bibitem[Bader et~al.(1996)Bader, Streitwieser, Neuhaus, Laidig, , and
  Speers]{Bader_DI_fermi_1996}
Richard F.~W. Bader, Andrew Streitwieser, Arnst Neuhaus, Keith~E. Laidig, , and
  Peter Speers.
\newblock Electron delocalization and the fermi hole.
\newblock \emph{Journal of the American Chemical Society}, 118\penalty0
  (21):\penalty0 4959--4965, 1996.

\bibitem[Howard and Krygowski(1997)]{Krygowski_aromaticity_ringCP_1997}
S.~T. Howard and T.~M. Krygowski.
\newblock Benzenoid hydrocarbon aromaticity in terms of charge density
  descriptors.
\newblock \emph{Canadian Journal of Chemistry}, 75\penalty0 (9):\penalty0
  1174--1181, 1997.

\bibitem[Savin et~al.(1996)Savin, Silvi, and
  Coionna]{Savin_ELF_delocalization_1996}
A.~Savin, B.~Silvi, and F.~Coionna.
\newblock Topological analysis of the electron localization function applied to
  delocalized bonds.
\newblock \emph{Canadian Journal of Chemistry}, 74\penalty0 (6):\penalty0
  1088--1096, 1996.

\bibitem[Santos et~al.(2005)Santos, Andres, Aizman, and
  Fuentealba]{Santos_ELF_aromaticity_2005}
Juan~C. Santos, Juan Andres, Arie Aizman, and Patricio Fuentealba.
\newblock An aromaticity scale based on the topological analysis of the
  electron localization function including σ and π contributions.
\newblock \emph{Journal of Chemical Theory and Computation}, 1\penalty0
  (1):\penalty0 83--86, 2005.

\bibitem[Krygowski et~al.(2004)Krygowski, Ejsmont, Stepie\'{n}, Cyra\'{n}ski,
  Poater, and Sol\`{a}]{Krygowski_aromaticity_substituents_2004}
Tadeusz~M. Krygowski, Krzysztof Ejsmont, Beata~T. Stepie\'{n}, Michal~K.
  Cyra\'{n}ski, Jordi Poater, and Miquel Sol\`{a}.
\newblock Relation between the substituent effect and aromaticity.
\newblock \emph{The Journal of Organic Chemistry}, 69\penalty0 (20):\penalty0
  6634--6640, 2004.

\bibitem[Poater et~al.(2004)Poater, Garcia-Cruz, Illas, and
  Sola]{Sola_aromaticity_discrepency_2004}
Jordi Poater, Isidoro Garcia-Cruz, Francesc Illas, and Miquel Sola.
\newblock Discrepancy between common local aromaticity measures in a series of
  carbazole derivatives.
\newblock \emph{Physical Chemistry Chemical Physics}, 6:\penalty0 314--318,
  2004.

\bibitem[Cyra{\~n}ski et~al.(2002)Cyra{\~n}ski, Krygowski, Katritzky, and von
  Ragu{\'e}~Schleyer‖]{Schleyer_aromaticity_unique_2002}
Michal~K. Cyra{\~n}ski, Tadeusz~M. Krygowski, Alan~R. Katritzky, and Paul von
  Ragu{\'e}~Schleyer‖.
\newblock To what extent can aromaticity be defined uniquely? .
\newblock \emph{The Journal of Organic Chemistry}, 67\penalty0 (4):\penalty0
  1333--1338, 2002.

\bibitem[Hecker et~al.(1978)Hecker, Rohr, and Stein]{Hecker_Ir_brittle_1978}
S.~S. Hecker, D.~L. Rohr, and D.~F. Stein.
\newblock Brittle fracture in iridium.
\newblock \emph{Metallurgical Transactions A}, 9\penalty0 (4):\penalty0
  481--488, 1978.

\bibitem[Fortes and Ralph(1967)]{Fortes_Ir_1967}
M.A Fortes and B~Ralph.
\newblock A field-ion microscope study of segregation to grain boundaries in
  iridium.
\newblock \emph{Acta Metallurgica}, 15\penalty0 (5):\penalty0 707--720, 1967.

\bibitem[Gornostyrev et~al.(2000)Gornostyrev, Katsnelson, Medvedeva, Mryasov,
  Freeman, and Trefilov]{Gornostyrev_Ir_2000}
Yu.~N. Gornostyrev, M.~I. Katsnelson, N.~I. Medvedeva, O.~N. Mryasov, A.~J.
  Freeman, and A.~V. Trefilov.
\newblock Peculiarities of defect structure and mechanical properties of
  iridium: Results of \textit{ab initio} electronic structure calculations.
\newblock \emph{Physical Review B}, 62:\penalty0 7802--7808, Sep 2000.

\bibitem[Gong(2011)]{Gong_Ir_2011}
H.R. Gong.
\newblock Ideal mechanical strength and interface cohesion property of ir-base
  superalloys from first principles calculation.
\newblock \emph{Materials Chemistry and Physics}, 126\penalty0 (1--2):\penalty0
  284--288, 2011.

\bibitem[Kamran et~al.(2008)Kamran, Chen, Chen, and Zhao]{Zhao_Ir_2008}
Sami Kamran, Kuiying Chen, Liang Chen, and Linruo Zhao.
\newblock Electronic origin of anomalously high shear modulus and intrinsic
  brittleness of fcc ir.
\newblock \emph{Journal of Physics: Condensed Matter}, 20\penalty0
  (8):\penalty0 085221, 2008.

\bibitem[\u{C}ern\'{y} and Pokluda(2010)]{Cerny_Ir_2010}
M.~\u{C}ern\'{y} and J.~Pokluda.
\newblock The theoretical shear strength of fcc crystals under superimposed
  triaxial stress.
\newblock \emph{Acta Materialia}, 58\penalty0 (8):\penalty0 3117--3123, 5 2010.

\bibitem[Liang et~al.(2012)Liang, Li, and Gong]{Gong_Ir_brittle_2012}
C.P. Liang, G.H. Li, and H.R. Gong.
\newblock Concerning the brittleness of iridium: An elastic and electronic
  view.
\newblock \emph{Materials Chemistry and Physics}, 133\penalty0 (1):\penalty0
  140--143, 2012.

\bibitem[Pauling(1946)]{Pauling_TSS_1946}
L.~Pauling.
\newblock Molecular architecture and biological reactions.
\newblock \emph{Chemical and Engineering News}, 24:\penalty0 1375--1377, 1946.

\bibitem[Jencks(1986)]{Jencks_RSD_1986}
W.~P. Jencks.
\newblock \emph{Catalysis in Chemistry and Enzymology}.
\newblock Dover Publication, New York, 1986.

\bibitem[Warshel et~al.(2006)Warshel, Sharma, Kato, Xiang, Liu, and
  Olsson]{Warshel_2006_preorganization}
Arieh Warshel, Pankaz~K. Sharma, Mitsunori Kato, Yun Xiang, Hanbin Liu, and
  Mats H.~M. Olsson.
\newblock Electrostatic basis for enzyme catalysis.
\newblock \emph{Chemical Reviews}, 106\penalty0 (8):\penalty0 3210--3235, 2006.

\bibitem[Prodan and Kohn(2005)]{Kohn_nearsightedness_2005}
E.~Prodan and W.~Kohn.
\newblock Nearsightedness of electronic matter.
\newblock \emph{Proceedings of the National Academy of Sciences of the United
  States of America}, 102\penalty0 (33):\penalty0 11635--11638, 2005.

\bibitem[Vanpoucke et~al.(2015)Vanpoucke, Ol{\'a}h, Proft, Speybroeck, and
  Roos]{DeProft_convergence_2015}
Danny E.~P. Vanpoucke, Julianna Ol{\'a}h, Frank~De Proft, Veronique~Van
  Speybroeck, and Goedele Roos.
\newblock Convergence of atomic charges with the size of the enzymatic
  environment.
\newblock \emph{Journal of Chemical Information and Modeling}, 55\penalty0
  (3):\penalty0 564--571, 2015.

\end{thebibliography}



\appendix{Functional Dependence}
\label{app:functionals}

Plots of the average kinetic energy per electron in gradient bundles for N$_2$ using various DFT methods are presented in \ref{fig:functionals}. 
We have tested LDA, BLYP, B3LYP, and M06L functionals, in addition to the PBE data presented in chapter \ref{cha:intrinsic}.
All plots are geometrically very similar. 
Local maxima and minima occur in the same gradient bundles (maxima at 80$^{\circ}$, minima at 0$^{\circ}$ and 180$^{\circ}$), and current investigation into the curvatures of the plots at local maxima and minima suggests these plots not to be significantly different.

\begin{figure}
	\begin{center}
		\subfigure{
			\resizebox{.48\textwidth}{!}{\includegraphics{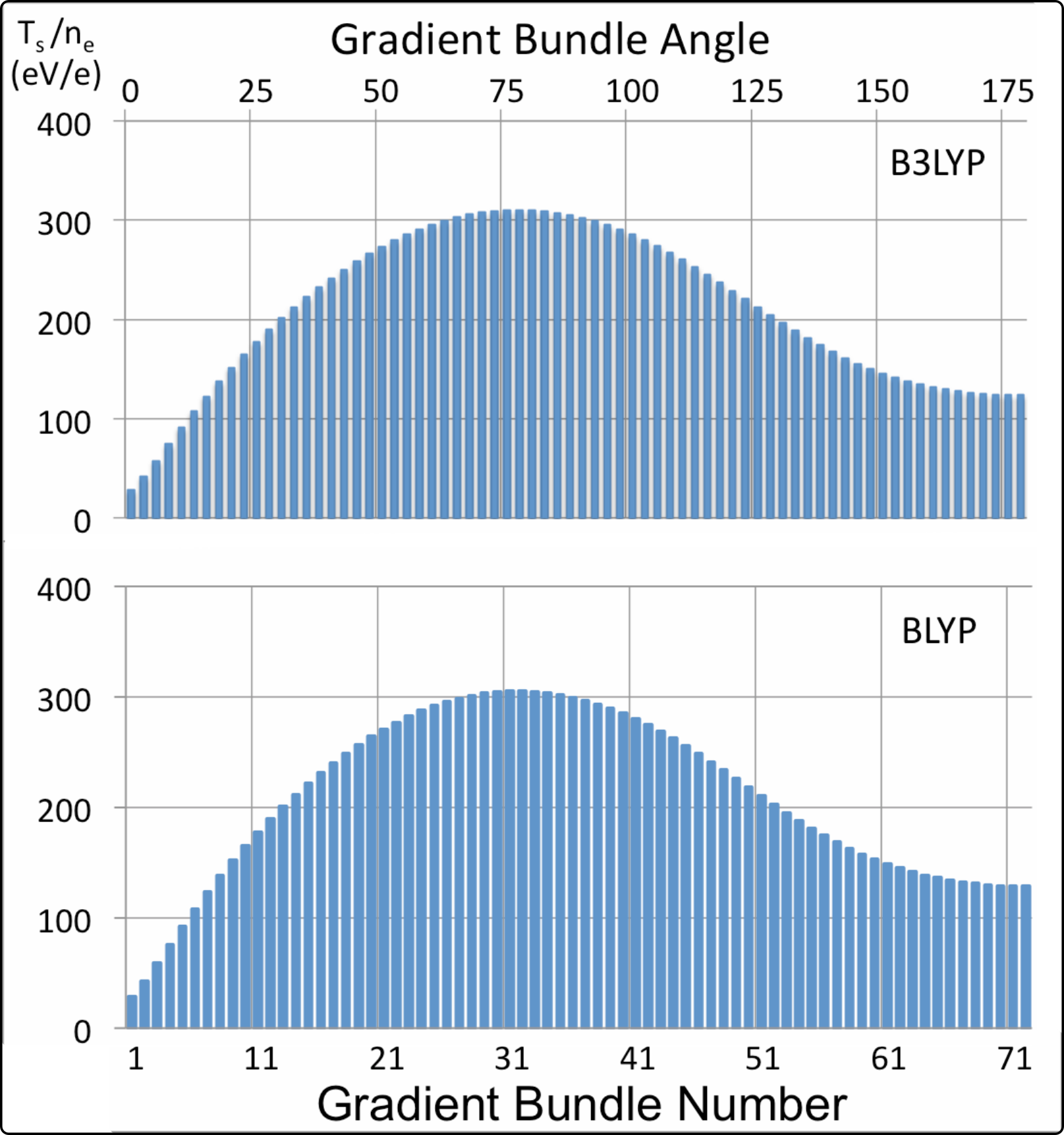}}
		} 
		\subfigure{
			\resizebox{.48\textwidth}{!}{\includegraphics{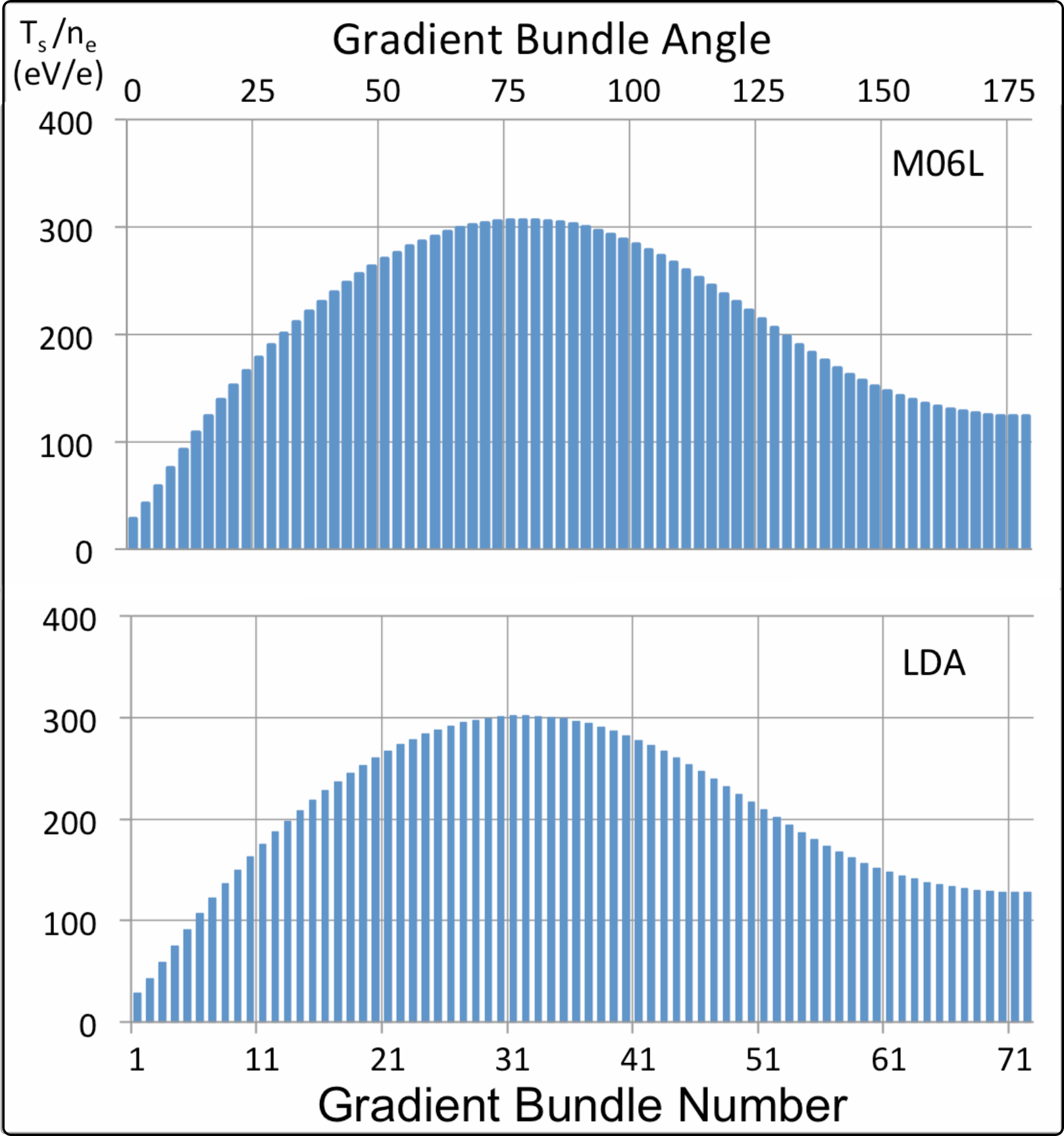}}
		}
		\caption{\label{fig:functionals} Average KE/e in gradient bundles seeded every 2.5$^{\circ}$ around a nucleus in N$_2$ using LDA, BLYP, B3LYP, and M06L functionals within the ADF software package. Gradient bundle angle is measured with respect to the bond path. Single point calculations were run using a TZP basis set using the geometry obtained from initial PBE calculations.}	
		\end{center}
\end{figure}

\appendix{HDAC8 Metal Swapping Data}
\label{app:Mswap}

\ref{tab:lit_G} provides experimental binding energies from \cite{Menergies} for various metal ions in DTPA and GEDTA used for metal swapping calculations for HDAC8.
\ref{tab:calc_G} lists calculated $\Delta$G values for metal swapping between DTPA and GEDTA, using Co$^{2+}$ as a reference point. 
\ref{tab:exp_G} contains the experimental $\Delta$G values for the same metal swapping reactions. 
\ref{tab:error_G} shows the difference between experimental and calculated values of $\Delta$G for metal swapping.

\begin{table}[ht]
\centering
\vspace{2mm}
 \caption{List of literature binding energies with regards to a series of chelating agents for all relevant metal ions studied in HDAC8.}
 \label{tab:lit_G}
\begin{tabular}{|c | c | c | c | c |}
\hline
 Metal & \multicolumn{2}{c |}{DTPA} & \multicolumn{2}{c |}{GEDTA} \\[1ex]
 \hline
 & Stability constant & $\Delta$G binding & Stability constant & $\Delta$G binding\\
 & & ($kcal/mol$) & & ($kcal/mol$)\\[1ex]
\hline
Co$^{2+}$ &18.4 & -25. & 12.5 & -17.\\[1ex]
Fe$^{2+}$ & 16.55 & -23. & 11.92 & -16.\\[1ex]
Mg$^{2+}$ & 9.3 & -13. & 5.21 & -7.1\\[1ex]
Mn$^{2+}$ & 15.6 & -21. & 12.3 & -17.\\[1ex]
Ni$^{2+}$ & 20.32 & -28. & 13.6 & -19.\\[1ex]
Zn$^{2+}$ & 18.75 & -26. & 14.5 & -20.\\[1ex]
\hline
\end{tabular}
\end{table}

\begin{table}[ht]
\centering
\vspace{2mm}
 \caption{Calculated $\Delta$G values for DTPA metal swapped with GEDTA ($kcal/mol$).}
 \label{tab:calc_G}
\begin{tabular}{| c | c c c c c c |}
 \hline
 & Co$^{2+}$ & Fe$^{2+}$ & Mg$^{2+}$ & Mn$^{2+}$ & Ni$^{2+}$ & Zn$^{2+}$ \\[1ex]
\hline
Co$^{2+}$ &0 & -2.25 & -3.41 & -1.96 & 1.31 & -2.37\\[1ex]
Fe$^{2+}$ &  & 0 & -1.16 & 0.29 & 3.56 & -0.12\\[1ex]
Mg$^{2+}$ & & & 0 & 1.45 & 4.72 & 1.03\\[1ex]
Mn$^{2+}$ & & & & 0 & 3.27 & -0.41\\[1ex]
Ni$^{2+}$ & & & & & 0 & -3.68\\[1ex]
Zn$^{2+}$ & & & & & & 0\\[1ex]
\hline
\end{tabular}
\end{table}

\begin{table}[ht]
\centering
\vspace{2mm}
 \caption{Experimental $\Delta$G values for DTPA metal swapped with GEDTA ($kcal/mol$).}
 \label{tab:exp_G}
\begin{tabular}{| c | c c c c c c |}
 \hline
 & Co$^{2+}$ & Fe$^{2+}$ & Mg$^{2+}$ & Mn$^{2+}$ & Ni$^{2+}$ & Zn$^{2+}$ \\[1ex]
\hline
Co$^{2+}$ &0 & -1.73 & -2.47 & -3.55 & 1.12 & -2.25\\[1ex]
Fe$^{2+}$ &  & 0 & -0.74 & -1.81 & 2.85 & -0.52\\[1ex]
Mg$^{2+}$ & & & 0 & -1.08 & 3.59 & 0.22\\[1ex]
Mn$^{2+}$ & & & & 0 & 4.66 & 1.30\\[1ex]
Ni$^{2+}$ & & & & & 0 & -3.37\\[1ex]
Zn$^{2+}$ & & & & & & 0\\[1ex]
\hline
\end{tabular}
\end{table}

\begin{table}[ht]
\centering
\vspace{2mm}
 \caption{Calculated $\Delta$G -- experimental $\Delta$G values for DTPA metal swapped with GEDTA ($kcal/mol$).}
 \label{tab:error_G}
\begin{tabular}{| c | c c c c c c |}
 \hline
 & Co$^{2+}$ & Fe$^{2+}$ & Mg$^{2+}$ & Mn$^{2+}$ & Ni$^{2+}$ & Zn$^{2+}$ \\[1ex]
\hline
Co$^{2+}$ &0 & -0.52 & -0.94 & 1.59 & 0.19 & -0.12\\[1ex]
Fe$^{2+}$ &  & 0 & -0.42 & 2.11 & 0.71 & 0.40\\[1ex]
Mg$^{2+}$ & & & 0 & 2.53 & 1.13 & 0.82\\[1ex]
Mn$^{2+}$ & & & & 0 & -1.40 & -1.71\\[1ex]
Ni$^{2+}$ & & & & & 0 & -0.31\\[1ex]
Zn$^{2+}$ & & & & & & 0\\[1ex]
\hline
\end{tabular}
\end{table}

\appendix{Permissions}

This appendix provides permissions from journals and co-authors of published material presented in this dissertation.

\subsection{Journal Permissions}
Excerpted from the licenses, copyright, and permissions of Royal Society of Chemistry journals, including \textit{Physical Chemistry Chemical Physics} and \textit{Physica Scripta}: ``You do not need to request permission to reuse your own figures, diagrams, etc, that were originally published in a Royal Society of Chemistry publication. However, permission should be requested for use of the whole article or chapter except if reusing it in a thesis. If you are including an article or book chapter published by us in your thesis please ensure that your co-authors are aware of this."

Excerpted from the permissions of Elsevier journals, including \textit{Computational and Theoretical Chemistry}: ``Authors can include their articles in full or in part in a thesis or dissertation for non-commercial purposes."

Excerpted from the permissions of American Chemical Society journals, including \textit{Journal of Physical Chemistry B}: ``ACS extends blanket permission to students to include in their theses and dissertations their own articles, or portions thereof, that have been published in ACS journals or submitted to ACS journals for publication, provided that the ACS copyright credit line is noted on the appropriate page(s)."

\subsection{Author Permissions}

Tim Wilson (Chapters 2 and 4), twilson@mines.edu\\
I'm sending this email indicating my authorization for our co-authored papers, ``In Search of an Intrinsic Chemical Bond" and ``The Influence of Zero-Flux Surface Motion on Chemical Reactivity" to appear in your dissertation.\\
Sincerely,\\
Tim Wilson

Jonathan Miorelli (Chapters 2 and 4), jmiorell@mines.edu\\
You have my permission to use both, ``In Search of an Intrinsic Chemical Bond" and ``The Influence of Zero-Flux Surface Motion on Chemical Reactivity" in your dissertation.\\
Jonathan Miorelli

Travis Jones (Chapter 2), trjones@mines.edu\\
You have my permission to use our paper ``In Search of Intrinsic Chemical Bond" in your dissertation.\\
And congratulations!\\
Travis

Charles Morgenstern (Chapter 4), cmorgens@mines.edu\\
You have my permission to include ``The Influence of Zero-Flux Surface Motion on Chemical Reactivity" in your thesis.\\
Charles Morgenstern 

Crystal Valdez (Chapter 5), cediev87@gmail.com\\
I give you permission to use parts of our co-authored submitted paper, ``Predictive methods for computational metalloenzyme redesign - a test case with carboxypeptiase A," in any parts of your dissertation.\\
Best,\\
Crystal

Quentin Smith (Chapter 5), schrodinator@gmail.com\\
No problem.

Nathan Gallup (Chapter 5), gallup@chem.ucla.edu\\
Yes, you may use it.

Michael Nechay (Chapter 5), michaelnechay@gmail.com\\
I grant you my permission to use parts of our co-authored published paper ``Histone Deacetylase 8: Characterization of Physiological Divalent Metal Catalysis" in your dissertation.\\
-Michael Nechay

\end{document}